\documentclass[a4paper,12pt]{book}
\usepackage{fullpage}
\usepackage{amssymb,amsmath}
\usepackage{graphicx}
\usepackage{epstopdf}
\usepackage{feynmp}
\usepackage{cite}
\usepackage{hyperref}
\usepackage{enumerate}
\usepackage{titlesec}
\usepackage{caption}          
\usepackage{subcaption}       

\def\be{\begin{equation}}
\def\ee{\end{equation}}
\def\bea{\begin{eqnarray}}
\def\eea{\end{eqnarray}}
\def\beal{\begin{equation}\begin{aligned}}
\def\eeal{\end{aligned}\end{equation}}
\def\nn{\nonumber}

\def\bra#1{\langle #1|}
\def\ket#1{|#1 \rangle}

\def\braket#1{\langle #1 \rangle}

\def\la{\lambda}
\def\lb{\tilde{\lambda}}

\def\Res_#1{\operatorname*{Res}_{#1}}
\def\Tr{\operatorname*{Tr}}

\def\d{\mathrm{d}}

\def\cA{\mathcal{A}}
\def\cM{\mathcal{M}}
\def\cN{\mathcal{N}}
\def\tf{\tilde{f}}

\def\ie{i.e. }
\def\eg{e.g. }

\def\eqn#1{Eq.~\eqref{#1}}
\def\eqns#1#2{Eqs.~\eqref{#1} and~\eqref{#2}}
\def\Eqn#1{Eq.~\eqref{#1}}
\def\Eqns#1#2{Eqs.~\eqref{#1} and~\eqref{#2}}
\def\fig#1{Fig.~{\ref{#1}}}

\def\tab#1{Table~{\ref{#1}}}

\def\chap#1{Chapter~{\ref{#1}}}
\def\sec#1{Section~{\ref{#1}}}
\def\secs#1#2{Section~{\ref{#1}} and~{\ref{#2}}}
\def\app#1{Appendix~{\ref{#1}}}
\def\rcite#1{Ref.~\cite{#1}}
\def\rcites#1{Refs.~\cite{#1}}

\def\citeSpinorHelicity{\cite{Berends:1981rb,DeCausmaecker:1981bg,
Gunion:1985vca,Kleiss:1985yh,Xu:1986xb,Gastmans:1990xh}}
\def\citeColorOrdering{\cite{Berends:1987cv,Mangano:1987xk,Mangano:1988kk,Bern:1990ux}}

\def\citePaper#1{\{#1\}}
\def\rcitePaper#1{Ref.~\{#1\}}

\def\fmfsettings{
    \fmfset{thin}{1.2pt}
    \fmfset{arrow_len}{7pt}
    \fmfset{arrow_ang}{20}
    \fmfset{wiggly_len}{7pt}
    \fmfset{curly_len}{6pt}
    \fmfset{dash_len}{8pt}
    \fmfset{dot_len}{3pt}
}

\title{Thesis}
\author{Alexander Ochirov}

\begin{document}

\frontmatter
\pagestyle{plain}

\renewcommand{\labelitemi}{$\bullet$}
\newpage\cleardoublepage

\thispagestyle{empty}

\begin{center}

\begin{minipage}{165mm}
\hspace{2cm}
\raisebox{-0.5\height}{\includegraphics[width=2.5cm]{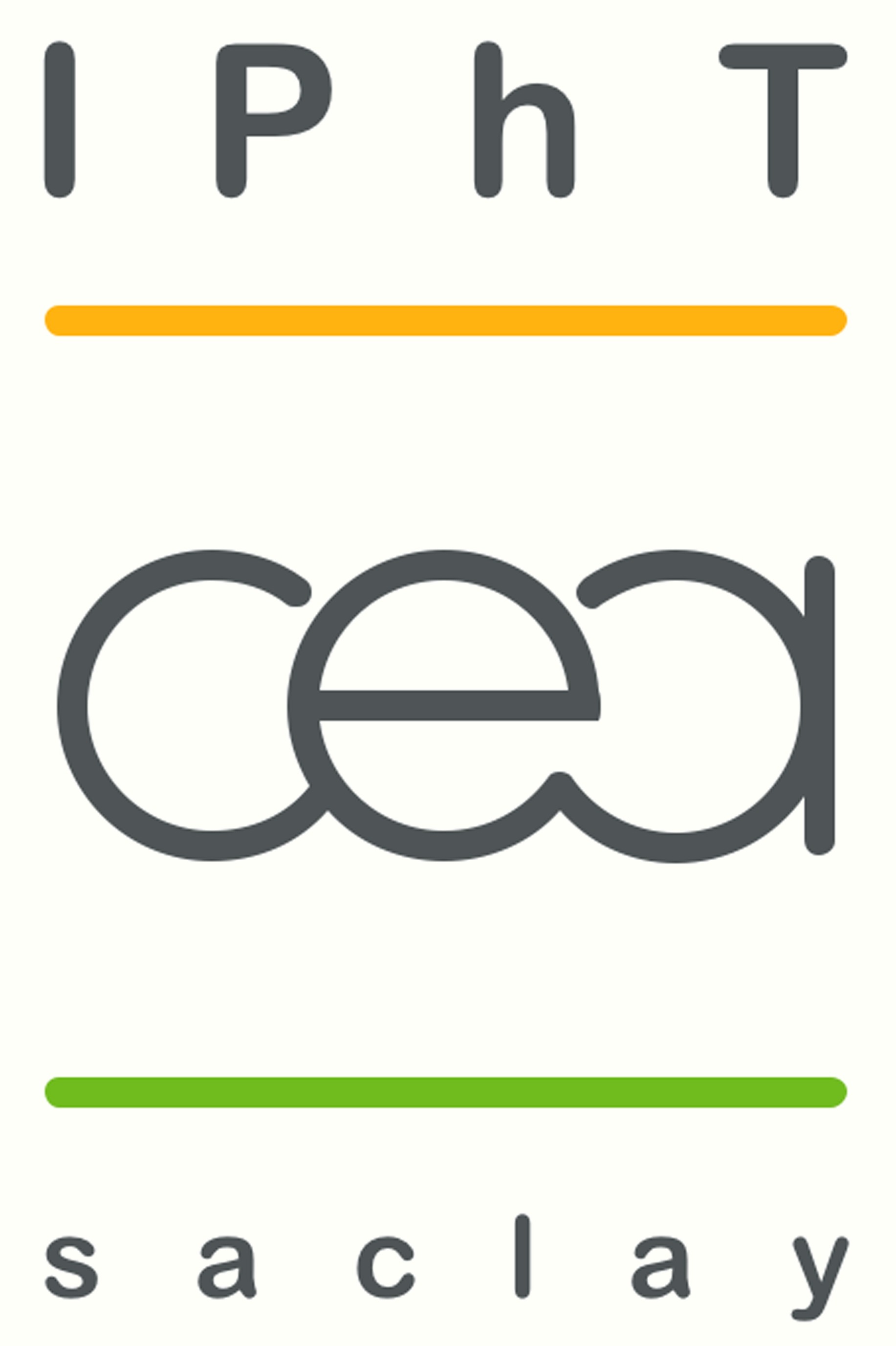}\hspace{2.7cm}}
\raisebox{-0.5\height}{\includegraphics[width=4.5cm]{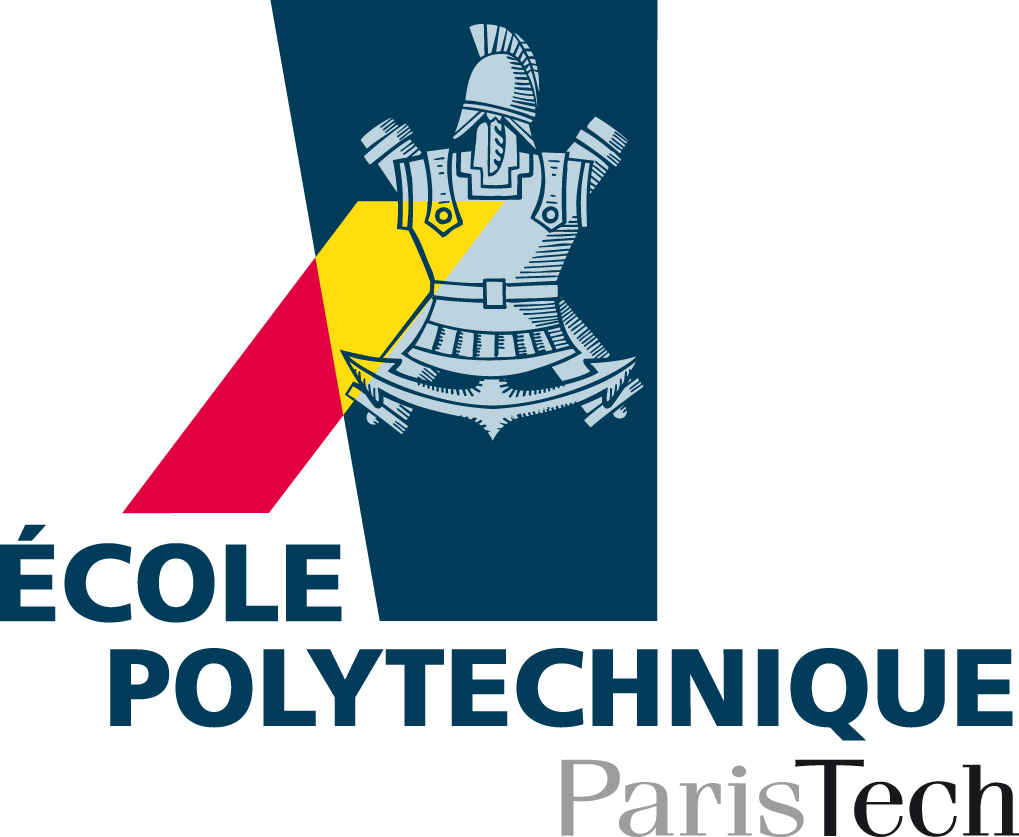}}
\end{minipage}

\vskip1.0cm

{\bf \'Ecole Polytechnique}
\vskip0.15cm
et
\vskip0.3cm
{\bf Institut de Physique Th\'eorique - CEA/Saclay}
\vskip0.4cm
\mbox{\'Ecole Doctorale de l'\'Ecole Polytechnique - ED 447}
\vskip1cm
{\bf \large Th\`ese de doctorat}
\vskip0.2cm
\mbox{Sp\'ecialit\'e: {\bf  Physique Th\'eorique}}
\vskip1cm
\hbox{\raisebox{0.4em}{\vrule depth 0pt height 0.4pt width 16cm}}
\vskip0.1cm
{\large \bf Scattering amplitudes in gauge theories}
\vskip0.2cm
{\large \bf with and without supersymmetry}
\vskip0.1cm
\hbox{\raisebox{0.4em}{\vrule depth 0pt height 0.4pt width 16cm}}
\vskip0.8cm
\normalsize
pr\'esent\'ee par {Alexander {\sc Ochirov}}
\vskip0.2cm
pour obtenir le grade de Docteur de l'\'Ecole Polytechnique
\vskip0.2cm
{Th\`ese pr\'epar\'ee sous la direction de Ruth {\sc Britto}}
\vskip1.6cm
Soutenue le 12 septembre 2014 devant le jury compos\'e de:
\normalsize
\vskip.6cm
\begin{tabular}{ll}
Ruth {\sc Britto} & Directeur de th\`ese\\
Zvi {\sc Bern} & Rapporteur\\
Gabriele {\sc Travaglini} & Rapporteur\\
David {\sc Kosower} & Examinateur\\
Marios {\sc Petropoulos} & Examinateur\\
Samuel {\sc Wallon} & Examinateur
\end{tabular} 

\end{center}

\section*{Abstract}

This thesis aims at providing better understanding of
the perturbative expansion of gauge theories with and without supersymmetry.
At tree level, the BCFW recursion relations are analyzed with respect to
their validity for general off-shell objects in Yang-Mills theory,
which is a significant step away from their established zone of applicability.
Unphysical poles constitute a new potential problem
in addition to the boundary behavior issue, common to the on-shell case as well.
For an infinite family of massive fermion currents,
both obstacles are shown to be avoided under certain conditions,
which provides a natural recursion relation.
At one loop, scattering amplitudes can be calculated from unitarity cuts
through their expansion into known scalar integrals with free coefficients.
A powerful method to obtain these coefficients, namely spinor integration,
is discussed and rederived in a somewhat novel form.
It is then used to compute analytically the infinite series of
one-loop gluon amplitudes in $\mathcal{N} = 1$ super-Yang-Mills theory
with exactly three negative helicities.
The final part of this thesis concerns
the intriguing relationship between gluon and graviton scattering amplitudes,
which involves a beautiful duality between
the color and kinematic content of gauge theories.
This BCJ duality is extended to include particles
in the fundamental representation of the gauge group,
which is shown to relieve the restriction of the BCJ construction
to factorizable gravities
and thus give access to amplitudes in generic (super-)gravity theories.

\section*{R\'esum\'e}

Cette th\`ese vise \`a assurer une meilleure compr\'ehension
de l'expansion perturbative des th\'eories de jauge avec et sans supersym\'etrie.
Au niveau des arbres, les relations de r\'ecurrence~BCFW sont analys\'ees
par rapport \`a leur validit\'e pour des objets g\'en\'eraux off-shell
en th\'eorie de Yang-Mills,
qui est un pas consid\'erable en dehors de leur zone d'application \'etablie.
Les p\^oles non physiques constituent un nouveau probl\`eme
en plus de celui du comportement limite, ce dernier commun au cas on-shell aussi.
Pour une famille infinie de courants de fermions massifs,
on presente certaines conditions qui garantissent que
ces deux obstacles sont \'evit\'es,
fournissant une relation de r\'ecurrence naturelle.
\`A une boucle, des amplitudes de diffusion peuvent \^etre calcul\'ees
\`a partir de coupes d'unitarit\'e gr\^ace \`a leur expansion
en int\'egrales scalaires connues avec des coefficients libres.
Une m\'ethode puissante pour obtenir ces coefficients,
\`a savoir l'int\'egration spinorielle,
est consider\'ee et rederiv\'ee sous une forme assez originale.
Elle est ensuite utilis\'ee pour calculer analytiquement
la s\'erie infinie des amplitudes des gluons \`a une boucle
dans la th\'eorie de $\mathcal{N}=1$ super-Yang-Mills
avec exactement trois h\'elicit\'es n\'egatives.
La derni\`ere partie de cette th\`ese concerne la relation intrigante
entre des amplitudes de diffusion des gluons et des gravitons,
qui implique une belle dualit\'e entre la couleur et le contenu cin\'ematique
de la th\'eorie de jauge.
On g\'en\'eralise cette dualit\'e~BCJ pour inclure des particules
dans la repr\'esentation fondamentale du groupe de jauge,
et on montre que cela l\`eve la restriction de la construction BCJ
aux gravit\'es factorisables et ainsi donne acc\`es \`a des amplitudes
dans des th\'eories de (super-)gravit\'e g\'en\'erales.

\section*{General audience abstract}

The scattering amplitudes subfield of particle physics occupies the space between collider phenomenology and string theory, thus making some mathematical insights from purely theoretical studies useful for the physics accessible by present-day experiments. Scattering amplitudes are excellent objects to investigate as in many cases they turn out to be calculable analytically. One method to compute them is recursive: once amplitudes for three particles are known, they can be used to construct those with any number of particles. The amplitudes that cannot be obtained by recursion can usually be cut into those that can. In this cut approach, the simpler ones are served as an input to compute more complicated ones. In this thesis, both kinds of methods are considered and used to obtain new analytic results for infinite families of scattering amplitudes.

\section*{R\'esum\'e g\'en\'eral}

Le domaine des amplitudes de diffusion se situe entre la ph\'enom\'enologie des collisionneurs et la th\'eorie des cordes. Gr\^ace \`a cela, certaines id\'ees math\'ematiques des \'etudes purement th\'eoriques deviennent utiles \`a la physique accessible par les exp\'eriences actuelles. Les amplitudes de diffusion sont d'excellents objets \`a \'etudier, car dans de nombreux cas, elles s'av\`erent \^etre analytiquement calculables. L'une des m\'ethodes pour les calculer est r\'ecursive : une fois les amplitudes pour trois particules connues, elles peuvent \^etre utilis\'ees pour construire celles avec un nombre quelconque de particules. Les amplitudes ne pouvant être obtenues par r\'ecurrence peuvent g\'en\'eralement \^etre coup\'ees en celles qui le peuvent. Avec cette approche-l\`a, les plus simples servent de donn\'ees d'entr\'ee pour calculer les plus compliqu\'ees. Dans cette th\`ese, les deux types de m\'ethode sont consid\'er\'es et utilis\'es pour obtenir de nouveaux r\'esultats analytiques pour des familles infinies d'amplitudes de diffusion.

\newpage

\section*{Acknowledgments}

I would like to thank my supervisor Ruth Britto for teaching me so much during these three years, as well as for her encouragement and advice always expressed with such amazing openness and optimism.

I am indebted to Henrik Johansson for his open attitude and patience
which lead to such a pleasant and fruitful collaboration.
I also thank Piotr Tourkine for all the exciting discussions on string theory,
which resulted not only in a paper but also in my ability to talk science in French.

I wish to thank all the members of IPhT Saclay for creating such a wonderful scientific environment. In particular, I am grateful to David Kosower and Gregory Korchemsky for being great mentors and setting an excellent example for me. I have also profited a lot from conversations with Pierre Vanhove, Leandro Almeida and Adriano Lo Presti. Moreover, I thank the administration team of the Institute for making the life of scientists much easier: Lo\"ic Bervas, Philippe Caresmel, Catherine Cataldi, Morgane Moulin, Laure Sauboy, Sylvie Zaffanella, as well as Audrey Lemarechal and Christine Ferret at Polytechnique.

I am grateful to many physicists outside Saclay for helpful conversations, including Zvi Bern, John Joseph Carrasco, David Dunbar, Einan Gardi, Warren Perkins, Radu Roiban, Emery Sokatchev, Simon Badger, Simon Caron-Huot, Tristan Dennen, Johannes Henn, Reinke Isermann, Edoardo Mirabella, Yang Zhang and Dima Chicherin,
as well as everybody I met at the most stimulating Carg\`ese school of 2012
and the Amplitudes conferences of 2013 and 2014.
Of course, I thank Gabriele Travaglini, Marios Petropoulos and Samuel Wallon for accepting to evaluate my thesis.

I am also indebted to my teachers and mentors from back home, especially Grigory Zalmanovich Mednikov, Wladimir von Schlippe, Andrei Arbuzov and Andrei Shuvaev.

I would like to thank the PhD students with whom I shared office however briefly: Jean-Marie, Ervand, Hendrik, Matthias, H\'el\`ene, Guilio and especially Alexandre, who was always eager to let me use his whiteboard as long as I occupied its upper parts.

I thank the senior PhD students, all doctors by now, for their hospitality: Kasper, Bruno, Julien, Romain, Igor Sh., Roberto and Francesco.
I am also grateful to the other fellow th\'esards
who witnessed me organizing our PhD student seminar
and even attended it not worse than before: Katya, Thomas E., Thomas A., Thiago, Misha, Antoine, Beno\^it, Yunfeng, Andrei, Ga\"elle, Samuel, J\'er\^ome, R\'emy, \'Eric, Yura, Guillaume, Maxime, Praxitelis, Soumya, Piotr W., Pierre and of course Hanna, who took on that honorable responsibility after me.

I want to thank many of my good friends scattered across the Paris area:
Olya, Gabriel, Adrien, Amine, Liz, Adam, Igor L., Igor R., Liza, Azat and George.
I would also like to thank my more distant but still the best friends Nikita, Misha, Nodar and \"Ozlem.
I am grateful to my parents for helping me to grow up and evolve into who I am now.
Finally, I thank Val\'erie for her friendship and love,
as well as for her confidence in the utility of fundamental science.

\newpage

~\,Papers written during the PhD study:

\begin{enumerate}[\{1\}]
   \item R.~Britto and A.~Ochirov, 
         \textit{On-shell recursion for massive fermion currents},
         \textit{JHEP} 1301 (2013) 002,
         \href{http://xxx.lanl.gov/abs/1210.1755}{\texttt{[arXiv:1210.1755]}}
   \item A.~Ochirov,
         \textit{All one-loop NMHV gluon amplitudes in $\mathcal{N} = 1$ SYM},
         \textit{JHEP} 1312 (2013) 080,
         \href{http://xxx.lanl.gov/abs/1311.1491}{\texttt{[arXiv:1311.1491]}}
   \item A.~Ochirov and P.~Tourkine,
         \textit{BCJ duality and double copy in the closed string sector},
         \textit{JHEP} 1405 (2014) 136,
         \href{http://xxx.lanl.gov/abs/1312.1326}{\texttt{[arXiv:1312.1326]}}
   \item H.~Johansson and A.~Ochirov,
         \textit{Pure gravities via color-kinematics duality for fundamental matter},
         \href{http://xxx.lanl.gov/abs/1407.4772}{\texttt{[arXiv:1407.4772]}}
\end{enumerate}

~
\tableofcontents

\mainmatter

\newpage

\chapter*{Introduction}
\addcontentsline{toc}{chapter}{Introduction}


In the last couple of decades, there has been impressive progress
in understanding perturbative expansion of gauge theories.
First came the realization that scattering matrix elements
not only constitute basic words of quantum field theoretic language,
but also turn out to be perfect objects to calculate analytically,
as was originally seen at tree level~\cite{Parke:1986gb}.
Then followed a number of beautiful insights
at one loop~\cite{Bern:1994zx,Britto:2004nc,Drummond:2008vq}
and beyond~\cite{Bern:2008qj,Bern:2010ue,Kosower:2011ty,ArkaniHamed:2012nw},
not to mention tree level again~\cite{Cachazo:2004kj,Britto:2004ap,Britto:2005fq}.

These developments resulted in taming gauge theory amplitudes analytically
for increasing and in some cases arbitrary number of particles.
Such remarkable progress was most welcome due its relevance
to the phenomenology of modern colliders,
as precise predictions of the Standard Model are crucial
for extracting new physics from the experimental data.
This resulted in many process-specific high-multiplicity calculations
(see, for example, the review~\cite{Butterworth:2014efa}),
as well as a number of semi-automated programs
for more general next-to-leading order computations~\cite{Berger:2008sj,
Hirschi:2011pa,Bevilacqua:2011xh,Cullen:2011ac,Cascioli:2011va,
Actis:2012qn,Cullen:2014yla}.

Most collider-relevant calculations are numerical,
as the analytic complexity tends to be quite prohibitive
for many theories of phenomenological interest,
of which quantum chromodynamics is the prime example.
However, it is usually the analytic understanding
that allows for efficient methods to be found and then be used numerically.
Hence the interest in different supersymmetric versions of QCD,
which seem more complicated only from the Lagrangian point of view,
but in practice turn out to be much more suitable for analytic studies.

Apart from the collider-related physics,
there is a lot to learn about gravity amplitudes.
At first glance, they might seem like rather odd objects to study,
as measuring the cross-section of graviton scattering experimentally
is rather unrealistic, to put it mildly.
Nonetheless, amplitude calculations can help answer
such basic and yet unresolved questions
as the ultraviolet properties of quantum gravity.
Although pure Einstein gravity is known to diverge
starting from two loops~\cite{Goroff:1985sz,Goroff:1985th,vandeVen:1991gw},
the situation is still rather unclear
for its supersymmetric extensions~\cite{Grisaru:1976nn,Tomboulis:1977wd}:
the extent to which supersymmetry improves the UV behavior of gravity
is not well understood.
Outstandingly, the absence of the four-loop divergence
in the maximally-supersymmetric gravity was discovered
by a perturbative calculation~\cite{Bern:2009kd}
and only then followed by symmetry arguments~\cite{Green:2010sp,
Bjornsson:2010wm,Bossard:2010bd,Beisert:2010jx,Bossard:2011tq}.
The next possible divergence is expected at seven loops,
which presently seems out of reach,
but the expected divergences of the lower-supersymmetry cases~\cite{Bern:2012gh,Bern:2013uka} are accessible by modern methods
and are likely to provide more insight in the structure of quantum gravity.

In this thesis, we consider scattering amplitudes of various kinds:
in gauge theory and in gravity, with and without supersymmetry;
we also review and develop different techniques to deal with them.

By now, there are three most basic theoretical tools that came into universal use:
\begin{itemize}
\item decomposing full amplitudes into simpler color-ordered components
\cite{Berends:1987cv,Mangano:1988kk,Bern:1990ux};
\item using helicity spinors both for fermions and bosons
\cite{Berends:1981rb,DeCausmaecker:1981bg,Xu:1986xb};
\item supersymmetric Ward identities and superspace coordinates
\cite{Grisaru:1976vm,Grisaru:1977px,Parke:1985pn,Kunszt:1985mg,Nair:1988bq}.
\end{itemize}
When needed, we recall some aspects of these standard techniques.
They are also nicely explained in \rcite{Dixon:1996wi}
and in the recent reviews~\cite{Elvang:2013cua,Henn:2014yza}.

At tree level, it is recursive methods that provide an efficient way
to calculate amplitudes both numerically and analytically.
One such approach, the Berends-Giele recursion~\cite{Berends:1987me}
operates with intermediate objects, the so-called off-shell currents,
which correspond to amplitudes with one leg off the mass shell.
A more modern method,
the Britto-Cachazo-Feng-Witten~\cite{Britto:2004ap,Britto:2005fq} recursion,
deals exclusively with fully on-shell objects.
In \chap{chap:treelevel}, based on \rcitePaper{1},
we will combine features of both approaches
and discover that the on-shell recursion
can be applicable to off-shell objects.
In particular, we use it to compute analytically
an infinite family of massive fermion currents.
Though these currents are essentially
an off-shell continuation of tree-level amplitudes,
they can also serve~\cite{Britto:2011cr} as key ingredients
for unitarity-based calculations of one-loop amplitudes with massive quarks.

More generally, at loop level,
the BCFW recursion has more limited use~\cite{Bern:2005hh}.
Nevertheless, many impressive analytic results~\cite{Bern:1994zx,Bern:1994cg,
Bedford:2004nh,Berger:2006vq,Bern:2004bt,Britto:2005ha,BjerrumBohr:2007vu,Dunbar:2009ax,
Britto:2006sj,Xiao:2006vt} are known thanks to a variety of modern on-shell methods.
In \chap{chap:oneloop}, based on \rcitePaper{2}, we review the powerful method of
spinor integration~\cite{Britto:2005ha,Anastasiou:2006jv} and then employ it
to calculate analytically the infinite series of
one-loop gluon amplitudes in $\mathcal{N} = 1$ super-Yang-Mills theory
with exactly three negative helicities.
Remarkably, this is achieved thanks to the concise analytic
formulas~\cite{Drummond:2008vq,Drummond:2008bq}
for tree-level amplitudes in $\mathcal{N} = 4$ SYM,
which can be obtained through
a supersymmetric version of the BCFW recursion~\cite{Elvang:2008na,Dixon:2010ik}.

Our results add to the body of one-loop amplitudes
known analytically to any multiplicity.
Interestingly, despite being computed in a supersymmetric Yang-Mills theory,
they can be regarded as a part of phenomenologically-relevant QCD amplitudes.
Moreover, we also use them to briefly discuss the applicability
of the loop-level BCFW recursion
beyond the previously established cases~\cite{Bern:2005hh}.

Somewhat unexpectedly, the progress in gauge theory amplitudes
greatly boosted the understanding of perturbation theory of gravity:
the integrands for graviton scattering amplitudes seem to be elegantly related
to those for gluon scattering amplitudes~\cite{Bern:2008qj,Bern:2010ue}.
More precisely, the problem of constructing gravity integrands
becomes reduced to finding such a representation of the gauge theory integrands
that its color and kinematic ingredients satisfy the same algebraic identities.
Such representations are said to obey the
{Bern-Carrasco-Johansson color-kinematics duality}~\cite{Bern:2008qj,Bern:2010ue}.
They were found in numerous calculations
\cite{Carrasco:2011mn,Bern:2012uf,Vanhove:2010nf,
Bern:2011rj,BoucherVeronneau:2011qv,Bern:2012cd,Bern:2012gh,Bern:2013uka,
Carrasco:2012ca,Chiodaroli:2013upa, Huang:2012wr,Bern:2013qca}
up to four loops in $\cN=4$ SYM~\cite{Bern:2009kd}.
At tree level, this duality is usually perceived
in terms of the celebrated Kawai-Lewellen-Tye relations~\cite{Kawai:1985xq}
between tree-level amplitudes in open and closed string theory,
further improved by monodromy relations and the momentum kernel~\cite{Stieberger:2009hq,
BjerrumBohr:2009rd,Tye:2010dd,BjerrumBohr:2010zs,BjerrumBohr:2010hn},
but a first-principle understanding at loop level is still
missing.\footnote{With the exception of one-loop amplitudes
in the self-dual sector of Yang-Mills theory \cite{Boels:2013bi}.}
One attempt to shed light on the string-theoretic origin of the BCJ duality
was made in \rcitePaper{3}, where both gauge theory and gravity
were considered in the purely closed-string context.

The color-kinematics duality was originally formulated for gauge theories
with purely-adjoint field content.
Moreover, it generically gave graviton amplitudes which belong to gravity theories
with some hardwired matter content.
For example, the gravity that can be obtained from
the pure non-supersymmetric Yang-Mills theory is
Einstein's general relativity theory coupled to an antisymmetric tensor and a dilaton.

As it turned out,
these two limitations can be resolved in one go.
In \chap{chap:bcj}, based on \rcitePaper{4},
we discuss an extension of the color-kinematics duality
that includes particles in the fundamental representation
of the Yang-Mills gauge group.
This provides a construction for (super-)gravity theories
with an arbitrary number of matter supermultiplets
which do not interact with each other.
Interestingly, it results as well in a recipe for amplitude integrands
in pure gravity theories.
In the latter case, the fundamental-representation construction
must develop a ghost-like behavior
in order to correctly cancel the unwanted degrees of freedom,
originally present in the adjoint-representation construction.

\section*{General conventions}
\addcontentsline{toc}{section}{General conventions}


In this thesis, we understand amplitudes, denoted by $\cA$, as constructed using
the textbook Feynman rules~\cite{Peskin:1995ev}
for the invariant $S$-matrix elements $i\cM$,
with all states considered incoming, but their momenta outgoing.
In gauge theory, calculations can be greatly simplified by separating
color factors from kinematic information.
Therefore, one usually deals with color-ordered amplitudes
that are defined as the coefficients of the leading color traces~\citeColorOrdering\,
and for which there exists a color-stripped version of Feynman rules~\cite{Dixon:1996wi}.
Such amplitudes, denoted by $A$, include only planar diagrams
with the external leg numbered counterclockwise in a fixed order.
Hence, their expressions can contain momentum sums
only for color-adjacent external particles,
for which we introduce the following short-hand notation:
\be
      P_{i,j} \equiv p_i + p_{i+1} + \dots + p_{j-1} + p_j ,
\label{momentumsum}
\ee
where indices are taken modulo the number of external particles $n$.

In a more general context, we will use the calligraphic letter~$\cA$,
instead of the plain $A$, to indicate that the amplitude $\cA$
contains not only kinematic information, but also color
or, in case of supersymmetric gauge theories, superspace degrees of freedom.

Unlike cross sections, scattering amplitudes are usually best expressed
not in terms of particle momenta $\{p_i\}_{i=1}^n$
but as functions of helicity spinors $\{\la_i,\lb_i\}_{i=1}^n$ ~\citeSpinorHelicity,
for which we use the following bra-ket notation:
\beal
   \overline{u}_+(p_i) & = \bra{\la_i} = \bra{i} , ~~~~~~~~~~
   u_-(p_i) = \ket{\la_i} = \ket{i} , \\
   \overline{u}_-(p_i) & =\;\! [\lb_i| =\;\! [i| , ~~~~~~~~~~
   u_+(p_i) = |\lb_i] \;\!= |i] .
\label{braket}
\eeal
The little-group freedom in choosing spinor phases reflects the fact that
only the absolute value of an amplitude affects the corresponding physical cross section.
We will find it convenient to adopt the following relation
between spinors with opposite energy signs:
\beal
            \ket{\!-\!p} & = i \ket{p} , \\
            |\!-\!p] & = i\,|p] .
\label{flippingmomentum}
\eeal
Moreover, we will use the following sign convention for momentum invariants:
\be
   s_{ij} = (p_i+p_j)^2 = \braket{ij} [ji] .
\ee
Other than that, we will not make precise the actual dependence of spinors
on their momenta.

Massive spinors can also be introduced in the spinor-helicity formalism,
but the details of various definitions (\eg~\cite{Kleiss:1985yh,Schwinn:2005pi})
will be irrelevant.
In this thesis, we adopt a round bracket notation $u(p) = |p)$
for an on-shell spinor with an arbitrary spin.
The Dirac equation, for any $m$, can be written as
\be
   \big\{\!\!\not{\!p} - m \big\} |p) =
   \big\{ \ket{p}[p| + |p]\bra{p} - m \big\} |p) = 0 .
\label{Dirac}
\ee

The polarization vector for a gauge boson of momentum $p$ is~\citeSpinorHelicity,
depending on its helicity,
\be
   \varepsilon^\mu_{p+} =  \frac{1}{\sqrt{2}}
      \frac{\langle n_p|\gamma^\mu|p]}{\braket{n_p p}} , \qquad
   \varepsilon^\mu_{p-} = -\frac{1}{\sqrt{2}}
      \frac{[n_p|\gamma^\mu|p \rangle}{[n_p p]} ,
\label{pvectors}
\ee
where $n_p$ is an arbitrary but fixed ``reference'' momentum satisfying $n_p^2=0$
and either $\braket{n_p p} \neq 0$ or $[n_p p] \neq 0$,
so that the denominator is nonzero.
The null reference momenta are chosen independently for each gauge boson.
The set of reference momenta is what we refer to
as the gauge choice for a particular calculation,
within the Feynman gauge used throughout the spinor-helicity formalism.

In the four-point case, we adopt the standard notation for the Mandelstam variables:
\be
      s = (p_1+p_2)^2 , \qquad
      t = (p_2+p_3)^2 , \qquad
      u = (p_1+p_3)^2 .
\label{Mandelstam}
\ee

To conclude this section, we note that in the first two chapters,
in absence of photons in our considerations,
we will be using wiggly lines for gluons in order to keep the graphs easily readable.
In \chap{chap:bcj}, however, we will reserve the wiggly line notation for gravitons,
and so gluons will be depicted by curly lines.

\chapter{Gauge theory at tree level}
\label{chap:treelevel}

Modern recursive methods~\cite{Berends:1987me,Britto:2004ap,Britto:2005fq}
have immensely simplified the computation of tree-level amplitudes in gauge theory
with respect to the straightforward Feynman-rule approach.
In some contexts, it is interesting to also study amplitudes (or rather, currents)
in which one or more legs is continued {\em off shell},
since they carry even more information than on-shell amplitudes.
For example, the Berends-Giele recursion~\cite{Berends:1987me}
among gluon currents in Yang-Mills theory is not only computationally powerful
for numerical results, but was also the crucial stepping stone
to establishing the first formulas for gluon amplitudes with arbitrary numbers of legs,
in certain helicity configurations~\cite{Berends:1987me,Kosower:1989xy}.
It is still possible to consider the limits in which internal propagators go on-shell
and apply the BCFW construction to find recursion relations~\cite{Feng:2011twa}.
Compared to the recursion relations for on-shell amplitudes,
the ones for currents require committing to a gauge choice,
and summing over all internal polarization states, including unphysical polarizations.

In this chapter, we seek compact analytic forms for currents of $n-2$ gluons
and two massive quarks, where one of the quarks is off shell
and the remaining particles are on shell.
These currents not only can be used to produce analytic formulas for on-shell tree amplitudes of gluons with massive quarks, such as those in \rcites{Quigley:2005cu,
Schwinn:2006ca, Ozeren:2006ft, Ferrario:2006np, Schwinn:2007ee, Hall:2007mz, Chen:2011sba, Boels:2011zz, Huang:2012gs},
but they are also key ingredients of an on-shell method of
computing 1-loop amplitudes with external massive fermions~\cite{Britto:2011cr}.
These amplitudes are of particular interest
in the context of LHC searches for new physics,
where production of top quarks plays a large role in both signals and backgrounds.
Massive fermion currents can be computed
with the off-shell Berends-Giele recursion~\cite{Berends:1987me}.
In \rcite{Rodrigo:2005eu}, this was used to give a compact result in the case
where all gluons have the same helicity,
with a particular gauge choice relative to the massive spinors.

We study the validity of the BCFW construction~\cite{Britto:2005fq,Badger:2005jv}
for these massive fermion currents.
The construction begins by shifting
the momenta of a pair of on-shell external legs by $+zq$ and $-zq$ respectively,
where $z$ is a complex variable and $q$ is obtained by requiring that both legs
remain on shell after the shift.
Then, the residue theorem produces a recursion relation from poles in $z$ taking values where propagators go on shell, as recalled in \sec{sec:arecursion}.
The construction breaks down if there are poles from other sources.
In Yang-Mills theory, the only other possible source is a ``boundary term'',
from a nonvanishing limit as $z$ is taken to infinity.\footnote{If
the theory is sufficiently well understood,
it is possible to include a boundary term explicitly at each step of the recursion
\cite{Feng:2009ei,Feng:2010ku,Benincasa:2011kn,Benincasa:2011pg,Feng:2011twa}.}
For off-shell currents, which we define more precisely in \sec{sec:cconventions},
there is another problematic source of poles, which we call ``unphysical poles'',
see \sec{sec:crecursion}.
They are due to the gauge dependence, and they spoil the recursion relation,
since we have no information about how to calculate their residues independently.

We deal with the two types of obstructions in \secs{sec:boundary}{sec:unphysical}, respectively.
We identify conditions under which the boundary terms and unphysical poles vanish for massive fermion currents, so that the BCFW construction produces a recursion relation. Then, in \sec{sec:cresults}, we proceed to solve the recursion in the particular case where all gluons have equal helicities. Compared to the more compact result of \rcite{Rodrigo:2005eu}, our formula also requires all gluons to use the same reference spinor but preserves the genericity of its value.

Our analysis of boundary terms in \sec{sec:boundary} is based on grouping Feynman diagrams conveniently and applying the Ward identity and inductive arguments. The argument establishes the absence of boundary terms for {\em general} off-shell objects in  Feynman gauge, provided that there are two on-shell gluons available to construct the momentum shift. 

In our study of unphysical poles in \sec{sec:unphysical}, we use off-shell gluon currents of the type originally derived by Berends and Giele~\cite{Berends:1987me}. We are motivated to generalize the currents in which one gluon has opposite helicity to all the others, by taking its reference spinor to be arbitrary. When the opposite-helicity gluon is color-adjacent to the off-shell leg, we find a very compact form for the current. When it is centrally located among the other gluons, we prove that the current is, in fact, independent of the arbitrary reference spinor.



\section{Recursion for amplitudes}
\label{sec:arecursion}

In this section, we review
the classical Britto-Cachazo-Feng-Witten recursion derivation~\cite{Britto:2005fq}
for tree-level scattering amplitudes in gauge theory.

Consider an $n$-point tree-level amplitude $\cA_n$ as a rational function
of external helicity spinors $\{\la_i,\lb_i\}_{i=1}^n$.
If one makes a linear change of the $k$-th and $l$-th spinor variables:
\beal
   \hat{\ket{k}} & = \ket{k} , ~~~~~~~~~~~~~~~~~\,
   \hat{\ket{l}} = \ket{l} + z \ket{k} , \\
   \hat{|k]}\;\,\!\!& = |k] - z |l] , ~~~~~~~~~~
   \hat{|l]}\;\! = |l] ,
\label{kl-shift}
\eeal
then the amplitude becomes a rational function $\cA_n(z)$ of the complex variable $z$.
We denote the shift \eqref{kl-shift} by $[kl \rangle$.
The crucial property of such shifts is that the momenta
\be
   \widehat{p}_k  = p_k - z q , ~~~~~~~~~~~~ \widehat{p}_l = p_l + z q ,
\label{kl-momenta}
\ee
shifted by the complex-valued vector
\be
   q^\mu = \langle k|\gamma^\mu|l]/2 ,
\label{q}
\ee
do not break the general momentum conservation and stay on shell at the same time.
Hence, we can still regard $\cA_n(z)$ as a scattering amplitude.

Now let us look at the following integral over a contour
big enough to encircle all finite poles of $\cA_n(z)$:
\be
   \oint \frac{\d z}{2 \pi i} \frac{\cA_n(z)}{z} ~~=~~ \cA_n(0) ~~
      + \!\!\!\!\!\!\!\!\sum_{\text{poles }z_P\text{ of }\cA_n(z)}
        \frac{1}{z_P}\,\Res_{z=z_P} \cA_n(z) .
\label{Cauchy}
\ee
Here we extracted by hand the first evident residue of $\cA_n(z)/z$ at $z=0$,
which is precisely the original physical amplitude $\cA_n$.
On the other hand, there might be a pole at infinity:
\be
   \oint \frac{\d z}{2 \pi i} \frac{\cA_n(z)}{z}
      = \lim_{z \rightarrow \infty} \cA_n(z) .
\label{boundary}
\ee
From \eqns{Cauchy}{boundary}, we immediately obtain
\be
   \cA_n ~~=~~-\!\!\!\!\!\!\!\!\sum_{\text{poles }z_P\text{ of }\cA_n(z)}
               \frac{1}{z_P}\,\Res_{z=z_P} \cA_n(z)
         ~~+~~ \lim_{z \to \infty} \cA_n(z) .
\label{basicbcfw}
\ee

To obtain recursion relations among amplitudes, we must have no pole at infinity.
This is assured if $ \lim_{z \to \infty} \cA_n(z) = 0 $,
which we call \emph{good boundary behavior}.
We will discuss in detail how it is achieved for gauge theory amplitudes
in \sec{sec:boundary}.

The remaining poles can have two possible origins,
due to their construction from Feynman rules:
\begin{enumerate}
\item vanishing of the denominator of a polarization vector,
when written as in \eqn{pvectors}.
\item vanishing of the denominator of a propagator.
\end{enumerate}

The poles of the first type are unphysical.
They have zero residues thanks to the gauge invariance of the amplitude:
the reference spinor of the shifted gauge boson can be chosen
to eliminate the $z$ dependence in its denominator.

The poles of the second type are physical,
and their residues are can be evaluated
as products of two amplitudes with fewer legs,
since the vanishing of a propagator denominator is an on-shell condition.

First, let us consider a pole from a scalar propagator with momentum $P$.
It is only affected by the shift~\eqref{kl-momenta} for those Feynman diagrams
in which it connects two trees, each with one of the shifted external legs.
Then the propagator momentum sum becomes $\hat{P} = P \pm z q$,
depending on the direction of $P$.
Taking into account that $q^2=0$, we obtain
\be
   \frac{i}{P^2(z)-m^2} = \frac{i}{P^2 \pm 2 (P\!\cdot\!q) z - m^2}
                        = \frac{\pm i}{2 (P\!\cdot\!q) (z - z_P)} ,
\label{scalarpropagatorshifted}
\ee
where the pole value is
\be
   z_P = \mp \frac{(P^2-m^2)}{2(P\!\cdot\!q)} .
\label{polevalue}
\ee
If we take the residue of the shifted propagator~\eqref{scalarpropagatorshifted},
when combined with the prefactor $-1/z_P$ from \eqn{basicbcfw},
it turns out to produce simply the unshifted scalar propagator:
\be
  -\frac{1}{z_P}\,\Res_{z=z_P} \frac{i}{P^2(z)-m^2} =  \frac{i}{P^2-m^2} .
\label{scalarpole}
\ee
The sum of all diagrams with such a pole inevitably gives the full amplitudes
on both sides of the $P$-channel propagator.
If there were only scalar poles,
a relation to amplitudes with fewer legs would be already established:
\be
   \cA_n ~~=~~ \!\!\!\!\!\!\!\!\!\!\sum_{\text{cut channels }P\text{ of }\cA_n}
               \!\!\!\!\!\!\!\cA_{r+1}(z_P) \frac{i}{P^2-m^2}\,\cA_{n-r+1}(z_P) .
\label{bcfwscalar}
\ee

Let us now replace the scalar propagator by a gauge boson one.
In Feynman gauge, it has the metric tensor $-g_{\mu \nu}$ in the numerator.
On the $z_P$-pole kinematics, it can be expanded as follows:

\be
   -g^{\mu \nu} = \varepsilon^\mu_{\hat{P}+} \varepsilon^\nu_{\hat{P}-}
                + \varepsilon^\mu_{\hat{P}-} \varepsilon^\nu_{\hat{P}+}
                + \varepsilon^\mu_{\hat{P}L} \varepsilon^\nu_{\hat{P}T}
                + \varepsilon^\mu_{\hat{P}T} \varepsilon^\nu_{\hat{P}L} ,
\label{bcfwmetric}
\ee
where the polarization vectors with the null momentum $\hat{P}$
are defined as in~\eqref{pvectors}, and
\be
   \varepsilon^{\mu}_{\hat{P}L} = \hat{P}^{\mu} , ~~~~~~~
   \varepsilon^{\mu}_{\hat{P}T} = - \frac{ n_{\hat{P}}^{\mu} }
                                         { (\hat{P}\!\cdot\!n_{\hat{P}}) } .
\label{bcfwpvectors}
\ee
Around the $P$-channel gauge boson propagator, the Feynman diagrams on both sides
now sum up to two amplitudes, each with one amputated polarization vector.
When such amputated amplitudes are contracted with the tensor~\eqref{bcfwmetric},
the contribution from
$ \varepsilon^\mu_{\hat{P}L} \varepsilon^\nu_{\hat{P}T}
+ \varepsilon^\mu_{\hat{P}T} \varepsilon^\nu_{\hat{P}L} $
vanishes due to the Ward identities on both sides, while
$ \varepsilon^\mu_{\hat{P}+} \varepsilon^\nu_{\hat{P}-}
+ \varepsilon^\mu_{\hat{P}-} \varepsilon^\nu_{\hat{P}+} $
produces two usual gauge theory amplitudes.
Consequently, in case of a pure gauge theory,
the recursion relation contains a sum over helicities of the intermediate gauge boson:
\be
   \cA_n ~~=~~ \!\!\!\!\!\!\!\!\!\!\sum_{\text{cut channels }P\text{ of }\cA_n}
             ~ \sum_{h=\pm}~\cA_{r+1}^{h}(z_P) \frac{i}{P^2}\,\cA_{n-r+1}^{-h}(z_P) .
\label{bcfwgluon}
\ee

Finally, if the $P$-channel propagator is fermionic,
on the $z_P$-pole kinematics its numerator can be expanded into the following state sum:
\be
   \not{\!\hat{P}} + m = \sum_{s=1,2} u_s(\hat{P}) \overline{u}_s(\hat{P}) ,
\ee
and the Feynman diagrams on both sides add up to two spinor-amputated amplitudes.
In the massless case, we can flip the momentum sign
according to \eqn{flippingmomentum}:
\be
   \not{\!\hat{P}} = \ket{\hat{P}}[\hat{P}| + |\hat{P}]\bra{\hat{P}}
                   = - i \big\{ \ket{\hat{P}}[-\hat{P}| + |\hat{P}]\bra{-\hat{P}} \big\} ,
\ee
and thus obtain a recursion relation for massless QCD:
\beal
   \cA_n ~~=~~&\!\!\!\!\!\!\!\!\!\!\sum_{\text{gluon channels }P\text{ of }\cA_n}
             ~ \sum_{h=\pm}~\cA_{r+1}^{h}(z_P) \frac{i}{P^2}\,\cA_{n-r+1}^{-h}(z_P) \\
         ~~+~~&\!\!\!\!\!\!\!\!\!\!\sum_{\text{quark channels }P\text{ of }\cA_n}
             ~ \sum_{h=\pm}~\cA_{r+1}^{h}(z_P) \frac{1}{P^2}\,\cA_{n-r+1}^{-h}(z_P) .
\label{bcfwqcd}
\eeal
Here we can see how amplitude calculations are sensitive to sign conventions,
such as~\eqref{flippingmomentum}.
In fact, we have used it already in \eqn{bcfwgluon},
when we implicitly assumed its corollary:
\be
   \varepsilon^\mu_{(-\hat{P})\pm} = \varepsilon^\mu_{\hat{P}\pm} .
\ee
Therefore, the subtle relative phases of gluon and quark contributions
in the recursion~\eqref{bcfwqcd} are self-consistent.

It is important to note that the on-shell recursion relations
contain significantly less terms for color-ordered amplitudes $A_n$,
because they are only allowed to have poles in sums of color-adjacent momenta.

In conclusion of this review section, we also remark
that the BCFW recursion was found~\cite{Bedford:2005yy,Cachazo:2005ca,
BjerrumBohr:2005jr,Benincasa:2007xk,Bern:2007xj,BjerrumBohr:2006yw,ArkaniHamed:2008yf}
to work surprisingly well for graviton amplitudes.
Subsequently, the cancellations, necessary for the good boundary behavior,
were argued to be related~\cite{Boels:2011mn,Boels:2012sy,Isermann:2013yha}
to the BCJ color-kinematics duality~\cite{Bern:2008qj,Bern:2010ue},
the latter being the main topic of \chap{chap:bcj}.

\section{Conventions for fermion currents}
\label{sec:cconventions}

In this section, we fix the conventions for the main object of interest of this chapter
--- off-shell fermion currents.

Momenta of gluons are directed outward, while momenta of fermions are directed inward.
We will be considering color-ordered amplitudes
and off-shell currents with one massive fermion line,
for example, $ iJ \big( 1_{\bar{Q}}^*, 2_Q, 3_g, 4_g, \dots, n_g \big)$, 
where the star means that the indicated leg is considered off-shell,
while the remaining legs are on-shell.
Note that we do not include the propagator for the off-shell leg in our definition.
For this current, the quark line has its arrow pointing from leg 2 to leg 1.
Due to that unfortunate, but conventional \cite{Hall:2007mz} numbering,
when the quark line matrices are read against the arrow,
then the gluon indices are contracted in reverse numerical order,
because they are still color-ordered counterclockwise.
So in slightly different notation, we can write
\be
   iJ \big( 1_{\bar{Q}}^*, 2_Q^{}, 3_g^{h_3}, 4_g^{h_4}, \dots, n_g^{h_n} \big)
      = | n^{h_n} \dots 4^{h_4} 3^{h_3} | 2 ) ,
\label{altnotation}
\ee
where the round bracket $|2)$ can be equal to either $ |2 \rangle $ or $|2]$,
depending on its spin.
This notation emphasizes the fact that the current is a spinorial object.
For example, to obtain the corresponding amplitude,
one should first put $p_1$ on shell
and then contract the current with either $[1|$ or $ \langle 1| $:
\be
   A \big( 1_{\bar{Q}}^-, 2_Q^-, 3_g^{h_3}, 4_g^{h_4}, \dots, n_g^{h_n} \big)
            = \langle 1 | n^{h_n} \dots 4^{h_4} 3^{h_3} |2] .
\label{notationexample}
\ee

The color-ordered Feynman rules behind these conventions
differ from the standard ones \cite{Dixon:1996wi}
by an extra minus sign in front of the cubic vertex:
\begin{subequations} \begin{align}
      \parbox{18mm}{
      \begin{fmffile}{vertex3} \fmfframe(10,10)(10,0){
      \fmfsettings \begin{fmfgraph*}(30,30)
            \fmflabel{$p, \lambda$}{g1}
            \fmflabel{$q, \mu$}{g2}
            \fmflabel{$r, \nu$}{g3}
            \fmftop{g2,g3}
            \fmfbottom{g1}
            \fmf{wiggly}{g2,v,g3}
            \fmf{wiggly}{g1,v}
      \end{fmfgraph*} }
      \end{fmffile}
      }
      & = -\frac{i}{\sqrt{2}} \left[ g_{\lambda \mu} (p - q)_{\nu}
                                   + g_{\mu \nu} (q - r)_{\lambda}
                                   + g_{\nu \lambda} (r - p)_{\mu} \right] ,
\label{vertex3G} \\
      \parbox{20mm}{
      \begin{fmffile}{vertex4} \fmfframe(10,10)(10,0){
      \fmfsettings \begin{fmfgraph*}(40,30)
            \fmflabel{$\lambda$}{g1}
            \fmflabel{$\mu$}{g2}
            \fmflabel{$\nu$}{g3}
            \fmflabel{$\rho$}{g4}
            \fmfleft{g1,g2}
            \fmfright{g4,g3}
            \fmf{wiggly}{g1,v,g2}
            \fmf{wiggly}{g3,v,g4}
      \end{fmfgraph*} }
      \end{fmffile}
      }
      & = \frac{i}{2} \left[ 2 g_{\lambda \nu} g_{\mu \rho}
                             - g_{\lambda \rho} g_{\mu \nu}
                             - g_{\lambda \mu} g_{\nu \rho} \right] .
\label{vertex4G}
\end{align} \end{subequations}
\!\!\!\!\;
The polarization vectors, defined in \eqn{pvectors},
contain arbitrary but fixed ``reference'' momenta,
independently chosen for each gluon.
The set of reference momenta is what we refer to
as the gauge choice for a particular calculation,
within the Feynman gauge used throughout the spinor-helicity formalism.
Any current we construct with a specific gauge choice
is expected to fit into a larger calculation,
such as the one-loop computations of \cite{Britto:2011cr},
in which all external legs are on-shell, so that ultimately,
after being combined with other ingredients computed in the same gauge,
no trace of the gauge choice remains. Therefore,
the reference spinors can be chosen to maximize computational convenience.
We delay the choice as far as possible, so that convenience can be evaluated later
in the full context of a larger calculation.
   
The spinors for the massive fermions satisfy the Dirac equation~\eqref{Dirac}.
We do not require any further details of their definitions,
so any of various conventions (\eg \cite{Kleiss:1985yh,Schwinn:2005pi})
can be used. The massless limit is smooth.

\section{Recursion for currents}
\label{sec:crecursion}

Let us now look at what happens, if we apply a BCFW shift to a pair of gluons,
since they are always on-shell in this context.
The idea is still to apply the residue theorem on $iJ(z)/z$
to recover the current as
\be
   iJ(z=0) ~~=~~-\!\!\!\!\!\!\!\!\sum_{\text{poles }z_P\text{ of }iJ(z)}
                 \frac{1}{z_P}\,\Res_{z=z_P} iJ(z)
           ~~+~~ \lim_{z \to \infty} iJ(z) .
\ee

If $J(z)$ goes to zero in the limit $z \to \infty$,
\ie in the case of good boundary behavior,
we can hope to obtain recursion relations among currents.

Just as for the amplitudes in \sec{sec:arecursion},
the finite poles of off-shell currents can have same two possible origins
due to their construction from Feynman rules:
\begin{enumerate}
\item vanishing of the denominator of a polarization vector,
when written as in \eqn{pvectors}.
\item vanishing of the denominator of a propagator.
\end{enumerate}

The propagator-induced poles are familiar from the recursion relations
for on-shell amplitudes~\cite{Britto:2005fq}
and were discussed in detail in \sec{sec:arecursion}.
The corresponding residues are easy to evaluate
as the product of two currents or amplitudes with fewer legs,
because the vanishing of a propagator denominator
puts the corresponding internal line on shell.

The first type of pole will be called an \emph{unphysical pole}.
In \sec{sec:arecursion}, we argued that in an on-shell amplitude
the reference spinor in such a pole could be freely chosen
to eliminate the $z$ dependence in the denominator,
but now with a quark line off-shell,
we must fix all reference spinors from the start, and they play an explicit role.
These unphysical poles are problematic,
since their location has no natural physical meaning,
and we have no independent way of computing their residues.
Thus, we will find conditions on the currents and shift
that prevent the appearance of unphysical poles.

With good boundary behavior and no unphysical poles,
there will be a recursion relation that takes the schematic form
\be
   iJ_n ~~=~~ \!\!\!\!\!\!\!\!\!\!\sum_{\text{cut channels }P\text{ of }J_n}
             ~ \sum_{h=\pm}~iJ_{r+1}^{h}(z_P) \frac{i}{P^2}\,iJ_{n-r+1}^{-h}(z_P) ,
\label{cbcfwgluon}
\ee
where the $i{\hat J}$  are currents and amplitudes with fewer legs,
$P$ is the momentum flowing between them which goes on shell at the pole $z_P$,
and the second sum is over internal helicities.
Of course, the propagator acquires a mass if it is fermionic.
If all of the off-shell legs belong
to just one of the two currents on the right-hand side,
then the other is replaced by a shifted on-shell amplitude $A(z_P)$.

\section{Boundary behavior}
\label{sec:boundary}

In this section, the behavior as $z \to \infty$ of the fermionic current
$iJ \big( 1_{\bar{Q}}^*, 2_Q^*, 3_g, 4_g, \dots, n_g \big) $
under the  $ [ k l \rangle $ shift \eqref{kl-shift},
where $k$ and $l$ represent any of the gluons.
We will conclude that the boundary term vanishes in the helicity cases
$(k_g^-,l_g^+)$, $(k_g^-,l_g^-)$ and thus $(k_g^+,l_g^+)$ as well,
for a generic gauge choice.
The generality of this argument is greatly enhanced by the fact that
we will not require any of the unshifted gluons to be on shell,
nor explicitly constrain the number of fermion lines.

\subsection{Choice of shift: helicities and polarizations}
\label{sec:shiftchoice}

Consider the superficial boundary behavior of individual Feynman diagrams,
following the flow of the additional momentum $zq$. Without the polarization vectors,
the diagrams where the $zq$ momentum goes only through 3-gluon vertices
and gluon propagators behaves the worst --- as $ O(z) $.
If $zq$ runs through a 4-gluon vertex or through a fermion line,
then such a diagram already behaves as $ O(1) $ or better.
Now, if we contract the vector indices for the $[kl \rangle$-shifted gluons
with their polarization vectors,
\beal
   \hat{\varepsilon}_{k-}^{\mu} & = -\frac{1}{\sqrt{2}}
      \frac{ [n_k |\gamma^{\mu}| k \rangle }
           { [n_k k] - z [n_k l] } , \:~~~~~~~~~~~
   \hat{\varepsilon}_{k+}^{\mu} =  \frac{1}{\sqrt{2}}
      \frac{ \langle n_k| \gamma^{\mu}| k] -z \langle n_k |\gamma^{\mu} |l] }
           { \langle n_k k \rangle } , \\
   \hat{\varepsilon}_{l-}^{\nu} & = -\frac{1}{\sqrt{2}}
      \frac{ [n_l| \gamma^{\nu} |l \rangle + z [n_l |\gamma^{\nu}| k \rangle }
           { [n_l l] } , ~~~
   \hat{\varepsilon}_{l+}^{\nu} =  \frac{1}{\sqrt{2}}
      \frac{ \langle n_l |\gamma^{\nu} |l] }
           { \langle n_l l \rangle + z \langle n_l k \rangle } ,
\label{polvectors}
\eeal
we see that for a generic gauge choice the off-shell current superficially
has $ O\left( \frac{1}{z} \right) $ behavior in the $ (k_g^-,l_g^+) $ case,
$O(z)$ in the $ (k_g^-,l_g^-) $ and $ (k_g^+,l_g^+) $ cases,
and $O(z^3)$ in the  $ (k_g^+,l_g^-) $ case.
Note that this behavior can be altered by special gauge choices,
\ie if $ n_k = l $ or $ n_l = k $, then
$ \hat{\varepsilon}_{k-}^{\mu} $ and $ \hat{\varepsilon}_{l+}^{\nu} $
lose their $z$ dependence instead of being $ O\left( \frac{1}{z} \right) $ .

The helicity case $(k_g^-,l_g^+)$ is thus safe automatically,
for a generic gauge choice.
In \sec{sec:mixhelspecialgauge}, we will discuss boundary behavior
for a special gauge choice that will be needed in the following sections.
In the remainder of this section, we will prove that
in a generic gauge (where $n_k \neq l$)
the off-shell current with helicities $ (k_g^-,l_g^-) $ also vanishes at infinity,
at least as $ O\left( \frac{1}{z} \right) $.
To do that, let us multiply this current by the $z$-independent factor
$ - \sqrt{2} [n_l l] $, so that only the numerator
$ [n_l |\gamma^{\nu}| \hat{l}\rangle $ remains contracted with
the $l$-th gluon's Lorentz index $\nu$.
The resulting expression depends only linearly on $[n_l|$,
which is a 2-dimensional massless spinor and thus can be expressed as
a linear combination of any two independent spinors of the same kind:
\be
   [n_l| = \alpha [l| + \beta [n_k| .
\label{n_l}
\ee
Therefore, it is enough to show that we get $ O\left( \frac{1}{z} \right) $ behavior
for the two special cases $n_l = l$ and $n_l=n_k$. Let us examine them one by one.

\subsection{Like-helicity shift, first term}
\label{sec:firstterm}

The first term in \eqn{n_l} yields
$ [l|\gamma^{\nu}|\hat{l} \rangle = 2 \hat{l}^{\nu} $,
making it possible to use the Ward identity,
which diagrammatically can be expressed as follows:
\beal
   \includegraphics[scale=1.0]{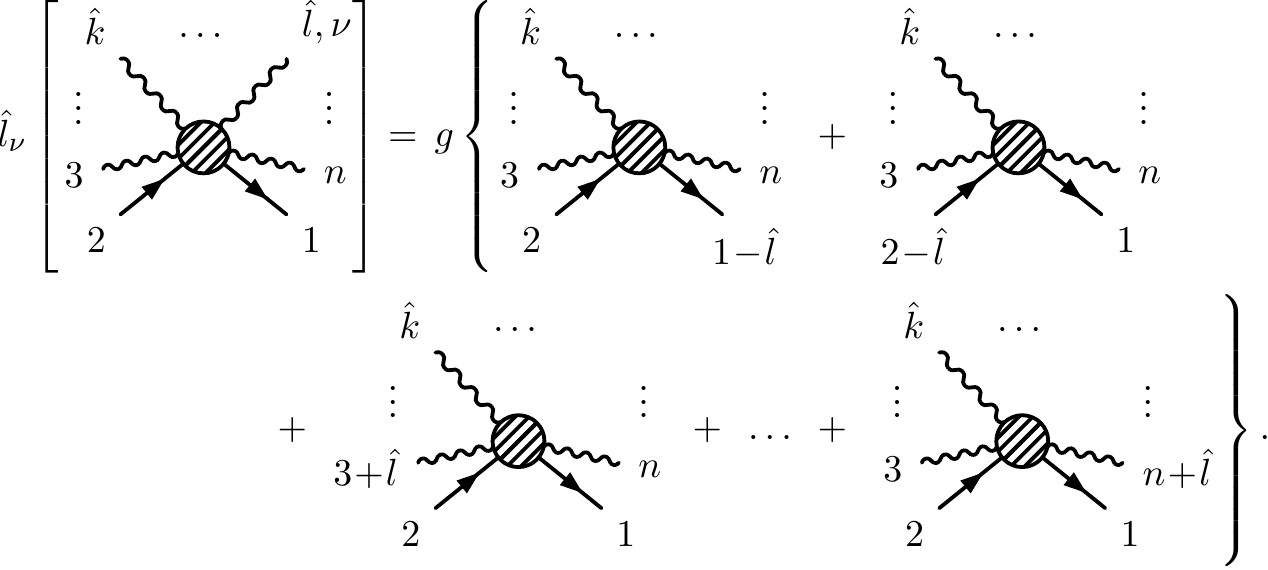}
\label{ward}
\eeal

Each diagram on the right-hand side of \eqref{ward} is supposed to have
an appropriate gauge group generator contracted with the leg
to which the momentum $ \hat{l} $ is added,
but that is irrelevant for the discussion of $ z \rightarrow \infty $ behavior.
For any leg that is initially on shell, such as the $k$-th gluon,
the corresponding right-hand-side term should naturally vanish,
because the resulting leg would go off shell and thus would be left out
when extracting the on-shell pole residue according to the standard LSZ procedure.

Now consider what the Feynman rules tell us about the diagrams
on the right-hand side. In case there are off-shell gluons,
if a diagram has the $zq$ momentum going from the $k$-th gluon to another
through 3-gluon vertices,
it must now behave no worse than $ O\left( \frac{1}{z} \right) $!
Indeed, it still has $ \hat{\varepsilon}_{k-}^{\mu} \sim O\left( \frac{1}{z} \right) $
on the $k$-th leg, and in addition to that the gluon propagator on the off-shell leg
is now $ O\left( \frac{1}{z} \right) $ as well.
If $zq$ runs through a 4-gluon vertex or through a fermion line
and ends up still on a gluon leg,
then such a diagram will behave at most as $  O\left( \frac{1}{z^2} \right) $.
A new ingredient here is the diagrams that have $zq$ momentum flowing through
3-point vertices to an off-shell fermion leg,
but it is easy to see that they will also behave
like $ O\left( \frac{1}{z} \right) $ or better.
In sum, applying the Ward identity in this way reduces
the maximal superficial power of $z$ at infinity by two.

There is one technical caveat about this argument: strictly speaking,
the Ward identity \eqref{ward} is valid for a ghostless gauge,
whereas in Feynman gauge it is necessary to introduce some extra terms
on the right-hand side.
We address this issue carefully in the next section
and find that the argument still holds in Feynman gauge.

\subsection{Ward identity argument in Feynman gauge}
\label{sec:wardid}


The supplementary terms we need to consider to fill the gap of the previous section
are due to the fact that the Noether current of the global gauge transformation
receives additional contributions from the gauge-fixing and ghost parts
of the effective Lagrangian. (Note that in the generalized axial gauge
the gauge-fixing term $ \propto \left( n^{\mu} A_{\mu}^a \right)^2 $
contains no derivatives and thus does not contribute to the Noether current.)
At tree level, however, the ghost part of the current does not produce
any non-vanishing diagrams, unless we consider a Green's function
with external ghost legs, which is not the case.
Therefore, the only non-trivial ingredient that we should worry about
in Feynman gauge is the gauge-fixing contribution to the Noether current.

To derive the Ward identities,
we consider an infinitesimal global gauge transformation,
\be
   A_{\mu}^a \rightarrow A_\mu^a - g f^{abc} \alpha^b A_{\mu}^c ,
\label{globalgauge}
\ee
which leaves invariant the $R_{\xi}$-gauge-fixing Lagrangian:
\be
   L_{\xi} = - \frac{1}{2 \xi} \left( \partial_{\mu} A^{a\mu} \right)^2 .
\label{lagrangiangf}
\ee
The latter generates the following contribution to the Noether gauge current:
\be
   J_{\xi}^{a\mu} = \frac{\partial L_{\xi}}{\partial (\partial_{\mu} A_{\nu}^b)}
                    g f^{abc} A_{\nu}^c
                  = \frac{g}{\xi} f^{abc} A^{b\mu} \partial_{\nu} A^{c\nu} .
\label{currentgf}
\ee
Taking a derivative gives
\be
   \partial_\mu J_{\xi}^{a\mu}(x)
      = \frac{g}{\xi} f^{abc} A^{b\mu}(x) \partial_{\mu} \partial_{\nu} A^{c\nu}(x) ,
\ee
so we retrieve the following momentum-space operator:
\be
   \hat{l}_\nu J_{\xi}^{b \nu} (\hat{l})
      = - \frac{i g}{\xi} f^{bcd} \int \frac{d^4 p}{(2\pi)^4}
          p_{\nu} p_{\rho} A^{c\nu}(p) A^{d\rho}(\hat{l}-p) .
\label{ccurrentgf}
\ee
This operator is to be inserted instead of the $\hat{l}$-th leg
and combined with the remaining $(n-1)$ legs of the off-shell current:
\be
      - \frac{i g}{\xi} f^{bcd} \int \frac{d^4 p}{(2\pi)^4}
                \hat{\varepsilon}_k^{\mu} p^{\nu} p^{\rho} 
      \left[ \parbox{152pt}{ \begin{fmffile}{wardsupplementary1}
      \fmfframe(24,18)(0,18){ \fmfsettings \begin{fmfgraph*}(80,60)
            \fmflabel{$1$}{q1}
            \fmflabel{$2$}{q2}
            \fmflabel{$3$}{g3}
            \fmflabel{$\vdots$}{d1}
            \fmflabel{$\hat{k},\mu,a\!\!\!$}{gk}
            \fmflabel{$\dots$}{d2}
            \fmflabel{$\!\!\!p,\nu,c$}{gp}
            \fmflabel{$\hat{l}\!-\!p,\rho,d$}{gl}
            \fmflabel{$\vdots$}{d3}
            \fmflabel{$n$}{gn}
            \fmfleft{q2,g3,,d1,,gk}
            \fmftop{d2}
            \fmfright{q1,gn,,d3,gl,gp}
            \fmf{plain_arrow}{c,q1}
            \fmf{plain_arrow}{q2,c}
            \fmf{wiggly}{g3,c}
            \fmf{wiggly}{gk,c}
            \fmf{wiggly,tension=0.5}{gp,c}
            \fmf{wiggly,tension=0.5}{gl,c}
            \fmf{wiggly}{gn,c}
            \fmfblob{0.25w}{c}
      \end{fmfgraph*} }
      \end{fmffile} } \right] .
\label{wardextra}
\ee

Note that in \eqn{wardextra} only the $\hat{k}$-th leg is considered
propagator-amputated, and we spell out its contraction
with the polarization vector $\hat{\varepsilon}_k^{\mu}$ explicitly.
In the following, we specialize to the Feynman gauge $\xi = 1$
and once again neglect all color information. 
Since \eqref{ccurrentgf} has two gluon legs and already contains one power of $g$,
then in order to construct a tree level contribution of order $O(g^{n-2})$
from an object with $(n+1)$ external legs we need to contract two of them together
without any interaction insertions.
Thus the other $(n-1)$ legs must form a normal connected tree-level diagram
of order $O(g^{n-3})$. An extra disconnected piece will naturally produce
a $\delta$-function which will annihilate the integration in \eqn{wardextra}.

We cannot contract together the two legs coming from \eqref{ccurrentgf},
because that would produce $ \delta^{(4)}(p+\hat{l}-p) = 0 $.
Moreover, if both directly contracted legs are not in \eqn{ccurrentgf},
say the $i$-th and $j$-th, then we will get $ \delta^{(4)}(p_i+p_j) = 0 $.
Obviously, we cannot contract directly a fermion with a gluon either.
Thus we are left only with the options of connecting one of the two legs
\eqref{ccurrentgf} with any of the remaining gluon legs ---
either the shifted one $\hat{k}$ or any of the unshifted legs.

Contraction of the first leg of \eqref{ccurrentgf} with the on-shell gluon $\hat{k}$
vanishes immediately due to the transversality of the polarization vector
$\hat{\varepsilon}_{k-}^{\mu}$,
whereas connecting the other leg to $\hat{k}$ results in $ p = \hat{k} + \hat{l} $ :
\beal
      \hat{\varepsilon}_k^{\mu} & p^{\nu} p^{\rho}
      \left[ \parbox{42mm}{
      \begin{fmffile}{wardsupplementary2}
      \fmfframe(38,0)(0,-2){ \fmfsettings \begin{fmfgraph*}(40,10)
            \fmflabel{$\hat{k},\mu$}{gk}
            \fmflabel{$\hat{l} \!-\! p,\rho$}{gl}
            \fmfleft{gk}
            \fmfright{gl}
            \fmf{wiggly}{gk,gl}
      \end{fmfgraph*} }
      \fmfframe(27,18)(0,18){ \fmfsettings \begin{fmfgraph*}(70,50)
            \fmflabel{$1$}{q1}
            \fmflabel{$2$}{q2}
            \fmflabel{$3$}{g3}
            \fmflabel{$\vdots$}{d1}
            \fmflabel{$k\!-\!1$}{gk1}
            \fmflabel{$k\!+\!1$}{gk2}
            \fmflabel{$\dots$}{d2}
            \fmflabel{$p,\nu$}{gp}
            \fmflabel{$\vdots$}{d3}
            \fmflabel{$n$}{gn}
            \fmfleft{q2,g3,,d1,gk1,gk2}
            \fmftop{d2}
            \fmfright{q1,gn,,d3,,gp}
            \fmf{plain_arrow}{c,q1}
            \fmf{plain_arrow}{q2,c}
            \fmf{wiggly}{g3,c}
            \fmf{wiggly,tension=0.5}{gk1,c}
            \fmf{wiggly,tension=0.5}{gk2,c}
            \fmf{wiggly}{gp,c}
            \fmf{wiggly}{gn,c}
            \fmfblob{0.25w}{c}
      \end{fmfgraph*} }
      \end{fmffile} } \right] \\ & =
      (2\pi)^4 \delta^{(4)}(\hat{l}-p+\hat{k}) \hat{\varepsilon}_k^{\mu}
                           (\hat{k}+\hat{l})_{\mu}
      \cdot (\hat{k}+\hat{l})^{\nu}
      \left[ \parbox{45mm}{
      \begin{fmffile}{wardsupplementary3}
      \fmfframe(27,18)(0,18){ \fmfsettings \begin{fmfgraph*}(70,50)
            \fmflabel{$1$}{q1}
            \fmflabel{$2$}{q2}
            \fmflabel{$3$}{g3}
            \fmflabel{$\vdots$}{d1}
            \fmflabel{$k\!-\!1$}{gk1}
            \fmflabel{$k\!+\!1$}{gk2}
            \fmflabel{$\dots$}{d2}
            \fmflabel{$\hat{k} \!+\! \hat{l},\nu$}{gp}
            \fmflabel{$\vdots$}{d3}
            \fmflabel{$n$}{gn}
            \fmfleft{q2,g3,,d1,gk1,gk2}
            \fmftop{d2}
            \fmfright{q1,gn,,d3,,gp}
            \fmf{plain_arrow}{c,q1}
            \fmf{plain_arrow}{q2,c}
            \fmf{wiggly}{g3,c}
            \fmf{wiggly,tension=0.5}{gk1,c}
            \fmf{wiggly,tension=0.5}{gk2,c}
            \fmf{wiggly}{gp,c}
            \fmf{wiggly}{gn,c}
            \fmfblob{0.25w}{c}
      \end{fmfgraph*} }
      \end{fmffile} } \right] = O\!\left( \frac{1}{z} \right) ,
\label{wardextra2}
\eeal
so that the only remaining element dependent on $z$ is
$ \hat{\varepsilon}_{k-}^{\mu} = O\left( \frac{1}{z} \right) $.
Now if we take an arbitrary unshifted gluon leg with momentum $p_j$
and Lorentz index $\lambda$ and contract it with the leg $p$ from \eqref{ccurrentgf},
we get
\beal
      \hat{\varepsilon}_k^{\mu} & p^{\nu} p^{\rho}
      \left[ \parbox{109pt}{
      \begin{fmffile}{wardsupplementary4}
      \fmfframe(28,0)(0,-2){ \fmfsettings \begin{fmfgraph*}(40,10)
            \fmflabel{$j,\lambda$}{gj}
            \fmflabel{$p,\nu$}{gp}
            \fmfleft{gj}
            \fmfright{gp}
            \fmf{wiggly}{gj,gp}
      \end{fmfgraph*} }
      \fmfframe(18,18)(0,18){ \fmfsettings \begin{fmfgraph*}(60,45)
            \fmflabel{$1$}{q1}
            \fmflabel{$2$}{q2}
            \fmflabel{$3$}{g3}
            \fmflabel{$\vdots$}{d1}
            \fmflabel{$\hat{k},\mu$}{gk}
            \fmflabel{$\dots$}{d2}
            \fmflabel{$\hat{l}\!-\!p,\rho$}{gl}
            \fmflabel{$\vdots$}{d3}
            \fmflabel{$n$}{gn}
            \fmfleft{q2,,,g3,,,,d1,,,gk}
            \fmftop{d2}
            \fmfright{q1,,,gn,,,,d3,,,gl}
            \fmf{plain_arrow}{c,q1}
            \fmf{plain_arrow}{q2,c}
            \fmf{wiggly}{g3,c}
            \fmf{wiggly}{gk,c}
            \fmf{wiggly}{gl,c}
            \fmf{wiggly}{gn,c}
            \fmfblob{0.25w}{c}
      \end{fmfgraph*} }
      \end{fmffile} } \right] \\ & =
      (2\pi)^4 \delta^{(4)}(p+p_j) \frac{-i g_{\lambda \nu}}{p_j^2} p_j^{\nu}
      \cdot \hat{\varepsilon}_k^{\mu} p_j^{\rho}
      \left[ \parbox{112pt}{
      \begin{fmffile}{wardsupplementary5}
      \fmfframe(18,18)(0,18){ \fmfsettings \begin{fmfgraph*}(60,45)
            \fmflabel{$1$}{q1}
            \fmflabel{$2$}{q2}
            \fmflabel{$3$}{g3}
            \fmflabel{$\vdots$}{d1}
            \fmflabel{$\hat{k},\mu$}{gk}
            \fmflabel{$\dots$}{d2}
            \fmflabel{$\hat{l}\!+\!p_j,\rho$}{gl}
            \fmflabel{$\vdots$}{d3}
            \fmflabel{$n$}{gn}
            \fmfleft{q2,,,g3,,,,d1,,,gk}
            \fmftop{d2}
            \fmfright{q1,,,gn,,,,d3,,,gl}
            \fmf{plain_arrow}{c,q1}
            \fmf{plain_arrow}{q2,c}
            \fmf{wiggly}{g3,c}
            \fmf{wiggly}{gk,c}
            \fmf{wiggly}{gl,c}
            \fmf{wiggly}{gn,c}
            \fmfblob{0.25w}{c}
      \end{fmfgraph*} }
      \end{fmffile} } \right] \frac{-i}{(\hat{l}+p_j)^2}
      = O\!\left( \frac{1}{z} \right) .
\label{wardextra3}
\eeal
In the second line, we have written out the propagator of the $(\hat{l}-p)$ leg,
making both $z$-dependent legs propagator-amputated,
so that the diagram in the brackets is a standard $O(z)$
and the overall expression obviously vanishes at infinity.

The last remaining type of contribution is not so straightforward.
Connecting an off-shell gluon leg to the $(\hat{l}-p)$ leg makes $p$ equal to
$\hat{l} + p_j$ and thus produces two more powers of $z$ in the numerator:
\beal
      \hat{\varepsilon}_k^{\mu} & p^{\nu} p^{\rho}
      \left[ \parbox{109pt}{
      \begin{fmffile}{wardsupplementary6}
      \fmfframe(28,0)(0,-2){ \fmfsettings \begin{fmfgraph*}(40,10)
            \fmflabel{$j,\lambda$}{gj}
            \fmflabel{$\hat{l} \!-\! p,\rho$}{gl}
            \fmfleft{gj}
            \fmfright{gl}
            \fmf{wiggly}{gj,gl}
      \end{fmfgraph*} }
      \fmfframe(18,18)(0,18){ \fmfsettings \begin{fmfgraph*}(60,45)
            \fmflabel{$1$}{q1}
            \fmflabel{$2$}{q2}
            \fmflabel{$3$}{g3}
            \fmflabel{$\vdots$}{d1}
            \fmflabel{$\hat{k},\mu$}{gk}
            \fmflabel{$\dots$}{d2}
            \fmflabel{$p,\nu$}{gp}
            \fmflabel{$\vdots$}{d3}
            \fmflabel{$n$}{gn}
            \fmfleft{q2,,,g3,,,,d1,,,gk}
            \fmftop{d2}
            \fmfright{q1,,,gn,,,,d3,,,gp}
            \fmf{plain_arrow}{c,q1}
            \fmf{plain_arrow}{q2,c}
            \fmf{wiggly}{g3,c}
            \fmf{wiggly}{gk,c}
            \fmf{wiggly}{gp,c}
            \fmf{wiggly}{gn,c}
            \fmfblob{0.25w}{c}
      \end{fmfgraph*} }
      \end{fmffile} } \right] \\ & =
      (2\pi)^4 \delta^{(4)}(\hat{l}-p+p_j) \frac{-i g_{\lambda \rho}}{p_j^2}
                           (\hat{l}+p_j)^{\rho}
      \cdot \hat{\varepsilon}_k^{\mu} (\hat{l}+p_j)^{\nu}
      \left[ \parbox{112pt}{
      \begin{fmffile}{wardsupplementary7}
      \fmfframe(18,18)(0,18){ \fmfsettings \begin{fmfgraph*}(60,45)
            \fmflabel{$1$}{q1}
            \fmflabel{$2$}{q2}
            \fmflabel{$3$}{g3}
            \fmflabel{$\vdots$}{d1}
            \fmflabel{$\hat{k},\mu$}{gk}
            \fmflabel{$\dots$}{d2}
            \fmflabel{$\hat{l}\!+\!p_j,\nu$}{gp}
            \fmflabel{$\vdots$}{d3}
            \fmflabel{$n$}{gn}
            \fmfleft{q2,,,g3,,,,d1,,,gk}
            \fmftop{d2}
            \fmfright{q1,,,gn,,,,d3,,,gp}
            \fmf{plain_arrow}{c,q1}
            \fmf{plain_arrow}{q2,c}
            \fmf{wiggly}{g3,c}
            \fmf{wiggly}{gk,c}
            \fmf{wiggly}{gp,c}
            \fmf{wiggly}{gn,c}
            \fmfblob{0.25w}{c}
      \end{fmfgraph*} }
      \end{fmffile} } \right] \frac{-i}{(\hat{l}+p_j)^2} .
\label{wardextra4}
\eeal
Fortunately, what we can see on the right-hand side apart from other $O(1)$ factors
is just what we started with --- a BCFW-shifted off-shell current
with both shifted legs propagator-amputated and contracted with
the polarization vector $\hat{\varepsilon}_{k-}^{\mu}$ on one side
and with the momentum $(\hat{l}+p_j)$ on the other.
The main difference is that it has one gluon leg less than before
and the momentum of the missing gluon is now added to the $\hat{l}$-th leg.

So we started with a superficially $O(z)$ object with $(n-4)$ unshifted gluons on
the left-hand side of \eqref{ward}, and we have managed to see that
the right-hand side of the Ward identity gives us $O(\frac{1}{z})$ contributions
plus $(n-4)$ superficially $O(z)$ objects of the same type,
but with $(n-5)$ unshifted gluons.
In the same fashion, we can apply the Ward identity repeatedly,
until there are no unshifted gluons left,
which proves that the initial object was indeed $O(\frac{1}{z})$.

In this way,
we have verified that the Ward identity argument is still valid in Feynman gauge.

\subsection{Like-helicity shift, second term}
\label{sec:secondterm}

Now we consider the case $ n_l = n_k = n $.
It turns out to be possible to deduce some interesting facts
about the boundary behavior of an off-shell current simply from the Feynman rules.

\subsubsection{Gluon trees, leading power of $z$}
\label{sec:gluonleading}

We start with the leading $ O(z) $ diagrams,
in which the $zq$ momentum flows only through 3-gluon vertices and gluon propagators,
which thus behave as $ O(z) $ at most.
Let us look closely at the part of such a diagram
that is directly adjacent to the $zq$ momentum flow,
\ie just a gluon tree with all but the $k$-th and $l$-th legs off-shell
and their propagators amputated, keeping in mind that any of the off-shell legs
can be extended by any sort of $z$-independent tree, including a fermion line.
      \begin{figure}[h]
      \centering
      \includegraphics[scale=1.0]{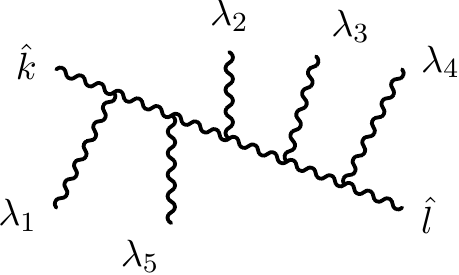}
      \vspace{-5pt}
      \caption{\small A generic gluon tree diagram with only 3-point vertices.
               \label{fig:gluontree}}
      \end{figure}

A tree with $n$ legs will have $(n-2)$ vertices $ \sim O(z) $,
$(n-3)$ internal propagators $ \sim O\left( \frac{1}{z} \right) $
and $(n-2)$ free indices.
The highest power of $z$ will be accumulated if we pick up $zq^{\lambda}$
from each vertex, a $ z [n |\gamma^{\nu}| k \rangle $ term
from the $l$-th gluon's polarization vector,
and another $ [n |\gamma^{\mu}| k \rangle $ coming from the $k$-th gluon:
$n$ vectors in total. Apart from that,
vertices and propagator numerators can only offer various combinations
of metric tensors, and the fact that there are only $(n-2)$ free indices means that
at least one contraction will take place among those vectors.
But any such contraction eliminates a power of $z$, since
\be
   [n|\gamma^{\mu}|k \rangle q_{\mu} = 0 , ~~~~~~~
   q^2 = 0 .
\ee
So the leading $ O(z) $ term vanishes and we are left only with $ O(1) $ at most.

\subsubsection{Fermion line insertion, leading power of $z$}
\label{sec:fermionleading}

Similarly, the leading  $ O(1) $ term vanishes for the diagrams
in which $zq$ flows through the fermion line.
To see this, consider once again only the terms directly adjacent
to the $zq$ momentum flow, \ie the relevant part of the fermion line
and mostly off-shell gluon trees on both of its sides.
      \begin{figure}[h]
      \centering
      \includegraphics[scale=1.0]{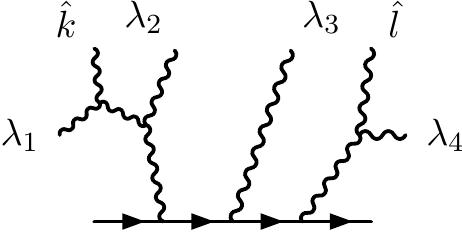}
      \vspace{-5pt}
      \caption{\small A generic diagram with $zq$ flowing through the fermion line and only 3-point gluon vertices. \label{fig:fermionline}}
      \end{figure}

The leading power of $z$ in a generic diagram with $n$ gluons will now be attained
by accumulating $(n-1)$ powers of $zq^{\lambda}$ from 3-gluon vertices
and the numerators of fermion propagators. As before,
$ [n |\gamma^{\mu}| k \rangle $ and $ z [n |\gamma^{\nu}| k \rangle $ will come from
the $k$-th and $l$-th gluons, respectively, so in total we have $(n+1)$ vectors
with only $(n-2)$ free indices to attribute to them.
Note that the relevant part of the fermion line consists of an odd number
of $\gamma$-matrices and thus can always be expressed as a linear combination
of eight basic matrices
$ \{\gamma^{\mu}\}_{\mu=0}^3 \cup \{\gamma^{\mu} \gamma^5\}_{\mu=0}^3 $
just by using the standard formula:
\be
\gamma^{\lambda} \gamma^{\mu} \gamma^{\nu}
            = g^{\lambda \mu} \gamma^{\nu}
            - g^{\lambda \nu} \gamma^{\mu}
            + g^{\mu \nu} \gamma^{\lambda}
            + i \epsilon^{\lambda \mu \nu \rho} \gamma_{\rho} \gamma^5 .
\label{threegamma}
\ee
The free index of the $ \gamma^{\mu} $ or $ \gamma^{\mu} \gamma^5 $
can either be left free as an off-shell gluon index
(leaving us with only $(n-3)$ free indices left for $(n+1)$ vectors)
or be contracted with one of the $(n+1)$ vectors.
So the number of free indices is smaller than the number of vectors at least by two.
The difference with the previous case is that now we have not only metric tensors
to do the index-contraction work,
but also the totally antisymmetric tensor coming from \eqref{threegamma}.
So lowering the number of free vector indices by two can be achieved
by either dotting one vector to another,
in which case we get zero just as in the gluon-tree case,
or by contracting three vectors to one antisymmetric tensor constructing terms like
$ \epsilon_{\lambda \mu \nu \rho}  \cdot z q^{\lambda} \cdot [n |\gamma^{\mu}| k \rangle \cdot z [n| \gamma^{\nu}| k \rangle $,
all of which vanish since we have copies of only two vectors
in the leading $O(z)$ term.
Thus all diagrams with $zq$ momentum flow through a fermion line
necessarily vanish at $ z \rightarrow \infty $.

\subsubsection{Gluon trees, next-to-leading power of $z$}
\label{sec:gluonsubleading}

What is left to consider is the possible $O(1)$ contribution from gluon trees.
To begin with, calculate the simplest gluon tree, \ie a single 3-gluon vertex
contracted with the two polarization vectors as given in \eqn{polvectors}:
\beal
      \parbox{35pt}{
      \begin{fmffile}{gluontree0} \fmfframe(5,0)(0,-10){
      \fmfsettings \begin{fmfgraph*}(30,30)
            \fmflabel{$\lambda$}{g}
            \fmflabel{$\hat{k}\!\!$}{gk}
            \fmflabel{$\!\hat{l}$}{gl}
            \fmftop{gk,gl}
            \fmfbottom{g}
            \fmf{wiggly}{gk,v,gl}
            \fmf{wiggly}{g,v}
      \end{fmfgraph*} }
      \end{fmffile}
      }
      & = \hat{\varepsilon}_{k-}^{\mu}
          \left( g_{\mu \nu} (\hat{k} - \hat{l})_{\lambda}
               + g_{\nu \lambda} (2 \hat{l} + \hat{k})_{\mu}
               - g_{\lambda \mu} (2 \hat{k} + \hat{l})_{\nu}
          \right)
          \hat{\varepsilon}_{l-}^{\nu} \\
      & = - \frac{1}{ 2 z [nl]^2 }
          \left( z \langle k |\gamma_{\lambda} |n] \cdot 2 \langle k | l | n] )
               + \langle k |\gamma_{\lambda}| n] \langle l |\gamma^{\nu} |n]
                 \cdot z \langle k |\gamma_{\nu}| l]
          \right) + O\!\left( \frac{1}{z} \right) \\
      & = - \frac{\langle k |\gamma_{\lambda} |n]}{2 [nl]^2 }
          \left( 2 \langle k l \rangle [ln]
               + 2 \langle l k \rangle [ln]
          \right) + O\!\left( \frac{1}{z} \right)
        = O\!\left( \frac{1}{z} \right) .
\label{gluonthree}
\eeal
Here the $O(z)$ terms vanish trivially in accord with our previous considerations,
but we see from the Fierz identity that the $O(1)$ term is canceled as well.
At four legs these cancellations continue to take place,
but start to involve $O(1)$ diagrams with a single quartic vertex insertion,
for instance:

\beal
   \includegraphics[scale=1.0]{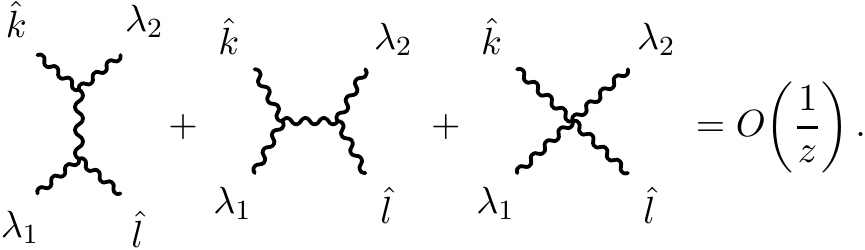}
\label{gluonfour}
\eeal

\begin{figure}[h]
      \centering
      \includegraphics[scale=1.0]{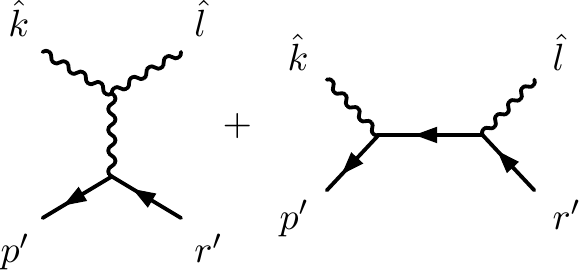}
      \vspace{-5pt}
      \caption{\small Diagrams for the amplitude of 2 gluons and 1 fermion line.
               \label{fig:gluonthreeappend}}
      \end{figure}

Evidently, such an intricate cancellation cannot be deduced
by examining Feynman diagrams separately.
Let us look again at the 3-gluon vertex \eqref{gluonthree}
from another point of view.
Attaching a gluon propagator to the off-shell line obviously does not change
the power of $z$, and the resulting off-shell 3-gluon current is a Lorentz vector.
If we contract it with a simple fermion line,
we obtain the first diagram in \fig{fig:gluonthreeappend},
which is a part of a scattering amplitude --- a gauge invariant object
that is well established to behave as $ O\left( \frac{1}{z} \right) $
for the $ (k_g^-,l_g^-) $ shift.

Moreover, the second diagram in \fig{fig:gluonthreeappend}
has $zq$ momentum flow through its fermion line,
so according to our previous discussion,
it is of order $ O\left( \frac{1}{z} \right) $ by itself.
Thus we can conclude that the first one is $ O\left( \frac{1}{z} \right) $ as well.
We obtained it by contracting the initial off-shell current vector
with a correctly defined fermion line.
The freedom of choosing the on-shell fermion momenta and helicities
spans the whole Minkowski space.
Therefore, the vector must have the same boundary behavior.

      \begin{figure}[h]
      \centering
      \includegraphics[scale=1.0]{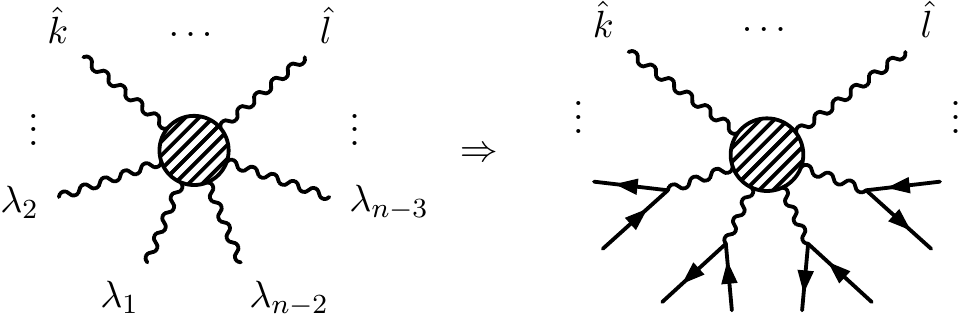}
      \vspace{-5pt}
      \caption{\small Contracting a gluon off-shell current with fermion lines.
               \label{fig:gluontensor}}
      \end{figure}

Along the same lines, we can now prove a very general statement:
\emph{An $n$-gluon off-shell current with only two shifted like-helicity legs on shell behaves as $ O\left( \frac{1}{z} \right) $.}
The current has a free Lorentz index for each off-shell leg,
so it is actually a tensor of rank $(n-2)$.
If we contract every index with its own fermion line (independent of $z$),
we will obtain an expression corresponding to a scalar amplitude
(\fig{fig:gluontensor}) which we know behaves as a whole
as $ O\left( \frac{1}{z} \right) $ under the $ (k_g^-,l_g^-) $ shift.
The freedom of choice of fermion momenta and helicities guarantees that
if the contracted expression vanishes at $ z \rightarrow \infty $,
then the initial tensor is bound to vanish too.

      \begin{figure}[t]
      \centering
      \includegraphics[scale=1.0]{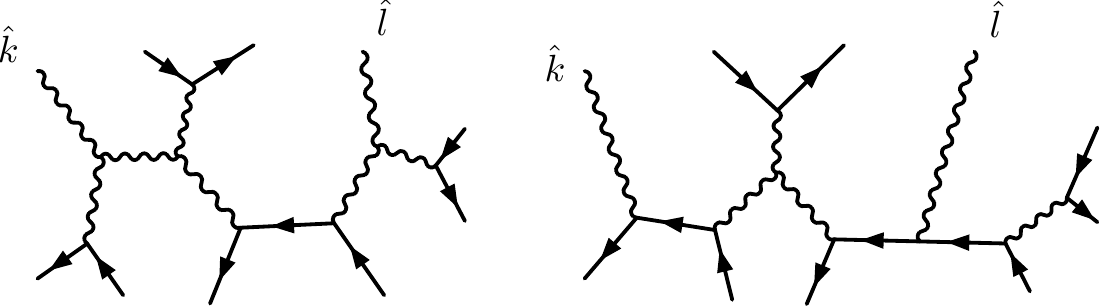}
      \vspace{-5pt}
      \caption{\small Diagrams for $ 4 \times \{ q \bar{q} \} \rightarrow \hat{g_k} \hat{g_l} $ with $zq$ momentum flow through fermion propagators. \label{fig:gluontensorlacking1}}
      \end{figure}

Of course, we will still lack some diagrams to build the full amplitude, but the lacking terms will in fact be those for which we have already proven the good behavior at $ z \rightarrow \infty $. Indeed, the result resembles an amplitude for a process where $(n-2)$ (distinct) quark-antiquark pairs go to 2 gluons, so it should contain not only the diagrams which are given by the right-hand side of \fig{fig:gluontensor}, but also those where some fermion lines have multiple fermion vertices and thus have fermion propagator insertions in them. Some of them look like the diagrams which are shown in \fig{fig:gluontensorlacking1}, \ie\ have $zq$ flow through at least one of those fermion lines and thus vanish at $ z \rightarrow \infty $.

\begin{figure}[t]
      \centering
      \includegraphics[scale=1.0]{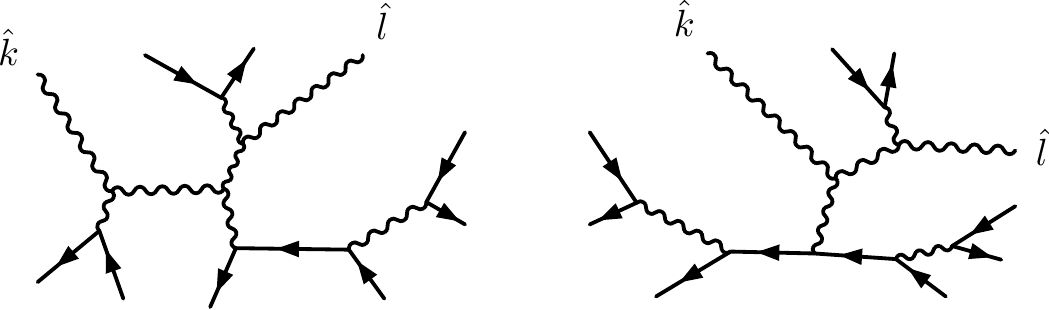}
      \vspace{-5pt}
      \caption{\small Diagrams for $ 4 \times \{ q \bar{q} \} \rightarrow \hat{g_k} \hat{g_l} $, that can be reduced to the case of smaller number of gluon legs. \label{fig:gluontensorlacking2}}
      \end{figure}

Others, however, may look like the diagrams shown in \fig{fig:gluontensorlacking2},
\ie contain some fermion lines connected to the shifted gluons only through their connection to other fermion lines. These diagrams can be reduced to the case of a smaller number of off-shell legs in the initial gluon current. 
Thus, we can construct an inductive argument, for which we have already verified the base case of $n=3$, to see that all the diagrams that we need to add to the contracted $n$-gluon current to form an amplitude behave as $ O\left( \frac{1}{z} \right) $ and the current itself is therefore bound to be $ O\left( \frac{1}{z} \right) $.

By the way, this inductive proof did not use the weaker $O(1)$ statement
of \sec{sec:gluonleading},
though we relied heavily on the $ O\left( \frac{1}{z} \right) $ statement
of \sec{sec:fermionleading}.
To conclude, let us recall the steps of our argument:

\begin{enumerate}

\item Any diagram with $zq$ momentum flow through at least one fermion propagator
behaves well.

\item The boundary behavior of the diagrams with $zq$ momentum flow
strictly through gluon propagators is the same as that of a gluon-only off-shell current.

\item Any off-shell current with 3 gluon legs vanishes as $ z \rightarrow \infty $
due to the cancellation which ensures the good behavior of the amplitude
$ q \bar{q} \rightarrow \hat{g} \hat{g} $.

\item Any off-shell current with $n$ gluon legs vanishes as $ z \rightarrow \infty $
to ensure the good behavior of the amplitude
$ (n-2) \times \{ q \bar{q} \} \rightarrow \hat{g} \hat{g} $,
provided the good behavior of the $(n-1)$--gluon current
and the diagrams with fermion propagator insertions.

\end{enumerate}

\subsection{Mixed-helicity shift in special gauges}
\label{sec:mixhelspecialgauge}

For a generic gauge, the boundary behavior of individual Feynman diagrams
under the $ [ k l \rangle $ shift is automatically $ O\left( \frac{1}{z} \right) $
in the mixed-helicity case $(k_g^-,l_g^+)$.
But in the following section, we will find that in order to avoid unphysical poles
we need to use special gauge choices $ n_k = p_l $ or $ n_l = p_k $.
Such gauges eliminate the $z$-dependence
from one of the polarization vector denominators
and thus turn the superficial behavior into $O(1)$.
However, we can easily rephrase our power-counting arguments
from Sections \ref{sec:gluonleading} and \ref{sec:fermionleading}
for the mixed-helicity case and find that the leading power of $z$
must always involve at least one contraction of two of the following three vectors:
$ zq^{\lambda} $, $ [n_k |\gamma^{\mu}| k \rangle $
and $ \langle n_l| \gamma^{\nu} |l] $,
with either $g_{\mu \nu}$ or $\epsilon_{\lambda \mu \nu \rho}$.
Either by imposing $ n_k = p_l $ or $ n_l = p_k $, we can guarantee that
any such contraction will give zero and thus ensure vanishing of the boundary term.

It is worth noting that if we take $ n_k = p_l $ and $ n_l = p_k $ simultaneously,
the superficial boundary behavior is worsened by two powers of $z$,
and the argument will in general be invalid.
Suppose that we first impose $ n_l = p_k $ and have $n_k$ unfixed.
Then we will be guaranteed to have no pole at infinity,
but might still have an unphysical pole at $ z_k = [n_k k]/[n_k l] $.
Now if we take $ n_k = p_l $, we can see that the pole $z_k$ goes smoothly to infinity.
In this way, the unphysical pole and the boundary term can be traded one for another,
and the problem is to find gauges in which neither survives.

\section{Avoiding unphysical poles}
\label{sec:unphysical}

In this section, we address the question of unphysical poles,
\ie the poles that come from polarization vectors \eqref{polvectors}
instead of propagators.
We construct explicit recursive proofs of the vanishing of the unphysical poles
for the following currents:

\begin{enumerate}

\item $ [ 3 4 \rangle $-shifted
$ iJ \big( 1_{\bar{Q}}^*, 2_Q, \hat{3}_g^-, \hat{4}_g^-, 5_g^-, \dots, n_g^- \big) $
with $ n_3 = n_4 = \dots = n_n $ ;

\item $ [ 4 3 \rangle $-shifted
$ iJ \big( 1_{\bar{Q}}^*, 2_Q, \hat{3}_g^+, \hat{4}_g^-, 5_g^-, \dots, n_g^- \big) $
with $ n_4 = n_5 = \dots = n_n = p_3 $ ;

\item $ [ 3 4 \rangle $-shifted
$ iJ \big( 1_{\bar{Q}}^*, 2_Q^*, \hat{3}_g^-, \hat{4}_g^+, 5_g^-, \dots, n_g^- \big) $
with $ n_3 = n_5 = \dots = n_n = p_4 $ .

\end{enumerate}

It is straightforward to prove analogous statements for the currents
with the opposite quark off shell or for flipped helicity assignments. For example,
$iJ \big( 1_{\bar{Q}}, 2_Q^*, 3_g^-, \dots, \widehat{(n\!-\!1)}_g^-, \hat{n}_g^+ \big)$ has no unphysical poles under the $ [ n\!-\!1 | n \rangle $ shift
if $ n_3 = \dots = n_{n-1} = p_n $.
One can also make a simultaneous flip of all gluon helicities trivially.

In short, the good gauge choices are:
\begin{itemize}
\item in the all-minus case, put all reference momenta equal to each other: $ n_i = q $;
\item in the one-plus cases, put reference momenta of negative-helicity gluons
equal to the momentum of the positive-helicity gluon: $ n_- = p_+ $.
\end{itemize}

Note that in the one-plus case with the positive-helicity gluon in central position
the unphysical poles vanish for a matrix-valued current $ | n^- \dots 5^- 4^+ 3^- | $
with \emph{both} fermions off shell,
\ie lacking spinors on both sides of the quark line. In fact, there is strong evidence (see the numerical results in the following section) that it will continue to be true for such one-plus currents $ | n^- \dots  (m\!+\!1)^- m^+ (m\!-\!1)^- \dots 3^- | $ irrespective of the position of the positive-helicity gluon with respect to the fermions.

In each case, our recursive argument is based on the following Berends-Giele
expansion~\cite{Berends:1987me}.
Consider constructing the $n$-particle current with one fermion line
by attaching the $n$-th gluon to the corresponding $(n\!-\!1)$-particle current.
Due to color ordering, it can be coupled directly to the off-shell quark,
to the $(n\!-\!1)$-th gluon, or to some more complicated gluon tree.
If we focus our attention on those gluon trees that include the $n$-th gluon
and attach to the quark line as a whole,
we can expand the current according to the number of legs in such trees,
as shown pictorially in \eqn{vertexrecursion}:
\beal
   \includegraphics[scale=1.0]{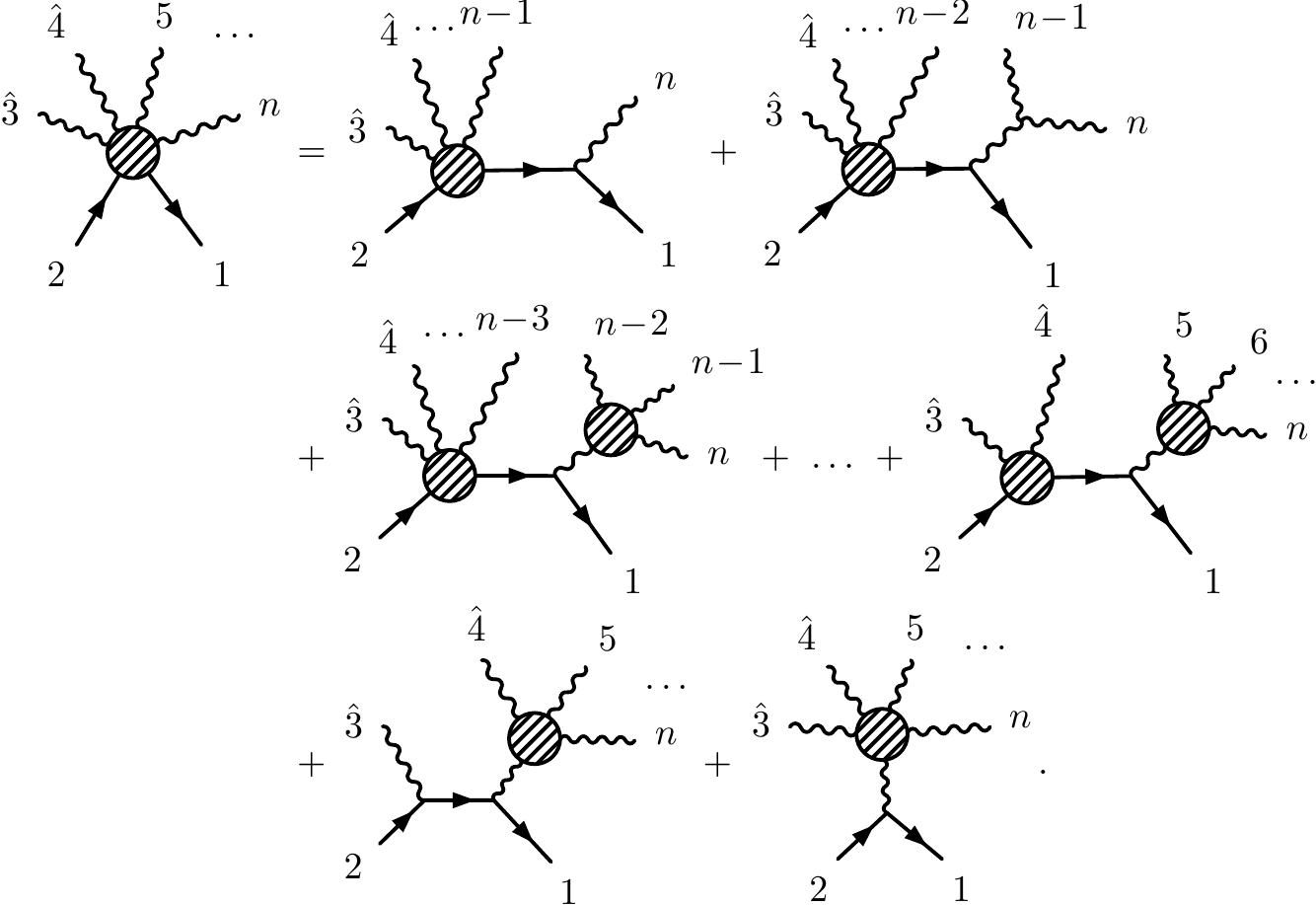}
\label{vertexrecursion}
\eeal

\subsection{All-minus currents}
\label{sec:allminus}

Let us prove that for the $ [ 3 4 \rangle $-shifted all-minus current $ iJ \big( 1_{\bar{Q}}^*, 2_Q^{}, \hat{3}_g^-, \hat{4}_g^-, 5_g^-, \dots, n_g^- \big) $ the residue at the unphysical pole $ z_3 = [n_3 3]/[n_3 4] $ vanishes when we make all the gluon reference momenta equal: $ n_3 = n_4 = \dots = n_n \equiv q $. 
      
Now if we already know that the residue at the unphysical pole $ z_3 = [q3]/[q4] $
vanishes for all the corresponding off-shell currents with fewer legs,
then only the last two diagrams in \eqn{vertexrecursion} remain to be calculated.
That can easily be done just by using color-ordered Feynman rules,
where we make use of the Berends-Giele formula \cite{Berends:1987me}
for currents of like-helicity gluons, which in our conventions is given by
\be
   iJ^{\mu} (1^-,2^-, \dots, n^-) =
      - \frac{[q|\gamma^{\mu} {\not}P_{1,n} |q]}
             {\sqrt{2} [q 1] [1 2] \dots [n\!-\!1~n] [n q] } .
\label{berendsgieleminus}
\ee
Evaluating the sum at the pole $z_3=0$, defined by $ [q \hat{3}] = 0 $, and
performing some manipulations using a Schouten identity between the two contributions,
we find the residue of the current~\eqref{vertexrecursion} at the unphysical pole:
\beal
      \left[q \hat{3}\right] \left\{
      \parbox{40mm}{ \begin{fmffile}{unphysallminus1}
      \fmfframe(20,12)(10,12){ \fmfsettings \begin{fmfgraph*}(75,60)
            \fmflabel{$1^*$}{q1}
            \fmflabel{$2$}{q2}
            \fmflabel{$\hat{3}^-$}{g3}
            \fmflabel{$\hat{4}^-$}{g4}
            \fmflabel{$5^-$}{g5}
            \fmflabel{$\dots$}{d}
            \fmflabel{$n^-$}{gn}
            \fmftop{,,g4,,g5,}
            \fmfbottom{q2,,q1,}
            \fmfleft{,,g3,}
            \fmfright{,,,gn,d,}
            \fmf{plain_arrow,tension=1.7}{q2,v2,v1}
            \fmf{plain_arrow,tension=1.2}{v1,q1}
            \fmf{wiggly,tension=1.2}{v1,c}
            \fmf{wiggly,tension=1.0}{g3,v2}
            \fmf{wiggly,tension=0.5}{g4,c}
            \fmf{wiggly,tension=0.5}{g5,c}
            \fmf{wiggly,tension=0.5}{gn,c}
            \fmfblob{0.20w}{c}
      \end{fmfgraph*} }
      \end{fmffile} }
      +
      \parbox{33mm}{ \begin{fmffile}{unphysallminus2}
      \fmfframe(15,12)(10,12){ \fmfsettings \begin{fmfgraph*}(60,60)
            \fmflabel{$1^*$}{q1}
            \fmflabel{$2$}{q2}
            \fmflabel{$\hat{3}^-$}{g3}
            \fmflabel{$\hat{4}^-$}{g4}
            \fmflabel{$5^-$}{g5}
            \fmflabel{$\dots$}{d}
            \fmflabel{$n^-$}{gn}
            \fmftop{,g4,,g5,d,}
            \fmfbottom{,q2,,q1,}
            \fmfleft{,,g3,}
            \fmfright{,,gn,}
            \fmf{plain_arrow,tension=1.7}{q2,v,q1}
            \fmf{wiggly,tension=2}{v,c}
            \fmf{wiggly}{g3,c}
            \fmf{wiggly}{g4,c}
            \fmf{wiggly}{g5,c}
            \fmf{wiggly}{gn,c}
            \fmfblob{0.25w}{c}
      \end{fmfgraph*} }
      \end{fmffile} }
      \right\} =
      \frac{ i |q] \langle 3|{\not}P_{3,n}|q] [q| ({\not}p_2 - m)  |2) }
           { \langle 3|2|q] [3 4] [4 5] \dots [n\!-\!1~n] [n q] } .
\label{minusn1}
\eeal
Here the right-hand side obviously vanishes
due to the presence of the on-shell spinor~$|2)$ next to $({\not}p_2 - m)$.

To conclude the proof, we do not even need to calculate the base of the recursion separately, because all the preceding formulas were general enough be valid for $  iJ \big( 1_{\bar{Q}}^*, 2_Q^{}, \hat{3}_g^-, \hat{4}_g^- \big) $ as well. Indeed, in that case the last two diagrams in \eqn{vertexrecursion} turn out to be the usual Feynman diagrams with the Berends-Giele current representing just the polarization vector of the shifted $4$th gluon.

\subsection{Currents with a single plus-helicity gluon in extreme position}
\label{sec:extremeposition}

For the $ [ 4 3 \rangle $-shifted current
$ iJ \big( 1_{\bar{Q}}^*, 2_Q^{}, \hat{3}_g^+, \hat{4}_g^-, 5_g^-, \dots, n_g^- \big) $,
we set $ n_4 = \dots = n_n = p_3 $ and rename $ n_3 \equiv q $.
Consider the same expansion \eqref{vertexrecursion}.
As in the previous case, for a recursive proof of the vanishing of the residue
at $ z_3 = - \langle q 3 \rangle / \langle q 4 \rangle $,
we only need to calculate the last two diagrams in \eqn{vertexrecursion}.
We use the following formula for the one-plus Berends-Giele current:
\be
   iJ^{\mu} (1^+,2^-, \dots, n^-) =
      - \frac{[1|\gamma^{\mu} {\not}P_{1,n} |1]}
             {\sqrt{2} [1 2] [2 3] \dots [n\!-\!1~n] [n 1] }
        \left\{ \sum_{l=3}^{n} \frac{ [1| {\not}P_{1,l} {\not}P_{1,l-1} |1] }
                                    { P_{1,l}^2 P_{1,l-1}^2 }
                             + \frac{ \braket{2q} }
                                    { \braket{21} \braket{1q} }
        \right\} ,
\label{berendsgieleplus1}
\ee
in which we retain dependence on the reference momentum $ n_1 \equiv q $
of the positive-helicity gluon.
This generalization is discussed in Appendix~B of \rcitePaper{1}.
It turns out that relaxing one reference momentum results in only one extra term
in \eqn{berendsgieleplus1}, which subsequently generates the pole for
$ i {\not}J ( \hat{3}^+, \hat{4}^-, \dots, n^-)$ at $ \langle \hat{3} q \rangle = 0 $.

Using the currents \eqref{berendsgieleminus} and \eqref{berendsgieleplus1},
we see again that the residue at $z_3$ vanishes due to the presence of the on-shell spinor $|2)$ next to $({\not}p_2 - m)$:
\beal
      \langle q \hat{3} \rangle \left\{
      \parbox{40mm}{ \begin{fmffile}{unphysextremeplus1}
      \fmfframe(20,12)(10,12){ \fmfsettings \begin{fmfgraph*}(75,60)
            \fmflabel{$1^*$}{q1}
            \fmflabel{$2$}{q2}
            \fmflabel{$\hat{3}^+$}{g3}
            \fmflabel{$\hat{4}^-$}{g4}
            \fmflabel{$5^-$}{g5}
            \fmflabel{$\dots$}{d}
            \fmflabel{$n^-$}{gn}
            \fmftop{,,g4,,g5,}
            \fmfbottom{q2,,q1,}
            \fmfleft{,,g3,}
            \fmfright{,,,gn,d,}
            \fmf{plain_arrow,tension=1.7}{q2,v2,v1}
            \fmf{plain_arrow,tension=1.2}{v1,q1}
            \fmf{wiggly,tension=1.2}{v1,c}
            \fmf{wiggly,tension=1.0}{g3,v2}
            \fmf{wiggly,tension=0.5}{g4,c}
            \fmf{wiggly,tension=0.5}{g5,c}
            \fmf{wiggly,tension=0.5}{gn,c}
            \fmfblob{0.20w}{c}
      \end{fmfgraph*} }
      \end{fmffile} }
      +
      \parbox{33mm}{ \begin{fmffile}{unphysextremeplus2}
      \fmfframe(15,12)(10,12){ \fmfsettings \begin{fmfgraph*}(60,60)
            \fmflabel{$1^*$}{q1}
            \fmflabel{$2$}{q2}
            \fmflabel{$\hat{3}^+$}{g3}
            \fmflabel{$\hat{4}^-$}{g4}
            \fmflabel{$5^-$}{g5}
            \fmflabel{$\dots$}{d}
            \fmflabel{$n^-$}{gn}
            \fmftop{,g4,,g5,d,}
            \fmfbottom{,q2,,q1,}
            \fmfleft{,,g3,}
            \fmfright{,,gn,}
            \fmf{plain_arrow,tension=1.7}{q2,v,q1}
            \fmf{wiggly,tension=2}{v,c}
            \fmf{wiggly}{g3,c}
            \fmf{wiggly}{g4,c}
            \fmf{wiggly}{g5,c}
            \fmf{wiggly}{gn,c}
            \fmfblob{0.25w}{c}
      \end{fmfgraph*} }
      \end{fmffile} }
      \right\} =
      \frac{-i |3] \braket{4q}  \langle q|{\not}P_{3,n}|3] [3| ({\not}p_2 - m)  |2) }
           { \langle q|2|3] \braket{43} [34] [\hat{4}5] \dots [n\!-\!1~n] [n3] } .
\label{plusn1}
\eeal

This evaluation is valid as well for $ iJ \big( 1_{\bar{Q}}^*, 2_Q^{}, \hat{3}_g^+, \hat{4}_g^- \big) $, thus ensuring the base of the recursive argument.

\subsection{Currents with a single plus-helicity gluon in next-to-extreme position}
\label{sec:nexttoextremeposition}

Here we consider a matrix-valued $[34 \rangle$-shifted current
$ iJ \big( 1_{\bar{Q}}^*, 2_Q^*, \hat{3}_g^-, \hat{4}_g^+, 5_g^-, \dots, n_g^- \big) $ with the positive-helicity gluon separated from the fermion line
by one negative-helicity gluon.
We set $ n_3 = n_5 = \dots = n_n = p_4 $ and $ n_4 \equiv q $.
In the expansion~\eqref{vertexrecursion}, we now need to examine the last three terms.
More specifically, we need to examine only their $q$-dependent parts,
since these are the ones that can have the unphysical pole at $\braket{q \hat{4}} = 0$.

The very last diagram in \eqn{vertexrecursion} contains the gluon current
$ iJ \big( \hat{3}^-, \hat{4}^+, 5^-, \dots, n^- \big) $,
for which we do not know a simple analytic formula.
Fortunately, according to the inductive argument outlined
in Appendix~B of \rcitePaper{1}, it does not depend on $q$,
so that diagram cannot contribute to the residue at $ \langle q \hat{4} \rangle  = 0 $.
We are thus left with the other two diagrams. 
As before,
\beal
      \parbox{33mm}{ \begin{fmffile}{unphysnexttoextremeplus1}
      \fmfframe(10,12)(10,12){ \fmfsettings \begin{fmfgraph*}(75,60)
            \fmflabel{$1^*$}{q1}
            \fmflabel{$2^*$}{q2}
            \fmflabel{$\hat{3}^-$}{g3}
            \fmflabel{$\hat{4}^+$}{g4}
            \fmflabel{$5^-$}{g5}
            \fmflabel{$\dots$}{d}
            \fmflabel{$n^-$}{gn}
            \fmftop{,,g4,,g5,}
            \fmfbottom{q2,,q1,}
            \fmfleft{,,g3,}
            \fmfright{,,,gn,d,}
            \fmf{plain_arrow,tension=1.7}{q2,v2,v1}
            \fmf{plain_arrow,tension=1.2}{v1,q1}
            \fmf{wiggly,tension=1.2}{v1,c}
            \fmf{wiggly,tension=1.0}{g3,v2}
            \fmf{wiggly,tension=0.5}{g4,c}
            \fmf{wiggly,tension=0.5}{g5,c}
            \fmf{wiggly,tension=0.5}{gn,c}
            \fmfblob{0.20w}{c}
      \end{fmfgraph*} }
      \end{fmffile} }
      =
      \frac{i}{\sqrt{2} [43]} i {\not}J (\hat{4}^+, 5^-, \dots, n^-)
      \frac{ {\not}p_2 - {\not}\widehat{p}_3 + m }{ (p_2 - \widehat{p}_3)^2 - m^2 }
      \Big( |4] \langle 3| + |3 \rangle [4| \Big) .
\label{plus43}
\eeal

We use the formula
\beal
            iJ \big( 1_{\bar{Q}}^*, 2_Q^*, 3_g^-, 4_g^+ \big)
            = \frac{i}{ [n_3 3] \langle n_4 4 \rangle } \Bigg\{
              \Big( |4] \langle n_4| \!+\! |n_4 \rangle [4| \Big)
              \frac{ {\not}p_2 \!-\! {\not}p_3 \!+\! m }{ (p_2 \!-\! p_3)^2 \!-\! m^2 }
            & \Big( |n_3] \langle 3| \!+\! |3 \rangle [n_3| \Big) \\
            + \frac{ [n_3 4] \langle n_4 3 \rangle }{ 2 [3 4] \langle 4 3 \rangle }
                                                       ({\not}p_3 \!-\! {\not}p_4)
            + \frac{ [n_3 4] }{ [3 4] } \Big( |4] \langle n_4| \!+\! |n_4 \rangle [4| \Big)
            - \frac{ \langle n_4 3 \rangle }{ \langle 4 3 \rangle }
                                      & \Big( |n_3] \langle 3| \!+\! |3 \rangle [n_3| \Big)
              \Bigg\} ,
\eeal
which is derived simply from Feynman rules, to evaluate the third-to-last diagram in expansion \eqref{vertexrecursion}:
\beal
      \parbox{40mm}{ \begin{fmffile}{unphysnexttoextremeplus2}
      \fmfframe(12,12)(12,0){ \fmfsettings \begin{fmfgraph*}(90,60)
            \fmflabel{$1^*$}{q1}
            \fmflabel{$2^*$}{q2}
            \fmflabel{$\hat{3}^-$}{g3}
            \fmflabel{$\hat{4}^+$}{g4}
            \fmflabel{$5^-$}{g5}
            \fmflabel{$6^-$}{g6}
            \fmflabel{$\dots$}{d}
            \fmflabel{$n^-$}{gn}
            \fmftop{,,,g4,,,g5,,g6,}
            \fmfbottom{q2,,,q1,}
            \fmfleft{,,g3,}
            \fmfright{,,,gn,d,}
            \fmf{plain_arrow,tension=2.3}{q2,c}
            \fmf{plain_arrow,tension=1.7}{c,v}
            \fmf{plain_arrow,tension=1.2}{v,q1}
            \fmf{wiggly,tension=0.75}{g3,c}
            \fmf{wiggly,tension=0.75}{g4,c}
            \fmf{wiggly,tension=2.0}{v,vg}
            \fmf{wiggly,tension=0.75}{g5,vg}
            \fmf{wiggly,tension=0.75}{g6,vg}
            \fmf{wiggly,tension=1.0}{gn,vg}
            \fmfblob{0.25h}{vg}
            \fmfblob{0.25h}{c}
      \end{fmfgraph*} }
      \end{fmffile} }
      = & -
      \frac{i}{\sqrt{2} \langle q \hat{4} \rangle [43]} i {\not}J (5^-, 6^-, \dots, n^-)
      \frac{ {\not}p_2 - {\not}p_3 - {\not}p_4 + m }{ (p_2 - p_3 - p_4)^2 - m^2 } \\ &
      \Bigg\{
            \Big( |4] \langle q| \!+\! |q \rangle [4| \Big)
            \frac{ {\not}p_2 - {\not}\widehat{p}_3 + m }{ (p_2 - \widehat{p}_3)^2 - m^2 }
          - \frac{ \langle q 3 \rangle }{ \langle 4 3 \rangle }
      \Bigg\}
      \Big( |4] \langle 3| + |3 \rangle [4| \Big) ,
\label{plus45}
\eeal

After using the formulas \eqref{berendsgieleminus} and \eqref{berendsgieleplus1}
for the Berends-Giele currents and substituting
\be
            |q \rangle = |\hat{4} \rangle \frac{\langle 3q \rangle}{\langle 34 \rangle} 
\label{g4plus}
\ee
in the residue of the unphysical pole,
we can combine the $q$-dependent terms of \eqref{plus43} and \eqref{plus45} into one term with the following spinor matrix in the middle:
\be
            \frac{ {\not}p_2 - {\not}\widehat{p}_3 - {\not}\widehat{p}_4 + m }
                 { (p_2 - \widehat{p}_3 - \widehat{p}_4)^2 - m^2 } ~ {\not}\widehat{p}_4 ~
            \frac{ {\not}p_2 - {\not}\widehat{p}_3 + m }
                 { (p_2 - \widehat{p}_3)^2 - m^2 } -
            \frac{ {\not}p_2 - {\not}\widehat{p}_3 - {\not}\widehat{p}_4 + m }
                 { (p_2 - \widehat{p}_3 - \widehat{p}_4)^2 - m^2 } +
            \frac{ {\not}p_2 - {\not}\widehat{p}_3 + m }
                 { (p_2 - \widehat{p}_3)^2 - m^2 } = 0 .
\label{g4plusmatrix}
\ee

Having established the induction, we turn back to $ iJ \big( 1_{\bar{Q}}^*, 2_Q^*, \hat{3}_g^-, \hat{4}_g^+, 5_g^- \big) $ and see that its expansion \eqref{vertexrecursion} contains precisely the three diagrams that we have just examined in a more general case. This provides the base of our inductive argument for $ iJ \big( 1_{\bar{Q}}^*, 2_Q^*, 3_g^-, 4_g^+, 5_g^-, \dots, n_g^- \big) $.

Our numerical results, discussed in \sec{sec:fermionnumerical},
indicate that the statement about the matrix-valued one-plus currents might be true
irrespective of the position of the positive-helicity gluon
as long as it is separated from the fermion by at least one negative-helicity gluon.
Unfortunately, it remains a challenge to show it.
Here we used an explicit formula for
$ iJ \big( 1_{\bar{Q}}^*, 2_Q^*, 3_g^-, 4_g^+ \big) $ to evaluate one of the diagrams.
To prove the vanishing of the unphysical pole for
$ iJ \big( 1_{\bar{Q}}^*, 2_Q^*, 3_g^-, \dots,
    (m\!-\!1)_g^-, m_g^+, (m\!+\!1)_g^-, \dots, n_g^- \big) $
in the same manner, one would need to have an explicit formula either for
$ iJ \big( 1_{\bar{Q}}^*, 2_Q^*, 3_g^-, \dots, (m\!-\!1)_g^-, m_g^+\big) $
or for
$ iJ \big( 1_{\bar{Q}}^*, 2_Q^*, m_g^+, (m\!+\!1)_g^-, \dots, n_g^- \big) $.

\section{Results for currents}
\label{sec:cresults}

In this section, we apply the constructions established in the previous section
to compute massive fermion currents from recursion relations.
First, we list 3- and 4-point currents as a starting point.
Next, we give a closed-form expression for currents
with an arbitrary number of gluons if their helicities are all alike.
For a fully non-recursive version and its derivation, we refer the reader to
Appendix~C of \rcitePaper{1}.
Finally, we state our numerical results for shifts producing recursion relations
in the case of one gluon of opposite helicity.

\subsection{3-point and 4-point currents}
\label{sec:3and4pt}

For completeness, we begin by listing the 3- and 4-point currents,
which are straightforward to derive from Feynman rules,
with full freedom of the choice of reference spinors.

\beal
   iJ \big( 1_{\bar{Q}}^*, 2_Q^*, 3_g^- \big) & = - i
      \frac{ |n_3] \langle 3| \!+\! |3 \rangle [n_3| }{ [n_3 3] } , \\
   iJ \big( 1_{\bar{Q}}^*, 2_Q^*, 3_g^+ \big) & = i
      \frac{ |3] \langle n_3| \!+\! |n_3 \rangle [3| }{ \langle n_3 3 \rangle } .
\label{iJ3}
\eeal

\beal
   iJ \big( 1_{\bar{Q}}^*, 2_Q,&3_g^-, 4_g^- \big)
      = - \frac{i}{ [n_3 3] [n_4 4] } \\ \times
          \Bigg\{\!\!&- \frac{1}{ \langle 3|2|3] }
                        \bigg[ |4 \rangle [n_4|2|3 \rangle [n_3|
                             - |n_4] \langle 4|1|n_3] \langle 3|
                           + m |4 \rangle [n_4 n_3] \langle 3|
                           + m |n_4] \langle 4 3 \rangle [n_3| \bigg] \\
                     &+ \frac{1}{[34]}
                        \bigg[ \frac{[n_3 n_4]}{2}({\not}p_3 \!-\! {\not}p_4)
                             + [n_3 4] \Big( |n_4] \langle 4|\!+\!|4 \rangle [n_4| \Big)
                             + [n_4 3] \Big( |n_3] \langle 3|\!+\!|3 \rangle [n_3| \Big)
                        \bigg]
          \Bigg\} |2) .
\label{iJ4mmq2}
\eeal

\beal
   iJ \big( 1_{\bar{Q}}^*, 2_Q^*, 3_g^-, 4_g^- \big)
      = - \frac{i}{ [n_3 3] [n_4 4] }
          \Bigg\{ \Big( |n_4] \langle 4| \!+\! |4 \rangle [n_4| \Big)
                  \frac{ {\not}p_2 \!-\! {\not}p_3 \!+\! m }{ (p_2\!-\!p_3)^2 \!-\! m^2 }
                & \Big( |n_3] \langle 3| \!+\! |3 \rangle [n_3| \Big) \\
                + \frac{1}{[34]}
                  \bigg[ \frac{[n_3 n_4]}{2}({\not}p_3 \!-\! {\not}p_4)
                       + [n_3 4] \Big( |n_4] \langle 4|\!+\!|4 \rangle [n_4| \Big)
                       + [n_4 3]&\Big( |n_3] \langle 3|\!+\!|3 \rangle [n_3| \Big)
                  \bigg]
          \Bigg\} ,
\label{iJ4mm}
\eeal
\beal
            iJ \big( 1_{\bar{Q}}^*, 2_Q^*, 3_g^-, 4_g^+ \big)
            = \frac{i}{ [n_3 3] \langle n_4 4 \rangle } \Bigg\{
              \Big( |4] \langle n_4| \!+\! |n_4 \rangle [4| \Big)
              \frac{ {\not}p_2 \!-\! {\not}p_3 \!+\! m }{ (p_2 \!-\! p_3)^2 \!-\! m^2 }
            & \Big( |n_3] \langle 3| \!+\! |3 \rangle [n_3| \Big) \\
            + \frac{ [n_3 4] \langle n_4 3 \rangle }{ 2 [3 4] \langle 4 3 \rangle }
                                                       ({\not}p_3 \!-\! {\not}p_4)
            + \frac{ [n_3 4] }{ [3 4] } \Big( |4] \langle n_4| \!+\! |n_4 \rangle [4| \Big)
            - \frac{ \langle n_4 3 \rangle }{ \langle 4 3 \rangle }
                                      & \Big( |n_3] \langle 3| \!+\! |3 \rangle [n_3| \Big)
              \Bigg\} ,
\eeal
\beal
            iJ \big( 1_{\bar{Q}}^*, 2_Q^*, 3_g^+, 4_g^- \big)
            = \frac{i}{ \langle n_3 3 \rangle [n_4 4] } \Bigg\{
              \Big( |n_4] \langle 4| \!+\! |4 \rangle [n_4| \Big)
              \frac{ {\not}p_2 \!-\! {\not}p_3 \!+\! m }{ (p_2 \!-\! p_3)^2 \!-\! m^2 }
            & \Big( |3] \langle n_3| \!+\! |n_3 \rangle [3| \Big) \\
            + \frac{ \langle n_3 4 \rangle [n_4 3] }{ 2 \langle 3 4 \rangle [4 3] }
                                                       ({\not}p_3 \!-\! {\not}p_4)
            + \frac{ \langle n_3 4 \rangle }{ \langle 3 4 \rangle }
                                        \Big( |n_4] \langle 4| \!+\! |4 \rangle [n_4| \Big)
            - \frac{ [n_4 3] }{ [4 3] }&\Big( |3] \langle n_3| \!+\! |n_3 \rangle [3| \Big)
              \Bigg\} .
\label{iJ4pm}
\eeal

\subsection{Closed form for all-minus currents}
\label{sec:allminussolution}

      \begin{figure}[h]
      \centering
      \includegraphics[scale=1.0]{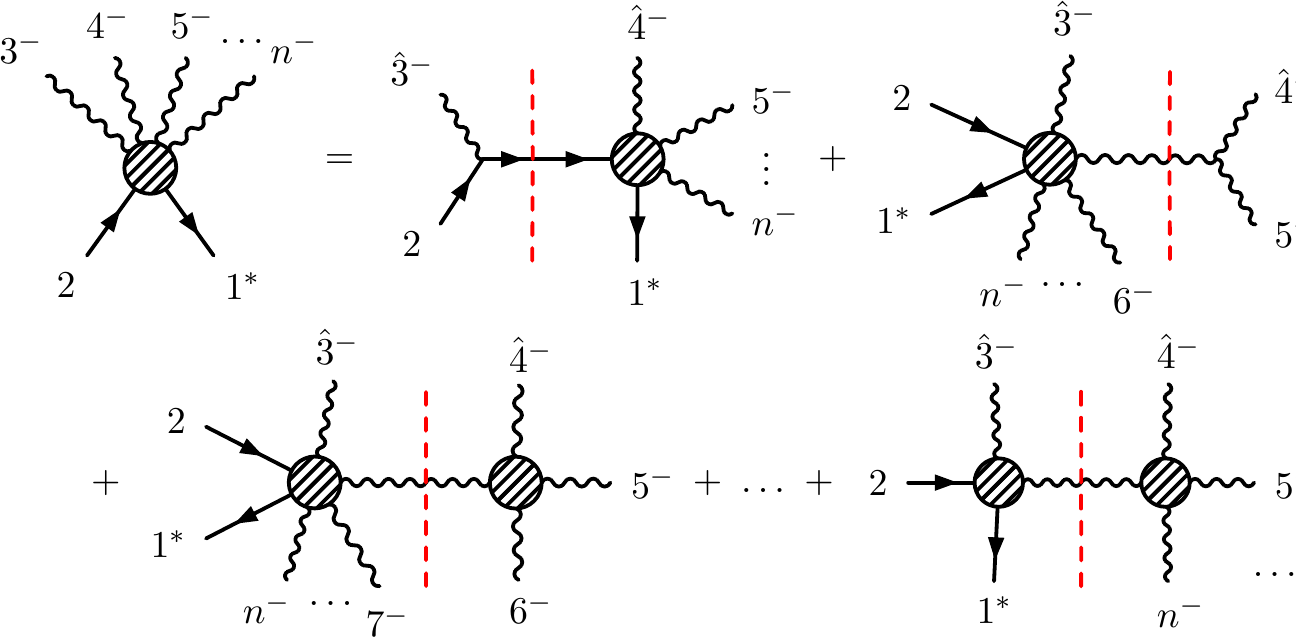}
      \vspace{-5pt}
      \caption{\small BCFW derivation of
               $iJ \big( 1_{\bar{Q}}^*, 2_Q^{}, 3_g^-, 4_g^-,\ldots, n_g^- \big)$.
               \label{fig:nmmmm}}
      \end{figure}

For $n \geq 5$, we compute the all-minus current $iJ \big( 1_{\bar{Q}}^*, 2_Q^{}, 3_g^-, 4_g^-,\ldots, n_g^- \big)$ by doing a $ [34 \rangle $ shift and setting all reference momenta equal to an arbitrary null vector $q$.
Since we have established the absence of boundary terms and unphysical poles
in the preceding sections, the BCFW expansion is given as follows:
\small
\beal
\label{bcfw-allminus} \!\!\!\!\!
iJ \big( 1_{\bar{Q}}^*, 2_Q, \hat{3}_g^-, \hat{4}_g^-,\ldots, n_g^- \big) =
iJ \big( 1_{\bar{Q}}^*, (2\!-\!\hat{3})_Q, \hat{4}_g^-,5_g^-,\ldots, n_g^- \big)
& \frac{ i ( {\not}p_2\!-\!{\not}\widehat{p}_3\!+\!m ) }
       { (p_2\!-\!{p}_3)^2\!-\!m^2 }
A \big(\!-\!(2\!-\!\hat{3})_{\bar{Q}} , 2_Q , \hat{3}_g^- \big) \\
+ \sum_{(h,\tilde{h})} \bigg[
iJ \big( 1_{\bar{Q}}^*, 2_Q, \hat{3}_g^-, (\hat{4}\!+\!5)_g^h,6_g^-,\ldots, n_g^- \big)
& \frac{i}{ (p_4\!+\!p_5)^2 }
A \big(\!-\!(\hat{4}\!+\!5)_g^{\tilde{h}}, \hat{4}_g^-, 5_g^- \big) \\
+ \sum_{k=6}^n
iJ \big( 1_{\bar{Q}}^*, 2_Q, \hat{3}_g^-, (\hat{P}_{4,k})_g^h, (k\!+\!1)_g^-,
\ldots, n_g^- \big)
& \frac{i}{ P_{4,k}^2 }
A \big(\!-\!(\hat{P}_{4,k})_g^{\tilde{h}}, \hat{4}_g^-, 5_g^- ,\ldots,k_g^-\big)
\bigg] .
\eeal
\normalsize
See \fig{fig:nmmmm}. 
Because we are working with off-shell currents, 
the sum over intermediate gluon polarization states $(h,\tilde{h})$
must now include the unphysical polarization state combinations $(L,T), (T,L)$,
which vanished automatically in the on-shell case due to the Ward identity.
This subtlety was first treated in a similar context in \rcite{Feng:2011twa}.
Specifically, the numerator of the Feynman propagator is decomposed as 
\be
   -g_{\mu \nu} = \varepsilon_{\mu}^+ \varepsilon_{\nu}^-
                + \varepsilon_{\mu}^- \varepsilon_{\nu}^+
                + \varepsilon_{\mu}^L \varepsilon_{\nu}^T
                + \varepsilon_{\mu}^T \varepsilon_{\nu}^L ,
\label{metric}
\ee
where 
\be
            \varepsilon_{\mu}^L = k_{\mu} , ~~~~~
            \varepsilon_{\mu}^T = - \frac{ q_{\mu} }{(k\!\cdot\!q)} .
\label{polvectorsextra}
\ee
	
It is clear that the gluon amplitude in the third line of \eqn{bcfw-allminus}
(the second line of \fig{fig:nmmmm}) vanishes identically,
due to the form of the all-minus Berends-Giele current~\eqref{berendsgieleminus}
contracted with any of the polarization vectors.
The three-point gluon amplitude in the second line of \eqn{bcfw-allminus}
(the last diagram of the first line of \fig{fig:nmmmm}) vanishes as well.
Therefore, the only contribution that is left is the single term in the first line,
involving a fermionic propagator.
The general expression for an $n$-point all-minus current can then be written as
\beal
   | n^- (n\!-\!1)^- \dots 4^- 3^- | 2 ) = \,
    & \frac{-i}
           { [q \hat{3}] [q \hat{4}] \dots [q ~\widehat{n\!-\!2}] [q ~n\!-\!1] [q n] } \\
      \times \Bigg\{
      \Big( |q] \langle n| \!+\! |n \rangle [q| \Big)
    & \frac{ {\not}p_2 \!-\! {\not}\widehat{P}_{3,n\!-\!1} \!+\! m }
           { (p_2 \!-\! P_{3,n\!-\!1})^2 \!-\! m^2 }
      \Big( |q] \langle \widehat{n\!-\!1}| \!+\! |\widehat{n\!-\!1} \rangle [q| \Big) \\
    + \frac{1}{[n\!-\!1~ n]}
      \bigg[ [q n] \Big( &|q] \langle n| + |n \rangle [q| \Big)
           + [q~n\!-\!1] \Big( |q] \langle \widehat{n\!-\!1}|
                                    \!+\! |\widehat{n\!-\!1} \rangle [q| \Big)
      \bigg] \Bigg\} \\ \times
    & \frac{ {\not}p_2 \!-\! {\not}\widehat{P}_{3,n\!-\!2} \!+\! m }
           { (p_2 \!-\! P_{3,n\!-\!2})^2 \!-\! m^2 }
      \Big( |q] \langle \widehat{n\!-\!2}| \!+\! |\widehat{n\!-\!2} \rangle [q| \Big)
      \times \cdots \\ \times
    & \frac{ {\not}p_2 \!-\! {\not}\widehat{P}_{3,4} \!+\! m }
           { (p_2 \!-\! P_{3,4})^2 \!-\! m^2 }
      \Big( |q] \langle \hat{4}| \!+\! |\hat{4} \rangle [q| \Big) \\ \times
    & \frac{ {\not}p_2 \!-\! {\not}\widehat{p}_3 \!+\! m }
           { (p_2 \!-\! p_3)^2 \!-\! m^2 }
      \Big( |q] \langle 3| \!+\! |3 \rangle [q| \Big) |2) ,
\label{iJminus}
\eeal
where the shifted momenta are defined recursively by
\be
   \left\{ \begin{aligned}
      z_k & = \frac{\langle \hat{k}|{\not}p_2\!-\!{\not}\widehat{P}_{3,k\!-\!1}|k] }
                   {\langle \hat{k}|{\not}p_2\!-\!{\not}\widehat{P}_{3,k\!-\!1}|k\!+\!1]} \\
      |\hat{k}] & = |k] - z_k |k\!+\!1] \\
      |\widehat{k\!+\!1} \rangle & = |k\!+\!1 \rangle + z_k |\hat{k} \rangle \\
      {\not}\widehat{P}_{3,k} & \equiv {\not} P_{3,k} - z_k
            \Big( |k\!+\!1] \langle \hat{k}| \!+\! |\hat{k} \rangle [k\!+\!1| \Big) ,
	\end{aligned} \right.
\label{shiftk}
\ee
with $ k = 3, 4, \dots, n - 2 $, and the initial values
\be
   z_2 = 0, \qquad  z_3 = \frac{\langle 3|2|3]}{\langle 3|2|4]}.
\label{z-init}
\ee

We have verified this formula numerically through $n=6$
by comparison with sums of Feynman diagrams.

The massless version ($m=0$) was found in \rcite{Berends:1987me}
for one helicity choice of the on-shell spinor,
namely $iJ \big( 1_{\bar{Q}}^*, 2_Q^{+}, 3_g^-,  \ldots,  n_g^- \big)$
in our reversed fermion momentum convention, so that $|2)=\ket{2}$.
In our calculation, rather than take the massless limit of \eqn{iJminus},
it would be more effective to return to the recursion relation as given in the nonvanishing first line of \eqn{bcfw-allminus},
so that the propagators can be replaced by simple spinor products
at each step of the recursion.
Recovering the compact form of \rcite{Berends:1987me} is not immediate for general $n$,
however, because we preserve a form of the current
in which the quark spinor $|2)$ is an explicit factor at the right of the expression,
free to take either helicity value.
This is important, because the shift of the quark momentum means that
the full internal helicity sum occurs at each stage of our recursion.
 
It is possible to solve the recursion~\eqref{shiftk} exactly
and write the shifted spinors for \eqn{iJminus} in a fully closed form.
For this non-recursive form and the outline of its derivation,
we refer the reader to Appendix~C of \rcitePaper{1}.

\subsection{Numerical results}
\label{sec:fermionnumerical}

Beyond the case of all gluons having the same helicity, we have found valid shifts numerically through $n=6$ in the case of one gluon of opposite helicity to the others (the ``one-plus'' case or its parity conjugate). A sufficient condition for a valid shift is to take the reference momenta of all the negative-helicity gluons equal to the momentum of the positive-helicity gluon: $n_- = p_+$.

For the choice of shifted gluons, we have identified two valid possibilities:
\begin{itemize}
      \item Shift the two gluons closest to the on-shell fermion;
if they both have negative helicities, choose the shift so that
the unphysical pole would come from the gluon adjacent to the on-shell fermion.
(These shifts are all valid in the all-minus case as well.)
      \item In the case with the plus-helicity gluon in central position,
shift the plus-gluon along with the any of the adjacent minus-gluons
irrespective of their position with respect to the fermions.
The unphysical poles then vanish, even with both fermions off-shell.
\end{itemize}

To be more precise, we found that for
$iJ \big( 1_{\bar{Q}}^*, 2_Q^*, 3_g^-, 4_g^+, 5_g^-, 6_g^- \big)$
in gauge $ n_3 = n_5 = n_6 = p_4 $ not only $[34\rangle$ shift produces no unphysical poles (as we have proved in Section 4.3), but $[54\rangle$ as well. Similarly,
$iJ \big( 1_{\bar{Q}}^*, 2_Q^*, 3_g^-, 4_g^-, 5_g^+, 6_g^- \big)$
in gauge $ n_3 = n_4 = n_6 = p_5 $ suffers from no unphysical poles both under $[65\rangle$ and $[45\rangle$ shifts.

In the 6-point case, we also have currents with two plus and two minus helicities,
but unfortunately we were unable to find a good gauge choice for them.

\section{Discussion}
\label{sec:treeleveldiscussion}

In this chapter, we have studied currents of $n-2$ gluons
of ``mostly-minus'' helicity and a massive quark-antiquark pair,
where the antiquark is off shell.  
Because of the off-shellness of the antiquark,
the choice of reference spinors plays an important role.
 
BCFW-type recursion relations are obtained under the following conditions,
which ensure the absence of a boundary term and unphysical poles.
The reference spinors of the negative-helicity gluons are all chosen to be equal.
If there is a single positive-helicity gluon,
its momentum is taken to be the reference spinor of the negative-helicity gluons.  
\begin{itemize}
\item In the case where all gluons have negative helicity, we have obtained both a recursive and a closed form for the current derived from recursion relations.
\item In the case where one gluon has positive helicity, and it is color-adjacent to the quark or antiquark, we have proven the validity of the recursion relation, but we do not have a closed form.
\item In the case where one gluon has positive helicity, and it is color-adjacent to two other gluons, we have found numerical evidence for the validity of the recursion relation in general, but were able to prove it only for the simplest configuration, with the positive gluon in next-to-extreme position.
\end{itemize}

In Yang-Mills theory, an on-shell alternative to the BCFW construction is the MHV diagram expansion \cite{Cachazo:2004kj},
in which maximally-helicity-violating (MHV) amplitudes play the role of interaction vertices, with a suitable on-shell prescription for the intermediate legs.
For off-shell currents, there is apparently no sensible expansion in MHV diagrams when the off-shell leg carries color charge, such as the Berends-Giele currents for gluons.

One might consider applying a BCFW shift to the massive fermion pair, but this construction fails off-shell.  
With a conventional definition of massive spinors \cite{Kleiss:1985yh,Schwinn:2005pi} in terms of a single reference vector, good boundary behavior is evident, but there are unavoidable, complicated unphysical poles, due to $\sqrt{z}$-dependence of the denominators of the massive spinors.
Even with both fermions on shell, the only shift known to be valid is quite specialized:  
each of the two massive fermion spinors has its reference vector constructed in terms of the other \cite{Boels:2011zz}.
This choice is not well suited for repeated application in an analytic recursion relation, because it is undesirable to keep track of the data of internal legs.  One would like the choice of reference spinor to be fixed once for all. Nevertheless, we looked at extending this construction off-shell. There is no $z$-dependence in the denominators, but when either of the on-shell massive spinors are stripped off, the miraculous cancellation reducing the boundary behavior from $O(1)$ to $O(1/z)$ no longer takes place.

In the course of studying boundary behavior in \sec{sec:boundary}, we have proven
the good boundary behavior of general off-shell objects in Feynman gauge,
as long as they contain at least two on-shell gluons that can be shifted.
This meshes with the argument of \rcite{Boels:2011mn}
in the light-cone gauge $q\!\cdot\!A=0$ specified by the BCFW-shift vector $q$
given in \eqn{q}.
Another analysis~\cite{Zhang:2013cha}, based on reduced Yang-Mills vertices,
found similar results as well.
Thus we could see that it is not the boundary behavior that hinders the BCFW recursion off shell, but the unphysical poles, coming from the polarization vectors.

Several questions arise for future exploration. Is there any choice of shift and reference spinors that eliminates boundary terms and unphysical poles for more general helicity configurations?  If so, can the recursion relation be solved neatly?  Do some shifts give more compact results than others?  Is there a neat solution for the current with a single gluon of opposite helicity, for which we have already proved the existence of recursion relations? In cases where unphysical poles are present: is there any way to understand them, so that their residues could be incorporated explicitly in the recursion relation?  Further results addressing these questions would certainly illuminate our understanding of the BCFW construction and its applicability to gauge-dependent objects.

\chapter{Gauge theory at one loop}
\label{chap:oneloop}

Recent developments in understanding scattering amplitudes
have lead to impressive achievements
in taming gauge theory amplitudes analytically
for increasing and in some cases arbitrary number of particles.
Table \ref{tab:known} provides a short summary of existing one-loop results
as of July 2014.
In it, ``maximally-helicity-violating'' (MHV) conventionally stands
for amplitudes with two minus-helicity gluons, whereas
the next-to-maximally-helicity-violating (NMHV) case
corresponds to three negative helicities.
In addition to that, general split-helicity color-ordered amplitudes
in $\mathcal{N}=1$ SYM are known as well due to their simple analytic behavior
which permits an elegant one-loop BCFW recursion~\cite{Bern:2005hh}.

\begin{table}[h]
\centering
    \begin{tabular}{| l | l | l | l |}
    \hline
    &
    $\mathcal{N}=4$ SYM & $\mathcal{N}=1$ SYM & QCD \\ \hline
    MHV &
    $n$-point in 1994 \cite{Bern:1994zx} &
    $n$-point in 1994 \cite{Bern:1994cg} &
    $n$-point in 2006 \cite{Bedford:2004nh,Forde:2005hh,Berger:2006vq} \\ \hline
    NMHV &
    $n$-point in 2004 \cite{Bern:2004bt} &
      6-point in 2005 \cite{Britto:2005ha}, &
      6-point in 2006 \cite{Britto:2006sj,Xiao:2006vt} \\
    & &
    7-point in 2009 \cite{BjerrumBohr:2007vu,Dunbar:2009ax} & \\
    & &
    \textbf{\textit{n}-point} in 2013 \citePaper{2} & \\ \hline
    \end{tabular}
\caption{\small Known analytic results for gluon amplitudes at one loop in gauge theories
         with and without supersymmetry, including the result of \citePaper{2}}
\label{tab:known} \end{table}

In this chapter, we discuss the completion of the lower middle cell
of table \ref{tab:known} with $n$-point analytic results.
To do that, we use spinor integration~\cite{Britto:2005ha,Anastasiou:2006jv}
which provides a sleek way to compute amplitude coefficients
of one-loop master integrals from unitarity cuts in a purely algebraic manner.
We briefly review its idea and recipes in \sec{sec:method}.
In fact, our bubble coefficient formula has a novel form with respect to those found in the literature~\cite{Britto:2006fc,Britto:2007tt,Britto:2008sw,Britto:2010xq},
so for completeness we provide a streamlined rederivation
of the spinor integration formalism in \sec{sec:spinorintderivation}.
Moreover, later in \sec{sec:simplifiedbubble},
we slightly adapt the bubble coefficient formula
to make full use of $\mathcal{N}=1$ supersymmetry.

Intuitively, the main difficulty in finding universal NMHV formulas is that
even at 7 points general patterns are not yet obvious,
because the numbers of minus and plus helicities are still comparable to each other,
whereas MHV amplitudes become ``saturated'' by positive helicities already for 6 external gluons.
So in \sec{sec:cutconstruction}, we construct a double cut
for an arbitrary multiplicity from the start,
for which we use the tree input from \rcite{Dixon:2010ik}.
Next, in \sec{sec:loopmomentum}, we carefully analyze
how the cut depends on loop momentum variables,
which is essential for getting to the master integral coefficients.

For the explicit formulas for the bubble and box coefficients,
we refer the reader to Sections~6 and~7 of \rcitePaper{2}.
We note that their further use is facilitated by their Mathematica implementation,
freely distributed along with it (see Appendix F therein for a brief description).
To verify our results, we performed a number of non-trivial checks,
which are summarized in \sec{sec:checks}.

We hope that our all-multiplicity results will provide a helpful testing ground
for further theoretic developments.
For instance, it is an interesting question whether any kind of on-shell recursion
relations can be established between the coefficients we have found.
We only took a quick peek into this, as is mentioned in \sec{sec:oneloopdiscussion}.

\section{Method: spinor integration}
\label{sec:method}

Scattering amplitudes in four dimensions are known to be reducible
\cite{Brown:1952eu,Melrose:1965kb,'tHooft:1978xw,Passarino:1978jh,vanNeerven:1983vr}
to the following basis of master integrals:
\be
   A^{\text{1-loop}} = \mu^{2\epsilon} \Big(
                       \sum C^{\text{box}} \, I_4
                     + \sum C^{\text{tri}} \, I_3
                     + \sum C^{\text{bub}} \, I_2 + R \Big) ,
\label{integralexpansion}
\ee
where the sums go over all distinct scalar integrals
and $R$ is the purely-rational part.
In this thesis, we adopt the conventional definition~\cite{Bern:1992em,Bern:1993kr}
for dimensionally-regularized massless scalar integrals:
\be
      I_n = (-1)^{n+1}(4\pi)^{\frac{d}{2}}i
            \int \frac{d^d \ell_1}{(2\pi)^d}
            \frac{ 1 }
                 { \ell_1^2 (\ell_1-K_1)^2 \dots (\ell_1-K_{n-1})^2 } ,
\label{integrals}
\ee
where $d = 4-2\epsilon$.
Due to the normalization, all the coefficient formulas we provide further
contain trivial prefactors $(4\pi)^{-d/2}$.
Analytic expressions for these integrals can be found in \rcite{Bern:1993kr}.

Since all one-loop integrals are well known,
the problem of finding the amplitude is equivalent
to finding their coefficients and rational part $R$.
Rather than try to construct one-loop amplitudes from Feynman rules,
it is much easier to construct their unitarity cuts from known tree amplitudes.
If we denote the unitarity cut operator by $\delta_2$, we obtain
\be
   \delta_2 A^{\text{1-loop}} = \sum C^{\text{box}} \, \delta_2 I_4
                              + \sum C^{\text{tri}} \, \delta_2 I_3
                              + \sum C^{\text{bub}} \, \delta_2 I_2 .
\label{cutintegralexpansion}
\ee
Of course, each cut is sensitive only to those coefficients
whose integrals' cuts do not vanish,
but by going through different cuts one can obtain all coefficients.
As is evident from \eqn{cutintegralexpansion},
$R$ is the only term in \eqn{integralexpansion}
completely invisible to four-dimensional cuts.
This can be cured by considering $d$-dimensional cuts,
(see, for instance, \rcites{Brandhuber:2005jw,Anastasiou:2006jv,Anastasiou:2006gt}).
However, $d$-dimensional tree amplitudes are much less available in the literature
than their four-dimensional versions, so other methods~\cite{Bern:2005hs,Bern:2005cq,Berger:2006ci,Berger:2006vq,Xiao:2006vt} are often preferable.
Fortunately,
amplitudes in supersymmetric theories have a vanishing rational part~\cite{Bern:1994cg},
so the four-dimensional cut methods are sufficient for the purposes of this thesis.

In the rest of this section,
we concentrate on the particular method of spinor integration
in four dimensions~\cite{Britto:2005ha,Britto:2006sj,Anastasiou:2006jv,Britto:2006fc,Britto:2007tt,Britto:2008sw,
Mastrolia:2009dr}.
We will summarize its ideas, write down the explicit formulas
for the coefficients of the master integrals,
and then give its streamlined derivation.

\subsection{General idea}
\label{sec:spinorintegrationsummary}

We start by constructing the standard unitarity cut, the double cut,
from two tree amplitudes.
Here, for simplicity, we write the four-dimensional $K$-channel cut
without any prefactors:
\be
   \text{Cut} = \sum_{h_1,h_2}
                \int \! d^4\ell_1 \, \delta(\ell_1^2) \delta(\ell_2^2)
                A^{\text{tree}}(-\ell_1^{\bar{h}_1},\dots,\ell_2^{h_2})
                A^{\text{tree}}(-\ell_2^{\bar{h}_2},\dots,\ell_1^{h_1}) .
\label{cut}
\ee
The most important step is then
to trade the constrained loop variables $\ell_1$ and $\ell_2=\ell_1-K$
for homogeneous spinor variables $\la$ and $\lb$ such that
\begin{subequations} \begin{align}
            \ell_1^{\mu} & =  \frac{K^2}{2}
                           \frac{ \bra{\la} \gamma^{\mu}|\lb] }
                                { \bra{\la}K|\lb] } , \\
            \ell_2^{\mu} & = -\frac{1}{2}
                           \frac{ \bra{\la}K|\gamma^{\mu}|K|\lb] }
                                { \bra{\la}K|\lb] } .
\end{align} \label{lvariables} \end{subequations}
\!\!\!\!\;The integration measure transforms as follows:
\be
            \int \! d^4\ell_1 \, \delta(\ell_1^2) \delta(\ell_2^2)
                      = -\frac{K^2}{4} \int_{\tilde{\la}=\bar{\la}} \!
                         \frac{ \braket{\la d\la} [\lb d\lb] }
                              { \bra{\la}K|\lb]^2 } .
\label{dLIPS4}
\ee

If one then expands these homogeneous variables in arbitrary basis spinors:
\be
            \la = \la_{p} + z \la_{q} , \quad
            \lb = \lb_{p} + \bar{z} \lb_{q} ,
\label{complexvariables}
\ee
then the connection to the integral over the complex plane becomes evident:
\be
            \int_{\tilde{\la}=\bar{\la}} \! \braket{\la d\la} [\lb d\lb]
                      = -(p+q)^2 \! \int \! dz \wedge d\bar{z} .
\label{complexmeasure}
\ee
So the phase space spinor integration can be treated as
a complex plane integration in disguise. 
In this spinorial language, it is possible to define
simple and self-consistent rules for taking residues.
For instance, we calculate the residue of simple pole $\braket{\zeta|\la}$ as follows:
\be
      \Res_{\lambda=\zeta} \; \frac{F(\la,\lb)}{\braket{\zeta|\la}}
            = F(\zeta,\tilde{\zeta}) .
\label{simplepolerule}
\ee
The full set of rules is given in detail in \app{app:residues}.

In essence, the method of \emph{spinor integration} uses
a spinorial version of Cauchy's integral theorem \cite{Mastrolia:2009dr}
to actually perform that complex plane integration
in a manner which exposes coefficients of different scalar integrals.

In short, once we rewrite the cut \eqref{cut} using homogeneous spinor variables
\be
            \text{Cut} = \int_{\tilde{\la}=\bar{\la}} \!
                         \braket{\la d\la} [\lb d\lb] \;
                         \mathcal{I}_{\text{spinor}} ,
\label{cutspinor}
\ee
integral coefficients are given by \emph{general algebraic formulas}
which are given below.
To write them we only need to introduce a short notation for the following vectors:
\be
            Q_i^{\mu}(K_i,K) = -K_i^{\mu}+ \frac{K_i^2}{K^2} K^{\mu} .
\label{Qi4dim}
\ee
These arise naturally because
all loop-dependent physical poles come from propagators
which can be rewritten in homogeneous variables as
\beal
            (\ell_1 - K_i)^2 = K^2 \frac{ \bra{\la}Q_i|\lb] }
                                     { \bra{\la}K|\lb] } .
\label{propagators4dim}
\eeal

\subsection{Box coefficient}
\label{sec:box}

The coefficient of the scalar box labeled by two uncut propagators $i$ and $j$
can be expressed as
\be
            C_{ij}^{\text{box}} =-\frac{2 K^2}{(4\pi)^{\frac{d}{2}}i}
                                  \mathcal{I}_{\text{spinor}}
                                  \bra{\la}Q_i|\lb] \bra{\la}Q_j|\lb]
                                  \bigg\{ \bigg|_{\substack{
                                             \la = \la^{ij}_+ \\
                                             \lb = \lb^{ij}_-
                                             }}
                                          +
                                          \bigg|_{\substack{
                                             \la = \la^{ij}_- \\
                                             \lb = \lb^{ij}_+
                                             }} 
                                  \bigg\} ,
\label{Cbox}
\ee
where spinors $\la = \la^{ij}_\pm$ and $\lb = \lb^{ij}_\pm$ correspond to
on-shell combinations of propagator momenta:
\be
            P^{ij}_{\pm}(K_i,K_j,K) = Q_i + x^{ij}_{\pm} Q_j ,
\label{Pbox}
\ee
\be
   x^{ij}_{\pm} = \frac{-(Q_i\!\cdot\!Q_j)
              \pm \sqrt{ (Q_i\!\cdot\!Q_j)^2 - Q_i^2 Q_j^2 } }{Q_j^2} .
\label{Xbox}
\ee

It is easy to see that these formulae are equivalent
to the well-understood quadruple cut method \cite{Britto:2004nc,Kosower:2011ty}.
Indeed, the sole purpose of factors $\bra{\la}Q_i|\lb]$ and $\bra{\la}Q_j|\lb]$
in \eqn{Cbox} is just to cancel the corresponding propagator factors
in the denominator of $\mathcal{I}_{\text{spinor}}$.
Now, by definition
\be
            \bra{\la^{ij}_\pm}Q_i|\lb^{ij}_\mp] =
            \bra{\la^{ij}_\pm}Q_j|\lb^{ij}_\mp] = 0 ,
\label{quadruplecut}
\ee
so formula \eqref{Cbox} effectively puts propagators $i$ and $j$ on shell,
thus converting the original double cut into a quadruple cut
and summing over the two solutions.

\subsection{Triangle coefficient}
\label{sec:triangle}

The coefficient of the scalar triangle labeled by one uncut propagator $i$
can be found to be equal to
\beal
   C_i^{\text{tri}} = &\,\frac{2}{(4\pi)^{\frac{d}{2}}i}
                         \frac{1}{(K^2 (x^i_+-x^i_-)^2)^{n-k+1}} \\ \times
                      &\,\frac{1}{(n\!-\!k\!+\!1)!}
                               \frac{\mathrm{d}^{(n-k+1)}}{\mathrm{d}t^{(n-k+1)}}
                               \mathcal{I}_{\text{spinor}}
                               \bra{\la}Q_i|\lb] \bra{\la}K|\lb]^{n-k+2}
                               \bigg\{ \bigg|_{\substack{
                                          \la = \la^i_+ - t \la^i_- \\
                                          \lb = x^i_+ \lb^i_- - t x^i_- \lb^i_+
                                          }}\!\!
                                          +
                                          \bigg|_{\substack{
                                          \la = \la^i_- - t \la^i_+ \\
                                          \lb = x^i_- \lb^i_+ - t x^i_+ \lb^i_-
                                          }}\!
                               \bigg\} \bigg|_{t=0} ,
\label{Ctri}
\eeal
where spinors $\la = \la^{i}_\pm$ and $\lb = \lb^{i}_\pm$ correspond to
the following on-shell momenta:
\be
            P^{i}_{\pm}(K_i,K) = Q_i + x^{i}_{\pm} K ,
\label{Ptri}
\ee
\be
   x^{i}_{\pm} = \frac{-(K\!\cdot Q_i)
             \pm \sqrt{ (K\!\cdot Q_i)^2 - K^2 Q_i^2 } }{K^2} .
\label{Xtri}
\ee
Here and below in this section,
$(n-k)$ is the difference between the numbers of $\la$-factors
in the numerator and the denominator of $\mathcal{I}_{\text{spinor}}$,
excluding the homogeneity-restoring factor $\bra{\la}K|\lb]^{n-k+2}$.

\subsection{Bubble coefficient}
\label{sec:bubble}

Finally, we find the coefficient of the $K$-channel scalar bubble
through the following general formula:
\small
\be
   C^{\text{bub}}\! = \!\frac{4}{\!(4\pi)^{\frac{d}{2}}i}\!
      \sum_{\text{residues}}\!
         \frac{1}{(n\!-\!k)!}\frac{\mathrm{d}^{(n-k)}}{\mathrm{d}s^{(n-k)}}
         \frac{1}{s} \ln\!\left(\!1\!+\!s\frac{\bra{\la}q|\lb]}
                                           {\bra{\la}K|\lb]} \right)\!\!
         \bigg[ \mathcal{I}_{\text{spinor}}
                \frac{\bra{\la}K|\lb]^{n-k+2}}
                     {\bra{\la}K|q\ket{\la}} \bigg|_{|\lb]=|K+s\,q\ket{\la}}
         \bigg] \bigg|_{s=0} ,
\label{Cbub}
\ee
\normalsize
where the derivative in $s$ is just a way to encode the extraction of the
$(n\!-\!k)$-th Taylor coefficient around $s = 0$.
Note that the formula contains an arbitrary light-like vector $q$.
Nonetheless, the answer does not depend on it
and thus can be simplified by an appropriate choice of $q$.

We point out the fact that \eqn{Cbub} looks different
from equivalent spinor integration formulas given earlier in
\rcites{Britto:2006fc,Britto:2007tt,Britto:2008sw,Britto:2010xq},
because here we chose to write it using as a sum over spinor residues
thus leaving the next step to be carried out afterwards
according to the conventions given in \app{app:residues}.
So in fact, \eqn{Cbub} can be considered as an intermediate step
in derivation of more involved formulas with all pole residues
already taken explicitly in full generality with the price of generating
extra sums and derivatives in another auxiliary parameter.
Further in \sec{sec:simplifiedbubble}, we provide another formula
which is even better suited for calculations with $\mathcal{N}=1$ supersymmetry.

\section{Derivation of spinor integration formulas}
\label{sec:spinorintderivation}

In this section, we derive the coefficient formulas~\eqref{Cbox},
\eqref{Ctri} and \eqref{Cbub} from the four-dimensional spinor integration.

First of all, we need the expressions for cut integrals
to identify their coefficients in the cut expressions.
Using the notation from the previous section, we can write~\cite{Britto:2005ha}: 
\begin{subequations} \begin{align}
            \delta_2 I^{\text{bub}} & = 
                - i (4\pi)^{-\epsilon}
                  2\pi i ,
\label{cutbubble} \\
            \delta_2 I^{\text{tri}}_i \:\, & = ~\;\,
                  i (4\pi)^{-\epsilon}
                  \frac{2\pi i}{K^2 ( x^i_+ - x^i_- )}
                  \ln \left( \frac{-x^i_-}{-x^i_+} \right) ,
\label{cuttriangle} \\
            \delta_2 I^{\text{box}}_{ij} & =
                - i (4\pi)^{-\epsilon}
                  \frac{2\pi i}{K^2 Q_j^2 ( x^{ij}_+ - x^{ij}_- )}
                  \ln \left( \frac{-x^{ij}_-}{-x^{ij}_+} \right) ,
\label{cutbox}
\end{align} \label{cutintegrals} \end{subequations}
\!\!\!\!\;where, strictly speaking, the operation $\delta_2$ acts on two loop propagators
as the following replacement rule:
\be
   \frac{1}{\ell^2} \rightarrow - 2 \pi i \delta_+(\ell^2) .
\label{cutoperation}
\ee

Formulas~\eqref{cutintegrals} can be immediately derived
from the two following identities:
\begin{subequations} \begin{align}
   \int_{\tilde{\la}=\bar{\la}} \!
   \frac{ \braket{\la d\la} [\lb d\lb] }{ \bra{\la}K|\lb]^2 }
      & = \frac{2 \pi i}{K^2} ,
\label{spinorintegrationexample1} \\
   \int_{\tilde{\la}=\bar{\la}} \!
   \frac{ \braket{\la d\la} [\lb d\lb] }{ \bra{\la}P|\lb] \bra{\la}Q|\lb] }
      & = \frac{2 \pi i}{2 \sqrt{(P\!\cdot\!Q)^2 - P^2 Q^2}}
          \ln \left(  \frac{(P\!\cdot\!Q) + \sqrt{(P\!\cdot\!Q)^2 - P^2 Q^2}}
                           {(P\!\cdot\!Q) - \sqrt{(P\!\cdot\!Q)^2 - P^2 Q^2}} \right) .
\label{spinorintegrationexample2}
\end{align} \label{spinorintegrationexamples} \end{subequations}
\!\!\!\!\;
As summarized in \sec{sec:spinorintegrationsummary},
we should start with a generic $K$-channel cut integrand~\eqref{cut}
and transform it to the spinor integral form~\eqref{cutspinor}.


In the most general $d$-dimensional case,
we can express the product of two amplitudes as follows:
\be
   \sum_{h_1,h_2} A^{\text{tree}}(-\ell_1^{\bar{h}_1},\dots,\ell_2^{h_2})
                  A^{\text{tree}}(-\ell_2^{\bar{h}_2},\dots,\ell_1^{h_1})
              =  \frac{ \prod_{i=1}^n (-2(P_i\!\cdot\!\ell_1)) }
                      { \prod_{i=1}^k (\ell_1 - K_i)^2 } ,
\label{cutinput}
\ee
for some fixed complex-valued $P_i$ constructed from external kinematics.
Then, by plugging in \eqns{lvariables}{dLIPS4},
we can rewrite the $K$-channel cut as
\beal
    \text{Cut} = & 
                   \int \! d^d\ell_1 \,
                   \delta_+(\ell_1^2) \, \delta_+( K^2 - 2 (K\!\cdot\!\ell_1) ) \,
                   \frac{ \prod_{i=1}^n (-2(P_i\!\cdot\!\ell_1)) }
                        { \prod_{i=1}^k (\ell_1 - K_i)^2 } \\
      = &\, \frac{\pi^{-\epsilon}}{\Gamma(-\epsilon)}
            \left( \frac{K^2}{4} \right)^{-\epsilon} \!\!\!
            \int_0^1 \! du \, u^{-1-\epsilon} \sqrt{1-u} \\ & \times
            \left( \! -\frac{(K^2)^{n-k+1}}{4} \right)
            \int_{\tilde{\la}=\bar{\la}}
            \frac{\!\!\!\!\!\! \braket{\la d\la} [\lb d\lb] }
                 { \bra{\la}K|\lb]^{n-k+2} }
            \frac{ \prod_{i=1}^n \bra{\la} R_i(P_i,K,u) |\lb] }
                 { \prod_{i=1}^k \bra{\la} Q_i(K_i,K,u) |\lb] } ,
\label{initialddim}
\eeal
where we introduced
\be
   R_i^{\mu}(P_i,K,u) = -\sqrt{1-u} \, P_i^{\mu}
                        -\frac{(1-\sqrt{1-u})(P_i\!\cdot\!K)}{K^2} K^{\mu} ,
\label{Ri}
\ee
\be
   Q_i^{\mu}(K_i,K,u) = -\sqrt{1-u} \, K_i^{\mu}
                        +\frac{K_i^2 - (1-\sqrt{1-u})(K_i\!\cdot\!K)}{K^2} K^{\mu} ,
\label{Qi}
\ee
and the $u$-integration part comes from $(-2\epsilon)$-dimensional phase-space measure.

In the cut-constructible case,
it is harmless to omit it,
so we rewrite (\ref{initialddim}) simply as
\be
    \text{Cut} = -\frac{(K^2)^{n-k+1}}{4}
            \int_{\tilde{\la}=\bar{\la}}
            \frac{\!\!\!\!\!\! \braket{\la d\la} [\lb d\lb] }
                 { \bra{\la}K|\lb]^{n-k+2} }
            \frac{ \prod_{i=1}^n \bra{\la} R_i(P_i) |\lb] }
                 { \prod_{i=1}^k \bra{\la} Q_i(K_i,K) |\lb] } ,
\label{initial4dim}
\ee
where vectors $Q_i$ now coincide with their definition~\eqref{Qi4dim}
and $R_i$ degenerate to
\be
            R_i^{\mu}(P_i) = -P_i^{\mu} .
\label{Ri4dim}
\ee

\subsection{Separating the bubble contribution}
\label{sec:separatingbub}

We now consider the spinor integral
\be
    \int_{\tilde{\la}=\bar{\la}}
            \frac{\!\!\!\!\!\! \braket{\la d\la} [\lb d\lb] }
                 { \bra{\la}K|\lb]^{n-k+2} }
            \frac{ \prod_{i=1}^n \bra{\la}R_i |\lb] }
                 { \prod_{i=1}^k \bra{\la}Q_i |\lb] } .
\label{startderivation}
\ee
The following derivation can be considered as the evaluation
of the core four-dimensional integral
in the $d$-dimensional unitarity cut~\eqref{initialddim},
meaning that the $u$-dependence of $R_i$ and $Q_i$
can in principle be retained all along.
Alternatively,
one can think of it in terms of the purely four-dimensional cut~\eqref{initial4dim}.
The first perspective lets us consider all $Q_i$ massive,
as is always true for their $d$-dimensional definition~\eqref{Qi}.
However, thanks to the cut-constructibility of the supersymmetric amplitudes,
which we aim to calculate in this chapter,
it is sensible to leave the question of $u$-integration~\cite{Anastasiou:2006jv,
Anastasiou:2006gt,Britto:2006fc,Britto:2007tt,Britto:2008sw,Feng:2008ju}
beyond the scope of this thesis.

We deform the denominator of~\eqref{startderivation}
using an arbitrary massless vector $n^{\mu}$
and a set of infinitesimal parameters $ \{s_i\}_{i=1}^{n-k+1} $:
\be
            \frac{1}{ \bra{\la}K|\lb]^{n-k+2} }
            \frac{ \prod_{i=1}^n \bra{\la}R_i |\lb] }
                 { \prod_{i=1}^k \bra{\la}Q_i |\lb] }
            = 
            \lim_{\{s\} \rightarrow 0}
            \frac{1}
                 { \bra{\la}K|\lb] }
            \frac{1}
                 { \prod_{i=1}^{n-k+1} \bra{\la}K\!+\!s_i n |\lb] }
            \frac{ \prod_{i=1}^n \bra{\la}R_i |\lb] }
                 { \prod_{i=1}^k \bra{\la}Q_i |\lb] } .
\label{deform}
\ee
Now that all the denominators are different,
we can use the following reduction formula:
\be
            \frac{ \prod_{i=1}^n \bra{\la} R_i |\lb] }
                 { \prod_{i=1}^{n+1} \bra{\la} Q_i |\lb] }
            = \sum_{i=1}^n
              \frac{1}{ \bra{\la} Q_i |\lb] }
              \frac{ \prod_{j=1}^n \bra{\la}R_j|Q_i\ket{\la} }
                   { \prod_{j \neq i}^{n+1} \bra{\la}Q_j|Q_i\ket{\la} } ,
\label{reduceformula}
\ee
to get
\beal
            \lim_{\{s\} \rightarrow 0}
            \frac{1}
                 { \bra{\la}K|\lb] }
            \bigg\{
                  \sum_{i=1}^{n-k+1}
                  \frac{1}{ \bra{\la}K\!+\!s_i n|\lb] } &
                  \frac{ \prod_{j=1}^n \bra{\la}R_j|K\!+\!s_i n\ket{\la} }
                       { \prod_{j \neq i}^{n-k+1}
                         \bra{\la}K\!+\!s_j n|K\!+\!s_i n\ket{\la} 
                         \prod_{j=1}^{k} \bra{\la}Q_j|K\!+\!s_i n\ket{\la} } \\
                  +
                  \sum_{i=1}^k
                  \frac{1}{ \bra{\la} Q_i |\lb] } &
                  \frac{ \prod_{j=1}^n \bra{\la}R_j|Q_i\ket{\la} }
                       { \prod_{j=1}^{n-k+1} \bra{\la}K\!+\!s_j n|Q_i\ket{\la} 
                         \prod_{j \neq i}^{k} \bra{\la}Q_j|Q_i\ket{\la} }
            \bigg\} .
\label{deformedintegrand}
\eeal

In the second line of \eqn{deform}, it is harmless to take the limit:
\beal
    \int_{\tilde{\la}=\bar{\la}}
            \frac{\!\!\!\!\!\! \braket{\la d\la} [\lb d\lb] }
                 { \bra{\la}K|\lb]^{n-k+2} }
            \frac{ \prod_{i=1}^n \bra{\la} R_i |\lb] }
                 { \prod_{i=1}^k \bra{\la} Q_i |\lb] }
            =
            \sum_{i=1}^k
            \int_{\tilde{\la}=\bar{\la}} \! &
            \frac{ \braket{\la d\la} [\lb d\lb] F_i(\la) }
                 { \bra{\la}K|\lb] \bra{\la} Q_i |\lb] } \\
            +
            \lim_{\{s\} \rightarrow 0} \!\!
            \sum_{i=1}^{n-k+1}  \!
            \frac{1}{ \prod_{j \neq i}^{n-k+1} (s_i\!-\!s_j) }
            \int_{\tilde{\la}=\bar{\la}} \! &
            \frac{ \braket{\la d\la} [\lb d\lb] G(\la,s_i) }
                 { \bra{\la}K|\lb] \bra{\la}K\!+\!s_i n|\lb] } ,
\label{coreintegral}
\eeal
where we denote
\be
            F_i(\la) = \frac{1}{ \bra{\la}K|Q_i\ket{\la}^{n-k+1} }
                       \frac{ \prod_{j=1}^n \bra{\la}R_j|Q_i\ket{\la} }
                            { \prod_{j \neq i}^k \bra{\la}Q_j|Q_i\ket{\la} } ,
\label{Fi}
\ee
\be
            G(\la,s_i) =
            \frac{1}{ \bra{\la}K|n\ket{\la}^{n-k} }
            \frac{ \prod_{j=1}^n \bra{\la}R_j|K\!+\!s_i n\ket{\la} }
                 { \prod_{j=1}^k \bra{\la}Q_j|K\!+\!s_i n\ket{\la} } .
\label{Gi}
\ee

Note that when $n-k+2=0$, the standard loop-momentum power-counting tells us
that the cut reduces only to boxes and there is no need to use \eqn{reduceformula}
to get to the box coefficient formula~\eqref{Cbox}.
If $n=k-1$, there are triangles, but still no bubbles.
They come into play when $n \geq k$
along with the second sum in \eqn{reduceformula}.
At $n \geq k+2$ the integrated expression can contain rational terms,
invisible for four-dimensional cuts.

We will first treat the first sum in \eqn{reduceformula}
which will be shown to contain only boxes and triangles.
The second sum is nothing but the bubble contribution,
which will be computed later.

\subsection{Boxes and triangles}
\label{sec:boxtri}

We start by introducing a Feynman parameter
to gather the two $\lb$-containing denominators under $F_i(\la)$ into one:
\be
            \int_{\tilde{\la}=\bar{\la}} \!
            \frac{ \braket{\la d\la} [\lb d\lb] F_i(\la) }
                 { \bra{\la}K|\lb] \bra{\la} Q_i |\lb] }
            =
            \int_0^1 \!\!\! dx
            \int_{\tilde{\la}=\bar{\la}} \!
            \frac{ \braket{\la d\la} [\lb d\lb] F_i(\la) }
                 { \bra{\la}\! (1-x) K \!+\! x \, Q_i|\lb]^2 }
            \equiv
            \int_0^1 \!\!\! dx
            \int_{\tilde{\la}=\bar{\la}} \!
            \frac{ \braket{\la d\la} [\lb d\lb] F_i(\la) }
                 { \bra{\la}Q_i(x) |\lb]^2 } .
\label{xFi}
\ee
Then we use the following formula:
\be
            \frac{ [\lb d\lb] [\tilde{\eta} \lb]^n }
                 { \bra{\la}K|\lb]^{n+2} }
            =
            \frac{ [\lb \partial_{\lb}] }{n+1} \;
            \frac{ [\tilde{\eta} \lb]^{n+1} }
                 { \bra{\la}K|\lb]^{n+1} \bra{\la}K|\tilde{\eta}] } ,
\label{toresidue}
\ee
in which we set $n=0$ to obtain
\be
            \int_0^1 \!\!\! dx
            \int \!
            \braket{\la d\la} [\lb \partial_{\lb}]
            \frac{ F_i(\la) [\tilde{\eta} \lb] }
                 { \bra{\la}Q_i(x) |\lb]
                   \bra{\la}Q_i(x) |\tilde{\eta}] } .
\label{Fitoresidue}
\ee
Here we transformed the spinor measure $ \braket{\la d\la} [\lb d\lb] $
which in standard complex variables corresponds to $ dz \wedge d\bar{z} $
to $ \braket{\la d\la} [\lb \partial_{\lb}] $.
The latter translates to $ dz \wedge d\bar{z} \frac{\partial}{\partial_{\bar{z}}} $
which, according to Cauchy's integral theorem [0905.2909],
is equivalent to summing the residues of all poles of the integrand.

      We are free to choose $\tilde{\eta} = \lb^i_+ $, for which
\be
            \bra{\la}Q_i(x) |\lb^i_+] = \frac{ \bra{\la}K|Q_i\ket{\la} }
                                                { \braket{\la|\la^i_+} }
                                           \left( x (x^i_+ + 1) - 1 \right) ,
\label{Qix}
\ee
so we get
\be
            \int_0^1 \!\!\! dx
            \int \!
            \braket{\la d\la} [\lb \partial_{\lb}]
            \frac{ F_i(\la) }
                 { \bra{\la}K|Q_i\ket{\la} }
            \frac{ \bra{\la}P^i_+|\lb] }
                 { \left( x (x^i_+ + 1) - 1 \right)
                   \bra{\la}Q_i(x) |\lb] } ,
\label{Fiintermediate}
\ee
where the apparent pole at $\la^i_+$ coming from $ \bra{\la}K|Q_i\ket{\la} $ in the denominator
has zero residue due to $ \bra{\la}P^i_+|\lb] $ in the numerator.
Next we split the last two denominators and take the integral in $x$:
\beal
            \int_0^1 \!\!\! dx &
            \int_{\text{not at }\la^i_+} \!\!\!\!\!\!\!\!\!\!\!\!\!\!\!\!
            \braket{\la d\la} [\lb \partial_{\lb}]
            \frac{ F_i(\la) }
                 { \bra{\la}K|Q_i\ket{\la} }
            \bigg\{
                  \frac{ x^i_+ + 1 }
                       { x (x^i_+ + 1) - 1 }
                - \frac{ \bra{\la}Q_i\!-\!K |\lb] }
                       { x \bra{\la}Q_i\!-\!K |\lb] +  \bra{\la}K|\lb] }
            \bigg\}
            \\ = &
            \int_{\text{not at }\la^i_+} \!\!\!\!\!\!\!\!\!\!\!\!\!\!\!\!
            \braket{\la d\la} [\lb \partial_{\lb}]
            \frac{ F_i(\la) }
                 { \bra{\la}K|Q_i\ket{\la} }
            \bigg\{
                  \ln(-x^i_+)
                - \ln \frac{ \bra{\la}Q_i|\lb] }
                           { \bra{\la}K|\lb] }
            \bigg\} .
\label{Fisplit}
\eeal

The first term in \eqn{Fisplit} is analytic in $\la$, so we can replace
the sum over all residues but $\la^i_+$ by simply
the residue at $\la^i_+$ with a minus sign.
In the second term, we can just split the only multiple pole $\la^i_-$ from the rest.
Even when taking the multiple pole according to \eqn{multiplespinorpole},
we replace $\lb$ by the pole value $\lb^i_-$,
which simplifies the argument of the logarithm to
\be
            \frac{ \bra{\la}Q_i|\lb^i_-] }
                 { \bra{\la}K|\lb^i_-] }
            = - x^i_- .
\label{atxim}
\ee
The remaining poles are simple and come from the following denominators in $F_i(\la)$:
\be
            \bra{\la}Q_j|Q_i\ket{\la}
                = \frac{\braket{\la|\la^{ij}_+}
                        [\lb^{ij}_+|\lb^{ij}_-]
                        \braket{\la^{ij}_-|\la} }
                       { x^{ij}_+ - x^{ij}_- } .
\label{boxpoles}
\ee

After restoring the full form of $F_i(\la)$ and the original sum over $i$, we obtain:
\beal
            \sum_{i=1}^k
            \int_{\tilde{\la}=\bar{\la}} \!
            \frac{ \braket{\la d\la} [\lb d\lb] F_i(\la) }
                 { \bra{\la}K|\lb] \bra{\la} Q_i |\lb] }
            =
            - 2\pi i \sum_{i=1}^k & \Res_{\la=\la^i_+}
            \frac{1}{ \bra{\la}K|Q_i\ket{\la}^{n-k+2} }
                       \frac{ \prod_{l=1}^n \bra{\la}R_l|Q_i\ket{\la} }
                            { \prod_{l \neq i}^k \bra{\la}Q_l|Q_i\ket{\la} }
            \ln(-x^i_+) \\
            - 2\pi i \sum_{i=1}^k & \Res_{\la=\la^i_-}
            \frac{1}{ \bra{\la}K|Q_i\ket{\la}^{n-k+2} }
                       \frac{ \prod_{l=1}^n \bra{\la}R_l|Q_i\ket{\la} }
                            { \prod_{l \neq i}^k \bra{\la}Q_l|Q_i\ket{\la} }
            \ln(-x^i_-) \\
            + 2\pi i \sum_{i \neq j}^k & \Res_{\la=\la^{ij}_{\pm}}
            \frac{1}{ \bra{\la}K|Q_i\ket{\la}^{n-k+2} }
                       \frac{ \prod_{l=1}^n \bra{\la}R_l|Q_i\ket{\la} }
                            { \prod_{l \neq i}^k \bra{\la}Q_l|Q_i\ket{\la} }
            \ln \frac{ \bra{\la}K|\lb] }
                     { \bra{\la}Q_i|\lb] } .
\label{Filn}
\eeal

We concentrate on the last double sum in order to isolate box contributions.
For that we denote
\be
            F_{ij}(\la)
                  = \frac{1}{ \bra{\la}K|Q_i\ket{\la}^{n-k+2} }
                    \frac{ \prod_{l=1}^n \bra{\la}R_l|Q_i\ket{\la} }
                         { \prod_{l \neq i,j}^k \bra{\la}Q_l|Q_i\ket{\la} } ,
\label{Fij}
\ee
and change the summing
\beal
            2\pi i \sum_{i \neq j}^k \Res_{\la=\la^{ij}_{\pm}}
            \frac{ F_{ij}(\la) }
                 { \bra{\la}Q_j|Q_i\ket{\la} }
            \ln \frac{ \bra{\la}K|\lb] }
                     { \bra{\la}Q_i|\lb] }
            =
            2\pi i \sum_{i < j}^k \Res_{\la=\la^{ij}_{\pm}}
            \bigg\{
            \frac{ F_{ij}(\la) }
                 { \bra{\la}Q_j|Q_i\ket{\la} }
            \ln \frac{ \bra{\la}K|\lb] }
                     { \bra{\la}Q_i|\lb] } & \\
            +
            \frac{ F_{ji}(\la) }
                 { \bra{\la}Q_i|Q_j\ket{\la} }
            \ln \frac{ \bra{\la}K|\lb] }
                     { \bra{\la}Q_j |\lb] } &
            \bigg\}
\label{Fijln}
\eeal

Now we can actually take the residues
according to the rules~\eqref{simplespinorpole}.
It is easy to prove that due to
\beal
            |Q_i\ket{\la^{ij}_{\pm}} &
                = \frac{ \pm x^{ij}_{\pm} }{ x^{ij}_+ - x^{ij}_- }
                  |\lb^{ij}_{\mp}] \braket{\la^{ij}_{\mp}|\la^{ij}_{\pm}} , \\
            |Q_j\ket{\la^{ij}_{\pm}} &
                = \frac{ \mp 1 }{ x^{ij}_+ - x^{ij}_- }
                  |\lb^{ij}_{\mp}] \braket{\la^{ij}_{\mp}|\la^{ij}_{\pm}} ,
\label{Qjla}
\eeal
$F_{ij}$ and $F_{ji}$ actually coincide at the poles:
\be
            F_{ij}(\la^{ij}_{\pm}) = F_{ji}(\la^{ij}_{\pm})
                 = \frac{1}{ \bra{\la^{ij}_{\pm}}\!K|\lb^{ij}_{\mp}]^{n-k+2} }
                   \frac{ \prod_{l=1}^n \bra{\la^{ij}_{\pm}}R_l|\lb^{ij}_{\mp}] }
                        { \prod_{l \neq i,j}^k 
                                        \bra{\la^{ij}_{\pm}}Q_l|\lb^{ij}_{\mp}] } .
\label{Fijla}
\ee
This lets us group together the two contributions in \eqn{Fijln}:
\beal 
            2\pi i \sum_{i < j}^k
            \bigg\{
                  \frac{ ( x^{ij}_+ - x^{ij}_- ) F_{ij}(\la^{ij}_+) }
                       { [\lb^{ij}_+|\lb^{ij}_-]
                         \braket{\la^{ij}_-|\la^{ij}_+} }
                  \ln \frac{ \bra{\la^{ij}_+} Q_i |\lb^{ij}_+] }
                           { \bra{\la^{ij}_+} Q_j |\lb^{ij}_+] }
                - \frac{ ( x^{ij}_+ - x^{ij}_- ) F_{ij}(\la^{ij}_-) }
                       { \braket{\la^{ij}_-|\la^{ij}_+}
                         [\lb^{ij}_+|\lb^{ij}_-] }
                  \ln \frac{ \bra{\la^{ij}_-} Q_i |\lb^{ij}_-] }
                           { \bra{\la^{ij}_-} Q_j |\lb^{ij}_-] } &
            \bigg\} \\
            = - \sum_{i < j}^k
            \frac{ 2\pi i }
                 { Q_j^2 ( x^{ij}_+ - x^{ij}_- ) }
            \bigg\{
                  \ln(-x^i_+)
                  F_{ij}(\la^{ij}_+)
                - \ln(-x^i_-)
                  F_{ij}(\la^{ij}_-) &
            \bigg\} ,
\label{Fijresidues}
\eeal
where, among other things, we used that
      \begin{eqnarray} \begin{aligned}
            Q_i & = \frac{ x^{ij}_+ P^{ij}_- - x^{ij}_- P^{ij}_+ }
                         { x^{ij}_+ - x^{ij}_- } , \\
            Q_j & = \frac{ P^{ij}_+ - P^{ij}_- }
                         { x^{ij}_+ - x^{ij}_- } .
\label{QiQj}
	\end{aligned} \end{eqnarray}
We then regroup the logarithms and rewrite \eqn{Filn} as
\beal
            - & \sum_{i=1}^k 2\pi i
            \Big\{
                  \ln(-x^i_+) \Res_{\la=\la^i_+}
                + \ln(-x^i_-) \Res_{\la=\la^i_-}
            \Big\}
            \frac{1}{ \bra{\la}K|Q_i\ket{\la}^{n-k+2} }
            \frac{ \prod_{l=1}^n \bra{\la}R_l|Q_i\ket{\la} }
                 { \prod_{l \neq i}^k \bra{\la}Q_l|Q_i\ket{\la} } \\
            + & \sum_{i < j}^k
            \frac{2\pi i}{ Q_j^2 (x^{ij}_+ - x^{ij}_-) }
            \ln{\left( \frac{-x^{ij}_-}{-x^{ij}_+} \right)}
            \frac{ F_{ij}(\la^{ij}_+) + F_{ij}(\la^{ij}_-) }{2} \\
            - & \sum_{i < j}^k
            \frac{2\pi i}{ Q_j^2 (x^{ij}_+ - x^{ij}_-) } \,
            \ln{\left( x^{ij}_+ x^{ij}_- \right)} \,
            \frac{ F_{ij}(\la^{ij}_+) - F_{ij}(\la^{ij}_-) }{2} .
\label{Fijboxready}
\eeal

The second line of \eqn{Fijboxready} already contains
the right cut box expression.
To see that the third line belongs to cut triangles we note that
\be
            \ln \left( x^{ij}_+ x^{ij}_- \right)
                = \ln \left( \frac{Q_i^2}{Q_j^2} \right)
                = \ln \left( \frac{Q_i^2}{K^2} \right)
                - \ln \left( \frac{Q_j^2}{K^2} \right)
                = \ln \left( x^i_+ x^i_- \right)
                - \ln \left( x^j_+ x^j_- \right) .
\label{trianglelog}
\ee
It is also straightforward to prove that
\beal
            2\pi i \Res_{\la=\la^{ij}_{\pm}}
            \frac{1}{ \bra{\la}K|Q_i\ket{\la}^{n-k+2} }
            \frac{ \prod_{l=1}^n \bra{\la}R_l|Q_i\ket{\la} }
                 { \prod_{l \neq i}^k \bra{\la}Q_l|Q_i\ket{\la} }
            & = -
            2\pi i \Res_{\la=\la^{ij}_{\pm}}
            \frac{1}{ \bra{\la}K|Q_j\ket{\la}^{n-k+2} }
            \frac{ \prod_{l=1}^n \bra{\la}R_l|Q_j\ket{\la} }
                 { \prod_{l \neq j}^k \bra{\la}Q_l|Q_j\ket{\la} } \\
            & =
            \frac{2\pi i}{ Q_j^2 (x^{ij}_+ - x^{ij}_-) } \,
            \big[ F_{ij}(\la^{ij}_+) - F_{ij}(\la^{ij}_-) \big] ,
\label{boxlemma}
\eeal
so the third line of \eqn{Fijboxready} becomes
\beal
            - \sum_{i < j}^k
            \frac{2\pi i}{ Q_j^2 (x^{ij}_+ - x^{ij}_-) }
            \Big\{
                  \ln \left( x^i_+ x^i_- \right)
                - \ln \left( x^j_+ x^j_- \right) &
            \Big\}
            \frac{ F_{ij}(\la^{ij}_+) - F_{ij}(\la^{ij}_-) }{2} \\
            =
            - \frac{1}{2} \sum_{i < j}^k
            \bigg\{
                  \ln \left( x^i_+ x^i_- \right)
                  2\pi i \Res_{\la=\la^{ij}_{\pm}} &
                  \frac{1}{ \bra{\la}K|Q_i\ket{\la}^{n-k+2} }
                       \frac{ \prod_{l=1}^n \bra{\la}R_l|Q_i\ket{\la} }
                            { \prod_{l \neq i}^k \bra{\la}Q_l|Q_i\ket{\la} } \\
                  +
                  \ln \left( x^j_+ x^j_- \right)
                  2\pi i \Res_{\la=\la^{ij}_{\pm}} &
                  \frac{1}{ \bra{\la}K|Q_j\ket{\la}^{n-k+2} }
                       \frac{ \prod_{l=1}^n \bra{\la}R_l|Q_j\ket{\la} }
                            { \prod_{l \neq j}^k \bra{\la}Q_l|Q_j\ket{\la} }
            \bigg\} \\
            =
            - \frac{1}{2} \sum_{i = 1}^k
            \ln \left( x^i_+ x^i_- \right)
            \sum_{j \neq i}^k
            2\pi i \Res_{\la=\la^{ij}_{\pm}} &
            \frac{1}{ \bra{\la}K|Q_i\ket{\la}^{n-k+2} }
            \frac{ \prod_{l=1}^n \bra{\la}R_l|Q_i\ket{\la} }
                            { \prod_{l \neq i}^k \bra{\la}Q_l|Q_i\ket{\la} } \\
            =
            + \frac{1}{2} \sum_{i = 1}^k
            \ln \left( x^i_+ x^i_- \right)
            2\pi i \Res_{\la=\la^i_{\pm}} &
            \frac{1}{ \bra{\la}K|Q_i\ket{\la}^{n-k+2} }
            \frac{ \prod_{l=1}^n \bra{\la}R_l|Q_i\ket{\la} }
                { \prod_{l \neq i}^k \bra{\la}Q_l|Q_i\ket{\la} } ,
\label{thirdline}
\eeal
where in the last line thanks to analyticity of the integrands
we replaced the sum over all simple poles $ \la=\la^{ij}_{\pm} $
by the sum over the two multiple poles $ \la=\la^i_{\pm} $ with a minus sign.

We can now group the first and the third lines of \eqn{Fijboxready}
into the correct cut triangle logarithm:
\beal
            \sum_{i=1}^k
            \int_{\tilde{\la}=\bar{\la}} \!
            \frac{ \braket{\la d\la} [\lb d\lb] F_i(\la) }
                 { \bra{\la}K|\lb] \bra{\la} Q_i |\lb] }
            =
            \sum_{i<j}^k
            \frac{2\pi i}{ Q_j^2 (x^{ij}_+ - x^{ij}_-) }
            \ln{\left( \frac{-x^{ij}_-}{-x^{ij}_+} \right)}
            \frac{ F_{ij}(\la^{ij}_+) + F_{ij}(\la^{ij}_-) }{2} & \\
            +
            \sum_{i=1}^k
            2\pi i
            \ln{\left( \frac{-x^i_-}{-x^i_+} \right)}
            \frac{1}{2}
            \Big\{
                  \Res_{\la=\la^i_+}
                - \Res_{\la=\la^i_-}
            \Big\}
            \frac{1}{ \bra{\la}K|Q_i\ket{\la}^{n-k+2} }
            \frac{ \prod_{l=1}^n \bra{\la}R_l|Q_i\ket{\la} }
                 { \prod_{l \neq i}^k \bra{\la}Q_l|Q_i\ket{\la} } & .
\label{Fijtriangleready}
\eeal

We now use \eqn{multiplespinorpole} to take the residues of the multiple poles.
When taking the residue $ \la^i_+ $,
we pick the reference spinor to be equal to $ \la^i_- $ and vice versa.
This lets us make use of the following properties:
      \begin{eqnarray} \begin{aligned}
            |K\ket{\la^i_{\pm}} &
                = \frac{ \mp 1 }{ x^i_+ - x^i_- }
                  |\lb^i_{\mp}] \braket{\la^i_{\mp}|\la^i_{\pm}} , \\
            |Q_i\ket{\la^i_{\pm}} &
                = \frac{ \pm x^i_{\pm} }{ x^i_+ - x^i_- }
                  |\lb^i_{\mp}] \braket{\la^i_{\mp}|\la^i_{\pm}} .
\label{Qila}
	\end{aligned} \end{eqnarray}
When the smoke clears, we use \eqn{Fijla} and finally rewrite:
\beal
            \sum_{i=1}^k
            \int_{\tilde{\la}=\bar{\la}} \! &
            \frac{ \braket{\la d\la} [\lb d\lb] F_i(\la) }
                 { \bra{\la}K|\lb] \bra{\la} Q_i |\lb] } \\
            =
            \sum_{i<j}^k &
            \frac{2\pi i}{ Q_j^2 (x^{ij}_+ - x^{ij}_-) }
            \ln{\left( \frac{-x^{ij}_-}{-x^{ij}_+} \right)}
            \frac{1}{2}
            \frac{1}{ \bra{\la}K|\lb]^{n-k+2} }
            \frac{ \prod_{l=1}^n \bra{\la}R_l|\lb] }
                 { \prod_{l \neq i,j}^k \bra{\la}Q_l|\lb] }
            \bigg\{ \bigg|_{\substack{
                       \la = \la^{ij}_+ \\
                       \lb = \lb^{ij}_-
                       }}
                  + \bigg|_{\substack{
                       \la = \la^{ij}_- \\
                       \lb = \lb^{ij}_+
                       }}
                    \bigg\} \\
            +
            \sum_{i=1}^k &
            \frac{2\pi i}{ K^2 (x^i_+ - x^i_-) }
            \ln{\left( \frac{-x^i_-}{-x^i_+} \right)}
            \frac{1}{2}
            \frac{1}{(K^2(x^i_+-x^i_-)^2)^{(n-k+1)}} \\ \times &
            \frac{1}{(n\!-\!k\!+\!1)!}
            \frac{\mathrm{d}^{(n-k+1)}}{\mathrm{d}t^{(n-k+1)}}
            \frac{ \prod_{l=1}^n \bra{\la}R_l|\lb] }
                 { \prod_{l \neq i}^k \bra{\la}Q_l|\lb] }
            \bigg\{ \bigg|_{\substack{
                       \la = \la^i_+ - t \la^i_- \\
                       \lb = x^i_+ \lb^i_- - t x^i_- \lb^i_+
                       }}
                  + \bigg|_{\substack{
                        \la = \la^i_- - t \la^i_+ \\
                        \lb = x^i_- \lb^i_+ - t x^i_+ \lb^i_-
                      }} 
            \bigg\} \bigg|_{t=0} .
\label{Fijfinal}
\eeal

\subsection{Bubble}
\label{sec:bub}

We return to the second line in \eqn{coreintegral} and,
much in the same way as in the previous section,
we gather the two $\lb$-containing denominators into one
by introducing a Feynman parameter and
then reduce the spinor measure with the formula~\eqref{toresidue}:
\beal
            \int_{\tilde{\la}=\bar{\la}} \!
            \frac{ \braket{\la d\la} [\lb d\lb] G(\la,s_i) }
                 { \bra{\la}K|\lb] \bra{\la}K\!+\!s_i n|\lb] }
            = &
            \int_0^1 \!\!\! dx
            \int_{\tilde{\la}=\bar{\la}} \!
            \frac{ \braket{\la d\la} [\lb d\lb] G(\la,s_i) }
                 { \bra{\la}K\!+\!x s_i n|\lb]^2 } \\
            = &
            \int_0^1 \!\!\! dx
            \int \!
            \braket{\la d\la} [\lb \partial_{\lb}]
            \frac{ G(\la,s_i) [\tilde{\eta} \lb] }
                 { \bra{\la}K\!+\!x s_i n|\lb]
                   \bra{\la}K\!+\!x s_i n|\tilde{\eta}] } .
\label{xGi}
\eeal
We can now simplify the integral in $x$
by picking the reference spinor $|\tilde{\eta}] = |n]$,
so after restoring the original sum in $s_i$ we get
\beal
            \lim_{\{s\} \rightarrow 0} \!\!
            \sum_{i=1}^{n-k+1}  \!
            \frac{1}{ \prod_{j \neq i}^{n-k+1} (s_i\!-\!s_j) }
            \int \!
            \braket{\la d\la} [\lb \partial_{\lb}]
            \left( - \frac{1}{s_i} \right)
            \ln \left( 1 + s_i \frac{ \bra{\la}n|\lb] }
                                    { \bra{\la}K|\lb] }
                \right)
            \frac{ G(\la,s_i) }
                 { \bra{\la}K|n\ket{\la} } .
\label{Giln}
\eeal

Now we use the following non-trivial lemma:
\beal
            \lim_{\{s\} \rightarrow 0}
            \sum_{i=1}^{m}
            \frac{f(s_i)}{ \prod_{j \neq i}^{m} (s_i\!-\!s_j) }
            =
            \frac{1}{(m-1)!}
            \frac{\mathrm{d}^{(m-1)}}{\mathrm{d}s^{(m-1)}}
            f(s) \bigg|_{s=0} ,
\label{lemma}
\eeal
to reduce the singular limit in $\{s_i\}_{i=1}^{n-k+1}$
to a well-defined Taylor coefficient extraction procedure
encoded by a derivative in a single parameter $s$:
\beal
          - \frac{1}{(n\!-\!k)!}
            \frac{\mathrm{d}^{(n-k)}}{\mathrm{d}s^{(n-k)}}
            \int \!
            \braket{\la d\la} [\lb \partial_{\lb}]
            \frac{1}{s}
            \ln \left( 1 + s \frac{\bra{\la}n|\lb]}
                                  {\bra{\la}K|\lb]}
                \right)
            \frac{ G(\la,s) }
                 { \bra{\la}K|n\ket{\la} } \bigg|_{s=0} .
\label{tobubble}
\eeal
Note that the logarithm expansion starts already with a factor of $s$,
which cancels the overall $1/s$ and that is the only role it plays in this contribution.
Thus it has nothing to do with neither boxes nor triangles.
As discussed earlier, the spinor measure $ \braket{\la d\la} [\lb \partial_{\lb}] $
here means is equivalent to the operation of taking the residues of the integrand.
This means that all operations in \eqn{tobubble} are already purely algebraic,
so we prefer to implement it in this form.

Finally, we rewrite the entire cut integral as
the sum of box, triangle and bubble contributions:
\beal
      \int_{\tilde{\la}=\bar{\la}} &
            \frac{\!\!\!\!\!\! \braket{\la d\la} [\lb d\lb] }
                 { \bra{\la}K|\lb]^{n-k+2} }
            \frac{ \prod_{i=1}^n \bra{\la}R_i|\lb] }
                 { \prod_{i=1}^k \bra{\la}Q_i|\lb] } \\
            =
            \sum_{i<j}^k &
            \frac{2\pi i}{ Q_j^2 (x^{ij}_+ - x^{ij}_-) }
            \ln{\left( \frac{-x^{ij}_-}{-x^{ij}_+} \right)}
            \frac{1}{2}
            \frac{1}{ \bra{\la}K|\lb]^{n-k+2} }
            \frac{ \prod_{l=1}^n \bra{\la}R_l|\lb] }
                 { \prod_{l \neq i,j}^k \bra{\la}Q_l|\lb] }
            \bigg\{ \bigg|_{\substack{
                       \la = \la^{ij}_+ \\
                       \lb = \lb^{ij}_-
                       }}
                  + \bigg|_{\substack{
                       \la = \la^{ij}_- \\
                       \lb = \lb^{ij}_+
                       }}
                    \bigg\} \\
            +
            \sum_{i=1}^k &
            \frac{2\pi i}{ K^2 (x^i_+ - x^i_-) }
            \ln{\left( \frac{-x^i_-}{-x^i_+} \right)}
            \frac{1}{2}
            \frac{1}{(K^2(x^i_+-x^i_-)^2)^{(n-k+1)}} \\ \times &
            \frac{1}{(n\!-\!k\!+\!1)!}
            \frac{\mathrm{d}^{(n-k+1)}}{\mathrm{d}t^{(n-k+1)}}
            \frac{ \prod_{l=1}^n \bra{\la}R_l|\lb] }
                 { \prod_{l \neq i}^k \bra{\la}Q_l|\lb] }
            \bigg\{ \bigg|_{\substack{
                       \la = \la^i_+ - t \la^i_- \\
                       \lb = x^i_+ \lb^i_- - t x^i_- \lb^i_+
                       }}
                  + \bigg|_{\substack{
                        \la = \la^i_- - t \la^i_+ \\
                        \lb = x^i_- \lb^i_+ - t x^i_+ \lb^i_-
                      }}
            \bigg\} \bigg|_{t=0} \\
            -
            2 \pi i & \sum_{\text{residues}}
            \frac{1}{(n\!-\!k)!}
            \frac{\mathrm{d}^{(n-k)}}{\mathrm{d}s^{(n-k)}}
            \frac{1}{s}
            \ln \left(1 + s \frac{\bra{\la}n|\lb]}
                                 {\bra{\la}K|\lb]}
                \right)
            \frac{1}{ \bra{\la}K|n\ket{\la}^{n-k+1} }
            \frac{ \prod_{j=1}^n \bra{\la}R_j|K\!+\!s n\ket{\la} }
                 { \prod_{j=1}^k \bra{\la}Q_j|K\!+\!s n\ket{\la} } \bigg|_{s=0} .
\label{boxtribub}
\eeal

To get the final coefficient formulas
\eqref{Cbox}, \eqref{Ctri}, \eqref{Cbub}, one just needs to
restore the original prefactors from the double cut and
compare the result with the cut integral expressions
\eqref{cutbox}, \eqref{cuttriangle}, \eqref{cutbubble}, respectively.

\section{Setup: $\mathcal{N}=1$ SYM at one loop}
\label{sec:setup}

In this section, we describe the general traits
of $\mathcal{N}=1$ super-Yang-Mills theory at one loop,
which will lead us to a simplified version~\eqref{CbubN2}
of the bubble coefficient formula~\eqref{Cbub},
better suited to that setup.

\subsection{Supersymmetry expansion}
\label{sec:susyexpansion}

The concept of supersymmetry proved to be directly useful
for understanding even non-supersymmetric gauge theories.
Whereas gluon tree amplitudes for pure quantum chromodynamics
equal those of supersymmetric Yang-Mills theory,
their one-loop analogues obey a simple expansion~\cite{Bern:1993mq,Bern:1994zx}:
\be
      A^{\text{1-loop}}_{\text{QCD}} =
      A^{\text{1-loop}}_{\mathcal{N}=4 \text{ SYM}} - 4 \,
      A^{\text{1-loop}}_{\mathcal{N}=1 \text{ chiral}} + 2 \,
      A^{\text{1-loop}}_{\mathcal{N}=0 \text{ scalar}} ,
\label{amplitudeQCD}
\ee
which splits the calculation of direct phenomenological interest
into three problems of increasing difficulty.\footnote{For more information
about various gauge theory supermultiplets, see \tab{Multiplets}
in \sec{sec:supermultiplets}.}
Taking into account that
\be
      A^{\text{1-loop}}_{\mathcal{N}=1 \text{ SYM}} =
      A^{\text{1-loop}}_{\mathcal{N}=4 \text{ SYM}} - 3 \,
      A^{\text{1-loop}}_{\mathcal{N}=1 \text{ chiral}} ,
\label{amplitudeN1}
\ee
it becomes clear that calculations in $\mathcal{N}=4,1$ SYM are important steps
to full understanding of QCD.

In this thesis, we deal with
one-loop NMHV amplitudes in $\mathcal{N}=1$ SYM.
More precisely, we concentrate on $n$-point one-loop contributions
from the $\mathcal{N}=1$ chiral multiplet in the adjoint representation,
which consists of a complex scalar and a Majorana fermion.
In fact, its effective number of supersymmetries is two,
which is reflected in its alternative name, $\mathcal{N}=2$ hyper multiplet,
and can be easily seen from its relation to $\mathcal{N}=2$ SYM:
\be
      A^{\text{1-loop}}_{\mathcal{N}=2 \text{ SYM}} =
      A^{\text{1-loop}}_{\mathcal{N}=4 \text{ SYM}} - 2 \,
      A^{\text{1-loop}}_{\mathcal{N}=1 \text{ chiral}} .
\label{amplitudeN2}
\ee

Moreover, amplitudes in four dimensions are known to be reducible
\cite{Brown:1952eu,Melrose:1965kb,'tHooft:1978xw,Passarino:1978jh,vanNeerven:1983vr}
to the following basis of master integrals:
\be
      A^{\text{1-loop}} = \mu^{2\epsilon} \Big(
                          \sum C^{\text{box}} \, I_4
                        + \sum C^{\text{tri}} \, I_3
                        + \sum C^{\text{bub}} \, I_2 + R \Big) ,
\label{amplitude}
\ee
where the sums go over all distinct scalar integrals
and $R$ is the purely rational part.
However, we know supersymmetry can constrain the general expansion \eqref{amplitude}:
the strongest, $\mathcal{N}=4$, supersymmetry leaves nothing but boxes $\{I_4\}$
\cite{Bern:1994zx},
while $\mathcal{N}=1,2$ supersymmetries eliminate the rational part $R$
\cite{Bern:1994cg}.
Since $R$ is the only term in \eqn{amplitude} invisible to four-dimensional cuts,
supersymmetric amplitudes can be characterized as cut-constructible.

\subsection{UV and IR behavior}
\label{sec:singular}


      Now we review a useful result from \rcite{Britto:2005ha},
where it was derived that
one can include all the infrared divergent one-mass and two-mass triangles
into the definition of new, finite, boxes
and thus leave only three-mass triangles in expansion \eqref{amplitude}.
Moreover, the only remaining divergent integrals are the bubbles:
\be
      I_2 = \frac{1}{\epsilon} + O(1) ,
\label{bubdiv}
\ee
so they alone must produce the remaining singular behavior of the amplitude.
As the latter is proportional to the tree amplitude
\be
      A^{\text{1-loop}}_{\mathcal{N}=1 \text{ chiral}}
    = \frac{1}{\epsilon} \sum C^{\text{bub}} + O(1)
    = \frac{1}{(4\pi)^\frac{d}{2} \epsilon} A^{\text{tree}} + O(1) ,
\label{amplitudeUV}
\ee
we retrieve a non-trivial relation among bubble coefficients:
\be
      \sum C^{\text{bub}} = \frac{1}{(4\pi)^\frac{d}{2}} A^{\text{tree}} ,
\label{bubblesum}
\ee
which we use as the first consistency check for our analytic results.

Having considerably reduced our problem,
we now summarize how we deal with the rest.
The best and immediately algebraic method to compute box coefficients
is from quadruple cuts, first introduced in \rcite{Britto:2004nc}.
Three-mass triangle coefficients can be found from triple cuts
\cite{Bern:1997sc,Forde:2007mi},
and it was done in full generality in \rcite{BjerrumBohr:2007vu,Dunbar:2009ax}.
In the following, we will thus concentrate mostly on bubbles,
for which we use the spinor integration technique~\cite{Britto:2005ha},
described in \sec{sec:method}.

For all calculations in this work, we pick the cut-channel momentum
to be $P_{1,k}$, defined according to \eqn{momentumsum}.
For brevity, we will spell it simply as $P_{1k}$.
If one wishes to compute another channel cut $P_{r,s}$,
one should simply cyclically relabel the legs $i \rightarrow (i\!-\!r\!+\!1)$
and set $k=s-r+1$.
As described in Appendix F of \rcitePaper{2},
the functions provided in the attached Mathematica notebook
have input arguments that are adapted for such relabeling.


\subsection{Example: $\overline{\text{MHV}}$-MHV bubbles in $\mathcal{N}=1$ SYM}
\label{sec:simplestbubbles}

      In this section, we employ the spinor integration technique
to derive explicitly a simple but non-trivial family of bubble coefficients
in $\mathcal{N}=1$ SYM.
To be more precise, we consider the contribution of
the $\mathcal{N}=1$ chiral multiplet in the loop.
To get bubble coefficients in pure $\mathcal{N}=1$ SYM
one just needs to multiply our results by $-3$.

      \begin{figure}[h]
      \centering
      \includegraphics[scale=1.0]{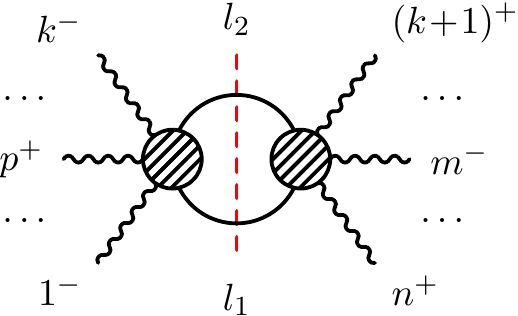}
      \vspace{-5pt}
      \caption{\small $P_{1k}$-channel cut for
               $ A_{\mathcal{N}=1 \text{ chiral}}^{\text{1-loop}}
                   (p^+\!\in\!\{1,\dots,k\}, m^-\!\in\!\{k\!+\!1,\dots,n\}) $
               \label{cut1ksimplest}}
      \end{figure}
      Consider the $P_{1k}$-channel cut of $\mathcal{N}=1$ chiral one-loop amplitude
with one plus-helicity gluon $p^+$ to the left of the cut
and one minus-helicity gluon $m^-$ to the right, see \fig{cut1ksimplest}.
The amplitude to the right of the cut is then just MHV,
whereas the one to the left is then
$(k+2)$-point $\text{N}^{k-2}$MHV=$\overline{\text{MHV}}$.
A nice property of this cut is that it is omnipresent
as a two-particle cut in MHV amplitudes,
a three-particle cut in NMHV amplitudes,
a four-particle cut in NNMHV amplitudes and so on.
At the same time, it is very simple to write down:
\beal
            \sum_{h_1,h_2} A(-\ell_1^{\bar{h}_1},\dots,\ell_2^{h_2})
                           A(-\ell_2^{\bar{h}_2},\dots,\ell_1^{h_1})
                = & \bigg(
                         - \frac{ [\ell_1 p] \braket{\ell_1 m} }
                                { [\ell_2 p] \braket{\ell_2 m} }
                         + 2
                         - \frac{ [\ell_2 p] \braket{\ell_2 m} }
                                { [\ell_1 p] \braket{\ell_1 m} }
                    \bigg) \\ \times
                    \frac{ (-1)^k i [\ell_1 p]^2 [\ell_2 p]^2 }
                         { [\ell_1 1] [12] \dots [k\!-\!1|k] [k \ell_2] [\ell_2 \ell_1] } &
                    \frac{ i \braket{\ell_1 m}^2 \braket{\ell_2 m}^2 }
                         { \braket{\ell_2|k\!+\!1} \braket{k\!+\!1|k\!+\!2}
                                    \dots \braket{n\!-\!1|n}
                           \braket{n \ell_1} \braket{\ell_1 \ell_2} } .
\label{simplestcut}
\eeal
The second line is just a product of tree amplitudes with two scalar legs
and the factor in the first line sums supersymmetric Ward identity
(SWI) factors \cite{Grisaru:1976vm,Grisaru:1977px,Parke:1985pn}
due to two scalars and two helicities of the Majorana fermion circulating in the loop.
Due to supersymmetry, instead of complicating the cut integrand,
that sum helps to simplify it:
\beal
            \sum_{h_1,h_2} \! A_1 A_2
                & = \frac{ (-1)^{k} \bra{m}P_{1k}|p]^2 }
                         { P_{1k}^2 [12] \dots [k\!-\!1|k]
                           \braket{k\!+\!1|k\!+\!2} \dots \braket{n\!-\!1|n} }
                    \frac{ \braket{\ell_1 m} \braket{\ell_2 m} [\ell_1 p] [\ell_2 p] }
                         { \braket{\ell_1 n} \braket{\ell_2|k\!+\!1} [\ell_1 1] [\ell_2 k] } \\
                  & \equiv \frac{F}{P_{1k}^2}
                    \frac{ \braket{\ell_1 m} \braket{\ell_2 m} [\ell_1 p] [\ell_2 p] }
                         { \braket{\ell_1 n} \braket{\ell_2|k\!+\!1} [\ell_1 1] [\ell_2 k] } \\
                & = \frac{F}{P_{1k}^2}
                    \frac{ \braket{\ell_1 m} \bra{\ell_1}P_{1k}|p]
                           \bra{m}P_{1k}|\ell_1] [p \ell_1] }
                         { \braket{\ell_1 n} \bra{\ell_1}P_{1k}|k]
                           \bra{k\!+\!1}P_{1k}|\ell_1] [1 \ell_1] } ,
\label{simplestcut2}
\eeal
where in the last line by $F$ we denoted a kinematic factor
independent of loop momenta
and then we eliminated $\ell_2$ in favor of $\ell_1$.
Now the introduction of the homogeneous variables is trivial,
so after restoring the integration measure \eqref{dLIPS4} we get:
\beal
            \text{Cut}(P_{1k}) = &
                  - \frac{F}{4}
                    \int_{\tilde{\la}=\bar{\la}} \!
                    \frac{ \braket{\la d\la} [\lb d\lb] }
                         { \bra{\la}P_{1k}|\lb]^2 }
                    \frac{ \braket{\la|m} \bra{\la}P_{1k}|p]
                                          \bra{m}P_{1k}|\lb] [p|\lb] }
                         { \braket{\la|n} \bra{\la}P_{1k}|k]
                                          \bra{k\!+\!1}P_{1k}|\lb] [1|\lb] } .
\label{simplestcut3}
\eeal
We then plug the spinorial integrand into \eqref{Cbub}
to obtain the bubble coefficient:
\beal
            C^{\text{bub},P_{1k}}_{\mathcal{N}=1 \text{ chiral}} = & -
                    \frac{F}{(4\pi)^{\frac{d}{2}}i}
                    \sum_{\text{residues}}
                    \frac{1}{s}
                    \ln \left( 1 + s \frac{\bra{\la}q|\lb]}
                                          {\bra{\la}P_{1k}|\lb]}
                        \right) \\ & \times
                    \bigg\{
                        \frac{ \braket{\la|m} \bra{\la}P_{1k}|p]
                                              \bra{m}P_{1k}|\lb] [p|\lb] }
                             { \braket{\la|n} \bra{\la}P_{1k}|k]
                                              \bra{k\!+\!1}P_{1k}|\lb] [1|\lb] }
                        \frac{1}{\bra{\la}P_{1k}|q\ket{\la}}
                        \bigg|_{|\lb]=|P_{1k}+s\,q\ket{\la}}
                    \bigg\} \bigg|_{s=0} .
\label{simplestbubble}
\eeal
Here we used the fact that the integrand \eqref{simplestcut2}
was homogeneous in $\ell_1$,
so the power of the derivative in $s$ is zero.
Therefore, only the first term in the expansion of
the logarithm $ \frac{1}{s} \ln(1+st) = t + O(s) $
survives in the limit $s \rightarrow 0$:
\beal
            C^{\text{bub},P_{1k}}_{\mathcal{N}=1 \text{ chiral}} =
                    \frac{F}{(4\pi)^{\frac{d}{2}}i}
                    \sum_{\text{residues}}
                    \frac{[q|\lb]}
                         {\bra{\la}P_{1k}|\lb]}
                    \frac{ \braket{\la|m}^2 \bra{\la}P_{1k}|p]^2 }
                         { \braket{\la|k\!+\!1} \braket{\la|n} 
                           \bra{\la}P_{1k}|1] \bra{\la}P_{1k}|k]
                           \bra{\la}P_{1k}|q] } .
\label{simplestbubble2}
\eeal
We see 5 poles in the denominator: $\ket{\la}=\ket{k\!+\!1}$, $\ket{\la}=\ket{n}$,
$\ket{\la}=|P_{1k}|1]$, $\ket{\la}=|P_{1k}|k]$ and $\ket{\la}=|P_{1k}|q]$.
Note that the factor $\bra{\la}K|\lb]$ never contains any poles
because in complex variable representation \eqref{complexvariables}
it becomes proportional to $(1+z\bar{z})$.
The sum of the residues produces the final answer:
\beal
   C^{\text{bub},P_{1k}}_{\mathcal{N}=1 \text{ chiral}}
     (p^+\!\in\!\{1,\dots,k\}, m^-\!\in\!\{k\!+\!1,\dots,n\}) \hspace{109pt} & \\
      = \frac{ (-1)^k }{ (4\pi)^{\frac{d}{2}}i }
        \frac{ \bra{m}P_{1k}|p]^2 }
             { [12] \dots [k\!-\!1|k]
               \braket{k\!+\!1|k\!+\!2} \dots \braket{n\!-\!1|n} } \hspace{89pt} & \\
             \times \bigg\{
                    \frac{ \braket{m|k\!+\!1}^2
                           \bra{k\!+\!1}P_{1k}|p]^2 [k\!+\!1|q] }
                         { \braket{k\!+\!1|n}
                           \bra{k\!+\!1}P_{1k}|1] \bra{k\!+\!1}P_{1k}|k] 
                           \bra{k\!+\!1}P_{1k}|k\!+\!1] \bra{k\!+\!1}P_{1k}|q] }
                    & \\ +
                    \frac{ \braket{mn}^2 \bra{n}P_{1k}|p]^2 [nq] }
                         { \braket{n|k\!+\!1}
                           \bra{n}P_{1k}|1] \bra{n}P_{1k}|k] 
                           \bra{n}P_{1k}|n] \bra{n}P_{1k}|q] }
                    & \\ +
                    \frac{1}{P_{1k}^2}
                    \bigg(
                    \frac{[1p]^2}{[1k][1q]}
                    \frac{ \bra{m}P_{1k}|1]^2 \bra{1}P_{1k}|q] }
                         { \bra{1}P_{1k}|1]
                           \bra{k\!+\!1}P_{1k}|1] \bra{n}P_{1k}|1] }
                    & \\ +
                    \frac{[kp]^2}{[k1][kq]}
                    \frac{ \bra{m}P_{1k}|k]^2 \bra{k}P_{1k}|q] }
                         { \bra{k}P_{1k}|k]
                           \bra{k\!+\!1}P_{1k}|k] \bra{n}P_{1k}|k] }
                    & \\ +
                    \frac{[pq]^2}{[1q][kq]}
                    \frac{ \bra{m}P_{1k}|q]^2 }
                         { \bra{k\!+\!1}P_{1k}|q] \bra{n}P_{1k}|q] } &
                    \bigg)
                    \bigg\} .
\label{simplestbubbleq}
\eeal

      Each term in \eqn{simplestbubble} can be generically eliminated
by an appropriate choice of reference spinor $|q]$.
Moreover, specific helicity configurations can further simplify the formula.
For instance, if in a $P_{1,3}$-channel NMHV bubble
the plus-helicity leg gluon $p^+$ is $3^+$
followed by the minus-helicity gluon $m^- = 4^-$ and we pick $|q]=|3]$,
then only two terms survive:
\beal
            C^{\text{bub},P_{1,3}}_{\mathcal{N}=1 \text{ chiral}}
              (1^-,2^-,3^+,4^-,5^+,\dots,n^+) = & \\
              \frac{1}{(4\pi)^{\frac{d}{2}} i}
              \frac{ \bra{4}P_{1,3}|3]^2 }
                   { [12] \braket{45} \dots \braket{n\!-\!1|n} \bra{n}P_{1,3}|1] } &
              \bigg\{
                \frac{ \bra{4}n|3] }
                     { [23] \bra{n}P_{1,3}|n] }
              + \frac{ \bra{2}1|P_{1,3}\ket{4} }
                     { P_{1,3}^2 \bra{1}P_{1,3}|1] }
              \bigg\} .
\label{N1chBubP13}
\eeal

      We checked on various examples that our result numerically coincides
with the equivalent all-$n$ formula found earlier in \rcite{Dunbar:2009ax}.
More than that, we found that we can reproduce their formula term by term
by choosing in \eqn{simplestbubbleq} $|q]=|P_{1k}\ket{m}$:
\footnotesize
\beal
   C^{\text{bub},P_{1k}}_{\mathcal{N}=1 \text{ chiral}}
     (p^+\!\in\!\{1,\dots,k\}, m^-\!\in\!\{k\!+\!1,\dots,n\})
      = \frac{ (-1)^k }{ (4\pi)^{\frac{d}{2}}i }
        \frac{ \bra{m}P_{1k}|p]^2 }
             { [12] \dots [k\!-\!1|k]
               \braket{k\!+\!1|k\!+\!2} \dots \braket{n\!-\!1|n} } & \\
             \times \bigg\{
                    \frac{1}{P_{1k}^2 \braket{k\!+\!1|n}}
                    \bigg(
                    \frac{ \bra{k\!+\!1}P_{1k}|p]^2 \braket{m|P_{1k}|k\!+\!1|m} }
                         { \bra{k\!+\!1}P_{1k}|1] \bra{k\!+\!1}P_{1k}|k] 
                           \bra{k\!+\!1}P_{1k}|k\!+\!1] }
                    -
                    \frac{ \bra{n}P_{1k}|p]^2 \braket{m|P_{1k}|n|m} }
                         { \bra{n}P_{1k}|1] \bra{n}P_{1k}|k] 
                           \bra{n}P_{1k}|n] }
                    \bigg) & \\
                    +
                    \frac{1}{[1k]}
                    \bigg(
                    \frac{ [1p]^2 \braket{m|P_{1k}|1|m} }
                         { \bra{1}P_{1k}|1] \bra{k\!+\!1}P_{1k}|1]
                           \bra{n}P_{1k}|1] }
                    -
                    \frac{ [kp]^2 \braket{m|P_{1k}|k|m} }
                         { \bra{k}P_{1k}|k] \bra{k\!+\!1}P_{1k}|k]
                           \bra{n}P_{1k}|k] }
                    \bigg) &
                    \bigg\} .
\label{simplestbubbledpw}
\eeal
\normalsize

      Needless to say, any bubble with an $\overline{\text{MHV}}$-MHV cut
can be obtained from the $P_{1k}$-channel bubble \eqref{simplestbubbleq} by appropriate relabeling.

\subsection{Modified bubble formula}
\label{sec:simplifiedbubble}

      We can already learn a more general lesson from the calculation
in \sec{sec:simplestbubbles}.
The supersymmetric helicity sum is well known~\cite{Bern:1994cg}
to simplify cut integrands instead of complicating them.

      We consider the $\mathcal{N}=1$ chiral multiplet
in the adjoint representation of the gauge group
which in fact has an effective $\mathcal{N}=2$ supersymmetry.
Thanks to that,
for all $\mathcal{N}=1$ chiral double cuts
the numerator and the denominator
have the same number of loop-momentum-dependent factors
and after introducing homogeneous variables $\la,\lb$
the only overall factor $\bra{\la}K|\lb]^{-2}$ comes
from the cut measure \eqref{dLIPS4}.
This means that when plugging the cut integrand
into the general bubble formula \eqref{Cbub}
we will always have a zero power of the derivative in $s$,
so we can set $s$ to zero from the start:
\be
            C^{\text{bub}}_{\mathcal{N}=1 \text{ chiral}} =
                  \frac{4}{(4\pi)^{\frac{d}{2}}i} \!
                  \sum_{\text{residues}}
                  \frac{\bra{\la}q|\lb]}
                       {\bra{\la}K|\lb]}
                  \bigg[ \mathcal{I}_{\text{spinor}}
                         \frac{\bra{\la}K|\lb]^2}
                              {\bra{\la}K|q\ket{\la}}
                  \bigg|_{|\lb]=|K\ket{\la}}
                  \bigg] .
\label{CbubN1}
\ee
Taking into account that
\be
            \mathcal{I}_{\text{spinor}} =
                - \frac{K^2}{4} \frac{1}{\bra{\la}K|\lb]^2}
                  \sum_{h_1,h_2} \! A_1 A_2 ,
\label{IspinorN1}
\ee
where by $\sum_{h_1,h_2} \! A_1 A_2$ we just mean the double cut
after loop-variable change, 
we retrieve a more direct formula for $\mathcal{N}=1$ chiral bubble coefficients:
\be
            C^{\text{bub}}_{\mathcal{N}=1 \text{ chiral}} =
                - \frac{K^2}{(4\pi)^{\frac{d}{2}}i} \!
                  \sum_{\text{residues}}
                  \frac{ [\lb|q] }
                       { \bra{\la}K|\lb] \bra{\la}K|q] }
                  \bigg[ \sum_{h_1,h_2} \! A_1 A_2
                  \bigg|_{|\lb]=|K\ket{\la}}
                  \bigg] .
\label{CbubN2}
\ee

      Incidentally, a close analogue of \eqref{CbubN2} has already been discovered
in \rcite{Elvang:2011fx} with the help of $\mathcal{N}=1$ superspace.

\section{Cut integrand construction}
\label{sec:cutconstruction}

      \begin{figure}[h]
      \centering
      \includegraphics[scale=1.0]{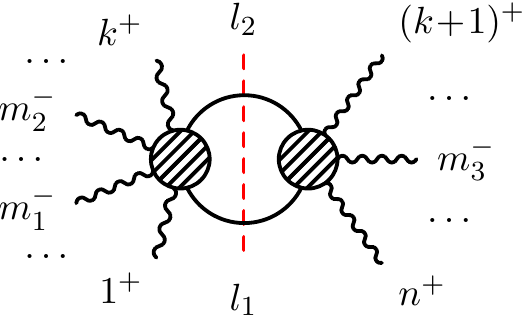}
      \vspace{-5pt}
      \caption{\small $P_{1k}$-channel cut for
               $ A_{\mathcal{N}=1 \text{ chiral}}^{\text{1-loop,NMHV}}
                   (m_1^-,m_2^-\!\in\!\{1,\dots,k\},
                          m_3^-\!\in\!\{k\!+\!1,\dots,n\}) $
               \label{cut1k}}
      \end{figure}

      Constructing appropriate cut integrands is crucial for using spinor integration
and getting clean analytic expressions.
In short, what we do is we sew tree amplitudes
and sum over the $\mathcal{N}=1$ chiral multiplet circulating in the cut.
An NMHV amplitude has 3 minus-helicity gluons,
so all its non-zero double cuts have an MHV amplitude on one side of the cut
and an NMHV one on the other side, as shown in \fig{cut1k}.
For some 3-particle cuts the NMHV amplitude happens to be $\overline{\text{MHV}}$
and the computation is greatly simplified which we exploited
in \sec{sec:simplestbubbles}.
But for all other integrands one needs to sew NMHV tree amplitudes,
which we describe in detail in the following section.

\subsection{NMHV tree amplitudes}
\label{sec:nmhvtrees}

      NMHV tree amplitudes are known to be encoded in the $\mathcal{N}=4$ SYM
$n$-point superamplitude \cite{Drummond:2008vq,Drummond:2008bq,Elvang:2008na}:
\be
   \mathcal{A}_n^{\text{NMHV}} = \mathcal{A}_n^{\text{MHV}}
      \sum_{s=r+2}^{r+n-3} \, \sum_{t=s+2}^{r+n-1} \mathcal{R}_{rst} ,
\label{NMHVtree}
\ee
where the overall prefactor is the well-known MHV superamplitude:
\be
   \mathcal{A}_n^{\text{MHV}} =
      \frac{ i \, \delta^{(8)}(Q) }
           { \braket{12} \braket{23} \dots \braket{n\!-\!1|n} \braket{n1} } ,
\label{MHVtree}
\ee
with the delta-function of the supermomentum defined
in terms of auxiliary Grassmann variables $\eta^A_i$ as
\be
   \delta^{(8)}(Q)
      = \prod_{A=1}^{4} \sum_{i<j}^m \eta^A_i \braket{ij}  \eta^A_j .
\label{delta8Q}
\ee
Moreover, the rest of the formula~\eqref{NMHVtree} involves the following object:
\be
   \mathcal{R}_{rst} =
      \frac{ -\braket{s\!-\!1|s} \braket{t\!-\!1|t} \, \delta^{(4)}(\Xi_{rst}) }
           { P_{\mathcal{S}}^2 \braket{s\!-\!1|P_{\mathcal{S}}|P_{\mathcal{T}}|r}
                               \braket{s|P_{\mathcal{S}}|P_{\mathcal{T}}|r}
                               \braket{t\!-\!1|P_{\mathcal{S}}|P_{\mathcal{R}}|r}
                               \braket{t|P_{\mathcal{S}}|P_{\mathcal{R}}|r} } ,
\label{susyRrst}
\ee
where the Grassmann delta-function is a product of the components of
\be
   \Xi_{rst}^A =
      - \sum_{i=t}^{r-1} \eta^A_i \braket{i|P_{\mathcal{S}}|P_{\mathcal{R}}|r}
      + \sum_{i=r}^{s-1} \eta^A_i \braket{i|P_{\mathcal{S}}|P_{\mathcal{T}}|r} .
\label{Xrst}
\ee

\begin{figure}[t]
      \centering
      \includegraphics[scale=1.0]{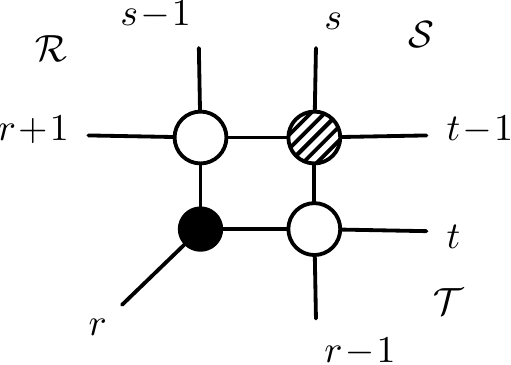}
      \vspace{-5pt}
\caption{\small Cut box diagram for determining the values of $r$, $s$ and $t$}
\label{RST}
\end{figure}

      Remarkably, in \eqn{NMHVtree} $r$ can be chosen arbitrarily.
The possible values of $s$ and $t$ (mod $n$) are already given
in the explicit double sum in \eqn{NMHVtree},
but we also find insightful the following graphic approach
from \rcite{Drummond:2008bq}.
After picking the $r$, one draws all cut-box-like diagrams
with one vertex having only one external leg $r$,
the opposite vertex with at least two external legs $s,\dots,t-1$ and
the other two vertices having at least one external legs, see \fig{RST}.
For brevity, in \eqns{susyRrst}{Xrst}
we denote the three collections of external legs as
$\mathcal{R} = \{r+1,\dots,s-1\}$,
$\mathcal{S} = \{s,\dots,t-1\}$ and
$\mathcal{T} = \{t,\dots,r-1\}$.

      Component amplitudes can then be extracted
from the super-amplitude \eqref{NMHVtree} using, for example,
the package GGT \cite{Dixon:2010ik} in the following representation:
      \begin{subequations} \begin{align}
            A^{\text{tree}}(1_g^+, \overset{+}{\dots},
                            a_g^-, \overset{+}{\dots},
                            b_g^-, \overset{+}{\dots}, n_g^-) &
            = \frac{i}{ \braket{12} \braket{23} \dots \braket{n1} }
              \sum_{s=2}^{n-3} \sum_{t=s+2}^{n-1}\!\!R_{nst} D_{nst;ab}^4 , \\
            A^{\text{tree}}(1_g^+, \overset{+}{\dots},
                   a_\Lambda^{-A}, \overset{+}{\dots},
                 b_\Lambda^{+BCD}, \overset{+}{\dots},
                            c_g^-, \overset{+}{\dots}, n_g^-) &
            = \frac{i \epsilon^{ABCD} }{ \braket{12} \braket{23} \dots \braket{n1} }
              \sum_{s=2}^{n-3} \sum_{t=s+2}^{n-1}\!\!
              R_{nst} D_{nst;ac}^3 D_{nst;bc} , \\
            A^{\text{tree}}(1_g^+, \overset{+}{\dots},
                         a_S^{AB}, \overset{+}{\dots},
                         b_S^{CD}, \overset{+}{\dots},
                            c_g^-, \overset{+}{\dots}, n_g^-) &
            = \frac{i \epsilon^{ABCD} }{ \braket{12} \braket{23} \dots \braket{n1} }
              \sum_{s=2}^{n-3} \sum_{t=s+2}^{n-1}\!\!
              R_{nst} D_{nst;ac}^2 D_{nst;bc}^2 ,
	\end{align} \label{nmhvtrees} \end{subequations}
\!\!\!\!\;where $R_{rst}$ is just the bosonic part of $\mathcal{R}_{rst}$:
\be
            R_{rst} = \frac{-\braket{s\!-\!1|s} \braket{t\!-\!1|t} }
                           { P_{\mathcal{S}}^2
                             \braket{s\!-\!1|P_{\mathcal{S}}|P_{\mathcal{T}}|r}
                             \braket{s|P_{\mathcal{S}}|P_{\mathcal{T}}|r}
                             \braket{t\!-\!1|P_{\mathcal{S}}|P_{\mathcal{R}}|r}
                             \braket{t|P_{\mathcal{S}}|P_{\mathcal{R}}|r} } ,
\label{Rrst}
\ee
whereas $D_{rst;ab}$ arise from differentiating the product of super-delta
functions inside $\mathcal{A}_n^{\text{MHV}}$ and $\mathcal{R}_{rst}$ :
\beal
            D_{rst;ab} =
            \begin{cases}
                  \braket{ab} \braket{r|P_{\mathcal{S}}|P_{\mathcal{T}}|r} &
                  \text{if } a,b \in \mathcal{S} \\
                - \braket{br} \braket{a|P_{\mathcal{S}}|P_{\mathcal{T}}|r} &
                  \text{if } a \in \mathcal{S}, b \in \mathcal{R} \\
                  \braket{ar} \braket{b|P_{\mathcal{S}}|P_{\mathcal{T}}|r} &
                  \text{if } a \in \mathcal{R}, b \in \mathcal{S} \\
                  \braket{br} \braket{a|P_{\mathcal{S}}|P_{\mathcal{R}}|r} &
                  \text{if } a \in \mathcal{S}, b \in \mathcal{T} \\
                - \braket{ar} \braket{b|P_{\mathcal{S}}|P_{\mathcal{R}}|r} &
                  \text{if } a \in \mathcal{T}, b \in \mathcal{S} \\
                - P_{\mathcal{S}}^2 \braket{ar} \braket{br} &
                  \text{if } a \in \mathcal{R}, b \in \mathcal{T} \\
                  P_{\mathcal{S}}^2 \braket{ar} \braket{br} &
                  \text{if } a \in \mathcal{T}, b \in \mathcal{R} \\
                  0 & \text{otherwise} .
            \end{cases}
\label{Drstab}
\eeal
By the derivation in Grassmann variables,
$D_{rst;ab}$ is antisymmetric in $a$ and $b$.

      From \eqn{nmhvtrees}, it is clear that,
much like ratios of spinor products relate
MHV amplitudes with fermions and scalars to purely gluonic ones
through standard supersymmetric Ward identities (SWI),
ratios of different $D_{rst;ab}$ do the same job for NMHV amplitude contributions.
We note here that one could in principle try to encode this information
using $\mathcal{N}=1$ superfields \cite{Bern:2009xq,Lal:2009gn,Elvang:2011fx}.
More than that, as we have already noted, the effective number of supersymmetries of
the $\mathcal{N}=1$ chiral multiplet in the adjoint representation is two,
so one can imagine even defining $\mathcal{N}=2$ hyper superspace.
However, even if Grassmann variables are undoubtedly
an indispensable tool for describing the theory in general,
sometimes they seem to put us farther away from calculating explicit formulas.
In this chapter, we find it direct enough
to assemble the cut without using the superspace.

\subsection{Cut integrand}
\label{sec:cutintegrand}

      Now we are ready to write down the cut integrand in full generality.
Consider the $P_{1k}$-channel cut shown on \fig{cut1k}.
It has two minus-helicity gluons labeled $m_1^-$ and $m_2^-$ on the left of the cut
and one such gluon $m_3^-$ on the right.
Evidently, all other cuts can be obtained from this one by appropriate relabeling.

      A scalar cut would be just a product of
the right-hand side scalar MHV amplitude
and left-hand side scalar NMHV amplitude.
As explained above, to account for the fact that there are two scalars
and two helicities of the Majorana fermion circulating in the loop,
we multiply it further by a sum of SWI factors:
\beal
            \sum_{h_1,h_2} A(-\ell_1^{\bar{h}_1},\dots,\ell_2^{h_2})
                           A(-\ell_2^{\bar{h}_2},\dots,\ell_1^{h_1})
                  = \frac{ i \braket{-\ell_2|m_3}^2 \braket{\ell_1|m_3}^2 }
                         { \braket{-\ell_2|k\!+\!1}
                           \braket{k\!+\!1|k\!+\!2} \dots \braket{n\!-\!1|n}
                           \braket{n|\ell_1} \braket{\ell_1|\!-\!\ell_2} } &\\\times
                    \frac{ i }
                         { \braket{-\ell_1|1} \braket{12} \dots \braket{k\!-\!1|k}
                           \braket{k|\ell_2} \braket{\ell_2|\!-\!\ell_1} }
                    \sum_{s=m_1+2}^{m_1-3} \, \sum_{t=s+2}^{m_1-1}
                    R_{m_1 st} D_{m_1 st; m_2(-\ell_1)}^2 D_{m_1 st; m_2 \ell_2}^2 &\\\times
                    \bigg(
                           \frac{ \braket{-\ell_2|m_3} D_{m_1 st; m_2(-\ell_1)} }
                                { \braket{ \ell_1|m_3} D_{m_1 st; m_2  \ell_2 } }
                         + 2
                         + \frac{ \braket{ \ell_1|m_3} D_{m_1 st; m_2  \ell_2 } }
                                { \braket{-\ell_2|m_3} D_{m_1 st; m_2(-\ell_1)} }
                    \bigg) & ,
\label{genericcuta}
\eeal
where both sums in the second line go cyclically
over labels $\{-\ell_1,1,,\dots,k,\ell_2\}$.

      Note that in \eqn{genericcuta} we picked $m_1$ to be the first argument
of $R_{rst}$ and $D_{rst;ab}$ and $m_2$ to be the last,
but in principle $m_1$ and $m_2$ can be interchanged
due to the arbitrariness of the choice of $r$ in the NMHV expansion \eqref{NMHVtree},
which is a non-trivial property of tree amplitudes.
It comes from the BCFW recursion that underlies formulas
\eqref{NMHVtree}-\eqref{Drstab} \cite{Drummond:2008cr}
and is related to the freedom of choosing BCFW shifts.
Anyway, the roles of $m_1$ and $m_2$ can also be interchanged
by a vertical flip of the amplitude.

      To make full use of the effective $\mathcal{N}=2$ supersymmetry
of the $\mathcal{N}=1$ chiral multiplet
in the adjoint representation of the gauge group
we rewrite it as follows:
\beal
            \sum_{h_1,h_2} \! A_1 A_2
                 = -\frac{ 1 }
                         { \braket{12} \dots \braket{k\!-\!1|k}
                           \braket{k\!+\!1|k\!+\!2} \dots \braket{n\!-\!1|n} }
                    \frac{ \braket{\ell_1|m_3} \braket{m_3|-\!\ell_2} }
                         { \braket{\ell_1|1} \braket{\ell_1|n} \braket{\ell_1|\ell_2}^2
                           \braket{k|\ell_2} \braket{k\!+\!1|\ell_2} } & \\ \times \!\!\!
                    \sum_{s=m_1+2}^{m_1-3} \, \sum_{t=s+2}^{m_1-1}
                    R_{m_1 st} D_{m_1 st; m_2(-\ell_1)} D_{m_1 st; m_2 \ell_2}
                    \big(
                      \!\braket{-\ell_2|m_3} D_{m_1 st; m_2(-\ell_1)}
                      + \braket{ \ell_1|m_3} D_{m_1 st; m_2  \ell_2 }
                    \big)^2 & ,
\label{genericcut}
\eeal
where the last factor squared is typically subject to non-trivial simplifications
involving the Schouten identity.

      The most important thing for applying spinor integration
is the dependence of the integrand on the loop variables.
Thus we need to do a case-by-case analysis of \eqn{genericcut} to expose them.
But first we consider a helicity configuration
for which there is only one case that contributes.

\subsection{Simpler bubble coefficients}
\label{sec:simplebubbles}

      \begin{figure}[h]
      \centering
      \includegraphics[scale=1.0]{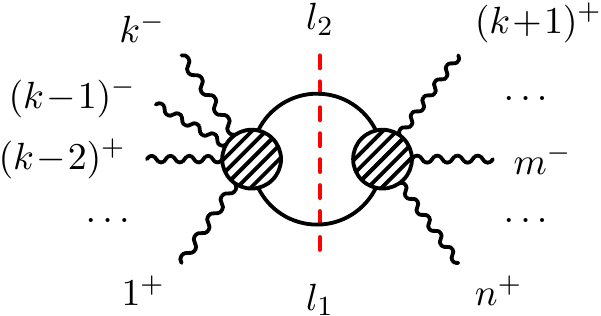}
      \vspace{-5pt}
      \caption{\small $P_{1k}$-channel cut for
               $ A_{\mathcal{N}=1 \text{ chiral}}^{\text{1-loop,NMHV}}
                   ((k\!-\!1)^-,k^-\!\in\!\{1,\dots,k\},
                          m^-\!\in\!\{k\!+\!1,\dots,n\}) $
               \label{cut1ksimple}}
      \end{figure}

      Consider a $P_{1k}$-channel cut
with minus-helicity gluons $k^-$ and $(k\!-\!1)^-$ adjacent to the cut.
The other negative helicity leg $m^-$ is at an arbitrary position
on the other side of the cut, see \fig{cut1ksimple}.
It turns out that the general formula \eqref{genericcut}
simplifies greatly in this case.
Indeed, if we take $r=m_1=(k\!-\!1)$, $m_2=k$ and $m_3=m$,
we can see from definition \eqref{Drstab} that
$ D_{(k-1)st; k \ell_2} $ is non-zero only for $s=\ell_2$,
because for all subsequent values of $s \in \{-\ell_1,\dots,k\!-\!4\} $
both $a=k$ and $b=\ell_2$ will belong to $\mathcal{R} = \{k\!-\!1,k,\dots,s\!-\!1\} $.
So the double sum in $s \in \{\ell_2,-\ell_1,\dots,k\!-\!4\} $
and $t \in \{s\!+\!2,\dots,k\!-\!2\} \subset \{1,\dots,k\!-\!2\} $
collapses to a single sum in $t \in \{1,\dots,k\!-\!2\} $:
\beal
   \sum_{h_1,h_2} \! A_1 A_2
      & = \frac{ 1 }
               { \braket{12} \dots \braket{k\!-\!1|k}
                 \braket{k\!+\!1|k\!+\!2} \dots \braket{n\!-\!1|n} }
          \frac{ \braket{\ell_1|m} \braket{-\ell_2|m} }
               { \braket{n|\ell_1} \braket{\ell_1|1} \braket{\ell_1|\ell_2}^2
                 \braket{k|\ell_2} \braket{\ell_2|k\!+\!1} } \\
      \times \sum_{t=1}^{k-2} &
          R_{(k-1) \ell_2 t} D_{(k-1) \ell_2 t; k(-\ell_1)} D_{(k-1) \ell_2 t; k \ell_2}
          \big(\!\braket{-\ell_2|m} D_{(k-1) \ell_2 t; k(-\ell_1)}
               + \braket{ \ell_1|m} D_{(k-1) \ell_2 t; k \ell_2}
          \big)^2 ,
\label{simplecut}
\eeal
where we compute
\be
            R_{(k-1) \ell_2 t}
                  = \frac{ \braket{k|\ell_2} \braket{t\!-\!1|t} }
                         { P_{t,k-1}^2 P_{t,k}^2 \braket{k\!-\!1|k}^3
                           \braket{\ell_2|P_{t,k}|P_{t,k-1}|k\!-\!1}
                           \bra{t\!-\!1}P_{t,k}|k] \bra{t}P_{t,k}|k] } ,
\label{Rsimple}
\ee
      \begin{subequations} \begin{align}
            D_{(k-1) \ell_2 t; k(-\ell_1)} & = \braket{k\!-\!1|k}
                                         \braket{-\ell_1|P_{t,k}|P_{t,k-1}|k\!-\!1} , \\
            D_{(k-1) \ell_2 t; k \ell_2}   & = \braket{k\!-\!1|k}
                                         \braket{ \ell_2|P_{t,k}|P_{t,k-1}|k\!-\!1} ,
      \end{align} \label{Dsimple} \end{subequations}
\!\!\!\!\;and the chiral sum is simplified by a Schouten identity:
\be
                   \braket{-\ell_2|m} D_{(k-1) \ell_2 t; k(-\ell_1)}
                 + \braket{ \ell_1|m} D_{(k-1) \ell_2 t; k \ell_2}
                 = \braket{k\!-\!1|k} \braket{\ell_1|\ell_2}
                   \braket{m|P_{t,k}|P_{t,k-1}|k\!-\!1} .
\label{chsumsimple}
\ee
Putting all these ingredients together, we observe numerous cancellations and find 
\beal
   \sum_{h_1,h_2} \! A_1 A_2 &
   = \frac{ 1 }
          { \braket{12} \dots \braket{k\!-\!2|k\!-\!1}
            \braket{k\!+\!1|k\!+\!2} \dots \braket{n\!-\!1|n} } \\ &
     \times \sum_{t=1}^{k-2}
     \frac{ \braket{m|P_{t,k}|P_{t,k-1}|k\!-\!1}^2 \braket{t\!-\!1|t} }
          { P_{t,k-1}^2 P_{t,k}^2 \bra{t}P_{t,k}|k] \bra{t\!-\!1}P_{t,k}|k] }
     \frac{ \braket{\ell_1|m} \braket{\ell_1|P_{t,k}|P_{t,k-1}|k\!-\!1}
            \braket{m|\ell_2} }
          { \braket{\ell_1|1} \braket{\ell_1|n} \braket{k\!+\!1|\ell_2} } \\ &
   \equiv \sum_{t=1}^{k-2}
     \frac{ F_t \braket{t\!-\!1|t} }
          { \bra{t\!-\!1}P_{t,k}|k] }
     \frac{ \braket{\ell_1|m}
            \braket{\ell_1|P_{t,k}|P_{t,k-1}|k\!-\!1}
            \braket{m|\ell_2} }
          { \braket{\ell_1|1} \braket{\ell_1|n}
            \braket{k\!+\!1|\ell_2} } ,
\label{simplecut2}
\eeal
where in the last line for brevity
we denoted the common factor independent of the loop momenta by $F_t$.
Note that, as expected, the number of loop momentum spinors is the same
for the numerator and the denominator.
Moreover, one should not miss the fact
that the $(t\!-\!1)$-th leg can become $(-\ell_1)$,
so poles are different for $t = 1$ and $t \neq 1$.
We then trade $\ell_2$ for $\ell_1$,
introduce the homogeneous variables
to find the following expression for the cut:
\beal
   \sum_{h_1,h_2} \! A_1 A_2
      & = F_1 \frac{ \braket{\la|m}
              \braket{ \la|P_{1k}|P_{1,k-1}|k\!-\!1}
                       \bra{m}P_{1k}|\lb] }
                     { \bra{\la}P_{1k}|k] \braket{\la|n}
                       \bra{k\!+\!1}P_{1k}|\lb] } \\
      & + \sum_{t=2}^{k-2}
              \frac{ F_t \braket{t\!-\!1|t} }
                   { \bra{t\!-\!1}P_{t,k}|k] }
              \frac{ \braket{\la|m}
                     \braket{\la|P_{t,k}|P_{t,k-1}|k\!-\!1}
                     \bra{m}P_{1k}|\lb] }
                   { \braket{\la|1} \braket{\la|n}
                     \bra{k\!+\!1}P_{1k}|\lb] } .
\label{simplecut3}
\eeal
To obtain the bubble coefficient, we plug this expression directly into
our simplified formula \eqref{CbubN2}:
\beal
            C^{\text{bub},P_{1k}}_{\mathcal{N}=1 \text{ chiral}} =
                    \frac{P_{1k}^2}{(4\pi)^{\frac{d}{2}}i}
                    \sum_{\text{residues}}
                    \frac{[q|\lb]}
                         {\bra{\la}P_{1k}|\lb]}
                    \bigg\{
                    F_1
                    \frac{ \braket{\la|m}^2
                           \braket{\la|P_{1k}|P_{1,k-1}|k\!-\!1} }
                         { \braket{\la|k\!+\!1} \braket{\la|n} 
                           \bra{\la}P_{1k}|k] \bra{\la}P_{1k}|q] } & \\
                    + \sum_{t=2}^{k-2}
                    \frac{ F_t \braket{t\!-\!1|t} }
                         { \bra{t\!-\!1}P_{t,k}|k] }
                    \frac{ \braket{\la|m}^2
                           \braket{\la|P_{t,k}|P_{t,k-1}|k\!-\!1} }
                         { \braket{\la|1} \braket{\la|k\!+\!1}
                           \braket{\la|n} \bra{\la}P_{1k}|q] } &
                    \bigg\} .
\label{simplebubble2}
\eeal
      We see 5 poles in the denominators:
$\ket{\la}=\ket{1}$, $\ket{\la}=\ket{k\!+\!1}$, $\ket{\la}=\ket{n}$,
$\ket{\la}=|P_{1k}|k]$ and $\ket{\la}=|P_{1k}|q]$.
The answer is then given by the sum of their residues:
\footnotesize
\beal
            C^{\text{bub},P_{1k}}_{\mathcal{N}=1 \text{ chiral}} =
                  \frac{1}{(4\pi)^{\frac{d}{2}}i}
                  \frac{ 1 }
                       { \braket{12} \dots \braket{k\!-\!2|k\!-\!1}
                         \braket{k\!+\!1|k\!+\!2} \dots \braket{n\!-\!1|n} }
                  \hspace{165pt} \\ \times
                  \bigg\{
                  \sum_{t=2}^{k-2} \!
                  \frac{ P_{1k}^2 \! \braket{m|P_{t,k}|P_{t,k-1}|k\!-\!1}^2 \!
                         \braket{t\!-\!1|t} \! }
                       { P_{t,k}^2 P_{t,k-1}^2 \!
                         \bra{t\!-\!1}P_{t,k}|k] \bra{t}P_{t,k}|k] }
                  \bigg(
                  \frac{ \braket{1m}^2 \!
                         \braket{1|P_{t,k}|P_{t,k-1}|k\!-\!1} [1q] }
                       { \braket{1n} \! \braket{1|k\!+\!1} \!
                         \bra{1}P_{1k}|1] \! \bra{1}P_{1k}|q] }
             \!+\!\frac{ \braket{nm}^2 \!
                         \braket{n|P_{t,k}|P_{t,k-1}|k\!-\!1} [nq] }
                       { \braket{n1} \! \braket{n|k\!+\!1} \!
                         \bra{n}P_{1k}|n] \! \bra{n}P_{1k}|q] } & \\
                + \frac{ \braket{k\!+\!1|m}^2
                         \braket{k\!+\!1|P_{t,k}|P_{t,k-1}|k\!-\!1} [k\!+\!1|q] }
                       { \braket{k\!+\!1|1} \! \braket{k\!+\!1|n} \!
                         \bra{k\!+\!1}P_{1k}|k\!+\!1] \!
                         \bra{k\!+\!1}P_{1k}|q] }
             \!+\!\frac{ \bra{m}P_{1k}|q]^2
                         \bra{k\!-\!1}P_{t,k-1}|P_{t,k}|P_{1,k}|q] }
                       { P_{1k}^2 \bra{1}P_{1k}|q] \!
                         \bra{k\!+\!1}P_{1k}|q] \! \bra{n}P_{1k}|q] } &
                  \bigg) \\
                + \frac{ \braket{m|P_{1k}|P_{1,k-1}|k\!-\!1}^2 }
                       { P_{1,k-1}^2 \! \bra{1}P_{1k}|k] }
                  \bigg(
                  \frac{1}{P_{1k}^2}
                  \bigg(
                  \frac{ \bra{m}P_{1k}|k]^2 \bra{k\!-\!1}P_{1k}|k]
                         \bra{k}P_{1k}|q] }
                       { \bra{n}P_{1k}|k] \bra{k\!+\!1}P_{1k}|k]
                         \bra{k}P_{1k}|k] [kq] }
             \!-\!\frac{ \bra{m}P_{1k}|q]^2 \bra{k\!-\!1}P_{1k}|q] }
                       { \bra{n}P_{1k}|q] \! \bra{k\!+\!1}P_{1k}|q] [kq] }
                  \bigg) & \\
                + \frac{ \braket{k\!+\!1|m}^2
                         \braket{k\!+\!1|P_{1k}|P_{1,k-1}|k\!-\!1} [k\!+\!1|q] }
                       { \braket{k\!+\!1|n} \! \bra{k\!+\!1}P_{1k}|k] \!
                         \bra{k\!+\!1}P_{1k}|k\!+\!1] \! \bra{k\!+\!1}P_{1k}|q] }
             \!+\!\frac{ \braket{nm}^2
                         \braket{n|P_{1k}|P_{1,k-1}|k\!-\!1} [nq] }
                       { \braket{n|k\!+\!1} \! \bra{n}P_{1k}|k] \!
                         \bra{n}P_{1k}|n] \! \bra{n}P_{1k}|q] } &
                  \bigg)
                  \bigg\} .
\label{simplebubbleq}
\eeal
\normalsize
This expression can be further simplified by an appropriate choice
of the arbitrary spinor $|q]$.
For example, setting it equal to $|P_{1k}\ket{m}$ gives
the following formula:
\beal
            C^{\text{bub},P_{1k}}_{\mathcal{N}=1 \text{ chiral}} =
                  \frac{1}{(4\pi)^{\frac{d}{2}}i}
                  \frac{ 1 }
                       { \braket{12} \dots \braket{k\!-\!2|k\!-\!1}
                         \braket{k\!+\!1|k\!+\!2} \dots \braket{n\!-\!1|n} }
                  \hspace{77pt} \\ \times
                  \bigg\{
                  \sum_{t=2}^{k-2}
                  \frac{ \braket{m|P_{t,k}|P_{t,k-1}|k\!-\!1}^2
                         \braket{t\!-\!1|t} }
                       { P_{t,k}^2 P_{t,k-1}^2
                         \bra{t\!-\!1}P_{t,k}|k] \bra{t}P_{t,k}|k] }
                  \bigg(
                  \frac{ \braket{m|P_{1k}|k\!+\!1|m}
                         \braket{k\!+\!1|P_{t,k}|P_{t,k-1}|k\!-\!1} }
                       { \braket{k\!+\!1|1} \braket{k\!+\!1|n}
                         \bra{k\!+\!1}P_{1k}|k\!+\!1] } & \\ +
                  \frac{ \braket{m|P_{1k}|1|m}
                         \braket{1|P_{t,k}|P_{t,k-1}|k\!-\!1} }
                       { \braket{1n} \braket{1|k\!+\!1}
                         \bra{1}P_{1k}|1] } +
                  \frac{ \braket{m|P_{1k}|n|m}
                         \braket{n|P_{t,k}|P_{t,k-1}|k\!-\!1} }
                       { \braket{n1} \braket{n|k\!+\!1}
                         \bra{n}P_{1k}|n] } &
                  \bigg) \\ +
                  \frac{ \braket{m|P_{1k}|P_{1,k-1}|k\!-\!1}^2 }
                       { P_{1k}^2 P_{1,k-1}^2
                         \bra{1}P_{1k}|k] }
                  \bigg(
                  \frac{ \braket{m|P_{1k}|k\!+\!1|m}
                         \braket{k\!+\!1|P_{1k}|P_{1,k-1}|k\!-\!1} }
                       { \braket{k\!+\!1|n}
                         \bra{k\!+\!1}P_{1k}|k]
                         \bra{k\!+\!1}P_{1k}|k\!+\!1] } \hspace{49pt} & \\ + \,
                  \frac{ \braket{m|P_{1k}|n|m}
                         \braket{n|P_{1k}|P_{1,k-1}|k\!-\!1} }
                       { \braket{n|k\!+\!1}
                         \bra{n}P_{1k}|k] \bra{n}P_{1k}|n] } \, + \,
                  \frac{ P_{1k}^2 \braket{m|P_{1k}|k|m}
                         \bra{k\!-\!1}P_{1k}|k] }
                       { \bra{n}P_{1k}|k] \bra{k\!+\!1}P_{1k}|k]
                         \bra{k}P_{1k}|k] } &
                  \bigg)
                  \bigg\} .
\label{simplebubble}
\eeal
In the following sections, we choose to provide only formulas with $q$ left arbitrary.

\section{Loop momentum dependence}
\label{sec:loopmomentum}

In this section, we carefully study the dependence
of the cut expression \eqref{genericcut} on the cut loop momenta $\ell_1$ and $\ell_2$.
They can be subsequently replaced by homogeneous variables $\la$ and $\lb$
in order to find the bubble coefficient corresponding to that cut.

\subsection{Case-by-case analysis}
\label{sec:caseanalysis}

\begin{figure}[h]
      \centering
      \includegraphics[scale=0.67]{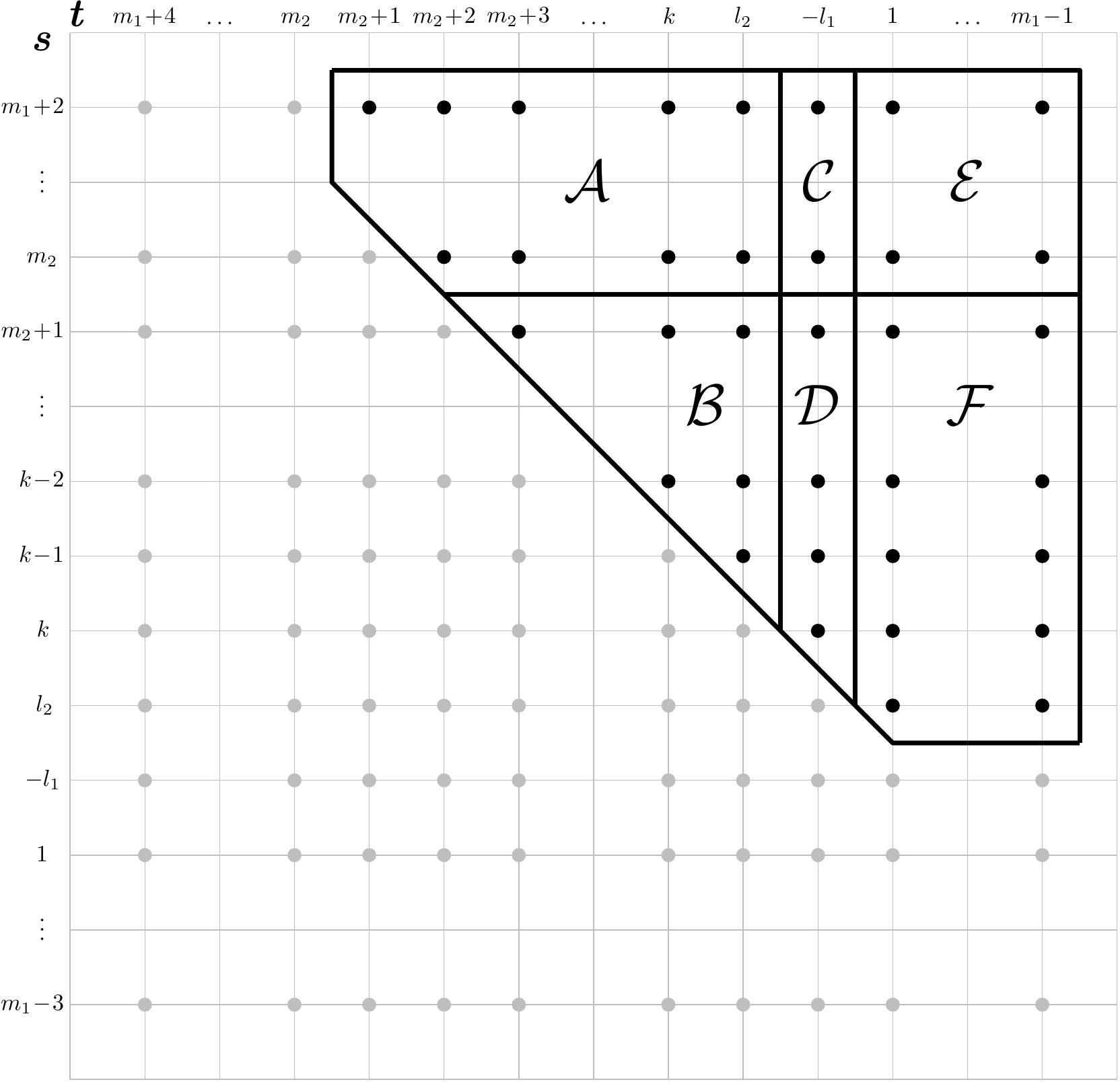}
      \vspace{-5pt}
      \caption{\small Values of $s$ and $t$ corresponding to non-zero contributions to the
               $P_{1k}$-channel cut.
               \label{stmesh}}
      \end{figure}

First of all, we find that in \eqn{genericcut}
the factor most frequently equal to zero is
\beal
   D_{m_1 st; m_2 \ell_2} \! = \!
   \begin{cases}
      \quad \braket{m_1|\ell_2} \braket{m_1|P_{m_1+1,s-1}|P_{s,t-1}|m_2} &
            \text{if } \{s,t\} \in \mathcal{A} \\
      - P_{s,t-1}^2 \braket{m_1|m_2} \braket{m_1|\ell_2} &
            \text{if } \{s,t\} \in \mathcal{B} \\
      \quad \braket{m_2|\ell_2}
            \big( \bra{m_1}P_{m_1+1,s-1}|\ell_1] \braket{\ell_1|m_1} &
            \text{if } \{s,t\} \in \mathcal{C} \\ ~~~~~~\quad\;
                - \braket{m_1|P_{m_1+1,s-1}|P_{1,m_1-1}|m_1} \big) \\
      - \braket{m_1|m_2}
            \big( \braket{m_1|\ell_1} [\ell_1|P_{s,k}\ket{\ell_2}
                - \braket{m_1|P_{1,m_1-1}|P_{s,k}|\ell_2} \big) &
            \text{if } \{s,t\} \in \mathcal{D} \\
      \quad \braket{m_2|\ell_2} \braket{m_1|P_{t,m_1-1}|P_{m_1+1,s-1}|m_1} &
            \text{if } \{s,t\} \in \mathcal{E} \\
      - \braket{m_1|m_2} \braket{m_1|P_{t,m_1-1}|P_{t,s-1}|\ell_2} &
            \text{if } \{s,t\} \in \mathcal{F} \\
      \quad 0 & \text{otherwise} ,
   \end{cases}
\label{D2cases}
\eeal
where we define the non-zero cases:
\begin{subequations} \begin{align}
                  \mathcal{A}: \; & s \in \{m_1\!+\!2,\dots,m_2\},
                             \;\;\; t \in \{m_2\!+\!1,\dots,\ell_2\} \\
                  \mathcal{B}: \; & s \in \{m_2\!+\!1,\dots,k\!-\!1\},
                               \,   t \in \{m_2\!+\!3,\dots,\ell_2\} \\
                  \mathcal{C}: \; & s \in \{m_1\!+\!2,\dots,m_2\},
                               \;\; t = -\ell_1 \\
                  \mathcal{D}: \; & s \in \{m_2\!+\!1,\dots,k\},
                            \quad\; t = -\ell_1 \\
                  \mathcal{E}: \; & s \in \{m_1\!+\!2,\dots,m_2\},
                               \;   t \in \{1,\dots,m_1\!-\!1\} \\
                  \mathcal{F}: \; & s \in \{m_2\!+\!1,\dots,\ell_2\},
                             \;\;\; t \in \{1,\dots,m_1\!-\!1\} .
\label{cases}
\end{align} \end{subequations}
\!\!\!\!\;Thus, we need to consider all other factors solely in these six cases.
For clearness, we depict them on a two-dimensional mesh in \fig{stmesh}.

Next, we expose the loop-momentum dependence of $D_{m_1 st; m_2(-\ell_1)}$:
\beal
   D_{m_1 st; m_2(-\ell_1)} \! = \!
   \begin{cases}
      - \braket{-\ell_1|m_1} \braket{m_1|P_{m_1+1,s-1}|P_{s,t-1}|m_2} &
            \text{if } \{s,t\} \in \mathcal{A} \\
      P_{s,t-1}^2 \braket{-\ell_1|m_1} \braket{m_1|m_2} &
            \text{if } \{s,t\} \in \mathcal{B} \\
      - \braket{-\ell_1|m_1}
        \big( \bra{m_1}P_{m_1+1,s-1}|\ell_1] \braket{\ell_1|m_2} &
            \text{if } \{s,t\} \in \mathcal{C} \\ ~~~~~~~~~~~\:
            - \braket{m_2|P_{1,s-1}|P_{s,m_1-1}|m_1} \big) \\
      P_{-\ell_1,s-1}^2 \braket{-\ell_1|m_1} \braket{m_1|m_2} &
            \text{if } \{s,t\} \in \mathcal{D} \\
      - \braket{-\ell_1|m_2} \braket{m_1|P_{t,m_1-1}|P_{m_1+1,s-1}|m_1} &
            \text{if } \{s,t\} \in \mathcal{E} \\
      \braket{m_1|m_2} \braket{-\ell_1|P_{t,s-1}|P_{t,m_1-1}|m_1} &
            \text{if } \{s,t\} \in \mathcal{F} .
   \end{cases}
\label{D1cases}
\eeal
Then we combine the $D$-terms together,
and, after applying Schouten identities where necessary, we find:
\beal
          & \braket{-\ell_2|m_3} D_{m_1 st; m_2(-\ell_1)}
          + \braket{ \ell_1|m_3} D_{m_1 st; m_2  \ell_2 } \\ &
          = \begin{cases}
            \quad \braket{\ell_1 \ell_2} \braket{m_1 m_3}
                                   \braket{m_1|P_{m_1+1,s-1}|P_{s,t-1}|m_2} &
                  \text{if } \{s,t\} \in \mathcal{A} \\
            \quad P_{s,t-1}^2 \braket{\ell_1 \ell_2} \braket{m_1 m_2} \braket{m_3 m_1} &
                  \text{if } \{s,t\} \in \mathcal{B} \\
                - \big( \braket{\ell_1 \ell_2} \braket{m_2 m_3}
                        \braket{m_1|P_{1,s-1}|P_{s,m_1-1}|m_1} \\ \,\,\,
                      + \braket{m_1 m_2} \braket{m_3 \ell_2}
                        \braket{\ell_1|P_{1,s-1}|P_{s,m_1-1}|m_1} &
                  \text{if } \{s,t\} \in \mathcal{C} \\ \,\,\,
                      - \braket{m_2 m_3} \braket{\ell_1 m_1}
                        \braket{m_1|P_{m_1+1,s-1}|P_{1,k}|\ell_2}
                  \big) \\
                - \braket{m_1 m_2} \big( \braket{\ell_1 m_1}
                                    \braket{m_3|P_{m_1,s-1}|P_{s,k}|\ell_2} &
                  \text{if } \{s,t\} \in \mathcal{D} \\ ~~~~~~~~~~
                      + \braket{m_3 m_1}
                        \braket{\ell_1|P_{1,m_1-1}|P_{s,k}|\ell_2} \big) \\
                - \braket{\ell_1 \ell_2} \braket{m_2 m_3}
                                   \braket{m_1|P_{t,m_1-1}|P_{m_1+1,s-1}|m_1} &
                  \text{if } \{s,t\} \in \mathcal{E} \\
            \quad \braket{\ell_1 \ell_2} \braket{m_1 m_2}
                                   \braket{m_3|P_{t,s-1}|P_{t,m_1-1}|m_1} &
                  \text{if } \{s,t\} \in \mathcal{F} .
            \end{cases}
\label{chsum}
\eeal

Finally, we write three distinct cases for the $R_{m_1 st}$ factor:
\beal
          R_{m_1 st} = \!
            \begin{cases}
                - \braket{s\!-\!1|s} \braket{t\!-\!1|t} \!/
                  \big( P_{s,t-1}^2 \!\!\!\!\!\!
                      & \braket{m_1|P_{m_1+1,t-1}|P_{s,t-1}|s\!-\!1}
                        \braket{m_1|P_{m_1+1,t-1}|P_{s+1,t-1}|s} \\ \!\!\!\!\!
                      & \braket{m_1|P_{m_1+1,s-1}|P_{s,t-2}|t\!-\!1}
                        \braket{m_1|P_{m_1+1,s-1}|P_{s,t-1}|t} \!
                  \big) \\ & ~~~~~~~~~~~~~~~~~~~~~~~~~~~~~~~~~~\,
                  \text{if } \{s,t\} \in \mathcal{A} \cup \mathcal{B} \\ \quad
                  \braket{s\!-\!1|s} \braket{\ell_1 \ell_2} \!/
                  \big( P_{-\ell_1,s-1}^2 \!\!\!\!\!\!
                      & \braket{m_1|P_{-\ell_1,m_1-1}|P_{-\ell_1,s-2}|s\!-\!1}
                        \braket{m_1|P_{-\ell_1,m_1-1}|P_{-\ell_1,s-1}|s} \\ \!\!\!\!\!
                      & \braket{\ell_1|P_{1,s-1}|P_{m_1+1,s-1}|m_1}
                        \braket{m_1|P_{m_1+1,s-1}|P_{s,k}|\ell_2} \!
                  \big) \\ & ~~~~~~~~~~~~~~~~~~~~~~~~~~~~~~~~~~\,
                  \text{if } \{s,t\} \in \mathcal{C} \cup \mathcal{D} \\ \quad
                  \braket{s\!-\!1|s} \braket{t|t\!-\!1} \!/
                  \big( P_{t,s-1}^2 \!\!\!\!\!\!
                      & \braket{m_1|P_{t,m_1-1}|P_{t,s-2}|s\!-\!1}
                        \braket{m_1|P_{t,m_1-1}|P_{t,s-1}|s} \\ \!\!\!\!\!
                      & \braket{m_1|P_{m_1+1,s-1}|P_{t+1,s-1}|t}
                        \braket{m_1|P_{m_1+1,s-1}|P_{t,s-1}|t\!-\!1} \!
                  \big) \\ & ~~~~~~~~~~~~~~~~~~~~~~~~~~~~~~~~~~\,
                  \text{if } \{s,t\} \in \mathcal{E} \cup \mathcal{F} .
            \end{cases}
\label{Rcases}
\eeal
Here the first and the third cases can develop simple loop dependence
on the borders of their respective domains:
in the first case $\ket{t}$ can be become includes $\ket{\ell_2}$, whereas
the third case includes $s=\ell_2$ and $t=1 \Rightarrow t-1=-\ell_1$
which have even a non-trivial overlap.
These subcases can only lead to loop spinors appearing
on the edges of spinor products and we will deal with these cases along the way.

Of course, in some particular lower-point cases these formulae
can be simplified further using momentum conservation and Schouten identities,
but they are simple enough for us to proceed in full generality.

\subsection{NMHV pole structure}
\label{sec:poles}

In principle, to obtain explicit bubble coefficients formulas,
all that remains to do is
to make loop-variable change in the cut integrand \eqref{genericcut}
and plug it into our simplified master formula \eqref{CbubN2}
in which the only non-trivial operation is taking spinor residues
with respect to $\la$.
We need to do it separately for different cases $\mathcal{A}$ through $\mathcal{F}$
and their subcases with slightly modified loop dependence and then sum over the cases.
Thus, we write \emph{a frame formula for a generic NMHV bubble coefficient}:
\begin{equation} \begin{aligned}
      C^{\text{bub},P_{1k}}_{\mathcal{N}=1 \text{ chiral}}
        (m_1^-,m_2^-\!\in\!\{1,\dots,k\}, m_3^-\!\in&\{k\!+\!1,\dots,n\}) \\
          = \sum_{ \{s,t\} \in \mathcal{A} } R_{\mathcal{A}}^{s,t}
          + \sum_{ \{s,t\} \in \mathcal{B} } R_{\mathcal{B}}^{s,t}
 +\!\!\!\!\!\!\sum_{ \{s,t=-\ell_1\} \in \mathcal{C} } \!\!\!\!\! & R_{\mathcal{C}}^s
 +\!\!\!\!\!\!\sum_{ \{s,t=-\ell_1\} \in \mathcal{D} } \!\!\!\!\! R_{\mathcal{D}}^s
          + \sum_{ \{s,t\} \in \mathcal{E} } R_{\mathcal{E}}^{s,t}
          + \sum_{ \{s,t\} \in \mathcal{F} } R_{\mathcal{F}}^{s,t} ,
\label{CbubN1frame}
\end{aligned} \end{equation}
where we introduced a shorthand notation for residue sums
of each individual contribution to the cut \eqref{genericcut}.

However, it is well known
\cite{Feng:2008ju,Hodges:2009hk,ArkaniHamed:2010gg,ArkaniHamed:2012nw}
that, in contrast to the Parke-Taylor MHV amplitudes \cite{Parke:1986gb},
the tree-level NMHV amplitudes derived from BCFW recursion contain spurious poles,
\ie poles that do not correspond to any physical propagator.
They can be viewed as an artifact of the on-shell derivation,
or as a price to pay to have more compact expressions
than what one would obtain from Feynman diagram calculations.
These poles obtain a geometrical meaning in (momentum) twistor variables
\cite{Hodges:2009hk,ArkaniHamed:2010gg,ArkaniHamed:2012nw}.

Fortunately, by definition \emph{spurious poles have zero residues},
so we can just omit them in our calculation of bubble coefficients.
To do this, we need to tell them apart from physical poles.
As already mentioned, the common MHV prefactor of \eqref{NMHVtree}
contains only physical poles.
Evidently, spurious poles come from denominators
of different $\mathcal{R}$-invariants.
Each term can have a non-zero spurious residue,
but they are bound to cancel in a sum over $s$ and $t$.

\begin{figure}[h]
      \centering
      \includegraphics[scale=1.1]{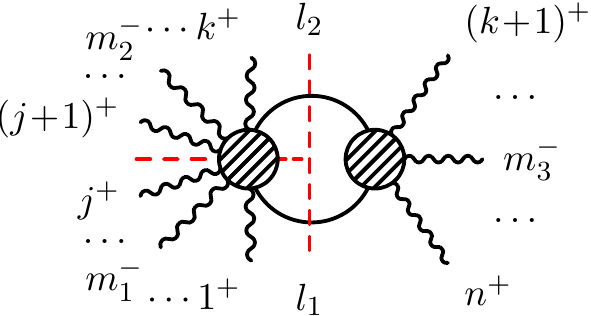}
      \vspace{-5pt}
      \caption{\small $P_{1k}$-channel cut for
               $ A_{\mathcal{N}=1 \text{ chiral}}^{\text{1-loop,NMHV}}
                   (m_1^-,m_2^-\!\in\!\{1,\dots,k\},
                          m_3^-\!\in\!\{k\!+\!1,\dots,n\}) $
               has non-vanishing three-mass triple cuts
               only between $m_1$ and $m_2$.
               \label{double2triple}}
      \end{figure}

Of course, for our one-loop calculation
we are only concerned by telling apart poles that depend on the loop momentum.
The common MHV denominator in \eqn{genericcut} already captures
four massless physical poles:
$\braket{\ell_1|1} \Rightarrow (\ell_1-p_1)^2$,
$\braket{k|\ell_2} \Rightarrow (\ell_2+p_k)^2$,
$\braket{k\!+\!1|\ell_2} \Rightarrow (\ell_2-p_{k+1})^2$,
$\braket{\ell_1|n} \Rightarrow (\ell_1+p_n)^2$.
So what we seek is a physical massive pole
that has to look like
\be
      P_{-\ell_1,j}^2 = (\ell_1 - P_{1,j})^2 = (\ell_2 + P_{j+1,k})^2 = P_{j+1,\ell_2}^2 .
\label{massivepole}
\ee
Moreover, the presence of such a pole means that one can cut it and obtain
a non-zero three-mass triple cut. This can only occur
if one cuts between the two minus-helicity gluons
on the left-hand side of the original double cut, as shown in \fig{double2triple}.
Therefore, $j \in \{m_1,\dots,m_2\!-\!1\} \cap \{2,\dots,k\!-\!2\} $.

Let us then examine one-by-one each of the five denominators
of $R_{m_1 st}$:

\begin{enumerate}
\item $ P_{\mathcal{S}}^2 = P_{s,t-1}^2 $
      produces massive physical poles, unless it is canceled by the numerator;
      it develops the desired loop-momentum dependence \eqref{massivepole}
      for either $\{s=-\ell_1, t=j\!+\!1\}$ or $\{s=j\!+\!1, t=-\ell_1\}$.
      However, the position of $s$ in the former also constrains
      $t$ to be in $\{2,\dots,m_1\!-\!1\}$,
      which is inconsistent with $j = t\!-\!1 \in \{m_1,\dots,m_2\!-\!1\}$,
      so only the latter case is meaningful.
\item $ \braket{s\!-\!1|P_{\mathcal{S}}|P_{\mathcal{T}}|r}
      = \braket{m_1|P_{t,m_1}|P_{t,s-1}|s\!-\!1} $
      can obviously produce a non-zero momentum square only
      if the two spinor arguments become adjacent.
      With $s \in \{m_1\!+\!2,\dots,m_1\!-\!3\}$
      it is only possible in case $s\!-\!1=m_1\!+\!1$.
      Moreover, to obtain the right loop dependence \eqref{massivepole},
      we need to have $t=-\ell_1$, for which this denominator becomes
      $ \braket{m_1|m_1\!+\!1} P_{-\ell_1,m_1}^2 $.
\item $ \braket{s|P_{\mathcal{S}}|P_{\mathcal{T}}|r}
      = \braket{m_1|P_{t,m_1}|P_{t,s}|s} $
      cannot produce a momentum square
      as $s$ is never adjacent to $m_1$.
\item $ \braket{t\!-\!1|P_{\mathcal{S}}|P_{\mathcal{R}}|r}
      = \braket{m_1|P_{s,m_1}|P_{s,t-1}|t\!-\!1} $
      cannot produce a momentum square
      because $t\!-\!1$ is never adjacent to $m_1$.
\item $ \braket{t|P_{\mathcal{S}}|P_{\mathcal{R}}|r}
      = \braket{m_1|P_{s,m_1}|P_{s,t}|t} $
      can be factorized with a momentum square as
      $ \braket{m_1|m_1\!-\!1} P_{s,m_1-1}^2 $ for $t=m_1\!-\!1$,
      but it cannot result in
      the desired loop-momentum dependence \eqref{massivepole} for any $s$.
\end{enumerate}

Thus we have only two potential sources of physical massive poles:
the first one, $P_{-\ell_1,m_1}^2$, comes from factorizing
$ \braket{s\!-\!1|P_{\mathcal{S}}|P_{\mathcal{T}}|r} $ for $s=m_1\!+\!2$,
while all subsequent poles come simply from $P_{\mathcal{S}}^2$
for $s \in \{m_1\!+\!2,\dots,\min(m_2,k\!-\!1)\}$.
In both cases $t$ remains equal to $-\ell_1$,
which corresponds to cases $\mathcal{C}$ and $\mathcal{D}$.
Moreover, the only way a massive pole can occur in case $\mathcal{D}$
is having the minus-helicity gluons adjacent to each other: $m_2=m_1\!+\!1$,
so that $s=m_2\!+\!1=m_1\!+\!2$.

To sum up, for a generic helicity configuration,
case $\mathcal{C}$ contains all physical massive poles:
\begin{itemize}
\item $R_{m_1(m_1\!+\!2)(-\ell_1)}$ generate two poles
      $P_{-\ell_1,m_1}^2$ and $P_{-\ell_1,m_1\!+\!1}^2$;
\item subsequent $R_{m_1 s (-\ell_1)}$ with $s \in \{m_1\!+\!3,\dots,m_2\}$
      each have only one pole $P_{-\ell_1,s-1}^2$.
\end{itemize}
The configuration with two adjacent minus-helicity gluons
generates a single physical massive pole $P_{-\ell_1,m_1}^2$
through $R_{m_1(m_2\!+\!1)(-\ell_1)}$ which belongs to case $\mathcal{D}$.
All other non-MHV-like loop-dependent poles are spurious
and thus can be omitted in the sum over residues.

\subsection{Massive pole residues}
\label{sec:massivepoles}

In this section, we specify how we take residues of massive poles.
If we have such a pole
\be
      (\ell_1-P_{1,i})^2 =  P_{1k}^2 \frac{ \bra{\la}Q_i|\lb] }
                                       { \bra{\la}P_{1k}|\lb] } ,
\label{pole1i}
\ee
after using \eqref{CbubN2}, it becomes proportional to $\braket{\la|Q_i|K|\la}$ .
Then from the definitions of $Q_i$,
$P^i_\pm$ and $x^i_\pm$ in \eqn{Qi4dim}, \eqref{Ptri} and \eqref{Xtri}, respectively,
one can deduce that
\be
            \braket{\la|Q_i|K|\la}
              = - \frac{ \braket{\la|\la^i_+}[\lb^i_+|\lb^i_-]
                         \braket{\la^i_-|\la} }
                       { x^i_+ - x^i_- } .
\label{trianglepoles}
\ee
This lets us split a massive pole into two massless ones,
which is why we introduce momenta $P^i_\pm$ in the first place.
So after taking the residues in the standard way \eqref{simplespinorpole}
and doing some simplifications we obtain the following simple prescription:
\be
            \Res_{\lambda=\la^i_\pm} \frac{ F(\la,\lb) }
                                          { \bra{\la}P_{1k}|\lb]
                                            \braket{\la|Q_i|P_{1k}|\la} }
                  = - \frac{  F(\la^i_\pm,\lb^i_\pm) }
                           { 4 \left( ( P_{1,i-1}\!\cdot\!P_{i,k} )^2
                                      - P_{1,i-1}^2 P_{i,k}^2
                               \right) } .
\label{massivepoleres}
\ee

The drawback of our method is that it introduces a superficial non-rationality
in otherwise rational coefficient formulas.
Indeed, massless momenta $P^i_\pm$ are defined in \eqn{Ptri} through $x^i_\pm$
which contain a non-trivial square root $\sqrt{ (K\!\cdot\!Q_i)^2 - K^2 Q_i^2 } $,
equal to $\sqrt{ ( P_{1,i-1}\!\cdot\!P_{i,k} )^2 - P_{1,i-1}^2 P_{i,k}^2 }$.
However, this square root dependence is guaranteed to effectively cancel
in the sum over $\pm$-solutions.

Other methods may produce explicitly rational expressions, such as
the three-mass triangle formula from \rcite{BjerrumBohr:2007vu,Dunbar:2009ax},
where our approach \eqref{Ctri} would generate superficially non-rational results.
We leave dealing with this minor issue for future work.

\section{Checks}
\label{sec:checks}

Having explained all the ingredients of our calculation,
for the explicit results we refer the reader to Sections 6 and 7 of \citePaper{2}.
Their further use is aided by the Mathematica implementation of the lengthy formulas,
which is distributed along with that paper
(see Appendix F therein for more information).
Now we summarize the checks that we we used to ensure the validity of our results.

First of all, we verified that the sum of all bubble coefficients~\eqref{bubblesum}
coincides numerically with the tree amplitude,
as discussed in \sec{sec:singular}.
We ensured this for all distinct helicity configurations at 6, 7 and 8 points.

As another strong and independent cross-check,
we compared our results with numerical data
kindly produced with the help of the powerful NGluon package
\cite{Badger:2010nx} by one of its authors.
To simulate the $\mathcal{N}=1$ chiral multiplet in the loop,
we had to add separate contributions from the fermion and the scalar loop.
Moreover, to remove the discrepancies
due to different implementation of spinor-helicity formalism,
we compared ratios of the master integral coefficients to the tree amplitude.
In this way, we witnessed agreement
for all types of coefficients
up to machine precision of 13 digits for 8-point amplitudes
and 12 digits for 17-point amplitudes.

Producing numerical tests for a large number (such as 25) of external gluons
becomes more involved, as their kinematics gets more and more singular.
There are numerical instabilities at the level of coefficient/tree ratios
which we believe to come from the spurious poles in $R_{rst}$ \eqref{Rrst}.
They cancel in the sum over $s$ and $t$, but can contaminate the numerical accuracy.
In fact, this issue occurs for the tree amplitude itself.

\section{Discussion}
\label{sec:oneloopdiscussion}

In this chapter, we have studied one-loop NMHV amplitudes
in $\mathcal{N}=1$ super-Yang-Mills theory for any number of external gluons
and showed a way find general analytic formulas
for all missing scalar integral coefficients:
\begin{itemize}
\item bubbles with arbitrary helicity assignment;
\item two-mass-easy and one-mass boxes with two minus-helicity gluons attached to one of the massive corners, but otherwise arbitrary.
\end{itemize}
We have also numerically verified the remaining all-$n$ formulas
calculated previously in \rcite{Dunbar:2009ax}
which are provided in Appendices~B and~C of \rcitePaper{2}
for completeness.

Our principal method was spinor integration
\cite{Britto:2005ha,Anastasiou:2006jv}.
It is a general one-loop method which combines mathematical elegance
with simplicity of computer implementation.
Even though we adapted it to the case of bubbles
with massless $\mathcal{N}=1$ chiral supermultiplet in the loop,
the method is general and can also be applied to
theories with massive particle content
\cite{Britto:2006fc}
and arbitrary loop-momentum power-counting
\cite{Anastasiou:2006gt}.

For all our results, we performed numerical tests at 8 and 17 points
and found agreement with numerical data produced by other methods.

Thus, NMHV amplitudes $\mathcal{N}=1$ SYM
add to the body of one-loop amplitudes known for all $n$.
Of course, such amplitudes were already numerically accessible
for phenomenological studies for multiplicities of order 20
\cite{Giele:2008bc,Badger:2010nx}.
Hopefully, new analytic results will prove useful
for the search of general mathematical structure of amplitudes,
such as recursion relations between separate coefficients
or their meaning in (momentum) twistor space \cite{Hodges:2009hk}.

\begin{figure}[t]
      \centering
      \includegraphics[scale=1.0]{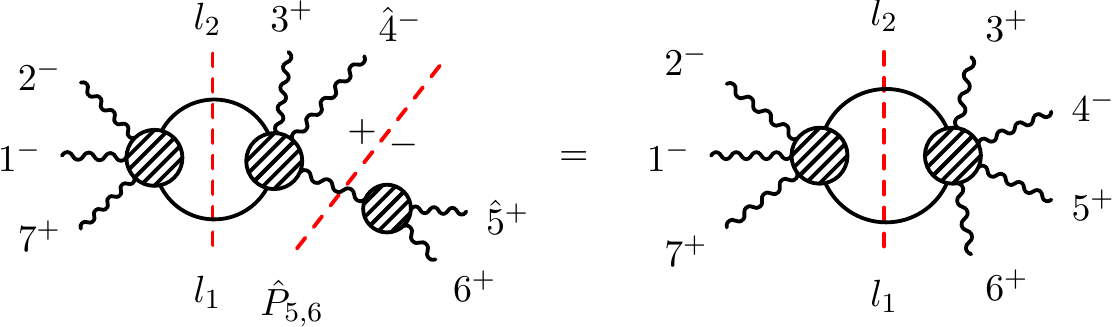}
      \vspace{-5pt}
      \caption{\small Recursion for the $P_{7,2}$-channel cut for
               $ A_{\mathcal{N}=1 \text{ chiral}}^{\text{1-loop,NMHV}}
                   (1^-,2^-,3^+,4^-,5^+,6^+,7^+) $
               \label{recursion}}
      \end{figure}

To illustrate one possible train of thought for further developments,
we have found several series of bubble coefficients that obey
simple BCFW recursion relations inherited from the tree amplitudes
which constitute the corresponding unitarity cuts.
For example, consider the coefficients of $P_{n,2}$-channel bubbles
in amplitudes of the form $A(1^-,2^-,3^+,4^-,5^+,\dots,n^+)$.
From our solution~\eqref{simplestbubbleq}, we can obtain:
\beal
   C^{\text{bub},P_{n,2}}_{\mathcal{N}=1 \text{ chiral}}
      (1^-,2^-,3^+,&4^-,5^+,\dots,n^+) = \\
      - \frac{1}{(4\pi)^{\frac{d}{2}} i} \frac{ \bra{4}P_{n,2}|n]^2 }{ [12] }
        \bigg\{&
            \frac{ \bra{4}3|n] }
                 { \braket{45} \dots \braket{n\!-\!2|n\!-\!1} \braket{n\!-\!1|3}
                   \bra{3}P_{n,2}|2] \bra{3}P_{n,2}|3] [n1] } \\
          +&\frac{ \braket{12} \bra{4}P_{n,2}|2]^2 }
                 { P_{n,2}^2 \braket{34} \dots \braket{n\!-\!2|n\!-\!1}
                   \bra{2}P_{n,2}|2] \bra{3}P_{n,2}|2]
                   \bra{n\!-\!1}P_{n,2}|2] } \\
          +&\frac{ \braket{4|n\!-\!1} }
                 { \braket{34}\braket{45} \dots \braket{n\!-\!2|n\!-\!1}
                   \braket{n\!-\!1|3} }
            \frac{ \bra{4}n\!-\!1|n] }
                 { \bra{n\!-\!1}P_{n,2}|2]
                   \bra{n\!-\!1}P_{n,2}|n\!-\!1] [n1] }
        \bigg\} .
\label{N1chBubn2}
\eeal
It is easy to check that these coefficients
are recursively related through the $[45\rangle$-shift for $n>7$.
However, the recursion fails if one tries to derive the 7-point coefficient
from the 6-point one, even though the cuts still satisfy that relation,
which is drawn in \fig{recursion}.

Of course, such a relation would also fail
even if we flip the helicity of the 3rd gluon.
That would produce the split-helicity case in which the recursion
is well understood and takes place only if one packs adjacent scalar bubble integrals
into two-mass triangles with a Feynman parameter in the numerator
and then works with coefficients of that modified basis \cite{Bern:2005hh}.
The problem is that, unlike the split-helicity case,
the NMHV integral basis consists not only from bubbles,
but also from three-mass triangles and various boxes,
and it is not yet understood how to repackage the full set
of one-loop integrals to make the recursion work.

This brings about another example within the same NMHV amplitude family:
we witnessed the validity of the $[45\rangle$-shift relation
between three-mass triangles $(23,4567,81)$ and $(23,456,71)$,
but not between $(23,456,71)$ and $(23,45,61)$.

In both examples, the recursion seems to work, but for some reason later than expected,
which leaves it unreliable for any predictive calculations.
However, it seems to be the perfect tool
to obtain better understanding of the underlying structure
of the NMHV amplitudes beyond the tree level.
For instance, impressive developments in $\mathcal{N}=4$ SYM
at the all-loop integrand level~\cite{ArkaniHamed:2012nw}
also heavily rely on the BCFW construction implemented in super-twistor variables.
It then seems natural that the on-shell recursion might eventually prove helpful
to tame \emph{integrated} loop amplitudes as well, hopefully, 
for arbitrary configurations of negative and positive helicities.

\chapter{Gauge theory and gravity}
\label{chap:bcj}

Over the last decade,
the progress in the field of scattering amplitudes
has revealed beautiful mathematical  structures,
hidden from the Lagrangian point of view.
One of them, discovered by Bern, Carrasco Johansson~\cite{Bern:2008qj,Bern:2010ue},
is a duality between the kinematic and color content of gauge theories.
That is, amplitudes can be constructed using kinematic numerators
that satisfy Lie-algebraic relations mirroring those of the color factors.
This has both interesting practical and conceptual consequences.
It greatly simplifies the construction of loop amplitudes in (super-)Yang-Mills theory,
and more importantly,
the duality relates multiloop integrands of (super-)Yang-Mills amplitudes
to those of (super-)gravity via a double-copy procedure. 
This has triggered a number of calculations and novel results
in ${\cal N}\ge 4$ supergravity that are otherwise extremely difficult to
obtain~\cite{Carrasco:2011mn,Bern:2012uf,Bern:2011rj,BoucherVeronneau:2011qv,Bern:2012cd,
Bern:2012uc,Bern:2012gh,Carrasco:2012ca,Carrasco:2013ypa,Bern:2013qca,Bern:2013uka}.

While the duality is a conjecture for loop amplitudes,
it is supported by strong evidence
through four loops in ${\cal N}=4$ SYM~\cite{Bern:2010ue,Bern:2012uf,Boels:2012ew}
and through two loops in pure Yang-Mills theory~\cite{Bern:2010ue,Bern:2013yya}.
Additional evidence comes from the attempts
to understand the kinematic algebra~\cite{Monteiro:2011pc,Boels:2013bi,Monteiro:2013rya}
and the Lagranian formulation~\cite{Bern:2010yg,Tolotti:2013caa} of the duality.
The appearance of similar structures in string theory~\cite{BjerrumBohr:2009rd, Stieberger:2009hq,Tye:2010dd,Mafra:2011kj,Mafra:2011nw,Mafra:2012kh,
Broedel:2013tta,Stieberger:2014hba,Mafra:2014oia},\citePaper{3},
Chern-Simons-matter theories~\cite{Bargheer:2012gv, Huang:2012wr,Huang:2013kca},
and, more recently, scattering equations~\cite{Cachazo:2012uq,Cachazo:2012da,Cachazo:2013gna,Cachazo:2013iea,Monteiro:2013rya}
have shed light on the universality of the duality.

The color-kinematics duality was originally formulated for gauge theories
with all fields in the adjoint representation of the gauge group.
The gravity theories that can be obtained
from the corresponding double copy of adjoint fields
are known as ``factorizable''~\cite{Carrasco:2012ca,Chiodaroli:2013upa}.
For ${\cal N}<4$ supersymmetry, such factorizable gravities have 
fixed matter content that goes beyond the minimal multiplets required by supersymmetry,
thus they are non-pure supergravities.
While this limitation is of little consequence at tree level,
for loop amplitudes the inability to decouple or freely tune the extra matter content
is a severe obstruction.
Many interesting gravity theories, such as Einstein gravity, pure ${\cal N}<4$ supergravity, or supergravities with generic matter, have been inaccessible at loop level from the point of view of color-kinematics duality. In contrast to this, the loop-level construction of pure $4\le{\cal N}\le8$ supergravities is well known~\cite{Bern:2010ue,Bern:2011rj,BoucherVeronneau:2011qv}.

Recently, the color-kinematics duality
has been studied away from the case of purely-adjoint gauge theories.
In \rcites{Huang:2012wr,Huang:2013kca}, the duality was found to hold
for two- or three-dimensional theories
of SU($N_1$)$\times$SU($N_2$) bi-fundamental matter,
whose group-theory structure can be embedded a Lie three-algebra.
Moreover, in \rcite{Chiodaroli:2013upa},
the duality was considered in the context of quiver gauge theories
with fields both in the adjoint and bi-fundamental representations.
However, these instances of the color-kinematics duality
stay somewhat close to the adjoint case:
in the former case,
due to the mentioned embedding using adjoint-like indices in the former case,
and in the latter case,
thanks to the theory formulation through orbifold projections
of adjoint parent theories.
The resulting double-copy prescription of \rcite{Chiodaroli:2013upa}
gave many non-factorizable gravities, even if not pure supergravities.

In this chapter, we eliminate the discussed limitations
formulate a double-copy prescription for amplitude integrands in Einstein gravity,
pure supergravity theories, and supergravities with tunable
non-self-interacting\footnote{By ``non-self-interacting'' we mean
no interactions beyond those required by general coordinate invariance
and supersymmetry, so that matter multiplets do not interact with each other.}
 matter. This is achieved by introducing the color-kinematics duality to gauge theories with matter fields in the fundamental representation.
We note that even at tree level, for four or more fundamental particles,
the double copy that follows from the fundamental color-kinematics duality is distinct
from the field-theory KLT relations~\cite{Kawai:1985xq,Bern:1998sv}.

Our main task is to obtain amplitudes for pure ${\cal N}<4$ supergravities,
including Einstein gravity. To do this, we combine double copies
of the adjoint and fundamental representations with a ghost prescription.
The double copies of fundamental matter are promoted to opposite-statistics states, which then cancel the unwanted matter content
in the factorizable gravities or adjoint double copies.
This construction critically relies on the double copy
of fermionic fundamental matter or supersymmetric extensions thereof.
As a naive alternative in the ${\cal N}=0$ case,
the fundamental scalar double copies works at one loop,
but starting at two loops it produces interactions
that no longer cancel the dilaton-axion states.
Although fundamental-scalar amplitudes appear to nontrivially satisfy
the color-kinematics duality, the corresponding double copies indicate
that the gravity amplitudes are corrected by four-scalar terms,
and possibly higher-order interactions,
consistently with the analysis of \rcite{Johansson:2013nsa}.
In this thesis, we mostly limit ourselves to the scalars
that are paired up with fermions within supersymmetric multiplets.

While we do not provide a rigorous proof of our framework,
we show its validity on various example calculations through two loops,
and we give a multiloop argument using the effective R-symmetry
of the tree amplitudes present in unitarity cuts.
As a warm-up, we discuss non-supersymmetric tree-level amplitudes.
At one loop, we obtain all the four-point color-kinematics numerators
with internal fundamental matter, and then we use them to reproduce
all the four-point one-loop supergravity amplitudes with external graviton multiplets and with or without matter in the loop.
As a highly nontrivial check at two loops,
we show in great detail how our prescription cancels the dilaton and axion
in the unitarity cuts of the four- and five-point Einstein gravity.

Interestingly, our approach directly generalizes to amplitudes in (super-)gravity theories
with arbitrary non-self-interacting matter: abelian vectors, fermions and scalars. Indeed,
once we obtain the tools to correctly subtract unwanted matter from loops,
we can reverse the procedure and add more matter instead,
introducing tunable parameters to count the number of matter states.
This generalization has a simple intuitive understanding:
the limited set of factorizable gravities can be regarded as a ``straightjacket''
imposed by the purely-adjoint color-kinematics duality with its unique color representation,
and as soon as we relax this and ``complexify'' the duality to include the fundamental representation, we naturally gain access to a wider range of (super-)gravities.
Indeed, this observation is consistent with the quiver and orbifold constructions
of \rcite{Chiodaroli:2013upa}.\footnote{We note as well that
interesting gravitational matter amplitudes can also be obtained
through adjoint double copies~\cite{Bern:1999bx,CGJR}
that give an arbitrary amount of non-abelian vector multiplets.}


\section{Color-kinematics duality in the adjoint representation}
\label{sec:bcjreview}

In this section,
we briefly review the BCJ duality and the double copy construction,
in the purely-adjoint context.

To begin with, consider a $n$-point $L$-loop color-dressed amplitude in gauge theory
as a sum of Feynman diagrams.
The color factors of graphs with quartic gluon vertices,
written in terms of the structure constants $\tf^{abc}$,
can be immediately understood as sums of cubic color diagrams.
Their kinematic decorations can also be adjusted, in a non-unique way,
so that their pole structure would correspond to that of trivalent diagrams.
This can be achieved
by multiplying and dividing terms by the denominators of missing propagators.
Each four-point vertex can thus be interpreted as an $s$-, $t$- or $u$-channel trees,
of a linear combination of those.
By performing this ambiguous diagram-reabsorption procedure,
one can represent the amplitude as a sum of cubic graphs only:
\be
   {\cal A}^{\text{$L$-loop}}_n = i^{L-1} g^{n+2L-2}\!\!\!\!\!\!\!
                  \sum_{\text{cubic graphs} \; \Gamma_i}
                  \int \prod_{j=1}^{L} \frac{\d^d \ell_j}{(2\pi)^d}
                  \frac{1}{S_i}
                  \frac{c_i \, n_i(\ell)}{D_i(\ell)} ,
\label{Ageneral}
\ee
where the denominators $D_i$, symmetry factors $S_i$ and color factors $c_i$
are understood in terms of the Feynman rules of the adjoint scalar $\phi^3$-theory
(without factors of $i$)
and the numerators $n_i$ generically lose their Feynman rules interpretation.

      \begin{figure}[t]
      \centering
      \includegraphics[scale=1.0]{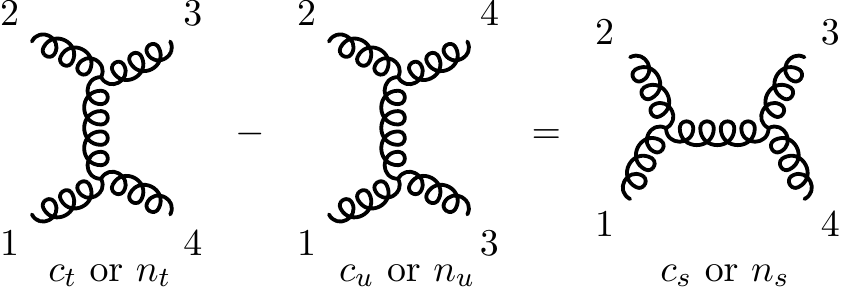}
      \vspace{-5pt}
      \caption{\small Basic Jacobi identity for the color factors or kinematic numerators.}
      \label{fig:jacobi0}
      \end{figure}

\begin{figure}[h]
      \centering
      \includegraphics[scale=1.0]{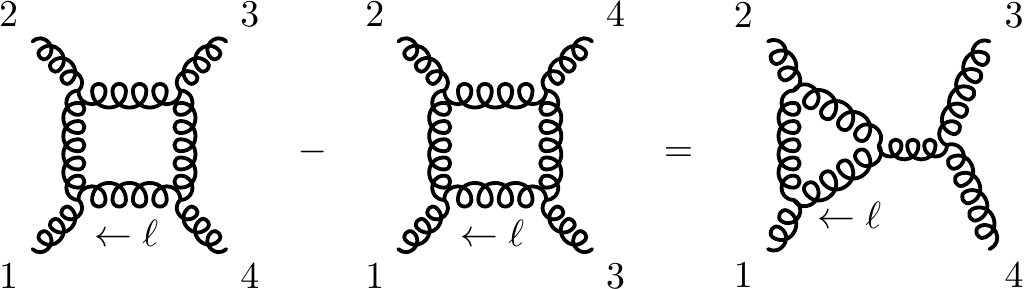}
      \vspace{-5pt}
      \caption{\small Sample Jacobi identity for one-loop numerators}
      \label{fig:jacobi1}
      \end{figure}

Note that the antisymmetry $\tf^{abc}=-\tf^{bac}$ and the Jacobi identity
\begin{equation}
      \tf^{a_2 a_3 b} \tf^{b a_4 a_1}
    - \tf^{a_2 a_4 b} \tf^{b a_3 a_1} =  \tf^{a_1 a_2 b} \tf^{b a_3 a_4} ,
\label{jacobi0}
\end{equation}
shown pictorially in \fig{fig:jacobi0},
induces numerous algebraic relations among the color factors,
such as the one depicted in \fig{fig:jacobi1}.

We are now ready to introduce the main constraint of
\emph{the BCJ color-kinematics duality} \cite{Bern:2008qj,Bern:2010ue}:
let the kinematic numerators $n_i$, defined so far very vaguely,
satisfy the same algebraic identities as their corresponding color factors $c_i$:
\begin{equation} \begin{aligned}
      c_i =-c_j ~~~&\Leftrightarrow~~~ n_i = -n_j  , \\
      c_i - c_j = c_k ~~~&\Leftrightarrow~~~ n_i - n_j = n_k .
\label{duality}
\end{aligned} \end{equation}
This reduces the freedom in the definition of $\{n_i\}$ substantially, but not entirely.
The numerators that obey this duality are usually called the BCJ numerators.
Note that even the basic Jacobi identity \eqref{jacobi0},
obviously true for the four-point tree-level color factors,
is much less trivial when written for the corresponding kinematic numerators.

Once imposed for gauge theory amplitudes, that duality results in
\emph{the BCJ double copy} construction for gravity amplitudes
in the following form:
\be
   {\cal M}^{\text{$L$-loop}}_n
      = i^{L-1} \Big(\frac{\kappa}{2}\Big)^{n+2L-2}\!\!\!\!\!\!\!
        \sum_{\text{cubic graphs} \; \Gamma_i} \int\!\!\frac{d^{Ld}\ell}{(2\pi)^{Ld}}
        \frac{1}{S_i} \frac{n_i n_i'}{D_i} ,
\label{BCJformGravityAdj}
\ee
where only one of the numerator sets, $\{n_i\}$ or $\{n'_i\}$,
needs to obey the color-kinematics duality \eqref{duality}.
In this way, gauge and gravity theories are related
at the integrand level in loop momentum space.
In this thesis, we loosely refer to \eqns{Ageneral}{BCJformGravityAdj},
related by the duality \eqref{duality}, as \emph{the BCJ construction}.

      A comment is due at loop level:
the loop-momentum dependence of numerators $n_i(\ell)$ should be traced with care.
For instance, in the kinematic Jacobi identity given in \fig{fig:jacobi1},
one permutes the legs $3$ and $4$, but keeps the momentum $\ell$ fixed,
because it is external to the permutation.
Indeed, if one writes that identity for the respective color factors,
the internal line $\ell$ will correspond to the color index
outside of the basic Jacobi identity of \fig{fig:jacobi0}.
In general, the correct loop-level numerator identities
correspond to those for the unsummed color factors
in which the internal-line indices are left uncontracted.

Formulas \eqref{Ageneral} and \eqref{BCJformGravityAdj}
are a natural generalization of the original discovery at tree level
\cite{Bern:2008qj}.
The double copy for gravity \eqref{BCJformGravityAdj} has been proven in
\rcite{Bern:2010yg} to hold to any loop order,
if there exists a BCJ representation \eqref{Ageneral}
for at least one of the gauge theory copies.
Such representations were found in numerous calculations
\cite{Bern:2010ue,Carrasco:2011mn,Bern:2012uf,Vanhove:2010nf,
Bern:2011rj,BoucherVeronneau:2011qv,Bern:2012cd,Bern:2012gh,Bern:2013uka,
Carrasco:2012ca,Chiodaroli:2013upa, Huang:2012wr,Bern:2013qca}
up to four loops in $\cN=4$ SYM~\cite{Bern:2009kd}.
A systematic way to find BCJ numerators is known
for Yang-Mills theory at tree level~\cite{Tolotti:2013caa},
and in $\cN=4$ SYM at one loop~\cite{Bjerrum-Bohr:2013iza}.
Moreover, for a restricted class of amplitudes
in the self-dual sectors of gauge theory and gravity,
one can trace the Lagrangian origin of the infinite-dimensional kinematic Lie algebra
\cite{Monteiro:2011pc,Boels:2013bi}.

The string-theoretic understanding of the double copy at tree level
dates back to the celebrated KLT relations \cite{Kawai:1985xq}
between tree-level amplitudes in open and closed string theory,
later improved with the discovery of monodromy relations and the momentum kernel
in
\rcites{Stieberger:2009hq,BjerrumBohr:2009rd,Tye:2010dd,BjerrumBohr:2010zs,
BjerrumBohr:2010hn}.
In the field theory limit, these relations implement the fact that
in amplitudes the degrees of freedom of a graviton can be
split off into those of two gauge bosons.
Recently, a new representation of the integrands open string amplitudes was
proposed by Mafra, Schlotterer and Stieberger~\cite{Mafra:2011kj}
where a new chiral block representation of the
open-string integrands, was introduced to construct BCJ numerators at $n$
points. All of this is applicable at tree level, whereas at loop level, the
relationship between open and closed string \emph{amplitudes} becomes obscure.

At the integrand level, five-point amplitudes were recently discussed
in \rcite{Green:2013bza} in open and closed string theory.
The authors of that work studied how the closed-string integrand
is related to the square of the open-string integrand,
and observed a detailed squaring behavior.
They also discussed the appearance of left-right mixing terms in this context.

Another attempt to shed light on the string-theoretic origin of the BCJ duality
was made in \rcitePaper{3}, where both gauge theory and gravity
were considered in the purely closed-string context.
It is shown that at tree level one can adapt the same mechanism
that was used for open-string integrands in \rcite{Mafra:2011kj},
\ie the MSS chiral block representation, to the closed string.
The loop-level analysis concentrated on the detailed comparison of
the one-loop four-point integrands in ${\cal N}=2$ SYM,
obtained in the BCJ form from field theory,
and in the worldline proper-time form from string theory.
On the one hand,
the BCJ representation is known to induce total derivatives
which integrate to zero in gauge theory,
but in the double copy produce the contributions
necessary to reproduce correct gravity amplitudes.
On the other hand,
in the heterotic string construction of Yang-Mills amplitudes,
the left- and right-moving sector do not communicate to each other
as they have different target spaces,
whereas in gravity amplitudes the two sectors mix due to left-right contractions.
The analysis in \rcitePaper{3} illustrated the relationship between
these left-right contractions
and the squares of the total derivatives in the loop momentum space.

\section{Color-kinematics duality in the fundamental representation}
\label{sec:bcjfund}

Now we give the general construction of the color-kinematics duality for scattering amplitudes in theories that have \emph{both} adjoint and fundamental particles.

Once again, we write the amplitudes of
$d$-dimensional (super-)Yang-Mills theory with fundamental matter as\footnote{We
use a different numerator normalization compared to \rcite{Bern:2010ue}.
Relative to that work, we absorb one factor of $i$ into the numerators,
giving a uniform overall $i^{L-1}$ to the gauge and gravity amplitudes.}
\be
   {\cal A}^{\text{$L$-loop}}_n = i^{L-1} g^{n+2L-2}\!\!\!\!\!\!\!
                  \sum_{\text{cubic graphs} \; \Gamma_i}
                  \int\!\!\frac{d^{Ld}\ell}{(2\pi)^{Ld}}
                  \frac{1}{S_i} \frac{n_i c_i}{D_i} ,
\label{BCJformYM}
\ee
where now the sum runs over all $n$-point $L$-loop graphs $\Gamma_i$ with trivalent vertices of two kinds, (adj., adj., adj.) and (adj., fund., antifund.), which correspond to particle lines of two types: adjoint vector and fundamental matter.
The color factors $c_i$ are built out of products of structure constants $\tf^{abc}$ and generators $T^{a}_{j \bar k}$ matching the respective vertices.
As before, each denominator $D_i$ is the product of the squared momenta of all internal lines of the graph, thus containing all the physical propagator poles.
$S_i$ are the usual combinatorial symmetry factors
that appear, for example, in Feynman diagrams.
The numerators $n_i$ are functions of momenta, polarizations
and other relevant quantum numbers, excluding color.

      \begin{figure}[t]
      \centering
      \includegraphics[scale=1.0]{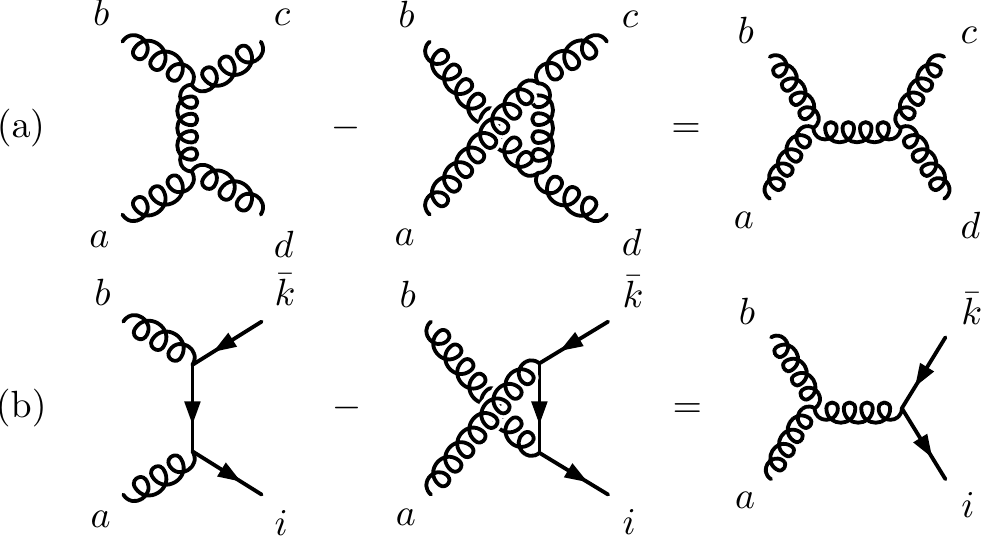}
      \vspace{-5pt}
\caption[a]{\small The algebra in the adjoint vector case~(a)
and in the fundamental matter case~(b).
Curly lines represent gluons or vector supermultiplets,
solid lines represents fermions, scalars or supersymmetric matter.
Alternatively, these diagrams describe standard Lie-algebra relations
for the color factors of the gauge group, as indicated by the external labels.}
      \label{fig:fjacobi}
      \end{figure}

The color factors in \eqref{BCJformYM} now satisfy linear relations of the schematic form
\be
   c_i - c_j = c_k ,
\label{Jacobi}
\ee
which originate from the Jacobi identity and the commutation relation for the generators of the gauge group:
\begin{subequations} \begin{align}
      \tf^{dae} \tf^{ebc} - \tf^{dbe} \tf^{eac} & = \tf^{abe} \tf^{dec}
      \label{jacobi} , \\
      T^{a}_{i \bar \jmath} \, T^{b}_{j \bar k}   -
      T^{b}_{i \bar \jmath} \, T^{a}_{j \bar k} & = \tf^{abe} \, T^{e}_{i \bar k}
      \label{commutation} .
\end{align} \label{chiraljacobi} \end{subequations}
\!\!\!\!\;These relations are depicted diagrammatically in \fig{fig:fjacobi}.
It is important to note the normalization convention in which the right-hand side of \eqn{commutation} is free of factors of $i$ and $\sqrt{2}$.
Additionally, we normalize the generators so that $\Tr(T^{a} T^{b})=\delta^{ab}$,
which, together with \eqn{commutation}, implies that
\be
   \tf^{abc} = \Tr([T^{a},T^{b}]T^{c}) = \sqrt{2}i f^{abc} ,
\ee
where $f^{abc}$ are the structure constants more commonly found in the literature.
These conventions imply that the generators are hermitian,
$(T^{a}_{i \bar \jmath})^*=T^{a}_{j \bar \imath}$, and the structure constants are imaginary\footnote{The choice of imaginary structure constants yields
hermitian generators in the adjoint representation:
$ (T^{a}_{\rm adj})_{bc} = \tf^{bac} ~\Rightarrow~
  (T^{a}_{\rm adj})_{bc} = (T^{a}_{\rm adj})_{cb}^* $.}
$(\tf^{abc})^*=-\tf^{abc}$.

The second property of the representation \eqref{BCJformYM} is the $Z_2$-freedom
in the definition of the color factors:
the interchange of two adjoint indices of the same vertex amounts to a flip in the overall sign of the color factor, \ie
\be
   c_j \rightarrow -c_j ~:~~~~~~  (\ldots \tf^{abc}\ldots) \rightarrow
     - (\ldots \tf^{abc}\ldots) = (\ldots \tf^{bac}\ldots) .
\label{antisymmetry}
\ee

To streamline the notation, we adopt the convention that the interchange of a fundamental and an antifundamental leg also induces a sign flip of the color factors.
This amounts to introducing new generators $T^{a}_{\bar \imath j}$
trivially related to the standard ones:
\be
   T^{a}_{\bar \imath j} = -T^{a}_{j \bar \imath} .
\label{antifundamentalgenerator}
\ee
When these generators have suppressed superscripts, we write them as 
$\overline{T}^{a}$.
With this convention we note that a color factor picks up a minus sign
whenever two legs attached to a cubic vertex of any representation are interchanged.
In practice, this means that we can identify the relative sign of a vertex with the cyclic orientation of its three legs, thus clockwise and counterclockwise orderings have different overall signs.

While we mainly consider matter particles to be in the fundamental representation of the gauge group, it is sometimes convenient to work with adjoint matter. The latter is obtained by simply swapping the generators as follows:
\be
      T^{a}_{i \bar \jmath} \rightarrow \tf^{bac}
      ~~{\rm with}~~i\rightarrow b,\, \bar \jmath \rightarrow c .
\label{representationswap}
\ee
By applying this rule, one goes from a complex representation to a real one,
which makes the color factors of matter and antimatter fields
indistinguishable from one another.
Let us illustrate how this affects a one-loop amplitude
in the form~\eqref{BCJformYM},
which is relevant for the explicit calculations of \sec{sec:YM}.

\begin{figure}[t]
      \centering
      \includegraphics[scale=0.22]{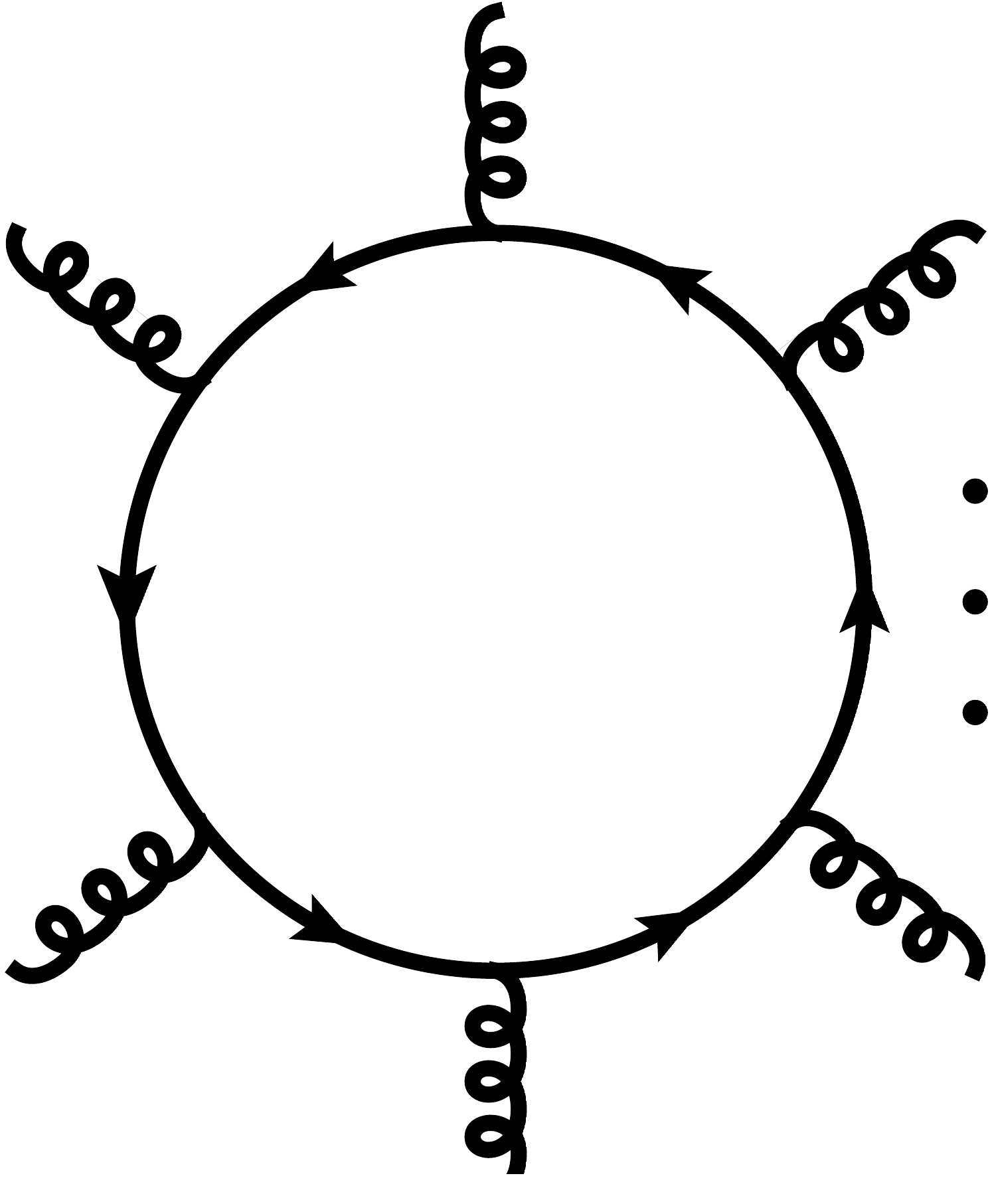}
      \vspace{-10pt}
\caption[a]{\small A typical one-loop graph
with external adjoint and internal fundamental particles}
\label{fig:NgonParentFund}
\end{figure}

Consider the one-loop ``ring diagram'' shown in \fig{fig:NgonParentFund}.
Its color factor is
\be
   c_{12 \cdots n} = \Tr(T^{a_1}T^{a_2} \dots T^{a_n})
   = T^{a_1}_{i \bar \jmath}T^{a_2}_{j \bar k} \dots T^{a_n}_{l \bar \imath} .
\label{ringdiagram}
\ee
Now consider the same diagram, but with the internal arrows reversed.
We denote the operation of reversing fundamental matter arrows (matter $\leftrightarrow$ antimatter) by a bar, hence the corresponding color factor is given by
\be
   \overline{c}_{12 \dots n}
   = \Tr(\overline{T}^{a_1}\overline{T}^{a_2} \dots \overline{T}^{a_n})
   = T^{a_1}_{\bar \imath j}T^{a_2}_{\bar j  k} \dots T^{a_n}_{\bar l i}
   = (-1)^n \, \Tr (T^{a_n} \dots T^{a_2} T^{a_1}) .
\label{ringdiagramconj}
\ee
These two diagrams have the same propagators but different color factors.
However, if we promote them to the adjoint representation using the rule \eqref{representationswap},
both color factors \eqref{ringdiagram} and \eqref{ringdiagramconj} are mapped to the same object
\be
   c^{\rm adj}_{12 \dots n}=\tf^{b a_1 c} \tf^{c a_2 d} \dots \tf^{e a_n b} .
\label{ringdiagramadjoint}
\ee
Since the two graphs have now identical color factors,
it becomes natural to repackage them as follows:
\be
   c_{12 \dots n} n_{12 \dots m}
   + \overline{c}_{12 \dots n} \overline{n}_{12 \dots n}
   ~~\longrightarrow ~~
   c^{\rm adj}_{12 \dots n} (n_{12\dots n}
   + \overline{n}_{12 \dots n}) = c^{\rm adj}_{12 \dots n} n^{\rm adj}_{12 \dots n} ,
\label{adjointPromotion}
\ee
where $\overline{n}_{12 \dots n}$ denotes the antimatter contribution,
and $n^{\rm adj}_{12 \dots n}$ defines an effective numerator
for adjoint matter inside the loop. As can be easily verified,
for generic one-loop numerators with adjoint external states and internal matter,
the same convenient relation exists between the fundamental and adjoint contributions,
\be
      n_i^{\rm adj} = n_i + \overline{n}_i .
\label{nplusnbar}
\ee
In the case that the matter multiplet is effectively non-chiral, implying that $n_i$ is effectively CPT-invariant and $n_i=\overline{n}_i$, the promotion in \eqn{adjointPromotion} gives a numerator from a non-minimal adjoint-matter multiplet.
Then the definition $n_i^{\rm adj} = n_i= (n_i+\overline{n}_i)/2$ may be more convenient to use, since it gives the minimal-multiplet contribution.

Returning to the general multiloop amplitude, we note that thus far the formula~\eqref{BCJformYM} is a trivial rewrite of standard (super-)Yang-Mills perturbation theory.
The only minor change is that we have implicitly absorbed the quartic interactions into the cubic graphs.
To expose a duality between color and kinematics~\cite{Bern:2008qj,Bern:2010ue},
we need to enforce nontrivial constraints on the kinematic numerators,
effectively making them behave as objects of a kinematic Lie algebra.
In particular, we demand that the numerator factors obey
the algebraic relations that are ubiquitous for color factors of any  Lie algebra.
Namely, the Jacobi/commutation relations~\eqref{Jacobi}
and the antisymmetry~\eqref{antisymmetry} under the interchange of legs attached to a single vertex.  These imply dual relations schematically written as follows:
\begin{subequations} \begin{align}
      c_i - c_j = c_k ~~~&\Leftrightarrow~~~ n_i - n_j = n_k , \\
      c_i = -c_j  ~~~&\Leftrightarrow~~~ n_i = -n_j .
\end{align} \label{fduality} \end{subequations}
\!\!\!\!\;For every such three-term color identity,
there is a corresponding kinematic numerator identity,
and for each sign flip in a color factor,
there is a corresponding sign flip in the kinematic numerator.

As before, the three-term identities for numerators take the same pictorial form
as the Lie-algebra identities for color factors, as shown in \fig{fig:fjacobi}.
However, they now represent the kinematic constraints
that can be consistently imposed on the interactions
involving not only adjoint but also fundamental particles.
That these constraints are consistent
with the amplitudes of various (super-)Yang-Mills theories is highly nontrivial.
Recall that in the purely-adjoint case the constraints~\eqref{duality}
lead to the BCJ relations~\cite{Bern:2008qj} between tree-level amplitudes,
which were proven in the context of
string~\cite{Stieberger:2009hq,BjerrumBohr:2009rd,BjerrumBohr:2010zs,
BjerrumBohr:2010hn,Barreiro:2013dpa}
and field theory~\cite{Feng:2010my,Chen:2011jxa,Du:2011js,Cachazo:2012uq}.
In the following, we will show that imposing the constraints~\eqref{fduality}
is sensible as well,
so we expect to find new relations for amplitudes with fundamental particles shortly.

\begin{figure}[t]
      \centering
      \includegraphics[scale=0.73]{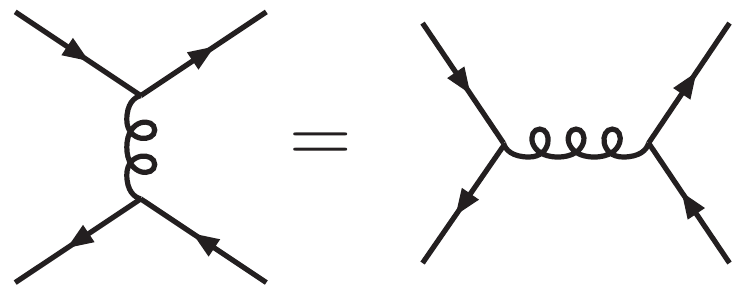}
      \vspace{-5pt}
\caption[a]{\small The additional two-term kinematic identity
that can be imposed on the kinematic numerators in particular situations.
However, it is not mandatory for obtaining a color-kinematics representation.
In general, the corresponding color factors do not satisfy this relation.}
\label{fig:TwoTermId}
\end{figure}

In addition to the three-term Jacobi/commutation identities~\eqref{chiraljacobi},
in \sec{sec:treeexamples} we will observe
that we can impose another nontrivial two-term constraint,
shown in \fig{fig:TwoTermId}, valid for fundamental matter numerators
in particular situations.
For the four-point pure-matter amplitude, the identity is simply
\be
   n_t = n_s ,
\label{4ptTwoTerm}
\ee
where $n_s$ and $n_t$ are the $s$- and $t$-channel kinematic numerators, respectively. 
It is consistent for indistinguishable complex scalars in any spacetime dimension,
if they have the following Lagrangian:
\be
   {\cal L}_{\rm scalar\;matter}
      = (\overline{D_\mu \phi})_{\bar \imath} (D^\mu \phi)_i
      - \frac{g^2}{4} (\overline{\phi}_{\bar \imath} T^a_{i \bar \jmath} \phi_j)
                      (\overline{\phi}_{\bar k} T^a_{k \bar l} \phi_l) .
\label{scalarlagrangian}
\ee
For indistinguishable minimal-Lorentz-representation fermions
in $d=3,4,6,10$ dimensions, the identity should also hold.
This can be explained by considering the real or non-chiral version
of the two-term identity,
\ie by converting between complex and real scalars, or Weyl and Majorana fermions.
For a real scalar, the identity becomes equivalent
to the adjoint Jacobi relation for kinematic numerators (with one term equal to zero),
which holds at tree level because the (YM + scalar)-theory is equivalent to pure Yang-Mills theory in the higher-by-one spacetime dimension.
For the non-chiral (Majorana) fermion, the identity again becomes equivalent
to the adjoint Jacobi relation for kinematic numerators,
which holds~\cite{Chiodaroli:2013upa} in $d=3,4,6,10$ due to a Fierz identity,
where the numerators and amplitudes belong to ${\cal N}=1$ SYM theory.

Although this identity can be consistently imposed (at tree level),
it is not necessary for obtaining a BCJ representation for the amplitude.
Without going into details, this can be seen from the fact that
the corresponding color factors will in general not satisfy this relation.
In terms of the generators, it would correspond to
\be
   T^a_{i \bar \jmath} T^a_{k \bar l} \stackrel{?}{=}
   T^a_{i \bar l} T^a_{k \bar \jmath} .
\label{ColorTwoTerm}
\ee
Although fundamental representations of generic gauge groups do not obey this relation,
the generator of U(1) do so, as well as generators of tensor representations
of U($N$)\footnote{For example, consider the $N\otimes N$
symmetric (antisymmetric) representation of U($N$) with generators 
$\widetilde{T}^{a}_{\beta \bar \gamma} = \Tr(M_{\beta} T^{a} M_{\gamma}^{\dagger})$,
where $M_{\beta}$ form a basis of symmetric (antisymmetric) $N$-by-$N$ matrices.}. 
The sufficient condition for the BCJ duality is that
the kinematic numerators obey the same {\it general} relations
as those of the color factors.
Since \eqn{ColorTwoTerm} is not a general identity,
imposing it on the numerators is optional.

An important feature of the amplitude representation in \eqn{BCJformYM} is that it is not unique. As the numerators of individual diagrams are not gauge-invariant, they have a shift freedom that leaves the amplitude invariant. It is known as generalized gauge invariance~\cite{Bern:2008qj, Bern:2010ue}. More precisely, a shift of the numerators $n_i \rightarrow n_i + \Delta_i$
does not change the amplitude, provided that $\Delta_i$ satisfy
\be
   \sum_{i} \int\!\!\frac{d^{LD}\ell}{(2\pi)^{DL}}
                    \frac{1}{S_i} \frac{ \Delta_i c_i }{D_i} =0 .
\ee
The duality conditions~\eqref{duality} constrain to some extent the freedom of the numerators, but not entirely.
The remaining freedom to move terms between different diagrams means that there usually exists many different amplitude representations that obey color-kinematics duality. This can be useful for finding representations that make various properties manifest. But more importantly, generalized gauge invariance provides a guidance for constructing gravity amplitudes out of the gauge-theory numerators.  Any prescription for this must preserve the generalized gauge invariance: the gravity amplitudes cannot depend on the arbitrariness of the gauge-theory numerators. We will use this fact in \sec{sec:ginvariance}.

The purely-adjoint BCJ duality is known~\cite{Bern:2010ue}
to provide the double copy construction~\eqref{BCJformGravityAdj}
for gravity amplitudes.
In \sec{sec:FundDoubleCopy},
we will present a generalization of \eqn{BCJformGravityAdj}
which will include fundamental matter particles.

\section{Tree-level examples}
\label{sec:treeexamples}

In this section, we give examples of the extended BCJ duality
for some simple tree-level amplitudes,
so as to make the formal description of \sec{sec:bcjfund} more concrete,
and to connect to the standard Feynman diagram expansion.

\subsection{Four-point amplitudes with fermions}
\label{sec:ftree4pt}

Consider the massless four-quark amplitude in QCD, $q \bar{q}q \bar{q}$.
If we take legs 1 and 3 to be quarks and legs 2 and 4 to be antiquarks,
and make no distinction between flavors,
then the tree-level amplitude is the sum of the following two Feynman diagrams:
\be
      \parbox{64pt}{
      \begin{fmffile}{qqQQ1} \fmfframe(12,12)(-12,12){
      \fmfsettings
      \begin{fmfgraph*}(50,40)
            \fmflabel{$1^-\!, i\!\!\!\!\!$}{q1}
            \fmflabel{$2^+\!, \bar \jmath\!\!\!\!\!$}{q2}
            \fmflabel{$\!\!\!\!\!\!\!\!3^-\!, k$}{q3}
            \fmflabel{$\!\!\!\!\!\!\!\!4^+\!, \bar l$}{q4}
            \fmfleft{q1,q2}
            \fmfright{q4,q3}
            \fmf{plain_arrow}{q2,v1,q1}
            \fmf{plain_arrow}{q4,v3,q3}
            \fmf{curly,tension=0.3}{v3,v1}
      \end{fmfgraph*} }
      \end{fmffile}
      }
      = -i\,T_{i \bar \jmath}^a T_{k \bar l}^a \,
         \frac{\braket{13} [24]}{s} = -i \frac{c_s n_s}{s} , \quad
      \parbox{60pt}{
      \begin{fmffile}{qqQQ2} \fmfframe(12,12)(-20,12){
      \fmfsettings
      \begin{fmfgraph*}(50,40)
            \fmflabel{$1^-\!, i\!\!\!\!\!$}{q1}
            \fmflabel{$2^+\!, \bar \jmath\!\!\!\!\!$}{q2}
            \fmflabel{$\!\!\!\!\!\!\!\!3^-\!, k$}{q3}
            \fmflabel{$\!\!\!\!\!\!\!\!4^+\!, \bar l$}{q4}
            \fmfleft{q1,q2}
            \fmfright{q4,q3}
            \fmf{plain_arrow}{q4,v1,q1}
            \fmf{plain_arrow}{q2,v2,q3}
            \fmf{curly,tension=0.3}{v2,v1}
      \end{fmfgraph*} }
      \end{fmffile}
      }
      = -i\,T_{i \bar l}^a T_{k \bar \jmath}^a \,
         \frac{\braket{13} [24]}{t} = -i \frac{c_t n_t}{t} .
\label{qqQQ}
\ee

An important property of that amplitude is that
the corresponding $u$-channel diagram cannot be constructed from Feynman rules,
as it is kinematically zero because of the external helicity choice. 
Incidentally, in these diagrams kinematics is dual to color.
That the Feynman rules land exactly on such a representation
is a somewhat accidental property of the particularly simple amplitude,
and it can be traced back to the gauge invariance of the individual diagrams.
Indeed, there exist no (generalized) gauge freedom that can shuffle terms
between the two numerators.
This is clear if we consider the quarks to have two different flavors,
\eg $q \bar{q} q' \bar{q}'$, in which case only the first of the diagrams~\eqref{qqQQ}
is allowed in the amplitude and thus must be gauge invariant by itself.

Let us take a step back and carefully check
that we have a BCJ representation of the form described in the previous section.
First of all, as the only adjoint particle is the intermediate gluon,
neither of the two identities of the kinematical algebra (shown in \fig{fig:fjacobi})
can be enforced.
Interestingly, the diagrams~\eqref{qqQQ} are clearly related by relabeling
$2 \leftrightarrow 4$,
and this is the only relation between the color factors
$T_{i \bar \jmath}^a T_{k \bar l}^a$ and $T_{i \bar l}^a T_{k \bar \jmath}^a$.
This relation\footnote{In the fully-adjoint case,
the relabeling relation also corresponds to the three-term Jacobi identity~\eqref{jacobi}
among the numerators, with one of them being zero.}
is also satisfied by the kinematic numerators $-i\braket{13}[24]$,
up to an extra minus sign due to Fermi-Dirac statistics.
In principle, on the gauge theory side such signs should be traced with care 
by considering the parity of the full permutation of external fermion states.
For example, the permutation $\{1,2,3,4\} \rightarrow \{1,4,3,2\}$
relating graphs~\eqref{qqQQ} is odd,
hence the aforementioned sign.

We also have the option to enforce the two-term identity.
Since the numerators are already unique, the identity should be automatically true. Indeed,
\be
   n_s = \braket{13} [24] = n_t .
\label{fermion4}
\ee

This example may seem trivial on the Yang-Mills side,
but it is highly nontrivial that one can now construct gravity amplitudes
using these numerators.
For example, the four-graviphoton amplitude\footnote{For brevity,
we use plain $M$ to denote gravity amplitudes ${\cal M}$
with omitted coupling constant, \ie with $\kappa=2$.} is given by
\be
   M_4^{\rm tree}(1^-_{\gamma},2^+_{\gamma},3^-_{\gamma},4^+_{\gamma})
      = -i \Big( \frac{n_s^2}{s} + \frac{n_t^2}{t} \Big)
      = -i \braket{13}^2 [24]^2 \Big(\frac{1}{s}+\frac{1}{t}\Big) .
\ee 
And if we have two types of U(1)-vectors, say $\gamma$ and $\gamma'$, we have the following amplitude
\be
   M_4^{\rm tree}(1^-_{\gamma},2^+_{\gamma},3^-_{\gamma'},4^+_{\gamma'})
      = -i \frac{n_s^2}{s} = -i \frac{\braket{13}^2 [24]^2}{s} ,
\label{TwoPhotonexample}
\ee
obtained by obvious removing the $t$-channel graph.

\begin{figure}[t]
      \centering
      \includegraphics[scale=1.00]{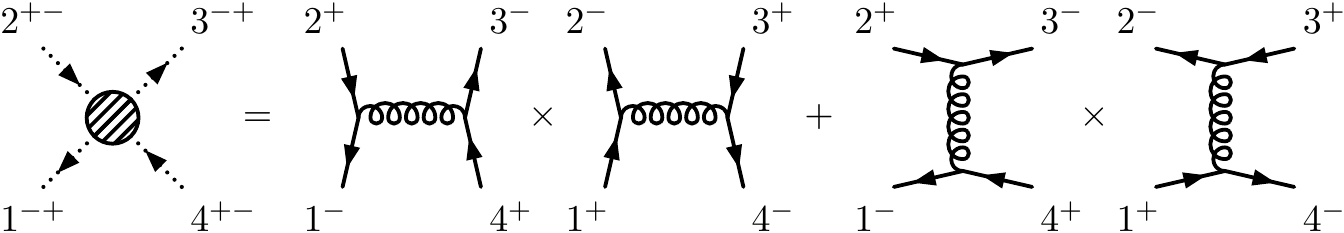}
\caption[a]{\small A double copy~\eqref{dilatonaxionexample} of diagrams
that obeys the color-kinematics duality in the fundamental representation.
It gives a four-scalar amplitude with gravitational interactions.}
\label{example4pt}
\end{figure}

There are many interesting generalizations of these amplitudes,
but let us now focus on the amplitudes
that will be relevant for obtaining pure Einstein gravity.
We are interested in the ``impurities'' that come from dilaton and axion amplitudes.
Since these particles are scalars the relevant amplitudes have no helicity weights,
the double copies should then be such that the external helicities are anticorrelated,
as shown in \fig{example4pt}. The amplitude in that figure is given by
\be
   M_4^{\rm tree}(1^{-+}_\varphi\!,2^{+-}_\varphi\!,3^{-+}_\varphi\!,4^{+-}_\varphi)
      = -i \Big( \frac{n_s \overline{n}_s}{s} + \frac{n_t \overline{n}_t}{t} \Big)
      = -i u^2 \Big( \frac{1}{s}+\frac{1}{t} \Big) = i\frac{u^3}{st} ,
\label{dilatonaxionexample}
\ee
where, as before, $\overline{n}_s$ and $\overline{n}_t$ denote
the numerators in \eqn{qqQQ} with inverted matter and antimatter.
They can be obtained by simple relabeling:
\be
   \overline{n}_s(1^+\!\!,2^-\!\!,3^+\!\!,4^-)
          = - n_t(2^-\!\!,3^+\!\!,4^-\!\!,1^+) , \qquad
   \overline{n}_t(1^+\!\!,2^-\!\!,3^+\!\!,4^-)
          = - n_s(2^-\!\!,3^+\!\!,4^-\!\!,1^+) .
\label{relabeling4}
\ee

Indeed, the states in the amplitude (\ref{dilatonaxionexample}) correspond to a complex scalar $\varphi^{-+}$ that can be understood as a linear combination of the dilaton and axion, $\varphi^{-+} \sim \phi + i a$, and similarly for the complex conjugate
$\varphi^{+-} \sim \phi - i a$.
If we have two different complex scalars, $\varphi$ and $\varphi'$, then we can have the following amplitude by removing the $t$-channel graph:
\be
   M_4^{\rm tree}
      (1^{-+}_\varphi\!,2^{+-}_\varphi\!,3^{-+}_{\varphi'}\!,4^{+-}_{\varphi'})
      = -i \frac{n_s \overline{n}_s}{s}= -i\frac{u^2}{s} .
\label{twoscalarexample}
\ee

\begin{figure}[t]
      \centering
      \includegraphics[scale=1.00]{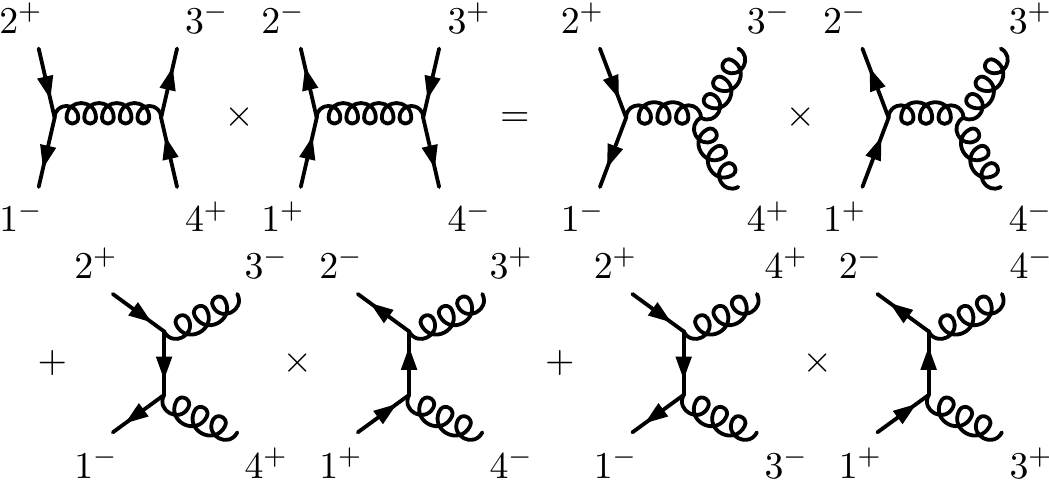}
\caption[a]{\small An equality between two different double copies: with two distinct fermion lines and with a single fermion line. Note that the propagators are implicitly included in this identity, and the $t$- and $u$-channel poles on the second line are spurious.}
\label{FDC4mixedFigure}
\end{figure}

As will be crucial for us in the following,
these dilaton-axion amplitudes were constructed
by taking double copies of fermion states.
This is not the conventional way,
as most often we think of dilaton-axion amplitudes
as arising in the double copy of amplitudes with only gluon states.
In that case, the relation between the gluon polarization vectors
and the polarization tensors of $\varphi^{+-}$ and $\varphi^{-+}$ is more transparent:
\be
   \varepsilon_{\mu\nu}^{+-} = \varepsilon_{\mu}^+ \varepsilon_{\nu}^- , \qquad
   \varepsilon_{\mu\nu}^{-+} = \varepsilon_{\mu}^- \varepsilon_{\nu}^+ .
\label{poltensors}
\ee
In particular, in the KLT approach~\cite{Kawai:1985xq},
we would construct the amplitude in \eqn{dilatonaxionexample}
as a double copy of color-ordered gluon amplitudes:
\be
   M_4^{\rm tree}
      (1^{-+}_\varphi\!,2^{+-}_\varphi\!,3^{-+}_\varphi\!,4^{+-}_\varphi)
      = -i s A_4^{\rm tree}(1^-_{g},2^+_{g},3^-_{g},4^+_{g})
             A_4^{\rm tree}(2^-_{g},1^+_{g},3^+_{g},4^-_{g})
      =  i \frac{u^3}{st} ,
\label{KLT4example}
\ee
which gives the same result as in \eqn{dilatonaxionexample}.
Similarly, one can construct the amplitude in \eqn{twoscalarexample}
through the double copy of gluons and a single fermion pair,
either using the KLT prescription or the BCJ duality for adjoint particles.
The KLT formula is
\be
   M_4^{\rm tree}
      (1^{-+}_\varphi\!,2^{+-}_\varphi\!,3^{-+}_{\varphi'}\!,4^{+-}_{\varphi'})
      = -i s A_4^{\rm tree}(1^-_{f},2^+_{f},3^-_{g},4^+_{g})
             A_4^{\rm tree}(2^-_{f},1^+_{f},3^+_{g},4^-_{g})
      = -i \frac{u^2}{s} ,
\ee
which again agrees with \eqn{twoscalarexample}.
Since the KLT approach is equivalent to the tree-level adjoint BCJ double copy,
we have an interesting equality
between the double copies of a single diagram~\eqref{twoscalarexample}
and a triplet of diagrams, as is illustrated in \fig{FDC4mixedFigure}.

For amplitudes with a single fermion pair,
one can put the fermions either in the adjoint or in the fundamental representation,
and the BCJ duality and double copy work in the same way for both cases.
This, together with the identity in \fig{FDC4mixedFigure},
explains why we were able to obtain the four-point amplitudes from
either the KLT approach or fundamental color-kinematics duality.
However, more generally, at higher points,
the KLT formula~\cite{Kawai:1985xq,Bern:1998sv}
will not be able to reproduce the wide range of gravitational interactions
that can be obtained from the double copy of diagrams
satisfying the extended BCJ duality.
This is because the latter approach
should admit an unlimited number of distinguishable (chiral) matter particles
(see \sec{sec:numerators}),
whereas the KLT formula should be limited to the double-copy spectrum
of the adjoint states naturally occurring in SYM theories.

\subsection{Five-point amplitudes with fermions}
\label{sec:ftree5pt}

The ease of constructing four-point gravity amplitudes is no coincidence.
Let us add another particle to the picture, a gluon,
and observe how the double-copy structure generalizes.
For easier bookkeeping, we first assume that the quarks have distinct flavors,
so the process is of the following type: $q\bar{q}q'\bar{q}'g$.
There are only five nonvanishing tree-level diagrams
that can be constructed from Feynman rules.
For completeness, we give all five contributions,
though some of them are related by relabeling:
\begin{subequations} \begin{align}
      \parbox{91pt}{
      \begin{fmffile}{qqQQg1} \fmfframe(10,17)(0,17){
      \fmfsettings
      \begin{fmfgraph*}(60,40)
            \fmflabel{$\!\!1^-\!, i$}{q1}
            \fmflabel{$2^+\!, \bar \jmath\!\!\!\!\!$}{q2}
            \fmflabel{$3^-\!, k\!\!\!\!\!$}{q3}
            \fmflabel{$\!\!4^+\!, \bar l$}{q4}
            \fmflabel{$\!5, a$}{g5}
            \fmfleft{q2,q3}
            \fmfright{q1,g5,q4}
            \fmf{plain_arrow}{q2,v1,v5,q1}
            \fmf{plain_arrow}{q4,v3,q3}
            \fmf{curly,tension=0}{v3,v1}
            \fmf{curly,tension=0}{g5,v5}
      \end{fmfgraph*} }
      \end{fmffile}
      }
      & = \frac{i}{\sqrt{2}} \frac{1}{s_{15}s_{34}} \,
          T_{i \bar m}^a T_{m \bar \jmath}^b T_{k \bar l}^b \,
          \braket{1|\varepsilon_5|1\!+\!5|3} [24] = -i \frac{c_1 n_1}{D_1} ,
\label{qqQQg1} \\
      \parbox{91pt}{
      \begin{fmffile}{qqQQg2} \fmfframe(10,17)(0,17){
      \fmfsettings
      \begin{fmfgraph*}(60,40)
            \fmflabel{$\!\!1^-\!, i$}{q1}
            \fmflabel{$2^+\!, \bar \jmath\!\!\!\!\!$}{q2}
            \fmflabel{$3^-\!, k\!\!\!\!\!$}{q3}
            \fmflabel{$\!\!4^+\!, \bar l$}{q4}
            \fmflabel{$5, a\!$}{g5}
            \fmfleft{q2,g5,q3}
            \fmfright{q1,q4}
            \fmf{plain_arrow}{q2,v5,v1,q1}
            \fmf{plain_arrow}{q4,v3,q3}
            \fmf{curly,tension=0}{v1,v3}
            \fmf{curly,tension=0}{v5,g5}
      \end{fmfgraph*} }
      \end{fmffile}
      }
      & = - \frac{i}{\sqrt{2}} \frac{1}{s_{25}s_{34}} \,
          T_{i \bar m}^b T_{m \bar \jmath}^a T_{k \bar l}^b \,
          \braket{13} [2|\varepsilon_5|2\!+\!5|4] = -i \frac{c_2 n_2}{D_2} ,
\label{qqQQg2} \\
      \parbox{91pt}{
      \begin{fmffile}{qqQQg3} \fmfframe(10,17)(0,17){
      \fmfsettings
      \begin{fmfgraph*}(60,40)
            \fmflabel{$\!\!1^-\!, i$}{q1}
            \fmflabel{$2^+\!, \bar \jmath\!\!\!\!\!$}{q2}
            \fmflabel{$3^-\!, k\!\!\!\!\!$}{q3}
            \fmflabel{$\!\!4^+\!, \bar l$}{q4}
            \fmflabel{$5, a\!$}{g5}
            \fmfleft{q2,g5,q3}
            \fmfright{q1,q4}
            \fmf{plain_arrow}{q2,v1,q1}
            \fmf{plain_arrow}{q4,v3,v5,q3}
            \fmf{curly,tension=0}{v1,v3}
            \fmf{curly,tension=0}{g5,v5}
      \end{fmfgraph*} }
      \end{fmffile}
      }
      & = \frac{i}{\sqrt{2}} \frac{1}{s_{12}s_{35}} \,
          T_{i \bar \jmath}^b T_{k \bar m}^a T_{m \bar l}^b \,
          \braket{1|3\!+\!5|\varepsilon_5|3} [24] = -i \frac{c_3 n_3}{D_3} ,
\label{qqQQg3} \\
      \parbox{91pt}{
      \begin{fmffile}{qqQQg4} \fmfframe(10,17)(0,17){
      \fmfsettings
      \begin{fmfgraph*}(60,40)
            \fmflabel{$\!\!1^-\!, i$}{q1}
            \fmflabel{$2^+\!, \bar \jmath\!\!\!\!\!$}{q2}
            \fmflabel{$3^-\!, k\!\!\!\!\!$}{q3}
            \fmflabel{$\!\!4^+\!, \bar l$}{q4}
            \fmflabel{$\!5, a$}{g5}
            \fmfleft{q2,q3}
            \fmfright{q1,g5,q4}
            \fmf{plain_arrow}{q2,v1,q1}
            \fmf{plain_arrow}{q4,v5,v3,q3}
            \fmf{curly,tension=0}{v3,v1}
            \fmf{curly,tension=0}{v5,g5}
      \end{fmfgraph*} }
      \end{fmffile}
      }
      & = - \frac{i}{\sqrt{2}} \frac{1}{s_{12}s_{45}} \,
          T_{i \bar \jmath}^b T_{k \bar m}^b T_{m \bar l}^a \,
          \braket{13} [2|4\!+\!5|\varepsilon_5|4] = -i \frac{c_4 n_4}{D_4} ,
\label{qqQQg4} \\
      \parbox{91pt}{
      \begin{fmffile}{qqQQg5} \fmfframe(10,17)(0,17){
      \fmfsettings
      \begin{fmfgraph*}(60,40)
            \fmflabel{$\!\!1^-\!, i$}{q1}
            \fmflabel{$2^+\!, \bar \jmath\!\!\!\!\!$}{q2}
            \fmflabel{$3^-\!, k\!\!\!\!\!$}{q3}
            \fmflabel{$\!\!4^+\!, \bar l$}{q4}
            \fmflabel{$\!5, a$}{g5}
            \fmfleft{q2,q3}
            \fmfright{q1,g5,q4}
            \fmf{plain_arrow}{q2,v1,q1}
            \fmf{plain_arrow}{q4,v3,q3}
            \fmf{curly,tension=0.01}{v1,v5}
            \fmf{curly,tension=0.01}{v3,v5}
            \fmf{curly,tension=0}{g5,v5}
      \end{fmfgraph*} }
      \end{fmffile}
      } & = \!
      \begin{aligned} \\
          \frac{i}{\sqrt{2}} \frac{1}{s_{12}s_{34}} \,
          \tilde{f}^{abc} T_{i \bar \jmath}^b T_{k \bar l}^c \,
          \Big( \bra{1}\varepsilon_5|2] \bra{3}5|4]
              - \bra{1}5|2] \bra{3}\varepsilon_5|4] & \\
              -\,2 \braket{13} [24] ((p_1\!+\!p_2)\!\cdot\!\varepsilon_5) & \Big)
        = -i \frac{c_5 n_5}{D_5} .
      \end{aligned}\!\!\!\!\!\!\!\!\!\!\!\!\!\!\!\!\!\!\!\!
\label{qqQQg5}
\end{align} \label{qqQQg} \end{subequations}
\!\!\!\!\;The amplitude is the sum of these five diagrams.

In \eqn{qqQQg}, we defined five color factors $c_i$
that satisfy two Lie algebra identities:
\begin{subequations} \begin{align}
   c_1 - c_2 = - c_5 ~~~~~ \Leftrightarrow ~~~~~
   T_{i \bar m}^a T_{m \bar \jmath}^b T_{k \bar l}^b -
   T_{i \bar m}^b T_{m \bar \jmath}^a T_{k \bar l}^b & = -
   \tilde{f}^{abc} T_{i \bar \jmath}^b T_{k \bar l}^c , \\
   c_3 - c_4 = c_5 ~~~~~~\:\: \Leftrightarrow ~~~~~
   T_{i \bar \jmath}^b T_{k \bar m}^a T_{m \bar l}^b -
   T_{i \bar \jmath}^b T_{k \bar m}^b T_{m \bar l}^a & =
   \tilde{f}^{abc} T_{i \bar \jmath}^b T_{k \bar l}^c .
\end{align} \end{subequations}
For the amplitude to satisfy the color-kinematics duality,
it is necessary that the kinematic numerators $n_i$ obey similar identities.
As we are not yet probing the four-gluon vertex,
it turns out that these identities automatically hold for the above Feynman diagrams,
provided the gluon polarization vector is transverse.
This can be verified using some spinor algebra involving Schouten identities:
\small
\begin{subequations} \begin{align}
 & n_1 - n_2 = - n_5 ~~~~~ \Leftrightarrow ~~~~~ \\
 & \braket{1|\varepsilon_5|1\!+\!5|3} [24]
   + \braket{13} [2|\varepsilon_5|2\!+\!5|4]
   = - \bra{1}\varepsilon_5|2] \bra{3}5|4]
     + \bra{1}5|2] \bra{3}\varepsilon_5|4]
     +\,2 \braket{13} [24] ((p_1\!+\!p_2)\!\cdot\!\varepsilon_5) . \nn \\
 & n_3 - n_4 = n_5 ~~~~~~\:\: \Leftrightarrow ~~~~~ \\
 & \braket{1|3\!+\!5|\varepsilon_5|3} [24]
   + \braket{13} [2|4\!+\!5|\varepsilon_5|4]
   = - \bra{1}5|2] \bra{3}\varepsilon_5|4]
     + \bra{1}\varepsilon_5|2] \bra{3}5|4]
     +\,2 \braket{13} [24] ((p_3\!+\!p_4)\!\cdot\!\varepsilon_5) . \nn
\end{align} \end{subequations}
\!\!\!\!\;\normalsize

This five-point amplitude is an interesting example
of the interplay between the color-kinematics duality with fundamental particles
and the (generalized) gauge invariance.
The appearance of the kinematical Lie algebra identities is correlated with the fact
that the five-point numerators are allowed to be gauge dependent.
We can see this gauge dependence by parameterizing the polarization vectors
with a reference momentum $q^\mu$, as in \eqn{polvectors}.
If we specialize to the case of the plus-helicity gluon, so that
$\varepsilon^\mu_{p+} = \bra{q}\gamma^\mu|p] / (\sqrt{2} \braket{qp})$,
we obtain the following numerators:
\begin{subequations} \begin{align}
   n_1 & = - \braket{13} [24] \frac{\bra{q}1|5]}{\braket{q5}} , \\
   n_2 & = \,\braket{13} \Big( [24] \frac{\bra{q}2|5]}{\braket{q5}}
                               + [25][54] \Big) , \\
   n_3 & = -\braket{13} [24] \frac{\bra{q}3|5]}{\braket{q5}} , \\
   n_4 & = \,\braket{13} \Big( [24] \frac{\bra{q}4|5]}{\braket{q5}}
                               - [25][54] \Big) , \\
   n_5 & = - n_1 + n_2 = n_3 - n_4 ,
\end{align} \label{qqQQgnumerators} \end{subequations}
\!\!\!\!\;which carry a nontrivial dependence on $q^\mu$.

Note that if all four quarks had the same flavor, $q\bar{q}q\bar{q} g$,
as in \sec{sec:ftree4pt}, we would have to include five more diagrams
with fermion lines stretching between $4^+$ and $1^-$ and between $2^+$ and $3^-$.
Not surprisingly, they can be obtained from the diagrams in \eqn{qqQQg}
by relabeling $1\leftrightarrow3$.
More than that, if we use the two-term identity from \fig{fig:TwoTermId},
four of these numerators can be directly identified with those in \eqn{qqQQgnumerators}:
\beal
   n_6 & \equiv - n_1\big|_{1\leftrightarrow3} = n_3, \\ 
   n_7 & \equiv - n_2\big|_{1\leftrightarrow3} = n_2, \\
   n_8 & \equiv - n_3\big|_{1\leftrightarrow3} = n_1, \\
   n_9 & \equiv - n_4\big|_{1\leftrightarrow3} = n_4, \\
   n_{10} & \equiv - n_5\big|_{1\leftrightarrow3} ,
\label{TwoTermId5pt}
\eeal
where the negative signs come from swapping the fermions $1\leftrightarrow3$.
These relations turn out to be consistent for any value of $q$,
confirming the validity of the two-term identity.

Now, let us construct gravity amplitudes. As before, we can easily construct a four-photon one-graviton amplitude
\beal
   M_5^{\rm tree}(1^{-}_\gamma,2^{+}_\gamma,3^{-}_\gamma,4^{+}_\gamma,5^{++}_h)
      = -i \sum_{i=1}^{10} \frac{n_i^2}{D_i} ,
\label{5ptPhoton}
\eeal
as well as a four-scalar one-graviton amplitude
\be
   M_5^{\rm tree}
      (1^{-+}_\varphi,2^{+-}_\varphi,3^{-+}_\varphi,4^{+-}_\varphi,5^{++}_h)
      = -i \sum_{i=1}^{10} \frac{n_i \overline{n}_i}{D_i} .
\label{FDC5}
\ee

Moreover, one can verify that, in analogy with \eqn{TwoPhotonexample} and \eqn{twoscalarexample}, taking just the five first diagrams~\eqref{qqQQg} results in the gauge-invariant amplitude that corresponds to gravitational interactions of two distinguishable scalars:
\be
   M_5^{\rm tree}(1^{-+}_\varphi,2^{+-}_\varphi,3^{-+}_{\varphi'},4^{+-}_{\varphi'},5^{++}_h)
      = -i \sum_{i=1}^{5} \frac{n_i \overline{n}_i}{D_i} .
\label{FDC5mixed}
\ee
The same procedure can be done for the photon amplitude~\eqref{5ptPhoton}.

At five points, the number of distinct matter particles is still quite small,
so we can reproduce the above amplitudes by applying
the KLT relation~\cite{Kawai:1985xq} in different ways:
\beal
   M_5^{\rm tree}(1,2,3,4,5)
      = i \big( s_{12} s_{34} A_5^{\rm tree}(1,2,3,4,5) & A_5^{\rm tree}(2,1,4,3,5) \\
             +\,s_{13} s_{24} A_5^{\rm tree}(1,3,2,4,5) & A_5^{\rm tree}(3,1,4,2,5) \big) .
\label{KLT5}
\eeal
Using this formula, one can easily check
Eqs.~\eqref{5ptPhoton}, \eqref{FDC5} and~\eqref{FDC5mixed}.

\subsection{Four-point amplitudes with scalars}
\label{sec:stree4pt}

Let us now consider amplitudes with fundamental scalars.
We will show that, even though scalars have more complicated interactions with gluons,
their amplitudes can be represented in a color-kinematics form.

First, consider a free massless complex scalar minimally coupled to QCD
with the following Lagrangian:
\be
   {\cal L}_{\rm scalar\;matter}
      = (\overline{D_\mu \phi})_{\bar \imath} (D^\mu \phi)_i .
\label{freescalarlagrangian}
\ee
The simplest process is $\phi \overline{\phi} \phi \overline{\phi}$.
If we take legs 1 and 3 to be scalars and legs 2 and 4 to be antiscalars,
and make no distinction between flavors,
then the tree-level amplitude is the sum of the following two Feynman diagrams:
\be
      \parbox{64pt}{
      \begin{fmffile}{ssSS1} \fmfframe(12,12)(-12,12){
      \fmfsettings
      \begin{fmfgraph*}(50,40)
            \fmflabel{$1, i\!\!\!\!\!$}{q1}
            \fmflabel{$2, \bar \jmath\!\!\!\!\!$}{q2}
            \fmflabel{$\!\!\!\!3, k$}{q3}
            \fmflabel{$\!\!\!\!4, \bar l$}{q4}
            \fmfleft{q1,q2}
            \fmfright{q4,q3}
            \fmf{dashes_arrow}{q2,v1,q1}
            \fmf{dashes_arrow}{q4,v3,q3}
            \fmf{curly,tension=0.3}{v3,v1}
      \end{fmfgraph*} }
      \end{fmffile}
      }
      = i\,T_{i \bar \jmath}^a T_{k \bar l}^a
        \Big( \frac{u}{s} + \frac{1}{2} \Big)
      = -i \frac{c_s n_s^{\rm free}}{s} , \quad
      \parbox{60pt}{
      \begin{fmffile}{ssSS2} \fmfframe(12,12)(-20,12){
      \fmfsettings
      \begin{fmfgraph*}(50,40)
            \fmflabel{$1, i\!\!\!\!\!$}{q1}
            \fmflabel{$2, \bar \jmath\!\!\!\!\!$}{q2}
            \fmflabel{$\!\!\!\!3, k$}{q3}
            \fmflabel{$\!\!\!\!4, \bar l$}{q4}
            \fmfleft{q1,q2}
            \fmfright{q4,q3}
            \fmf{dashes_arrow}{q4,v1,q1}
            \fmf{dashes_arrow}{q2,v2,q3}
            \fmf{curly,tension=0.3}{v2,v1}
      \end{fmfgraph*} }
      \end{fmffile}
      }
      = i\,T_{i \bar l}^a T_{k \bar \jmath}^a \,
        \Big( \frac{u}{t} + \frac{1}{2} \Big) 
      = -i \frac{c_t n_t^{\rm free}}{t} .
\label{ssSS}
\ee

Obviously, both diagrams are gauge-invariant separately.
Therefore, for the same reason as in the fermion case,
they comprise a BCJ representation.

However, the two-term identity from \fig{fig:TwoTermId} is not satisfied:
\be
   n_s^{\rm free} = -\Big(u+\frac{s}{2}\Big) ~\neq~
   n_t^{\rm free} = -\Big(u+\frac{t}{2}\Big) .
\label{freescalar4}
\ee
Moreover, these numerators are evidently not related by supersymmetry
to those in \eqn{fermion4}.
Indeed, from supersymmetric Ward identities
\cite{Grisaru:1976vm,Grisaru:1977px,Parke:1985pn,Kunszt:1985mg,Nair:1988bq,Dixon:1996wi},
one would rather obtain the following numerators:
\be
   n_s = - u = n_t ,
\label{scalar4}
\ee
related to those in \eqn{fermion4} by the replacement
$\braket{13} \rightarrow \braket{24}$.

As advertised in \sec{sec:bcjfund}, to fix this,
we can consider the following scalar self-interaction term:
\be
   {\cal L}_{\text{self-interaction}} =
      - \frac{g^2}{4} (\overline{\phi}_{\bar \imath} T^a_{i \bar \jmath} \phi_j)
                      (\overline{\phi}_{\bar k} T^a_{k \bar l} \phi_l) ,
\label{selfinteractionlagrangian}
\ee
which adds another Feynman diagram to the process~\eqref{ssSS}:
\be
      \parbox{64pt}{
      \begin{fmffile}{ssSS3} \fmfframe(12,12)(-12,12){
      \fmfsettings
      \begin{fmfgraph*}(50,40)
            \fmflabel{$1, i\!\!\!\!\!$}{q1}
            \fmflabel{$2, \bar \jmath\!\!\!\!\!$}{q2}
            \fmflabel{$\!\!\!\!3, k$}{q3}
            \fmflabel{$\!\!\!\!4, \bar l$}{q4}
            \fmfleft{q1,q2}
            \fmfright{q4,q3}
            \fmf{dashes_arrow}{q2,v,q1}
            \fmf{dashes_arrow}{q4,v,q3}
      \end{fmfgraph*} }
      \end{fmffile}
      }
      = -\frac{i}{2} \big( T_{i \bar \jmath}^a T_{k \bar l}^a
                         + T_{i \bar l}^a T_{k \bar \jmath}^a \big)
      = -i \bigg( \frac{c_s (n_s - n_s^{\rm free})}{s}
                + \frac{c_t (n_t - n_t^{\rm free})}{s} \bigg) .
\label{ssSSselfint}
\ee
Obviously, by symmetrically absorbing this diagram into those in \eqn{ssSS},
one obtains the simpler scalar numerators~\eqref{scalar4}.
Taking into account that those are related to the fermion numerators~\eqref{fermion4},
we regard the four-scalar contact term as needed
for generalizing to supersymmetric theories. 
In the supersymmetric theories, such an interaction is consistent
with the statement that we only work with matter supermultiplets
that are ``non-self-interacting'', \ie they contain no self-interactions
beyond those required by supersymmetry and gauge invariance.

In the rest of this section,
we will consider both cases, with and without scalar self-interaction.
However, we will refrain from building gravity amplitudes out of them,
because in the former case they simply coincide with the fermion double copies,
as follows from supersymmetric Ward identities,
and in the latter case the resulting gravity theory has not been identified yet.

\subsection{Five-point amplitudes with scalars}
\label{sec:stree5pt}

Now let us consider five-point amplitudes with complex scalars.

First, we replace only one of the fermion pairs in \eqn{qqQQg} by complex scalars,
obtaining the following process: $\phi\overline{\phi}q\bar{q}g$.
Note that it is entirely insensitive
to the scalar self-interaction~\eqref{selfinteractionlagrangian}.
There are seven nonvanishing tree-level diagrams that can be constructed from Feynman rules,
two of which contain a four-point $\overline{\phi} A A \phi$ vertex.
To obtain a color-kinematics representation,
we should absorb the latter diagrams into two trivalent diagrams with the same color factors:
\begin{subequations} \begin{align}
    & \parbox{91pt}{
      \begin{fmffile}{ssqqg1} \fmfframe(10,17)(0,17){
      \fmfsettings
      \begin{fmfgraph*}(60,40)
            \fmflabel{$1, i$}{s1}
            \fmflabel{$2, \bar \jmath\!\!$}{s2}
            \fmflabel{$3^-\!, k\!\!\!\!\!$}{q3}
            \fmflabel{$\!\!\!4^+\!, \bar l$}{q4}
            \fmflabel{$\!5, a$}{g5}
            \fmfleft{s2,q3}
            \fmfright{s1,g5,q4}
            \fmf{dashes_arrow}{s2,v1,v5,s1}
            \fmf{plain_arrow}{q4,v3,q3}
            \fmf{curly,tension=0}{v3,v1}
            \fmf{curly,tension=0}{g5,v5}
      \end{fmfgraph*} }
      \end{fmffile}
      }
      +
      \parbox{81pt}{
      \begin{fmffile}{ssqqg1a} \fmfframe(0,17)(0,17){
      \fmfsettings
      \begin{fmfgraph*}(60,40)
            \fmflabel{$1, i$}{s1}
            \fmflabel{$2, \bar \jmath\!\!$}{s2}
            \fmflabel{$3^-\!, k\!\!\!\!\!\!$}{q3}
            \fmflabel{$\!\!4^+\!, \bar l$}{q4}
            \fmflabel{$\!5, a$}{g5}
            \fmfleft{s2,q3}
            \fmfright{s1,g5,q4}
            \fmf{dashes_arrow}{s2,v1,s1}
            \fmf{plain_arrow}{q4,v3,q3}
            \fmf{curly,tension=0}{v3,v1}
            \fmf{curly,tension=0}{g5,v1}
      \end{fmfgraph*} }
      \end{fmffile}
      } = \!
      \begin{aligned} \\ &
          \frac{i}{\sqrt{2}} \frac{1}{s_{15}s_{34}} \,
          T_{i \bar m}^a T_{m \bar \jmath}^b T_{k \bar l}^b \, \\ & \!\times\!
          \Big( 2(p_1\!\cdot\!\varepsilon_5) \bra{3}2|4]
               + (p_1\!\cdot\!p_5) \bra{3}\varepsilon_5|4] \Big)
        = -i \frac{c_1 n_1}{D_1} ,
      \end{aligned}\!\!\!\!\!\!\!\!\!\!\!\!\!\!\!\!\!\!\!\!\!\!\!\!\!\!\!\!\!\!
\label{ssqqg1} \\
    & \parbox{70pt}{
      \begin{fmffile}{ssqqg2} \fmfframe(10,17)(-10,17){
      \fmfsettings
      \begin{fmfgraph*}(60,40)
            \fmflabel{$1, i$}{s1}
            \fmflabel{$2, \bar \jmath\!\!$}{s2}
            \fmflabel{$3^-\!, k\!\!\!\!\!$}{q3}
            \fmflabel{$\!\!\!4^+\!, \bar l$}{q4}
            \fmflabel{$5, a\!$}{g5}
            \fmfleft{s2,g5,q3}
            \fmfright{s1,q4}
            \fmf{dashes_arrow}{s2,v5,v1,s1}
            \fmf{plain_arrow}{q4,v3,q3}
            \fmf{curly,tension=0}{v1,v3}
            \fmf{curly,tension=0}{v5,g5}
      \end{fmfgraph*} }
      \end{fmffile}
      }
      +
      \parbox{86pt}{
      \begin{fmffile}{ssqqg2a} \fmfframe(21,17)(-7,17){
      \fmfsettings
      \begin{fmfgraph*}(60,40)
            \fmflabel{$1, i$}{s1}
            \fmflabel{$2, \bar \jmath\!\!$}{s2}
            \fmflabel{$3^-\!, k\!\!\!\!\!\!$}{q3}
            \fmflabel{$\!\!4^+\!, \bar l$}{q4}
            \fmflabel{$5, a\!$}{g5}
            \fmfleft{s2,g5,q3}
            \fmfright{s1,q4}
            \fmf{dashes_arrow}{s2,v1,s1}
            \fmf{plain_arrow}{q4,v3,q3}
            \fmf{curly,tension=0}{v1,v3}
            \fmf{curly,tension=0}{v1,g5}
      \end{fmfgraph*} }
      \end{fmffile}
      } = \!
      \begin{aligned} \\ &
          \frac{i}{\sqrt{2}} \frac{1}{s_{25}s_{34}} \,
          T_{i \bar m}^b T_{m \bar \jmath}^a T_{k \bar l}^b \, \\ & \!\times\!
          \Big( 2(p_2\!\cdot\!\varepsilon_5) \bra{3}1|4]
               + (p_2\!\cdot\!p_5) \bra{3}\varepsilon_5|4] \Big)
        = -i \frac{c_2 n_2}{D_2} ,
      \end{aligned}\!\!\!\!\!\!\!\!\!\!\!\!\!\!\!\!\!\!\!\!
\label{ssqqg2} \\
    & \parbox{83pt}{
      \begin{fmffile}{ssqqg3} \fmfframe(10,17)(0,17){
      \fmfsettings
      \begin{fmfgraph*}(60,40)
            \fmflabel{$1, i$}{s1}
            \fmflabel{$2, \bar \jmath\!\!$}{s2}
            \fmflabel{$3^-\!, k\!\!\!\!\!$}{q3}
            \fmflabel{$\!\!4^+\!, \bar l$}{q4}
            \fmflabel{$5, a\!$}{g5}
            \fmfleft{s2,g5,q3}
            \fmfright{s1,q4}
            \fmf{dashes_arrow}{s2,v1,s1}
            \fmf{plain_arrow}{q4,v3,v5,q3}
            \fmf{curly,tension=0}{v1,v3}
            \fmf{curly,tension=0}{g5,v5}
      \end{fmfgraph*} }
      \end{fmffile}
      }
        = - \frac{i}{2\sqrt{2}} \frac{1}{s_{12}s_{35}} \,
          T_{i \bar \jmath}^b T_{k \bar m}^a T_{m \bar l}^b \,
          \bra{3}\varepsilon_5|3\!+\!5|1\!-\!2|4] = \frac{c_3 n_3}{D_3} ,
\label{ssqqg3} \\
    & \parbox{91pt}{
      \begin{fmffile}{ssqqg4} \fmfframe(10,17)(0,17){
      \fmfsettings
      \begin{fmfgraph*}(60,40)
            \fmflabel{$1, i$}{s1}
            \fmflabel{$2, \bar \jmath\!\!$}{s2}
            \fmflabel{$3^-\!, k\!\!\!\!\!$}{q3}
            \fmflabel{$\!\!4^+\!, \bar l$}{q4}
            \fmflabel{$\!5, a$}{g5}
            \fmfleft{s2,q3}
            \fmfright{s1,g5,q4}
            \fmf{dashes_arrow}{s2,v1,s1}
            \fmf{plain_arrow}{q4,v5,v3,q3}
            \fmf{curly,tension=0}{v3,v1}
            \fmf{curly,tension=0}{v5,g5}
      \end{fmfgraph*} }
      \end{fmffile}
      }
        = \frac{i}{2\sqrt{2}} \frac{1}{s_{12}s_{45}} \,
          T_{i \bar \jmath}^b T_{k \bar m}^b T_{m \bar l}^a \,
          \bra{3}1\!-\!2|4\!+\!5|\varepsilon_5|4] = -i \frac{c_4 n_4}{D_4} ,
\label{ssqqg4} \\
    & \parbox{91pt}{
      \begin{fmffile}{ssqqg5} \fmfframe(10,17)(0,17){
      \fmfsettings
      \begin{fmfgraph*}(60,40)
            \fmflabel{$\!\!1^-\!, i$}{s1}
            \fmflabel{$2^+\!, \bar \jmath\!\!\!\!\!$}{s2}
            \fmflabel{$3^-\!, k\!\!\!\!\!$}{q3}
            \fmflabel{$\!\!4^+\!, \bar l$}{q4}
            \fmflabel{$\!5, a$}{g5}
            \fmfleft{s2,q3}
            \fmfright{s1,g5,q4}
            \fmf{dashes_arrow}{s2,v1,s1}
            \fmf{plain_arrow}{q4,v3,q3}
            \fmf{curly,tension=0.01}{v1,v5}
            \fmf{curly,tension=0.01}{v3,v5}
            \fmf{curly,tension=0}{g5,v5}
      \end{fmfgraph*} }
      \end{fmffile}
      } = \!
      \begin{aligned} \\
          \frac{i}{\sqrt{2}} \frac{1}{s_{12}s_{34}} \,
          \tilde{f}^{abc} T_{i \bar \jmath}^b T_{k \bar l}^c \,
          \Big( ((p_1\!+\!p_2)\!\cdot\!\varepsilon_5) \bra{3}1\!-\!2|4] & \\
             -\,((p_1\!-\!p_2)\!\cdot\!p_5) \bra{3}\varepsilon_5|4]
              + ((p_1\!-\!p_2)\!\cdot\!\varepsilon_5) \bra{3}5|4] & \Big)
        = -i \frac{c_5 n_5}{D_5} .
      \end{aligned}
\label{ssqqg5}
\end{align} \label{ssqqg} \end{subequations}
\!\!\!\!\;Thanks to this reabsorption, $n_1$ and $n_2$ are associated to their purely trivalent graphs.
The amplitude is the sum of the resulting five contributions.
It can be verified that now both the color factors and the kinematics numerators
obey equivalent relations:
\begin{subequations} \begin{align}
   & c_1 - c_2 = - c_5 ~~~~~ \Leftrightarrow ~~~~~ n_1 - n_2 = - n_5 , \\
   & c_3 - c_4 = c_5 ~~~~~~\:\: \Leftrightarrow ~~~~~ n_3 - n_4 = n_5 .
\end{align} \end{subequations}
Moreover, as fundamental lines are of different origin,
no two-term identities can be imposed in this case,
so the color-kinematics duality is satisfied in its fullest version.

If we specialize to the case of plus-helicity gluon with a reference spinor $\ket{q}$,
we obtain the following explicit numerators:
\begin{subequations} \begin{align}
   n_1 & = - \bra{3}2|4] \frac{\bra{q}1|5]}{\braket{q5}}
           - (p_1\!\cdot\!p_5) \frac{\braket{3q}[54]}{\braket{q5}} , \\
   n_2 & = - \bra{3}1|4] \frac{\bra{q}2|5]}{\braket{q5}}
           - (p_2\!\cdot\!p_5) \frac{\braket{3q}[54]}{\braket{q5}} , \\
   n_3 & = \phantom{-} \frac{ \braket{3q} [5|3\!+\!5|1\!-\!2|4] }
                 { 2 \braket{q5} } , \\
   n_4 & = -\frac{ \braket{3|1\!-\!2|4\!+\!5|q} [54] }
                 { 2 \braket{q5} } , \\
   n_5 & = - n_1 + n_2 = n_3 - n_4 .
\end{align} \label{ssqqgnumerators} \end{subequations}
\!\!\!\!\;On the other hand, we may consider another set of numerators defined by
the SWI-replacement $ \braket{13} \rightarrow - \braket{23} $ in \eqn{qqQQgnumerators}.
In fact, for the gauge choice $\ket{q}=\ket{3}$,
they are equal to the numerators~\eqref{ssqqgnumerators}.
As the amplitude is gauge-independent,
we see that these two numerators sets are equivalent.

Now let us consider the process with no fermions at all,
but for simplicity let the scalars be distinct:
$\phi\overline{\phi}\phi'\overline{\phi}'g$.
If we start by looking at the case with no self-interaction,
then we can construct only five inequivalent tree-level diagrams from Feynman rules,
two of which contain a four-point $\overline{\phi} A A \phi$ vertex.
To obtain a color-kinematics representation, it suffices to include them
into two trivalent diagrams with the same color factors:
\small
\begin{subequations} \begin{align}
    & \parbox{97pt}{
      \begin{fmffile}{ssSSg1} \fmfframe(10,17)(0,17){
      \fmfsettings
      \begin{fmfgraph*}(60,40)
            \fmflabel{$1, i$}{s1}
            \fmflabel{$2, \bar \jmath\!\!$}{s2}
            \fmflabel{$3, k\!\!$}{q3}
            \fmflabel{$4, \bar l$}{q4}
            \fmflabel{$\!5, a$}{g5}
            \fmfleft{s2,q3}
            \fmfright{s1,g5,q4}
            \fmf{dashes_arrow}{s2,v1,v5,s1}
            \fmf{dashes_arrow}{q4,v3,q3}
            \fmf{curly,tension=0}{v3,v1}
            \fmf{curly,tension=0}{g5,v5}
      \end{fmfgraph*} }
      \end{fmffile}
      }
      +
      \parbox{87pt}{
      \begin{fmffile}{ssSSg1a} \fmfframe(0,17)(0,17){
      \fmfsettings
      \begin{fmfgraph*}(60,40)
            \fmflabel{$1, i$}{s1}
            \fmflabel{$2, \bar \jmath\!\!$}{s2}
            \fmflabel{$3, k\!\!$}{q3}
            \fmflabel{$4, \bar l$}{q4}
            \fmflabel{$\!5, a$}{g5}
            \fmfleft{s2,q3}
            \fmfright{s1,g5,q4}
            \fmf{dashes_arrow}{s2,v1,s1}
            \fmf{dashes_arrow}{q4,v3,q3}
            \fmf{curly,tension=0}{v3,v1}
            \fmf{curly,tension=0}{g5,v1}
      \end{fmfgraph*} }
      \end{fmffile}
      } =
      \begin{aligned}
          \frac{i}{\sqrt{2}} \frac{1}{s_{15}s_{34}} \,
          T_{i \bar m}^a T_{m \bar \jmath}^b T_{k \bar l}^b \, \qquad & \\ \!\times\!
          \Big( 2(p_1\!\cdot\!\varepsilon_5) (p_2\!\cdot\!(p_3\!-\!p_4)) & \\
              +\,(p_1\!\cdot\!p_5) (\varepsilon_5\!\cdot\!(p_3\!-\!p_4)) & \Big)
        = -i \frac{c_1 n_1^{\rm free}}{D_1} ,
      \end{aligned}\!\!\!\!\!\!\!\!\!\!\!\!\!\!\!\!\!\!\!\!
\label{ssSSg1} \\
    & \parbox{80pt}{
      \begin{fmffile}{ssSSg2} \fmfframe(10,17)(-10,17){
      \fmfsettings
      \begin{fmfgraph*}(60,40)
            \fmflabel{$1, i$}{s1}
            \fmflabel{$2, \bar \jmath\!\!$}{s2}
            \fmflabel{$3, k\!\!$}{q3}
            \fmflabel{$4, \bar l$}{q4}
            \fmflabel{$5, a\!$}{g5}
            \fmfleft{s2,g5,q3}
            \fmfright{s1,q4}
            \fmf{dashes_arrow}{s2,v5,v1,s1}
            \fmf{dashes_arrow}{q4,v3,q3}
            \fmf{curly,tension=0}{v1,v3}
            \fmf{curly,tension=0}{v5,g5}
      \end{fmfgraph*} }
      \end{fmffile}
      }
      +
      \parbox{104pt}{
      \begin{fmffile}{ssSSg2a} \fmfframe(24,17)(-7,17){
      \fmfsettings
      \begin{fmfgraph*}(60,40)
            \fmflabel{$1, i$}{s1}
            \fmflabel{$2, \bar \jmath\!\!$}{s2}
            \fmflabel{$3, k\!\!$}{q3}
            \fmflabel{$4, \bar l$}{q4}
            \fmflabel{$5, a\!$}{g5}
            \fmfleft{s2,g5,q3}
            \fmfright{s1,q4}
            \fmf{dashes_arrow}{s2,v1,s1}
            \fmf{dashes_arrow}{q4,v3,q3}
            \fmf{curly,tension=0}{v1,v3}
            \fmf{curly,tension=0}{v1,g5}
      \end{fmfgraph*} }
      \end{fmffile}
      } =
      \begin{aligned}
          \frac{i}{\sqrt{2}} \frac{1}{s_{25}s_{34}} \,
          T_{i \bar m}^b T_{m \bar \jmath}^a T_{k \bar l}^b \, \qquad & \\ \!\times\!
          \Big( 2(p_2\!\cdot\!\varepsilon_5) (p_1\!\cdot\!(p_3\!-\!p_4)) & \\
              +\,(p_2\!\cdot\!p_5) (\varepsilon_5\!\cdot\!(p_3\!-\!p_4)) & \Big)
        = -i \frac{c_2 n_2^{\rm free}}{D_2} ,
      \end{aligned}\!\!\!\!\!\!\!\!\!\!\!\!\!\!\!\!\!\!\!\!
\label{ssSSg2} \\
    & \parbox{97pt}{
      \begin{fmffile}{ssSSg5} \fmfframe(10,17)(0,17){
      \fmfsettings
      \begin{fmfgraph*}(60,40)
            \fmflabel{$1, i$}{s1}
            \fmflabel{$2, \bar \jmath\!\!$}{s2}
            \fmflabel{$3, k\!\!$}{q3}
            \fmflabel{$4, \bar l$}{q4}
            \fmflabel{$\!5, a$}{g5}
            \fmfleft{s2,q3}
            \fmfright{s1,g5,q4}
            \fmf{dashes_arrow}{s2,v1,s1}
            \fmf{dashes_arrow}{q4,v3,q3}
            \fmf{curly,tension=0.01}{v1,v5}
            \fmf{curly,tension=0.01}{v3,v5}
            \fmf{curly,tension=0}{g5,v5}
      \end{fmfgraph*} }
      \end{fmffile}
      } =
      \begin{aligned}
          \frac{i}{\sqrt{2}} \frac{1}{s_{12}s_{34}} \,
          \tilde{f}^{abc} T_{i \bar \jmath}^b T_{k \bar l}^c \,
          \Big( ((p_1\!+\!p_2)\!\cdot\!\varepsilon_5)
                ((p_1\!-\!p_2)\!\cdot\!(p_3\!-\!p_4)) & \\
             -\,((p_3\!-\!p_4)\!\cdot\!\varepsilon_5)
                ((p_1\!-\!p_2)\!\cdot\!p_5) & \\
             +\,((p_1\!-\!p_2)\!\cdot\!\varepsilon_5)
                ((p_3\!-\!p_4)\!\cdot\!p_5) & \Big)
        = -i \frac{c_5 n_5^{\rm free}}{D_5} .
      \end{aligned}
\label{ssSSg5}
\end{align} \label{ssSSg} \end{subequations}
\normalsize\!\!\!\!\;
Again, we can verify that the color factors and the kinematics numerators
obey equivalent relations:
\be
   c_1 - c_2 = - c_5 ~~~~~ \Leftrightarrow ~~~~~
   n_1^{\rm free} - n_2^{\rm free} = - n_5^{\rm free} .
\ee
The amplitude is the sum of these three and two more contributions,
which can be obtained from \eqns{ssSSg1}{ssSSg2} by relabeling
$\{1\leftrightarrow3,2\leftrightarrow4\}$.

The amplitude with four indistinguishable scalars can be further obtained
by relabeling, as in \eqn{TwoTermId5pt}, though without fermionic minus signs.
However, in contrast with the fermion case, the two-term identity is not respected:
\beal
   n_6^{\rm free} & \equiv n_1^{\rm free}\big|_{1\leftrightarrow3} \neq n_3^{\rm free} , \\ 
   n_7^{\rm free} & \equiv n_2^{\rm free}\big|_{1\leftrightarrow3} \neq n_2^{\rm free} , \\
   n_8^{\rm free} & \equiv n_3^{\rm free}\big|_{1\leftrightarrow3} \neq n_1^{\rm free} , \\
   n_9^{\rm free} & \equiv n_4^{\rm free}\big|_{1\leftrightarrow3} \neq n_4^{\rm free} , \\
   n_{10}^{\rm free} & \equiv n_5^{\rm free}\big|_{1\leftrightarrow3} .
\label{ScalarNoTwoTermId5pt}
\eeal
Moreover, the numerators~\eqref{ssSSg} are not supersymmetrically related
to those in \eqn{qqQQg}.

The situation is thus the same as for the four-scalar amplitudes.
Therefore, we can fix it in the same way.
Indeed, if we consider fundamental scalars with self-interaction
described by the Lagrangian~\eqref{scalarlagrangian},
we can find a way to reabsorb diagrams with the four-point interactions
$\overline{\phi}AA\phi$ and $\overline{\phi}\phi \overline{\phi}\phi$
into the trivalent graphs
and to obtain the following color-kinematics representation:
\begin{subequations} \begin{align}
    & \parbox{97pt}{
      \begin{fmffile}{ssSSg1} \fmfframe(10,17)(0,17){
      \fmfsettings
      \begin{fmfgraph*}(60,40)
            \fmflabel{$1, i$}{s1}
            \fmflabel{$2, \bar \jmath\!\!$}{s2}
            \fmflabel{$3, k\!\!$}{q3}
            \fmflabel{$4, \bar l$}{q4}
            \fmflabel{$\!5, a$}{g5}
            \fmfleft{s2,q3}
            \fmfright{s1,g5,q4}
            \fmf{dashes_arrow}{s2,v1,v5,s1}
            \fmf{dashes_arrow}{q4,v3,q3}
            \fmf{curly,tension=0}{v3,v1}
            \fmf{curly,tension=0}{g5,v5}
      \end{fmfgraph*} }
      \end{fmffile}
      }
      = - \frac{i \sqrt{2}}{s_{15}s_{34}} \,
          T_{i \bar m}^a T_{m \bar \jmath}^b T_{k \bar l}^b \,
          (p_1\!\cdot\!\varepsilon_5) s_{24}
      = -i \frac{c_1 n_1}{D_1} ,
\label{ssSSg1reduced} \\
    & \parbox{97pt}{
      \begin{fmffile}{ssSSg2} \fmfframe(10,17)(-10,17){
      \fmfsettings
      \begin{fmfgraph*}(60,40)
            \fmflabel{$1, i$}{s1}
            \fmflabel{$2, \bar \jmath\!\!$}{s2}
            \fmflabel{$3, k\!\!$}{q3}
            \fmflabel{$4, \bar l$}{q4}
            \fmflabel{$5, a\!$}{g5}
            \fmfleft{s2,g5,q3}
            \fmfright{s1,q4}
            \fmf{dashes_arrow}{s2,v5,v1,s1}
            \fmf{dashes_arrow}{q4,v3,q3}
            \fmf{curly,tension=0}{v1,v3}
            \fmf{curly,tension=0}{v5,g5}
      \end{fmfgraph*} }
      \end{fmffile}
      }
      = - \frac{i \sqrt{2}}{s_{25}s_{34}} \,
          T_{i \bar m}^b T_{m \bar \jmath}^a T_{k \bar l}^b
          \Big( (p_2\!\cdot\!\varepsilon_5) (s_{25}\!-\!s_{13})
              + (p_4\!\cdot\!\varepsilon_5) s_{25} \Big)
      = -i \frac{c_2 n_2}{D_2} ,
\label{ssSSg2reduced} \\
    & \parbox{97pt}{
      \begin{fmffile}{ssSSg5} \fmfframe(10,17)(0,17){
      \fmfsettings
      \begin{fmfgraph*}(60,40)
            \fmflabel{$1, i$}{s1}
            \fmflabel{$2, \bar \jmath\!\!$}{s2}
            \fmflabel{$3, k\!\!$}{q3}
            \fmflabel{$4, \bar l$}{q4}
            \fmflabel{$\!5, a$}{g5}
            \fmfleft{s2,q3}
            \fmfright{s1,g5,q4}
            \fmf{dashes_arrow}{s2,v1,s1}
            \fmf{dashes_arrow}{q4,v3,q3}
            \fmf{curly,tension=0.01}{v1,v5}
            \fmf{curly,tension=0.01}{v3,v5}
            \fmf{curly,tension=0}{g5,v5}
      \end{fmfgraph*} }
      \end{fmffile}
      }
      \begin{aligned}
        = \frac{i}{\sqrt{2}} \frac{1}{s_{12}s_{34}} \,
          \tilde{f}^{abc} T_{i \bar \jmath}^b T_{k \bar l}^c \,
          \Big(\,&((p_1\!+\!p_2)\!\cdot\!\varepsilon_5)
                  ((p_1\!-\!p_2)\!\cdot\!(p_3\!-\!p_4)) \\
              -\,&((p_3\!-\!p_4)\!\cdot\!\varepsilon_5)
                 (((p_1\!-\!p_2)\!\cdot\!p_5)\!-\!(p_3\!\cdot\!p_4)) \\
              +\,&((p_1\!-\!p_2)\!\cdot\!\varepsilon_5)
                 (((p_3\!-\!p_4)\!\cdot\!p_5)\!-\!(p_1\!\cdot\!p_2)) \\
              +\,&(p_4\!\cdot\!\varepsilon_5) s_{12}
                - (p_2\!\cdot\!\varepsilon_5) s_{34} \Big)
        = -i \frac{c_5 n_5}{D_5} ,
      \end{aligned}
\label{ssSSg5reduced}\!\!\!\!\!\!\!\!\!\!\!\!\!\!\!\!\!\!\!\!
\end{align} \label{ssSSgreduced} \end{subequations}
\!\!\!\!\;obeying
\be
   c_1 - c_2 = - c_5 ~~~~~ \Leftrightarrow ~~~~~ n_1 - n_2 = - n_5 .
\ee
Again, the numerators of the flipped trivalent diagrams are defined by simple relabeling:
\beal
   n_3 & = n_1\big|_{1\leftrightarrow3,2\leftrightarrow4} , \\
   n_4 & = n_2\big|_{1\leftrightarrow3,2\leftrightarrow4} .
\eeal
The amplitude with four indistinguishable scalars can be obtained by further relabeling,
as in \eqns{TwoTermId5pt}{ScalarNoTwoTermId5pt}.
Moreover, the two-term identity from \fig{fig:TwoTermId} is now satisfied,
just as in the fermion case:
\beal
   n_6 & \equiv n_1\big|_{1\leftrightarrow3} = n_3 , \\ 
   n_7 & \equiv n_2\big|_{1\leftrightarrow3} = n_2 , \\
   n_8 & \equiv n_3\big|_{1\leftrightarrow3} = n_1 , \\
   n_9 & \equiv n_4\big|_{1\leftrightarrow3} = n_4 , \\
   n_{10} & \equiv n_5\big|_{1\leftrightarrow3} .
\label{ScalarTwoTermId5pt}
\eeal

In addition to that, it can be easily checked that for any helicity of the gluon,
and for any choice of its reference momentum,
the scalar numerators related to the fermion ones~\eqref{qqQQg}
by the SWI-replacement $ \braket{13} \rightarrow \braket{24} $,
or to the mixed ones~\eqref{ssqqg}
by the SWI-replacement $ \braket{23} \rightarrow - \braket{24} $.

These five-point calculations are fully consistent with the four-point example
of \sec{sec:stree4pt}.
That is, the fundamental scalars described
by the particular self-interacting Lagrangian~\eqref{scalarlagrangian}
can be considered as the supersymmetry continuation of the massless fermions,
and respect the two-term identity in \fig{fig:TwoTermId}.

It is interesting to relate these calculations to an earlier attempt~\cite{Vera:2012ds} to impose the color-kinematics duality on fundamental scalars. In the spirit of that work, we take the limit corresponding to multi-Regge kinematics for the double copies of the numerators~\eqref{ssSSg} with both co- and anti-aligned scalar lines.  Indeed, in doing so, we recover the full multi-Regge gravitational vertex,
which includes the square of Lipatov's effective vertex of QCD~\cite{Lipatov:1976zz,Fadin:1975cb,Kuraev:1976ge,Kuraev:1977fs,Balitsky:1978ic}
and the correct subleading terms required by the Steinmann relations~\cite{Steinmann:1960a,Steinmann:1960b}. This is consistent with the behavior of a correct gravitational amplitude \cite{SabioVera:2011wy}, but at the same time somewhat surprising since the similar calculation of \rcite{Vera:2012ds} yielded a double-copy amplitude with missing subleading terms. While we do not resolve the mismatch here, we note that in later work~\cite{Johansson:2013nsa} the missing terms were recovered in the context of the adjoint-representation double copy.

Having considered these explicit tree-level examples of the color-kinematics duality for fundamental matter, we can now proceed to discussing formal aspects of how we treat supersymmetric Yang-Mills theories with fundamental matter multiplets.

\section{On-shell spectrum and supermultiplets}
\label{sec:supermultiplets}

In this section, we briefly describe the supersymmetric multiplets
to set the notation for the subsequent sections.
More precisely, we explain how on-shell superspace variables are used
to package the on-shell states of super-Yang-Mills theories
into their supermultiplets:
a vector multiplet ${\cal V}_{\cal N}$, as well as a chiral $ \Phi_{\cal N}$
and an antichiral $\overline{\Phi}_{\cal N}$ matter multiplets.
We put them respectively belong into the adjoint,
fundamental and antifundamental representations of the gauge group.

Note that from now on, it will be more convenient
to use the same letters for on-shell states and their fields,
for example, $A^\pm$ for the helicity states of gluons.
Moreover, we denote on-shell Weyl fermions by $\lambda^\pm$ or $\psi^\pm$,
depending on whether they belong to a vector or a matter multiplet, respectively.
Similarly, the vector-multiplet scalars are denoted by $\varphi$
and the matter-multiplet ones by $\phi$.

Perhaps the best starting point for understanding the spectrum of massless gauge theories
is the on-shell supermultipet of ${\cal N}=4$ SYM~\cite{Kunszt:1985mg,Nair:1988bq,Drummond:2008vq,Elvang:2011fx}.
Using standard auxiliary Grassmann variables $\eta^A$,
this vector multiplet can be expressed as a super-wave function
\be
      V_{{\cal N}=4} = A^+ + \lambda_A^+ \eta^A+ \frac{1}{2} \varphi_{AB} \eta^A\eta^B
                     + \frac{1}{3!} \epsilon_{ABCD}\lambda_-^A\,\eta^B\eta^C\eta^D
                     + A_- \eta^1\eta^2\eta^3\eta^4 ,
\ee
where the R-symmetry indices live in SU(4)
and $\epsilon_{ABCD}$ is its Levi-Civita tensor.
The ${\cal N}=4$ vector multiplet is necessarily non-chiral,
in contrast to the cases with lower supersymmetry.
However, the lower supersymmetric multiplets can be considered
as its direct truncations.

The on-shell vector multiplets for reduced supersymmetry can be divided into
a chiral multiplet $V_{\cal N}$ and an antichiral one $\overline{V}_{\cal N}$.
For ${\cal N}=2,1,0$ they are explicitly:
\beal
      V_{{\cal N}=2} & = A^+ + \lambda^+_A \eta^A + \varphi_{12} \eta^1 \eta^2 ,
      ~~~~~~
      \overline{V}_{{\cal N}=2} = \overline{\varphi}^{12} 
                                + \epsilon_{AB} \lambda_-^A \eta^B
                                + A_- \eta^1 \eta^2 , \\
      V_{{\cal N}=1} & = A^+ + \lambda^+ \eta^1 ,
      ~~~~~~~~~~~~~~~~~~~~\,
      \overline{V}_{{\cal N}=1} = \lambda_- + A_- \eta^1 , \\
      V_{{\cal N}=0} & = A^+ ,
      ~~~~~~~~~~~~~~~~~~~~~~~~~~~~~~\,
      \overline{V}_{{\cal N}=0} = A_- ,
\label{chiralVectorMult}
\eeal
where SU(2) indices $A,B=1,2$ are inherited from SU(4) R-symmetry
and are raised and lowered using $\epsilon_{AB}$.
We find it convenient to assemble these chiral vector multiplets
into a single non-chiral multiplet
\be
      {\cal V}_{\cal N} = V_{\cal N} + \overline{V}_{\cal N} \, \theta ,
\label{nonChiralMult}
\ee
where we introduced an auxiliary parameter $\theta$ defined as
\be
      \theta = \!\!\! \prod_{A={\cal N}+1}^4 \!\!\! \eta^A .
\label{theta}
\ee
It is nilpotent, $\theta^2=0$, and commuting or anticommuting
for an even or odd number of supersymmetries, respectively.

While the non-chiral multiplets~\eqref{nonChiralMult}
do not increase the supersymmetry with respect to their chiral constituents,
they allow the more uniform treatment of the components
of the ${\cal N}=2,1,0$ scattering amplitudes
by assembling its full state dependence into a single generating function
$ {\cal A}_n(p_i, \eta^A_i, \theta_i) $.
For example, the $n$-point MHV tree amplitude in ${\cal N}<4$ super-Yang-Mills theories
can be written as
\be
   {\cal A}^{\rm MHV, tree}_n(p_i, \eta^A_i, \theta_i)
      = \frac{ i \, \delta^{(2{\cal N})}(Q)
               \sum_{i<j}^m \theta_{i} \theta_{j} \braket{ij}^{4-{\cal N}} }
             { \braket{12} \braket{34} \dots \braket{n\!-\!1|n} \braket{n1} } ,
\label{genMHVtree}
\ee
where $\theta$'s mark the external legs that belong
to the $\overline{V}_{\cal N}$ multiplet
and $\eta$'s encode the on-shell supersymmetry inside both multiplets.
The delta-function of the supermomentum
$ Q^{A}_\alpha = \sum_i |i\rangle_\alpha \eta_i^A $
is defined analogously to \eqn{delta8Q} as 
\be
   \delta^{(2{\cal N})}(Q)
      = \prod_{A=1}^{\cal N} \sum_{i<j}^m \eta^A_i \braket{ij}  \eta^A_j .
\label{delta2NQ}
\ee

All vector multiplets discussed so far, whether chiral or non-chiral,
belong to the adjoint representation of the gauge group.
For the matter content, on the other hand,
one can choose between the adjoint, fundamental or antifundamental representations.
The last two choices naturally define
the chiral matter multiplet $\Phi_{\cal N}$
and the antichiral one $\overline{\Phi}_{\cal N}$.
For ${\cal N}=2,1$ they are explicitly
\beal
   (\Phi_{{\cal N}=2})_i &= \psi^+_i+\phi_{Ai} \eta^A + \psi^{-}_i \eta^1 \eta^2 ,
   ~~~~~~
   (\overline{\Phi}_{{\cal N}=2})_{\bar \imath}
      = \psi^+_{\bar \imath} + \epsilon_{AB}\,\overline{\phi}^A_{\bar \imath} \eta^B
                             + \psi^{-}_{\bar \imath} \eta^1 \eta^2 , \\
   (\Phi_{{\cal N}=1})_i &= \psi^+_i+\phi_i\,\eta^1 ,
   ~~~~~~~~~~~~~~~~~~~~\,\,
   (\overline{\Phi}_{{\cal N}=1})_{\bar \imath}
      = \overline{\phi}_{\bar \imath} + \psi^-_{\bar \imath} \eta^1 .
\label{chiralMatterMult}
\eeal
Note that although the notation is similar
for the states in the two ${\cal N}=2$ matter multiplets, the states are distinct,
as emphasized by the fundamental/antifundamental subscripts that distinguish the color representations. 

Non-supersymmetric matter can either be a chiral Weyl fermion or a complex scalar:
\beal
      (\Phi_{{\cal N}=0})_i \equiv     (\Phi_{{\cal N}=0}^{\rm fermion})_i &= \psi^+_i ,~~~~~~~~~~~~
      (\overline{\Phi}_{{\cal N}=0})_{\bar \imath}\equiv       (\overline{\Phi}_{{\cal N}=0}^{\rm fermion})_{\bar \imath}
            = \psi^-_{\bar \imath} , \\
      (\Phi_{{\cal N}=0}^{\rm scalar})_i &= \phi_i ,\hskip 3.85cm
      (\overline{\Phi}_{{\cal N}=0}^{\rm scalar})_{\bar \imath}
            = \overline{\phi}_{\bar \imath} .
\eeal
We consider the Weyl fermions to be the default ${\cal N}=0$ matter multiplets.
With this choice,
the formalism for constructing pure supergravities will be uniform for any value of ${\cal N}$.

For completeness, and for later use, we remark that
if the chiral and antichiral ${\cal N}=1$ multiplets in \eqn{chiralMatterMult}
are put into the adjoint representation of the gauge group (or any real representation),
then it is convenient to combine them into
a non-chiral minimal ${\cal N}=2$ multiplet.
After relabeling the scalar fields, we have
\be
      \Phi_{{\cal N}=1} + \overline{\Phi}_{{\cal N}=1} \, \eta^2
      ~~~~ \rightarrow ~~~~
      \Phi_{{\cal N}=2}^{\rm adj}=\psi^++\phi_{A}\eta^A+\psi^-\eta^1\eta^2 .
\label{adjointNeq2}
\ee
Note that in \chap{chap:oneloop},
as well as in the literature on scattering amplitudes,
this matter multiplet is conventionally referred to as ``${\cal N}=1$ chiral'',
even though it effectively has ${\cal N}=2$ supersymmetry.
In this chapter, to avoid confusion with the fundamental chiral multiplets,
we prefer to call it ``${\cal N}=2$ adjoint.''

Similarly,
the non-supersymmetric chiral fermion matter can be promoted to a Majorana fermion,
and the fundamental complex scalar becomes equivalent to two real scalars,
all in the adjoint representation.

Finally, the particle and helicity content of the supermultiplets discussed so far
is summarized in \tab{Multiplets}.

\begin{table*}
\centering
\begin{tabular}{|c|c|c|c|c|c||c|c|c|c||c|c|c|c|}
\hline 
Field & $+1$ & $\!+\frac{1}{2}$ & $\,\,0\,\,$ & $\!-\frac{1}{2}$ & $-1$ &
Field & $\!+\frac{1}{2}$ & $\,\,0\,\,$ & $\!-\frac{1}{2}$ &
Field & $\!+\frac{1}{2}$ & $\,\,0\,\,$ & $\!-\frac{1}{2}$ \\
\hline 
$V_{{\cal N}=4}$ &1&4&6&4&1 & $\Phi_{{\cal N}=2}$&1&2&1 & $\overline{\Phi}_{{\cal N}=2}$ &1&2&1 \\
\hline 
${\cal V}_{{\cal N}=2}$ &1&2&2&2&1& $\Phi_{{\cal N}=1}$&1&1&0 & $\overline{\Phi}_{{\cal N}=1}$ &0&1&1 \\
\hline 
${\cal V}_{{\cal N}=1}$ &1&1&0&1&1& $\Phi_{{\cal N}=0}$ &1&0&0& $\overline{\Phi}_{{\cal N}=0}$&0&0&1 \\
\hline
${\cal V}_{{\cal N}=0}$ &1&0&0&0&1& $\Phi^{\rm scalar}_{{\cal N}=0}$ &0&1&0&  $\Phi^{\rm adj}_{{\cal N}=2}$ &1&2&1 \\
\hline
\end{tabular}
\caption[a]{\small Summary of the particle and helicity content of the various on-shell YM supermultiplets considered in this thesis}
\label{Multiplets} 	
\end{table*}

\section{Construction of pure gravity theories}
\label{sec:FundDoubleCopy}

In this section, we address the problem of constructing pure (super-)gravity theories with ${\cal N}<4$ supersymmetry. Such theories are ``non-factorizable'', meaning that their loop amplitudes cannot be constructed by squaring, or double copying, numerators of pure (super-)Yang-Mills theories. However, remarkably, pure gravity amplitudes can be obtained from non-pure Yang-Mills theory, as we show next.

\subsection{Double copies of physical states}
\label{sec:tensorrules}

We define the following tensor products of the adjoint and fundamental on-shell states in SYM theories with ${\cal N}$- and ${\cal M}$-fold supersymmetry:
\begin{subequations} \begin{align}
   \text{factorizable graviton multiplet}: \qquad
      {\cal H}_{{\cal N}+{\cal M}} & \equiv
      {\cal V}_{\cal N} \otimes {{\cal V}'\!}_{\cal M},
\label{tensoringvector} \\
   \text{gravity matter}: \qquad
      {X}_{{\cal N}+{\cal M}} & \equiv
      \Phi_{\cal N} \otimes \overline{\Phi}'_{\cal M},
\label{tensoringmatter}  \\
   \text{gravity antimatter}: \qquad
      \overline{X}_{{\cal N}+{\cal M}} & \equiv
      \overline{\Phi}_{\cal N} \otimes \Phi'_{\cal M}.
\label{tensoringantimatter} 
\end{align} \label{supertensoring} \end{subequations}
\!\!\!\!\;where we take $0 \le {\cal N} \le 2$ and $0 \le {\cal M} \le 2$,
so that $X_{{\cal N}+{\cal M}}$ is either a vector multiplet $V_{{\cal N}+{\cal M}}$,
or a lower-spin matter multiplet $\Phi_{{\cal N}+{\cal M}}$.
These are the states that naturally appear in the double copies of amplitudes
that obey the color-kinematics duality for adjoint and fundamental particles.

\begin{table*}
\centering
\begin{tabular}{|c|c|c|}
\hline 
SUGRA & tensoring vector states &
$ \text{ghosts} = \text{matter} \otimes \overline{\text{matter}}$ \phantom{\big[} \\
\hline
\;\!\!\!${\cal N}\!=0\;\!\!+\;\!\!0$\!\!\!\; &
$A^\mu \;\!\!\otimes\:\!\! A^\nu \;\!\! = \;\!\! h^{\mu \nu} \oplus \phi \oplus a$ &
$ (\psi^+\!\otimes\:\!\! \psi^-) \oplus (\psi^-\!\otimes\:\!\! \psi^+)
            = \phi \oplus a $ \phantom{\Big[}\!\! \\
\hline
\;\!\!\!${\cal N}\!=1\;\!\!+\;\!\!0$\!\!\!\; &
${\cal V}_{{\cal N}=1} \;\!\!\otimes\:\!\! A^\mu \;\!\!
      = \;\!\! H_{{\cal N}=1} \oplus \Phi_{{\cal N}=2}$ &
$ (\Phi_{{\cal N}=1} \;\!\!\otimes\:\!\! \psi^{-})  \oplus
            (\overline{\Phi}_{{\cal N}=1} \;\!\!\otimes\:\!\! \psi^{+})
            = \Phi_{{\cal N}=2} $ \phantom{\Big[}\!\!\! \\
\hline
\;\!\!\!${\cal N}\!=2\;\!\!+\;\!\!0$\!\!\!\; &
${\cal V}_{{\cal N}=2} \;\!\!\otimes\:\!\! A^\mu \;\!\!
      = \;\!\! H_{{\cal N}=2} \oplus {\cal V}_{{\cal N}=2}$ &
$ (\Phi_{{\cal N}=2} \;\!\!\otimes\:\!\! \psi^{-}) \oplus
      (\overline{\Phi}_{{\cal N}=2} \;\!\!\otimes\:\!\! \psi^{+})
      = {\cal V}_{{\cal N}=2} $ \phantom{\Big[}\!\! \\
\hline
\;\!\!\!${\cal N}\!=1\;\!\!+\;\!\!1$\!\!\!\; &
\!${\cal V}_{{\cal N}=1} \;\!\!\otimes\! {\cal V}_{{\cal N}=1} \;\!\!
      = \;\!\! H_{{\cal N}=2} \oplus 2\Phi_{{\cal N}=2}$\!\! &
\!$ (\Phi_{{\cal N}=1} \;\!\!\otimes\:\!\! \overline{\Phi}_{{\cal N}=1}) \oplus
      (\overline{\Phi}_{{\cal N}=1} \;\!\!\otimes\:\!\! \Phi_{{\cal N}=1})
      = 2 \Phi_{{\cal N}=2} $ \phantom{\Big[}\!\!\!\!\!\!\!\!\; \\
\hline
\;\!\!\!${\cal N}\!=2\;\!\!+\;\!\!1$\!\!\!\; &
\!${\cal V}_{{\cal N}=2} \;\!\!\otimes\! {\cal V}_{{\cal N}=1} \;\!\!
      = \;\!\! H_{{\cal N}=3} \oplus {\cal V}_{{\cal N}=4}$~\,\!\! &
\!$ (\Phi_{{\cal N}=2} \;\!\!\otimes\:\!\! \overline{\Phi}_{{\cal N}=1}) \oplus
      (\overline{\Phi}_{{\cal N}=2} \;\!\!\otimes\:\!\! \Phi^+_{{\cal N}=1})
      = {\cal V}_{{\cal N}=4} $ \phantom{\Big[}\!\!\! \\
\hline
\;\!\!\!${\cal N}\!=2\;\!\!+\;\!\!2$\!\!\!\; &
\!${\cal V}_{{\cal N}=2} \;\!\!\otimes\! {\cal V}_{{\cal N}=2} \;\!\!
      = \;\!\! H_{{\cal N}=4} \oplus 2{\cal V}_{{\cal N}=4}$\!\! &
\!$ (\Phi_{{\cal N}=2} \;\!\!\otimes\:\!\! \overline{\Phi}_{{\cal N}=2}) \oplus
      (\overline{\Phi}_{{\cal N}=2} \;\!\!\otimes\:\!\! \Phi_{{\cal N}=2})
      = 2 {\cal V}_{{\cal N}=4} $ \phantom{\Big[}\!\!\!\!\!\! \\
\hline
\end{tabular}
\caption[a]{\small Pure gravities are constructed from states that are double copies of pure SYM vectors, and similarly ghosts from matter-antimatter double copies. For compactness, pairs of chiral vectors, or pairs of chiral matter multiplets, are combined into non-chiral real multiplets.}
\label{tab:DCconstructions}
\end{table*}

The factorizable graviton multiplet~\eqref{tensoringvector},
naturally obtained in the adjoint double copy~\eqref{BCJformGravityAdj},
is characteristic of some supergravity theory,
but it is not the spectrum of a pure supergravity theory.
Indeed, ${\cal H}$ is reducible into the (non-chiral) pure graviton multiplet $H$
and a (chiral) complex matter multiplet $X$:
\be
   {\cal H}_{{\cal N}+{\cal M}} \equiv {\cal V}_{\cal N} \otimes {{\cal V}'\!}_{\cal M}
      = H_{{\cal N}+{\cal M}} \oplus {X}_{{\cal N}+{\cal M}}
                              \oplus {\overline X}_{{\cal N}+{\cal M}} .
\label{tensordecomposition}
\ee
For example, in the bosonic case the double copy reduces to the graviton, dilaton and axion:
$A^\mu \otimes A^\nu = h^{\mu\nu} \oplus \varphi^{+-} \oplus \varphi^{-+} $
with $\varphi \sim \phi +i a$.
This and other cases are explicitly listed in \tab{tab:DCconstructions}.
While this reduction is easy to carry out for the on-shell asymptotic states,
the same is not true for the off-shell internal states
that are obtained from the double copy construction \eqref{BCJformGravityAdj}.
The reason is that the double copies of gluons $A^\mu \otimes A^\nu$ are difficult
to decompose since amplitudes have no free Lorentz indices,
or, alternatively, reducing the product using little-group indices
would break Lorentz invariance of the off-shell states.

In this thesis, we propose to use the fact that multiplets $X$ and $\overline{X}$ appear
not only in the product~\eqref{tensordecomposition} of the adjoint vectors
but also as the fundamental matter double copy
in \eqns{tensoringmatter}{tensoringantimatter}.
This will let us take the factorizable graviton multiplet~${\cal H}$
and mod out by~$X$ and~$\overline{X}$.
To do that, we will promote the matter double copies $X$ and $\overline{X}$ to be ghosts\footnote{This is superficially similar to the Faddeev-Popov method~\cite{Faddeev:1967fc} that removes the unphysical YM states from propagating in loops, although the details are quite different.}.

First of all, let us do some simple counting of the on-shell degrees of freedoms
to show that it is indeed the same $X$, $\overline{X}$
that appear in \eqns{supertensoring}{tensordecomposition}.
For this, we add the bosonic and fermonic counts of states in the multiplets.
To start with, note that all minimal (chiral) supermultiplets
with ${\cal N}$-fold supersymmetry have exactly $2^{\cal N}$ states
(\eg see \eqns{chiralVectorMult}{chiralMatterMult}).
Therefore, non-chiral pure vector and graviton multiplets have twice as many states,
$2^{{\cal N}+1}$, except for the maximally-supersymmetric multiplets that we do not consider here.
For example, pure YM theory and Einstein gravity both have ${\cal N}=0$,
so they both contain $2^{0+1}=2$ physical states:
the two on-shell gluons or gravitons.
For this bosonic case, \eqn{tensordecomposition} becomes
\be
   4 = 2 \otimes 2 = 2 \oplus 1 \oplus 1 ,
\ee
where the right-hand side represents the two gravitons, dilaton and axion,
or rather the mixed dilaton-axion states $\varphi^{+-}$ and $\varphi^{-+}$
introduced in \sec{sec:ftree4pt}. 
In that picture, $X$ can be thought of as $\varphi^{+-} = A^+ \otimes A^-$.
For a general $0\le{\cal N},{\cal M} \le 2$ supersymmetic theory,
\eqn{tensordecomposition} becomes
\be
   2^{{\cal N}+{\cal M}+2} = 2^{{\cal N }+1} \otimes 2^{{\cal M}+1}
                           = 2^{{\cal N}+{\cal M}+1} \oplus 2^{{\cal N}+{\cal M}}
                                                     \oplus 2^{{\cal N}+{\cal M}} .
\ee
Indeed, the right-hand side represents the pure graviton multiplet
in ${\cal N}+{\cal M}$ supergravity
plus two minimal (chiral) matter multiplets in the same theory.
Clearly, the matter multiplet $X$ defined in \eqn{tensoringmatter}
has the same state counting,
\be
   2^{{\cal N }+{\cal M }} =2^{{\cal N }} \otimes 2^{{\cal M }} ,
\ee
and similarly for $\overline{X}$.
In the bosonic case, we have  $1=1\otimes 1$,
which now implies that $X=\varphi^{+-}=\psi^+ \otimes \psi^-$,
where $\psi^+$, $\psi^-$ are on-shell Weyl fermions.

In this way, we have shown that simple counting is consistent
with our recipe for obtaining the pure graviton multiplet $H$
from ${\cal V} \otimes {{\cal V}'}$
by modding out by $\Phi \otimes \overline{\Phi}'$ and $\overline{\Phi} \otimes \Phi'$.

\subsection{Numerator double copies}
\label{sec:numerators}

In \sec{sec:bcjfund}, we considered (super-)Yang-Mills theories where adjoint and fundamental color representations are dual to vector and matter multiplets, respectively, in the sense that the kinematic structure of the amplitudes is governed by the adjoint and fundamental color-kinematics duality. In \sec{sec:tensorrules}, we explained how the on-shell SYM states  can be tensored to obtain the on-shell spectrum of pure supergravities, including Einstein gravity. To obtain gravity amplitudes, we now have to do the same double-copy construction for the interacting theories with internal off-shell states.

Pure (super-)gravity amplitudes are obtained by recycling the cubic diagrams~\eqref{BCJformYM} of the Yang-Mills theory and constructing the proper double copies of the kinematic numerators, similarly to \eqn{BCJformGravityAdj}. The numerator double copies should be constructed so that internal adjoint lines in gauge theory become an off-shell equivalent of ${\cal V} \otimes {\cal V}'$ lines in gravity. And similarly, fundamental and antifundamental gauge-theory lines become gravitational lines that are off-shell continuations of $\Phi \otimes \overline{\Phi}'$ and $\overline{\Phi} \otimes \Phi'$, respectively. That is, the numerator copies should {\it not} be taken between diagrams that produce cross terms between the vector and matter internal lines. Such double copies would in general not be consistent since the kinematic algebra is different for the two types of states. 

In fact, the above double-copy structure is already present in Yang-Mills theory. By definition, since \eqn{BCJformYM} describes gauge theory amplitudes, there are no products of numerators and color factors, $n_i c_i$, that involve cross-terms between internal vector lines and fundamental color lines. Thus the transition from Yang-Mills to gravity amplitudes should be as straightforward as replacing the color factors with kinematic numerators. For example, replacing $c_i \rightarrow n_i'$ in \eqn{BCJformYM} would give valid gravity amplitudes, albeit with undesired matter content
given by $\Phi \otimes \Phi'$ and $\overline{\Phi} \otimes \overline{\Phi}'$.
To get matter states such as $\Phi \otimes \overline{\Phi}'$, we can instead let $c_i \rightarrow \overline{n}'_i$, where the bar operation swaps matter and antimatter. However, that would give gravity amplitudes with extra physical matter. Since we are trying to remove the matter already present in the decomposition of ${\cal V} \otimes {\cal V}'$, what we want is matter that behaves as ghosts. This means that we should insert a minus sign for each closed loop of matter particles. Hence we do the following replacement in \eqn{BCJformYM},
\be
   c_i \rightarrow (-1)^{|i|}\bar{n}'_i
\ee
where $|i|$ counts the number of closed matter loops in the $i$'th trivalent graph.
More generally, we can do the replacement $c_i \rightarrow (N_X)^{|i|} \bar{n}'_i$, where $N_X+1$ is the number of complex matter multiplets in the desired gravity theory.

Thus we propose the following formula for amplitudes in pure (super-)gravity:
\be
   {\cal M}^{\text{$L$-loop}}_n
      = i^{L-1} \Big(\frac{\kappa}{2}\Big)^{n+2L-2}
        \sum_{i} \int\!\!\frac{d^{Ld}\ell}{(2\pi)^{Ld}}
        \frac{(-1)^{|i|}}{S_i} \frac{n_i \overline{n}_i'}{D_i} ,
\label{BCJformGravity}
\ee
where the sum runs over all $m$-point $L$-loop graphs with trivalent vertices
of two kinds: (fact. grav., fact. grav., fact. grav.) and (fact. grav., matter, antimatter),
corresponding to particle lines of two types: factorizable graviton multiplets and matter ghosts. 
The graphs are in one-to-one correspondence with the ones
in the gauge theory amplitude~\eqref{BCJformYM},
and the denominators $D_i$ and the symmetry factors $S_i$ are same.
The objects $n_i$ and $n_i'$ are the numerators of two gauge theory amplitudes
where at least one of the two copies obeys the color-kinematics duality.
The bar on top of one of the numerators denotes
the operation of conjugating the matter particles
and reversing the fundamental-representation flows.
Finally, the contribution of each graph enters the sum with a sign
determined by the number of closed matter-ghost loops $|i|$.

When there are no fundamental lines in the diagram,
both the matter conjugation the sign modifier acts trivially,
so eliminating all graphs with matter ghosts reduces \eqn{BCJformGravity}
to the adjoint double copy~\eqref{BCJformGravityAdj}.

\subsection{Gauge invariance of the gravity construction}
\label{sec:ginvariance}

Now let us show that the double-copy construction~\eqref{BCJformGravity}
for pure gravity amplitudes obeys generalized gauge invariance.
This means that the freedom to shift numerators in gauge theory
should correspond to a similar freedom in shifting one copy of the gravity numerators.
If this was not true, the result of the double copy would depend on the gauge
chosen for the Yang-Mills amplitude computation, and thus be necessarily wrong.
Moreover, while it is not obvious, experience tells us that the notion of generalized gauge invariance is perhaps as useful and constraining for amplitudes as the more familiar notion of gauge invariance is for Lagrangians. Therefore, showing that the gravity construction is generalized-gauge-invariant is a strong check of consistency.

Recall that the generalized gauge invariance in Yang-Mills theory allows us to freely shift $n_i \rightarrow n_i + \Delta_i$ provided that $\Delta_i$ satisfy
\be
   \sum_{i} \int\!\!\frac{d^{Ld}\ell}{(2\pi)^{Ld}}
                    \frac{1}{S_i} \frac{ \Delta_i c_i }{D_i} =0.
\ee
The origin of this freedom comes from the overcompleteness of the color factor basis.
Indeed, the Jacobi identities and commutation relations satisfied by the color factors
can indirectly enter the amplitude multiplied by arbitrary kinematic functions,
which we may refer to as pure gauge garbage.
For example, consider a shift of the numerators $n_i$, $n_j$ and $n_k$
of to some three diagrams
with identical graph structure except for a single edge:
\begin{subequations} \begin{align}
   {\cal A}^{\text{$L$-loop}}_n & \rightarrow {\cal A}^{\text{$L$-loop}}_n
      + 
        \int K_{\rm pure\,gauge}
            \big( \tf^{dac} \tf^{cbe} - \tf^{dbc} \tf^{cae} - \tf^{abc} \tf^{dce} \big)
            c^{abde}_{\rm rest} , \\
   {\cal A}^{\text{$L$-loop}}_n & \rightarrow {\cal A}^{\text{$L$-loop}}_n
      + 
        \int K_{\rm pure\,gauge}
            \big(T^{a}_{i \bar \jmath} \, T^{b}_{j \bar k}
               - T^{b}_{i \bar \jmath} \, T^{a}_{j \bar k}
               - \tf^{abc} \, T^{c}_{i \bar k} \big)
            c^{ab}_{k \bar \imath, {\rm rest}} ,
\label{Ashift}
\end{align} \end{subequations}
where the standard loop integration measure is suppressed for brevity.
Taking into account that the denominators and symmetry factors are included into $K_{\rm pure\,gauge}$,
the three numerators will be shifted by
\be
   \Delta_i = S_i D_i K_{\rm pure\, gauge} , ~~~
   \Delta_j = S_j D_j K_{\rm pure\, gauge} , ~~~
   \Delta_k = S_k D_k K_{\rm pure\, gauge} .
\ee
Of course, $K_{\rm pure\,gauge}$ is multiplied here by something that vanishes \eqref{chiraljacobi},
so the amplitude cannot depend on it.

Translating this example to gravity
through the replacement $c_i \rightarrow (-1)^{|i|} \bar{n}'_i$, we obtain\footnote{
This argument trivially extends to the more general replacement prescription
$c_i \rightarrow (N_X)^{|i|} \bar{n}'_i$.}
\be
   M^{\text{$L$-loop}}_n \rightarrow M^{\text{$L$-loop}}_n
      + (-1)^{|i|} \int K_{\rm pure\,gauge}
        \big( {\bar n}'_i - {\bar n}'_j - {\bar n}'_k \big)  .
\label{Mshift}
\ee
The crucial point here is that the commutation and Jacobi relations in \fig{fig:fjacobi}
leave the number of closed fundamental matter loops invariant,
so in \eqn{Mshift} we safely set $|i|=|j|=|k|$.
For the gravitational amplitude~\eqref{Mshift}
to be free of the pure-gauge garbage,
the numerators $\bar{n}'_i$ must satisfy the color-dual algebra: ${\bar n}_i' - {\bar n}_j' = {\bar n}_k' $.

In summary, shifting the gauge-theory numerators $n_i \rightarrow n_i + \Delta_i$
gives new gravity numerators $n_i \bar{n}'_i \rightarrow (n_i + \Delta_i) \bar{n}'_i$.
If $\bar{n}_i$ satisfy the same algebraic relations as the color factors~$c_i$,
then we are guaranteed to leave the gravity amplitude invariant, since the following identity is inherited from the Yang-Mills case:
\be
   \sum_{i} \int\!\!\frac{d^{Ld}\ell}{(2\pi)^{Ld}}
                    \frac{1}{S_i} \frac{ \Delta_i \bar{n}'_i }{D_i} =0.
\ee
As in the purely-adjoint case~\cite{Bern:2010yg},
this argument can be easily extended to show that,
to obtain a consistent double-copy amplitude,
it is sufficient for only
one of the two numerator sets ($n_i$ or $\bar{n}'_i$)
to satisfy the color-kinematics duality, 

From this line of arguments, one can also see that
the additional two-term algebraic identity~\eqref{4ptTwoTerm},
depicted in \fig{fig:TwoTermId}, is optional.
It is not needed for obtaining generalized-gauge-invariant amplitudes
since the corresponding color identity is not generically present for fundamental representations of classical groups, such as SU($N$).
In fact, in those spacetime dimensions where the two-term identity~\eqref{4ptTwoTerm}
can be imposed, the double-copy gravity amplitudes stay the same
with or without imposing it.
Indeed, one can think of the identity~\eqref{4ptTwoTerm}
as a special gauge choice in Yang-Mills theory,
hence the gravity amplitudes cannot depend on this choice.
We have also confirmed this property by constructing explicit tree-level amplitudes
through seven points.

\subsection{Checks of two-loop cuts}
\label{sec:cutchecks}

\begin{figure}[t]
      \centering
      \includegraphics[scale=1.00]{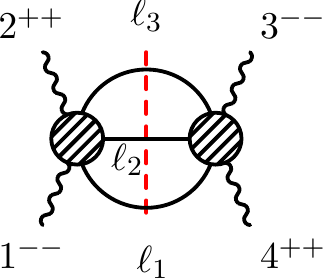}
      \hspace{30pt}
      \includegraphics[scale=1.00,trim=0 10pt 0 0]{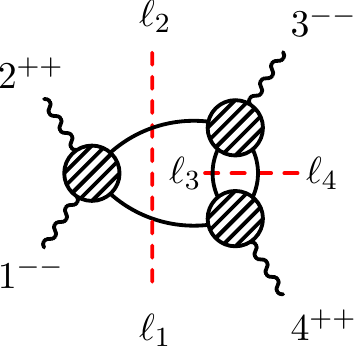}
\caption[a]{\small Two-loop cuts of the four-graviton amplitude
that can have ghost matter ciculating in only a single loop at the time.}
\label{fig:twoloopcut4ptAB}
\end{figure}

\begin{figure}[t]
      \centering
      \includegraphics[scale=1.00]{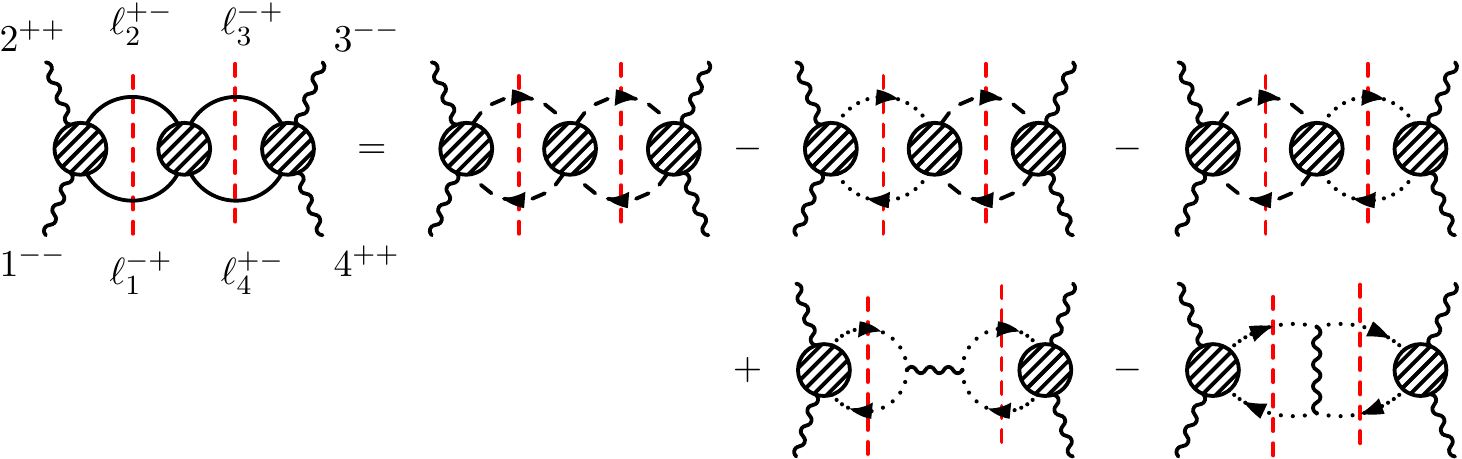}
      \vspace{-5pt}
\caption[a]{\small A two-loop cut of the four-graviton amplitude.
A particular internal helicity configuration is expanded using
the diagrammatic rules given in \fig{fig:doublecopylines}.
In the text, we explain that these dilaton-axion contributions
cancel among themselves, leaving the pure gravity cut.}
\label{fig:twoloopcut4pt}
\end{figure}

In this section, we consider an explicit multiloop example that provides a nontrivial check that the double-copy prescription correctly eliminates the dilaton and axion contributions from the unitarity cuts in Einstein gravity. We concentrate on two-loop cuts in four dimensions; the one loop case will be treated in full rigor in subsequent sections with complete four-point calculations in $d=4-2\epsilon$ dimensions. For the construction of the two-loop four-point amplitude it is sufficient to consider three types of unitarity cuts: the two shown in \fig{fig:twoloopcut4ptAB} and the first diagram shown in \fig{fig:twoloopcut4pt}. For the purpose of checking the cancellation of dilaton and axion contributions the latter cut is the most nontrivial. The two cuts in \fig{fig:twoloopcut4ptAB} can only have matter circulating in a single sub-loop at the time, so they can at most teach us about the dilaton-axion cancellations that take place in one-loop amplitudes, the topic of the following sections.

\begin{figure}[t]
      \centering
      \includegraphics[scale=1.00]{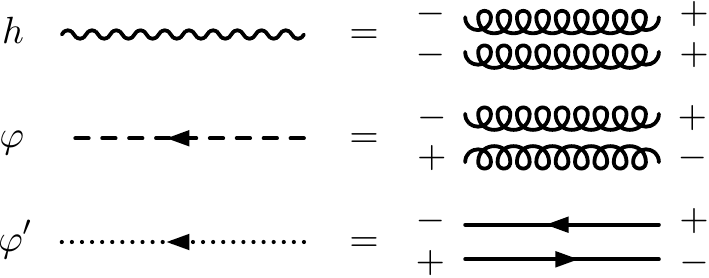}
\caption[a]{\small Diagrammatic rules for the gravity cut lines in \fig{fig:twoloopcut4pt}:
graviton lines correspond to double copies of gluon lines,
whereas mixed dilaton-axion lines can be realized either in the same way
or as fundamental double copies of fermions.}
\label{fig:doublecopylines}
\end{figure}

Now, consider the double $s$-channel cut of the four-graviton two-loop amplitude. To be explicit, consider the external configuration $M(1^{--}_h\!, 2^{++}_h\!, 3^{--}_h\!, 4^{++}_h) $, and focus on only the internal matter contributions as shown in \fig{fig:twoloopcut4pt}. If the theory is pure, these matter contributions should add up to zero. As shown, the cut has an expansion in terms of different particle and diagram contributions coming from our prescription. In order to not clutter the diagrams with explicit double-line notation that reflect the tensor structure of the gravity states, we translate to the particle-line notation given in \fig{fig:doublecopylines}. As explained in \sec{sec:ftree4pt}, the complexified dilaton-axion contributions arise as double copies of gluons with anticorrelated helicities, gravitons from double copies of correlated gluons, and matter ghosts from anticorrelated fermions (quarks).

The diagrammatic expansion in \fig{fig:twoloopcut4pt} is straightforward, except for the special treatment of the opposite-statistics nature of the ghosts. In particular, in last two diagrams the intermediate graviton channel needs to be resolved before the sign of the diagram can be determined due to the different number of ghost loops. 
 
To calculate the result of the cut we can take advantage of a convenient simplification. 
If we focus for a moment on the left- and the right-most four-point tree amplitudes (or blobs) appearing in the factorization of the cut, we see that they are in fact identical for all the five contributions in \fig{fig:twoloopcut4pt}. They are explicitly
\begin{subequations} \begin{align}
\begin{aligned}
   M_4^{\rm tree}
      (1^{--}_h\!,2^{++}_h\!,(-\ell_2)^{-+}_{\varphi}\!,(-\ell_1)^{+-}_{\varphi})
&= M_4^{\rm tree}
      (1^{--}_h\!,2^{++}_h\!,(-\ell_2)^{-+}_{\varphi'}\!,(-\ell_1)^{+-}_{\varphi'}) \\
&= \frac{i \bra{1}\ell_1|2]^4}{s (p_1\!-\!\ell_1)^2 (p_1\!-\!\ell_2)^2} ,
\end{aligned}
\label{exttree1} \\
\begin{aligned}
   M_4^{\rm tree}
      (3^{--}_h\!,4^{++}_h\!,(-\ell_4)^{+-}_{\varphi}\!,(-\ell_3)^{-+}_{\varphi})
&= M_4^{\rm tree}
      (3^{--}_h\!,4^{++}_h\!,(-\ell_4)^{+-}_{\varphi'}\!,(-\ell_3)^{-+}_{\varphi'}) \\
&= \frac{i \bra{3}\ell_3|4]^4}{s (p_3\!-\!\ell_3)^2 (p_3\!-\!\ell_4)^2} ,
\end{aligned}
\label{exttree2}
\end{align} \label{exttrees} \end{subequations}
\!\!\!\!\;where $\varphi$ denote the complexified dilaton-axion states arising from gluon double copies, and $\varphi'$ are the fundamental fermion double copies.

The amplitudes in \eqn{exttrees} only contribute to the cut as overall factors corresponding to the outer blobs in \fig{fig:twoloopcut4pt}, and thus we may ignore them for now and only check the possible cancelation among the five terms associated with the middle blob gravity tree amplitudes.
Conveniently, we already have all the ingredients at hand, from the results of \sec{sec:ftree4pt}.
The first term is  $M_4^{\rm tree} (\ell^{-+}_{1\,\varphi}\!,\ell^{+-}_{2\,\varphi}\!, \ell^{-+}_{3\,\varphi}\!,\ell^{+-}_{4\,\varphi})$, and second and the third terms give identical contributions
\be
 - M_4^{\rm tree} (\ell^{-+}_{1\,\varphi'},\ell^{+-}_{2\,\varphi'},
                   \ell^{-+}_{3\,\varphi}\!,\ell^{+-}_{4\,\varphi}) = 
 - M_4^{\rm tree} (\ell^{-+}_{1\,\varphi}\!,\ell^{+-}_{2\,\varphi}\!,
                   \ell^{-+}_{3\,\varphi'},\ell^{+-}_{4\,\varphi'}) .
\ee
For the fourth and fifth terms we recycle the numerators $n_s$ and $n_t$ from \eqn{qqQQ}, with $p_i\rightarrow l_i$. Finally, we get the following sum for the cut
\beal
     M_4^{\rm tree} (\ell^{-+}_{1\,\varphi}\!,\ell^{+-}_{2\,\varphi}\!,
                     \ell^{-+}_{3\,\varphi}\!,\ell^{+-}_{4\,\varphi})
 - 2 M_4^{\rm tree} (\ell^{-+}_{1\,\varphi}\!,\ell^{+-}_{2\,\varphi}\!,
                     \ell^{-+}_{3\, \varphi'},\ell^{+-}_{4\, \varphi'})
 - i \frac{n_s \overline{n}_s}{(\ell_1\!+\!\ell_2)^2}
 + i \frac{n_t \overline{n}_t}{(\ell_2\!+\!\ell_3)^2} & \\
 =-i \frac{n_s \overline{n}_s}{(\ell_1\!+\!\ell_2)^2}
 - i \frac{n_t \overline{n}_t}{(\ell_2\!+\!\ell_3)^2}
 +2i \frac{n_s \overline{n}_s}{(\ell_1\!+\!\ell_2)^2} 
 - i \frac{n_s \overline{n}_s}{(\ell_1\!+\!\ell_2)^2}
 + i \frac{n_t \overline{n}_t}{(\ell_2\!+\!\ell_3)^2} & = 0 .
\label{cancellation4}
\eeal
which is identically zero using \eqns{dilatonaxionexample}{twoscalarexample}.

Cancellations also happen among cuts where some of internal states are gravitons and some are matter states, that is, configurations such as $M_4^{\rm tree} (l^{++}_{1 h}\!,l^{--}_{2 h}\!, \ell^{-+}_{3\,\varphi}\!,\ell^{+-}_{4\,\varphi})$ for the middle blob. These are again effectively the same type of cancelation that has to happen for one-loop amplitudes. In summary, these cancelations  leaves the full four-dimensional cut corresponding to \fig{fig:twoloopcut4pt} completely free of dilaton and axion contributions.  This shows that our prescription~\eqref{BCJformGravity} for calculating pure gravity amplitudes works correctly for this two-loop four-point example. We have repeated the same calculation \eqref{cancellation4} in the various supersymmetric settings listed in \tab{tab:DCconstructions}.
In all cases, the unwanted matter, corresponding to $X$ and $\overline{X}$ multiplets,
cancels out.

\begin{figure}[t]
      \centering
      \includegraphics[scale=1.00]{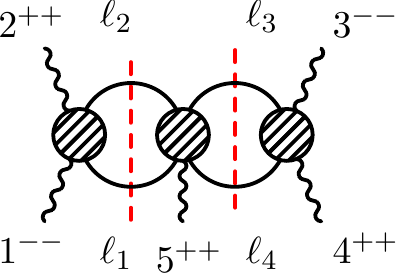}
      \vspace{-5pt}
\caption[a]{\small A two-loop cut of the five-graviton amplitude}
\label{fig:twoloopcut5pt}
\end{figure}

The four-point analysis can be repeated in exactly the same way for the two-loop cut
of the five-graviton amplitude $ M(1^{--}_h\!,2^{++}_h\!,3^{--}_h\!,4^{++}_h\!,5^{++}_h) $.
Its most nontrivial cut is shown in \fig{fig:twoloopcut5pt}.
To analyze it, we can use the tree-level input of \sec{sec:ftree5pt}.
At five points, the cancellation analogous to \eqn{cancellation4} takes the explicit form,
\small
\beal
     M_5^{\rm tree} (\ell^{-+}_{1\,\varphi}\!,\ell^{+-}_{2\,\varphi}\!,
                     \ell^{-+}_{3\,\varphi}\!,\ell^{+-}_{4\,\varphi}\!,5^{++}_h)
 - 2 M_5^{\rm tree} (\ell^{-+}_{1\,\varphi}\!,\ell^{+-}_{2\,\varphi}\!,
                     \ell^{-+}_{3\, \varphi'},\ell^{+-}_{4\, \varphi'},5^{++}_h) 
 - i \sum_{i=1}^5    \frac{n_i \overline{n}_i}{D_i}
 + i \sum_{i=6}^{10} \frac{n_i \overline{n}_i}{D_i} & = 0 ,
\label{cancellation5}
\eeal
\normalsize
where we used \eqns{FDC5}{FDC5mixed} to obtain zero.
We have also checked numerically that a similar cancellation occurs for cuts of the same topology with up to three more gravitons in the outer two blobs
and up to two more gravitons in the middle blob of this cut topology.

It is interesting to note that if we try to perform similar exercises and cut checks for double copies of fundamental scalars, instead of fermions, we encounter trouble.  The resulting ghost contributions will cancel the dilaton and axion in one-loop amplitudes,
however, at two loops the construction fails to produce pure theories. This happens already for the cut analogous to \fig{fig:twoloopcut4pt}, with fermion double copies replaced by scalar ones. In the next section, we give an indirect argument that confirms that scalars double copies are unsuitable for canceling the dilaton and axion.

\subsection{General multiloop argument}
\label{sec:multiloop}

In this section, we argue that our prescription cancels the dilaton and axion from four-dimensional unitarity cuts at any loop order. We expect the construction to also produce correct $(4-2\epsilon)$-dimensional unitarity cuts, but since such cuts can be very subtle and regularization dependent, we give no explicit argument in that case.

For simplicity, we restrict the argument to Einstein gravity as the supersymmetric cases should be straightforward generalizations thereof. We want to show that a generic four-dimensional unitarity cut,
\be
   \sum_{S~{\rm states}} M^{\rm tree}_{(1)} M^{\rm tree}_{(2)}
                   \dots M^{\rm tree}_{(k)} ,
\label{cuttrees}
\ee
constructed from double copies of YM diagrams according to our prescription, contains only gravitons in the internal states.

In the bosonic case, our prescription involves the product
of two copies of (YM + fund. fermion) theories, where the states are tensored according to their gauge-group representation.
Let us use the fact that at tree level the states and partial amplitudes in this YM theory can be mapped to those of ${\cal N}=1$ SYM simply by replacing the fundamental-representation color factors with those in the adjoint (cf. \eqn{representationswap}).
Indeed, from the works of \rcites{Drummond:2008cr,Dixon:2010ik,Reuschle:2013qna,Melia:2013bta,Melia:2013epa}, we know that the color-stripped tree-level amplitudes of ${\cal N}=1$ SYM (which are a subsector of ${\cal N}=4$ SYM)
can be combined to construct all the tree amplitudes in massless multiple-flavor QCD, and vice versa.
Therefore, for convenience, we map the states of (YM + fund. fermion)-theory
to a subsector of ${\cal N}=4$ SYM, and thus we can label the fields by the familiar SU(4) R-charge indices.

Let us look at the on-shell gravitational states
that the tensor product of two copies of (YM + fund. fermion)-theory gives us
according to our prescription.
If we reserve the R-charge indices $1,\dots,4$ for the left side of the double copy
and $5,\dots,8$ for the right side,
then the we have the tensor products for the gluons $(A^{+} \oplus A^{-}_{1234}) \otimes (A^{+} \oplus A^{-}_{5678})$,
as well as the matter states $\psi^{+}_1 \otimes \psi^{-}_{678}$
and $\psi^{-}_{234} \otimes \psi^{+}_{5}$, where the latter should be promoted to ghosts inside loop amplitudes.
This results in the following spectrum:
\be
   S=\{h^{+},~\varphi_{1234},~\varphi_{5678},~
         \hat{\varphi}_{1\,678},~\hat{\varphi}_{234\,5},~h^{-}_{1234\,5678}\} ,
\ee
where the hat notation marks the states that are treated as ghosts in the loop amplitudes.

Now let us consider the simplest possible R-charge rotation,
a permutation of the indices ${\cal R}_{25}=\{2 \leftrightarrow 5\}$,
which gives the following set of states:
\be
   S'=\{h^{+},~\varphi_{134\,5}, ~\varphi_{2\,678},~
          \hat{\varphi}_{1\,678},~\hat{\varphi}_{234\,5},~h^{-}_{1234\,5678}\} .
\ee
Although we will not prove it, we claim that the gravitational tree amplitudes, and thus the unitarity cut \eqref{cuttrees}, following from a double-copy construction should be invariant under this rotation. This can be argued because of the close relationship between the tree-level kinematical numerators of (YM + fund. fermion)-theory and those of ${\cal N}=1$ SYM,
which in turn can be obtained from ${\cal N}=4$ SYM.
If this claim is true, we can observe
that the rotated spectrum $S'$ corresponds to double copies $A^{+} \otimes A^{+}$,
$A^{-}_{1234} \otimes A^{-}_{5678}$,
$(\hat{\psi}^{+}_1 \oplus \psi^{+}_2) \otimes \psi^{-}_{678}$ and
$(\hat{\psi}^{-}_{234} \oplus \psi^{-}_{134}) \otimes \psi^{+}_{5}$.
This can be considered to belong to a color-kinematics construction
with adjoint and fundamental fields,
where on one side we have (YM + fund. fermion + fund. fermion ghost)
and on the other side (YM + fund. fermion).
If the fundamental fields are restricted to live on internal lines of the loop amplitudes,
as they are in our construction,
then the amplitudes in the (YM + fund. fermion + fund. fermion ghost)-theory receives no contributions from the fermions, because they must cancel each other entirely.
Since the numerators corresponding to fundamental particles
vanish on one side of the double copy,
the gravity amplitude receives no contributions from fermion double copies. 
We can then remove the scalars produced by these double copies,
and the spectrum~$S'$ is secretly equivalent to the pure gravity spectrum
\be
   S''=\{h^{+},~h^{-}_{1234\,5678}\} .
\ee

In other words, in our prescription,
the four-dimensional unitarity cuts should be equivalent to those of the pure gravity theory
\beal
   \sum_{S~{\rm states}}   M^{\rm tree}_{(1)} M^{\rm tree}_{(2)}
                     \dots M^{\rm tree}_{(k)} & =
   \sum_{S'~{\rm states}}  M^{\rm tree}_{(1)} M^{\rm tree}_{(2)}
                     \dots M^{\rm tree}_{(k)} \\ & =
   \sum_{S''~{\rm states}} M^{\rm tree}_{(1)} M^{\rm tree}_{(2)}
                     \dots M^{\rm tree}_{(k)} .
\eeal
Indeed, this is consistent with the explicit calculations in \sec{sec:cutchecks}.

Note that for the above argument to be valid something must fail if applied to scalar double copies,
since we observed that the calculations in \sec{sec:cutchecks} do not work for them.
To see this, consider the possible scalar prescription that relies on tensor products
$(A^{+} \oplus A^{-}_{1234}) \otimes (A^{+} \oplus A^{-}_{5678})$ and
$\phi_{12} \otimes \phi_{78}$ and $\phi_{34} \otimes \phi_{56}$.
This gives the following states:
\be
   S_\phi=\{h^{+},~\varphi_{1234},~\varphi_{5678},~
           \hat{\varphi}_{12\,78},~\hat{\varphi}_{34\,56},~h^{-}_{1234\,5678}\} .
\ee
Now perform, for example, the rotation
${\cal R}_{15;26} = \{1 \leftrightarrow 7, 2\leftrightarrow 8 \}$
of the R-charge indices:
\be
   S'_\phi=\{h^{+},~\varphi_{34\,78},~\varphi_{12\,56},
             ~\hat{\varphi}_{12\,78},~\hat{\varphi}_{34\,56},~h^{-}_{1234\,5678}\} .
\ee
Assuming that this is a valid double-copy spectrum the resulting matter can be thought of either as
$\phi_{12} \otimes (\phi_{56} \oplus \hat{\phi}_{78})$ and
$\phi_{34} \otimes (\hat{\phi}_{56} \oplus \phi_{78})$, or as
$(\phi_{12} \oplus \hat{\phi}_{34}) \otimes \phi_{56}$ and
$(\hat{\phi}_{12} \oplus \phi_{34}) \otimes \phi_{78}$, or even as $(\hat{\phi}_{12} \oplus \phi_{34}) \otimes (\hat{\phi}_{56} \oplus \phi_{78})$.
The first two cases look similar to the above fermionic situation, so we can attempt to interpret either of them as coming from a color-kinematics construction
with a (YM + fund. scalar + fund. scalar ghost)-theory on one side of the double copy. However, one can check that there will be an irregular four-scalar interaction that couples the ghosts to the non-ghost scalars, preventing a complete cancellation of the matter. Needless to say, this cannot happen to fermions. The third factorization $(\hat{\phi}_{12} \oplus \phi_{34}) \otimes (\hat{\phi}_{56} \oplus \phi_{78})$ is an even stranger case, since either side appears to contain a theory with one complex scalar whose antiparticle is a ghost, which is clearly not sensible.

Of course, one can consider doing other swaps of the R-symmetry indices,
but all possible permutations have similar problems.
While this does not conclusively prove that the scalar double-copy prescription fails for pure gravities, it does show that the argument that worked for fermion double copies
encounter obstructions when applied to scalars. In any case, the explicit cut checks in the previous section show that starting at two loops the scalar double copy fails to properly cancel the dilation and axion states.

\section{One-loop four-point gauge-theory amplitudes with internal matter}
\label{sec:YM}

In this section, we work out the duality-satisfying numerators
for massless one-loop four-point amplitudes with four external vector-multiplet legs and internal fundamental matter running in the loop.

\subsection{Reducing numerators to a master}
\label{sec:generalapproach}

\begin{figure}[t]
      \centering
      \includegraphics[scale=0.80]{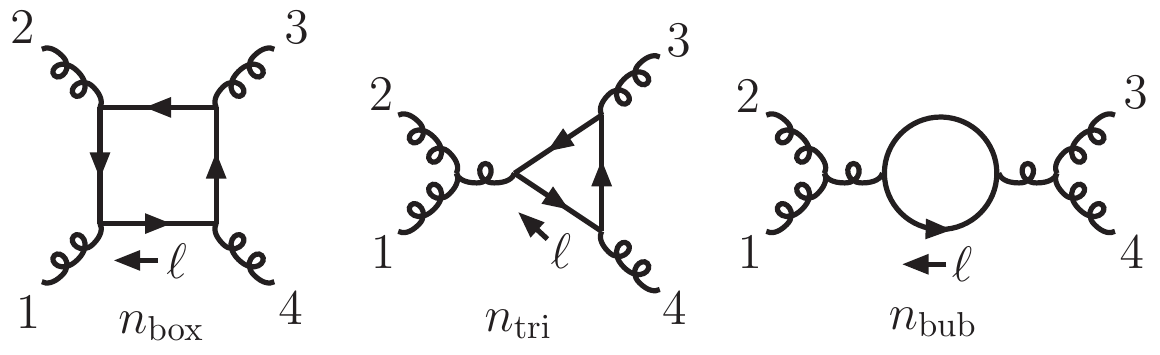}
      \vspace{-5pt}
\caption[a]{\small The three graphs that contribute to one-loop four-point amplitudes.
Curly lines correspond to adjoint vector multiplets
and solid lines represent fundamental matter
(or, in general, particles of any representation).
Our convention is that the external momenta are outgoing in all graphs.}
\label{fig:ChiralBoxes}
\end{figure}

\begin{figure}[t]
      \centering
      \includegraphics[scale=0.80]{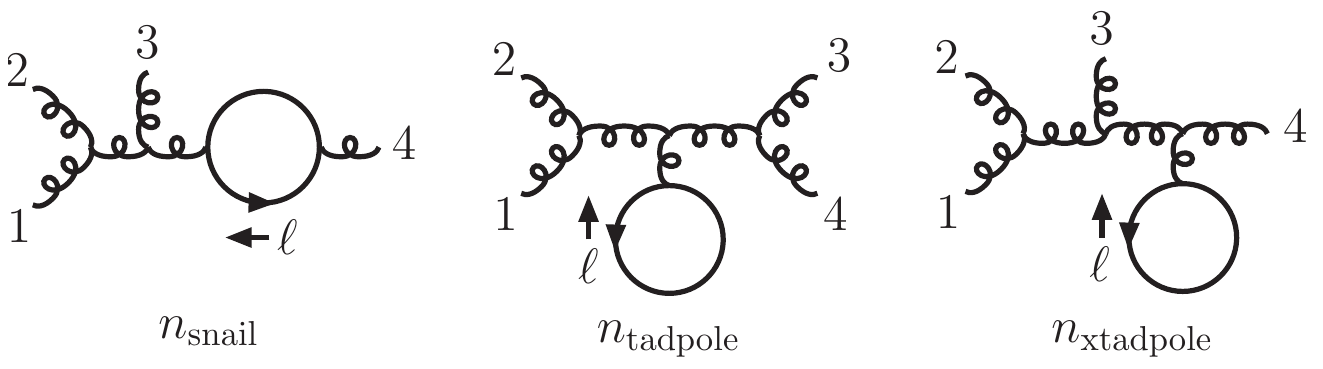}
      \vspace{-5pt}
\caption[a]{\small Three graphs that do not contribute to the amplitudes in dimensional regularization because they give scaleless integrals.}
\label{SnailTadpolesFigure}
\end{figure}

At four points, the master diagram that determines all other one-loop topologies is the box in \fig{fig:ChiralBoxes}, whose numerator we denote as
\be
   n_{\rm box}(1,2,3,4,\ell) ,
\ee
where $\ell$ is the loop momentum and
$1,\dots,4$ are collective labels for the momentum, helicity and particle type of the external legs.
Other numerators can be obtained from the master using a set of kinematic  identities that follow from the color-kinematics duality.
Thanks to our conventions for the color algebra,
\eqns{jacobi}{commutation},
the dual kinematic relations for both fundamental and adjoint loop states translate to antisymmetrization of the external legs, which we denote by a commutator bracket.
In this way, the triangle and the bubble numerators, shown in \fig{fig:ChiralBoxes},  are obtained from the box as
\beal
   n_{\rm tri}(1,2,3,4,\ell) & \equiv n_{\rm box}([1,2],3,4,\ell) , \\ 
   n_{\rm bub}(1,2,3,4,\ell) & \equiv n_{\rm tri}(1,2,[3,4],\ell)
                                    = n_{\rm box}([1,2],[3,4],\ell) .
\label{Jac1}
\eeal
More explicitly,
$ n_{\rm tri}(1,2,3,4,\ell) = n_{\rm box}(1,2,3,4,\ell) - n_{\rm box}(2,1,3,4,\ell) $,
etc.
Further antisymmetrization gives the numerators of snail- and tadpole-type graphs,
shown in \fig{SnailTadpolesFigure}:
\beal
   n_{\rm snail}(1,2,3,4,\ell) & \equiv n_{\rm box}([[1,2],3],4,\ell) , \\  
   n_{\rm tadpole}(1,2,3,4,\ell) & \equiv n_{\rm box}([[1,2],[3,4]],\ell) , \\
   n_{\rm xtadpole}(1,2,3,4,\ell) & \equiv n_{\rm box}([[[1,2],3],4],\ell) .
\label{Jac2}
\eeal
Here a ``snail'' is synonymous to a bubble on an external leg,
whereas ``tadpole'' and ``xtadpole'' stand for tadpoles
on an internal leg and an external leg, respectively.
These three diagrams will integrate to zero in dimensional regularization,
as they will be identified with scaleless integrals. 

All graphs that we considered so far have the counterclockwise orientation
of the fundamental color flow. At one loop, a symmetry trick can be used to obtain the numerators with the internal-matter arrow reversed.
The matter-conjugated box diagram is given by mirroring the graph,
so that it is equivalent to the master box with all arguments reversed,
\be
   \overline{n}_{\rm box}(1,2,3,4,\ell) = n_{\rm box} (4,3,2,1,-\ell) ,
\label{reverse}
\ee
and the others are obtained by conjugating \eqns{Jac1}{Jac2}.
In \sec{sec:symmetriescuts}, we give an alternative definition of $\overline{n}_i$
using the building blocks of the numerators.

In addition to the above relations, we need to constrain the numerators
with the theory-specific physical information.
This is done using the unitarity method~\cite{Bern:1994zx,Bern:1994cg}.
\Eqns{Jac1}{Jac2} together with the unitarity cuts are equivalent
to imposing the color-kinematics duality through functional equations.
In the absence of a direct way to obtain the duality-satisfying numerators,
we will find a solution of these equations using an ansatz for the master box numerator.

\subsection{The ansatz construction}
\label{sec:ansatz}

In this section, we explicitly construct a compact ansatz
for the master numerator $n_{\rm box}$,
suitable for a wide range of four-point amplitudes of various theories
(see alternative constructions
in \rcites{Carrasco:2012ca,Nohle:2013bfa,Chiodaroli:2013upa} and \citePaper{3}).

To begin with, we list the elementary building blocks that are allowed to appear in the numerators. These include the Mandelstam invariants $s$, $t$ and $u$,
of which only two are independent,
as well as the scalar products $\ell \cdot p_i$ of the loop and external momenta.
Here we introduce a shorthand notation
for the three independent combinations of such products
that we use in the rest of this chapter:
\be
      \ell_s = 2\, \ell \cdot \! (p_1+p_2) , \qquad
      \ell_t = 2\, \ell \cdot \! (p_2+p_3) , \qquad
      \ell_u = 2\, \ell \cdot \! (p_1+p_3) .
\label{lproducts}
\ee

Another invariant is the Lorentz square of the loop momentum $\ell^2$.
However, in dimensional regularization, we must distinguish between
the four-dimensional square and the $d$-dimensional one.
We choose to work with the latter and the difference between the two:
\be
\mu^2 = \ell^2_{d=4} - \ell^2 ,
\label{mu}
\ee
which corresponds to the square of the loop-momentum component
orthogonal to the four-dimensional spacetime
and is positive-valued in the metric signature $(+-\dots-)$.

Finally, we introduce the parity-odd Levi-Civita invariant
\be
      \epsilon(1,2,3,\ell) = {\rm Det}(p_1,p_2,p_3,\ell_{d=4}) ,
\ee
which integrates to zero in gauge theory,
but do contribute to gravity when squared in the double-copy construction~\eqref{BCJformGravity}.

These building blocks already make it possible to write down an ansatz
for each of the distinct helicity configurations of the external states.
For four external gluons, a sufficient set of configurations consists of the split-helicity case $(1^-\!,2^-\!,3^+\!,4^+)$ and the alternating-helicity case $(1^-\!,2^+\!,3^-\!,4^+)$. All other external-state configurations in SYM are obtainable via supersymmetric Ward identities
\cite{Grisaru:1976vm,Grisaru:1977px,Parke:1985pn,Mangano:1990by} and permutations of labels. However, instead of using such identities we choose to rely the extended on-shell superspace formalism of \sec{sec:supermultiplets}
to take care of the bookkeeping of external states.
Namely, for the four-point MHV amplitude, we introduce variables
\be
   \kappa_{ij} = \frac{ [12] [34] }{ \braket{12}\!\braket{34} }
   \delta^{(2{\cal N})}(Q) \braket{ij}^{4-{\cal N}} \theta_{i} \theta_{j}  ,
\label{kappa}
\ee
where the auxiliary parameter $\theta_{i}$ mark that the external leg $i$
belong to the anti-chiral vector multiplet, $\overline{V}_{\cal N}$, also containing the negative-helicity gluon. Implicitly the unmarked legs are of chiral type, $V_{\cal N}$, which contains the positive-helicity gluon.
The super momentum delta function $\delta^{(2{\cal N})}(Q)$ ensures that the
${\cal N}$-fold supersymmetry Ward identities are respected among the components.
The remaining kinematic factors guarantee that the gluonic components take on a simple familiar form,
\be
   \kappa_{ij} = istA_4^{\rm tree}(\dots i^- \!\!\!\dots j^- \!\!\!\dots)
                 (\eta_i^1 \eta_i^2 \eta_i^3 \eta_i^4)
                 (\eta_j^1 \eta_j^2 \eta_j^3 \eta_j^4) +\ldots .
\label{kappagluon}
\ee
For example, the pure-gluon component of $\kappa_{12}$ is explicitly  
$\braket{12}^2 [34]^2$ using the spinor-helicity variables.
Thus, for a given graph topology we can ignore the helicity labels and instead work with a single generating function, spanned by the six independent $\kappa_{ij}$
that encode all the component numerators.

Now we can write down an ansatz for the master numerator of a superamplitude with external vector multiplets and any type of internal massless particles,
\be
   n_{\rm box}(1,2,3,4,\ell) = \sum_{1 \le i<j \le 4}
      \frac{\kappa_{ij}}{s_{ij}^{N}}
      \Big( \sum_{k} a_{ij;k} M_{k}^{(N)}
          + i \epsilon(1,2,3,\ell) \sum_{k} \tilde a_{ij;k} M_{k}^{(N-2)}
      \Big) ,
\label{ansatz}
\ee
where $a_{ij;k}$ and $\tilde a_{ij;k}$ are free parameters of the ansatz,
$s_{ij} \equiv (p_i+p_j)^2$ are the Mandelstam invariants,
and $M^{(N)}_k$ denotes a monomial of dimension $2N$
built out of $N$ products of quadratic Lorentz invariants.
The monomials are drawn from the following set:
\be
   M^{(N)} = \Big\{ \prod_{i=1}^N m_i ~ \Big|~
                    m_i \in \{ s,\,t,\,\ell_s,\,\ell_t,\,\ell_u,\,\ell^2,\,\mu^2 \} 
             \Big\} ,
\label{monomialset}
\ee
which contains $C_{N+6}^6$ distinct elements,  $C_{n}^k$ being the binomial numbers.
Hence the ansatz have
$ 6 (C_{N+6}^6 + C_{N+4}^6) $ free coefficients in total.
We vary the parameter $N$ in \eqn{ansatz} depending on
the effective number of supersymmetries ${\cal N}_{\rm eff}$ of the amplitude, such that
\be
   N=4-{\cal N}_{\rm eff} ,
\ee
which is in accord with the classic loop-momentum power-counting argument of \rcite{Bern:1994cg} and the ansatz rules for color-kinematics numerators of \rcite{Bern:2012uf}.

As a simple and illuminating example that
the ansatz \eqref{ansatz} is finely tuned to the correct answer,
we can consider the case of ${\cal N}=3$ SYM,
which is equivalent to ${\cal N}_{\rm eff}=4$ SYM on shell.
Thus, using $N=0$, the ansatz becomes the six-fold parametrization
\be
   n_{\rm box}^{{\cal N}=3}(1,2,3,4,\ell)=\sum_{1\le i<j\le4}a_{ij;0}\,\kappa_{ij} ,
\ee
where in order to make $\kappa_{ij}$ well defined it is given by \eqn{kappa} using  ${\cal N}=3$.
As is easily checked, for $a_{ij;0}=1$ the sum collapses to the known answer
proportional to the tree superamplitude:
\be
   n_{\rm box}^{{\cal N}=3} = n_{\rm box}^{{\cal N}=4}
      = i s t A^{\rm tree}_4(p_i, \eta^A_i)
      = \frac{ [12] [34] }{ \braket{12} \braket{34} } \delta^{(8)}(Q) .
\label{N4Box}
\ee

Finally, note that the ansatz \eqref{ansatz} was chosen so that
the numerators with minus-helicity legs $i$ and $j$ will only contain
poles in the $s_{ij}$ channel.
This is a nontrivial highly-constraining aesthetic condition.
We found that such pole structure gives a necessary and sufficient class
of non-localities for the four-point one-loop numerators,
if they are built as rational functions of gauge-invariant building blocks.
Certainly, numerator non-localities can be completely avoided
by using gauge-dependent building blocks,
such as formal polarization vectors~\cite{Bern:2013yya, Nohle:2013bfa},
but we choose to work with the former type of ansatz.

\subsection{Imposing symmetries and cuts}
\label{sec:symmetriescuts}

Before proceeding with the calculations of specific theories, we impose some extra universal constraints on the box numerator ansatz. They are not always necessary for the color-kinematics duality to be satisfied, but they often simplify the construction of the amplitudes both in gauge theory and in gravity.

\begin{figure}
      \centering
      \includegraphics[scale=1.00]{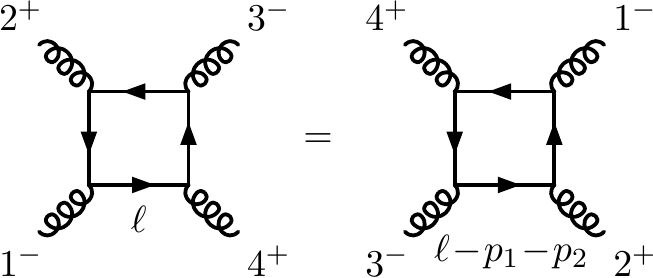}
\caption[a]{\small The cyclic symmetry of the alternating-helicity box diagram
                   that should be imposed on the numerator ansatz}
\label{fig:rotation}
\end{figure}

We choose to impose the cyclic symmetry natural to the box topology:
\be
   n_{\rm box}(1,2,3,4,\ell) = n_{\rm box}(2,3,4,1,\ell-p_1) . 
\label{rotatesymmetry}
\ee
Without this constraint the relabeling symmetry of the box numerator, and the decendant numerators \eqref{Jac1} and \eqref{Jac2}, would cease to be manifest.
In general, the constraint \eqref{rotatesymmetry} give identifications of various components in the supersymmetry expansion, and in particular it implies a two-site permutation symmetry of the alternating-helicity component $n_{\rm box}(1^-,2^+,3^-,4^+,\ell)$.

\begin{figure}[t]
      \begin{subfigure}{0.5\textwidth}
      \centering
      \includegraphics[scale=1.00]{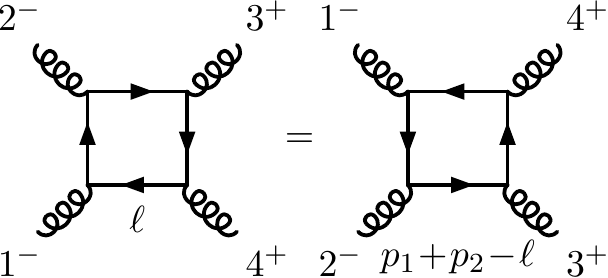}
      \caption[a]{\small \label{flip1}}
      \end{subfigure}
      \begin{subfigure}{0.5\textwidth}
      \centering
      \includegraphics[scale=1.00]{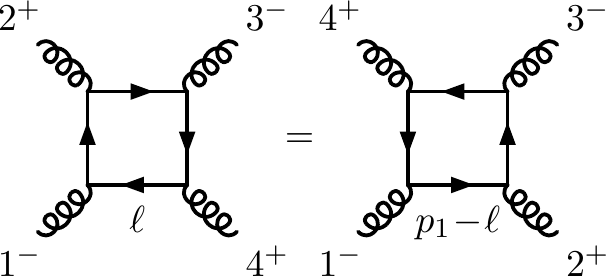}
      \caption[a]{\small \label{flip2}}
      \end{subfigure}
      \vspace{-5pt}
\caption[a]{\small The anti-fundamental boxes can be obtained
as the flipped fundamental boxes.
Through the matter-conjugation operation,
these relations impose constraints on the ansatz.}
\label{boxflip}
\end{figure}

Furthermore, we would like to extend the cyclic symmetry of the box numerator to a dihedral symmetry. The diagram flip relation \eqn{reverse} should then be imposed as a symmetry constraint on the ansatz. Recall that this equation reads
\be
   \overline{n}_{\rm box}(1,2,3,4,\ell) = n_{\rm box}(4,3,2,1,-\ell) .
\label{reverse2}  
\ee
The graph-symmetry origin of this equation is illustrated in \fig{boxflip}, where it is applied to the two distinct helicity components of the box numerator.
To use the above equation as a constraint, we need to precisely define how reversing the internal matter arrow operates on the ansatz \eqref{ansatz}. The matter conjugation should flip the chirality of the internal-loop matter particles while keeping the external chiral vectors unaltered. By using CPT invariance, we can alternatively regard this as a flip of the chirality of the external vector particles while keeping the internal matter unaltered. For consistent interactions, we also need to flip the sign of parity-odd momentum invariants.
The conjugated one-loop four-point MHV numerators are then
\be
   \overline{n}_i(1,2,3,4,\ell) = n_i(1,2,3,4,\ell)
      \big|_{ \kappa_{ij} \rightarrow \kappa^{\complement}_{ij} ,\,
              \epsilon(1,2,3,\ell)\rightarrow -\epsilon(1,2,3,\ell)} ,
\label{matterConjugation}
\ee
where $ \kappa^{\complement}_{ij} = \kappa_{kl} $ marks the pair of legs
$\{k,l\}=\{1,2,3,4\} \setminus \{i,j\}$ unmarked by $\kappa_{ij}$. Combining \eqns{matterConjugation}{reverse2} give the desired extra constraint.

\begin{figure}[t]
      \centering
      \begin{subfigure}{0.32\textwidth}
            \includegraphics[scale=1.00]{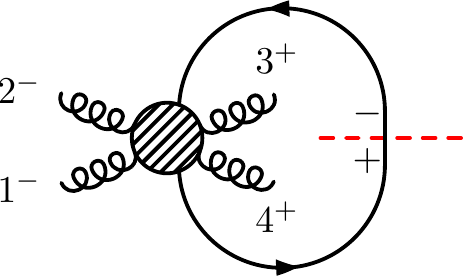}
      \caption[a]{\small \label{fig:singlecut1}}
      \end{subfigure}
      \begin{subfigure}{0.32\textwidth}
            \includegraphics[scale=1.00]{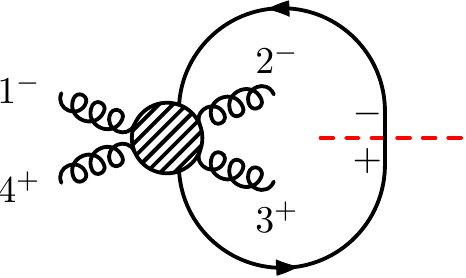}
      \caption[a]{\small \label{fig:singlecut2}}
      \end{subfigure}
      \begin{subfigure}{0.32\textwidth}
            \includegraphics[scale=1.00]{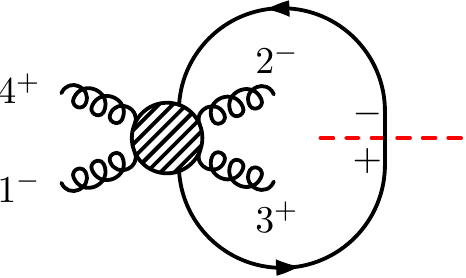}
      \caption[a]{\small \label{fig:singlecut3}}
      \end{subfigure}
      \vspace{-5pt}
\caption[a]{\small Non-planar single-line cuts used to compute the numerators and amplitudes}
\label{fig:singlecuts}
\end{figure}

The ansatz~\eqref{ansatz} is ultimately constrained by the unitarity cuts.
We choose to work with the non-planar single-line cuts shown in \fig{fig:singlecuts}. These cuts can be constructed by taking color-ordered six-point tree-level amplitudes
and identifying a conjugate pair of fundamental particles on opposite-site external legs. Because the tree amplitude is color ordered, the identification of momenta on opposite ends of the amplitude does not produce singularities corresponding to soft or collinear poles.

As the internal loop momenta are subject to only one constraint, $\ell^2=0$,
the single-line cuts are sensitive to most terms in the integrand of the amplitude. Undetected terms correspond to tadpoles and snails (external bubbles), which invariably integrates to zero in a massless theory. Indeed, the tadpole and snail graphs previously described do not contribute to this cut, instead these will be indirectly constrained through the kinematic relations of the numerators. 

Even after the symmetries and unitarity cuts are imposed on the numerators,
there remain free parameters in ansatz that can be interpreted as the residual generalized gauge freedom of the current amplitude representation.
For simplicity, we will fix this freedom by suitable aesthetic means,
as discussed in the next section.

\subsection{The amplitude assembly}
\label{sec:amplitudeconstruction}

In this section, we provide the precise details on how to assemble the full one-loop gauge-theory amplitude from the numerators and the color factors. This discussion will be valid for all of the YM theories and amplitudes discussed in the subsequent sections.

The complete MHV (super-)amplitude can be written as
\be
   {\cal A}^{\text{1-loop}}_4 = \sum_{S_4}
      \Big(
            \frac{1}{8}\, {\cal I}_{\rm box}
          + \frac{1}{4}\, {\cal I}_{\rm tri}
          + \frac{1}{16}\,{\cal I}_{\rm bub}
      \Big) ,
\label{ampl}
\ee
where the three non-scalar integrals correspond to the canonically labeled graphs
in \fig{fig:ChiralBoxes}, and the sum runs over the set $S_4$
of 24 permutations of the four external legs.
The rational numbers correspond to the symmetry factors that remove overcounting,
both from the permutation sum and the internal phase-space integration.
In the case of fundamental matter multiplets in the loop,
each integral is formed by combining two diagrams with opposite orientations of the internal-matter arrow:
\be
   {\cal I}_i = \int\!\!\frac{d^{d}\ell}{(2\pi)^{d}}
      \frac{(c_i n_i + \overline{c}_i \overline{n}_i)}{D_i} ,
\label{Ifund}
\ee
where $d=4-2\epsilon$ is the spacetime dimension in dimensional regularization,
and the denominators $D_i$ are products of the squared momenta
of the four internal lines in each graph in \fig{fig:ChiralBoxes},
thus accounting for the four propagators.
The color factors for the canonically labeled box, triangle and bubble graphs are
\beal
   c_{\rm box} & = \Tr(T^{a_1}T^{a_2}T^{a_3}T^{a_4}) , \\
   c_{\rm tri} \;\: & = \Tr([T^{a_1},T^{a_2}]T^{a_3}T^{a_4}) , \\
   c_{\rm bub} & = \Tr([T^{a_1},T^{a_2}][T^{a_3},T^{a_4}]) .
\label{fundColorFactors}
\eeal
The antifundamental color factors $\overline{c}_i$ are obtained from $c_i$ by replacing $T^a_{i\bar \jmath} \rightarrow T^a_{\bar \imath j}$ as discussed in \sec{sec:bcjfund}.

From \eqn{fundColorFactors}, it is obvious that the fundamental matter amplitude
can be rewritten in terms of only single traces,
thus giving color-ordered partial amplitudes in its decomposition
\be
   {\cal A}^{\text{1-loop,fund}}_4(1,2,3,4)
    = \!\!\sum_{\{i_1,i_2,i_3,i_4\} \in S_4/Z_4 \hspace{-30pt}}\!\!\!\!
      \Tr(T^{a_{i_1}}T^{a_{i_2}}T^{a_{i_3}}T^{a_{i_4}}\!)\,
      A^{\rm fund}_4(i_1,i_2,i_3,i_4) ,
\label{fundOrdering}
\ee
where the sum runs over the cyclically inequivalent subset of $S_4$.

As explained in \sec{sec:bcjfund}, the fully adjoint amplitude can be obtained
from the amplitude in \eqn{ampl} after swapping the generators inside the color factors~\eqref{representationswap}.
The integrals ${\cal I}_i$ then take the following form: 
\be
   {\cal I}_i = \int\!\!\frac{d^{d}\ell}{(2\pi)^{d}}
      \frac{c_i^{\rm adj} n_i^{\rm adj}}{D_i} ,
\label{Iadj}
\ee
where the adjoint color factors are given by
\beal
   c_{\rm box}^{\rm adj} & = \tf^{ba_1c}\tf^{ca_2d}\tf^{da_3e}\tf^{ea_4b} , \\
   c_{\rm tri}^{\rm adj} \: & = \tf^{a_1a_2c}\tf^{bcd}\tf^{da_3e}\tf^{ea_4b} , \\
   c_{\rm bub}^{\rm adj} & = \tf^{a_1a_2c}\tf^{bcd}\tf^{deb}\tf^{ea_3a_4} .
\label{adjColorFactors}
\eeal

In the color-ordering decomposition of the amplitude, the adjoint color factors~\eqref{adjColorFactors}
produce double traces, which in SU($N_c$) gives the following:
\beal
   {\cal A}^{\text{1-loop,adj}}_4(1,2,3,4) =\;
    & N_c\!\!\!\! \sum_{\{i_1,i_2,i_3,i_4\} \in S_4/Z_4 \hspace{-30pt}}\!\!\!\!\!\;
      \Tr(T^{a_{i_1}}T^{a_{i_2}}T^{a_{i_3}}T^{a_{i_4}}\!)\,
      A^{\rm adj}_4(i_1,i_2,i_3,i_4) \\
    & + \!\! \!\! \sum_{\{i_1,i_2,i_3,i_4\} \in S_4/S_{4;2}\hspace{-33pt}}\!\!\!\!
      \Tr(T^{a_{i_1}}T^{a_{i_2}}\!)\Tr(T^{a_{i_3}}T^{a_{i_4}}\!)\,
      A^{\rm adj}_{4;2}(i_1,i_2,i_3,i_4) ,
\label{adjOrdering}
\eeal
where $S_{4;2}$ is the subset of $S_4$ that leaves the double trace structure invariant.
The partial amplitudes $A^{\rm adj}_{4;2}$ in the subleading-color part are not independent but
related to the leading-color partial amplitudes $A^{\rm adj}_{4}$~\cite{Bern:1990ux,Bern:1994zx}.

A difference in  the structure of the one-loop color-ordered partial amplitudes
$A^{\rm fund}_{m}$ and $A^{\rm adj}_{m}$ is that the latter satisfy refection relations,
\be
      A^{\rm adj}_m(1,2,3,\dots,m) = (-1)^m A^{\rm adj}_m(m,\dots,,3,2,1) ,
\label{mirror}
\ee
whereas the former generally do not. However, for the cases where the matter is effectively non-chiral, and thus $n_i=\bar{n}_i$, the reflection relation is restored for the fundamental partial amplitudes. In this case, it is also true that the partial amplitudes $A^{\rm adj}_4$ and $A^{\rm fund}_4$ in \eqns{fundOrdering}{adjOrdering} are identical.

\subsection{The amplitude with ${\cal N}=2$ fundamental or adjoint matter}
\label{sec:N2}

Here we construct the four-point one-loop amplitude with a ${\cal N}=2$ fundamental matter multiplet $\Phi_{{\cal N}=2}$ circulating in the loop
and external adjoint vector multiplets ${\cal V}_{{\cal N}=2}$.
Because $\Phi_{{\cal N}=2}$ has maximal amount of supersymmetry for matter, it is effectively a non-chiral multiplet. Only the color representation distinguish it from its conjugate multiplet. Similarly, the color-stripped partial amplitudes are insensitive to the representation. To see this note that the particle content of the multiplets $\Phi_{{\cal N}=2}$,
$\overline{\Phi}_{{\cal N}=2}$ in \eqn{chiralMatterMult}
and $\Phi_{{\cal N}=2}^{\rm adj}$ in \eqn{adjointNeq2} coincide
up to the gauge-group representation indices.
Using this fact, we can equate the one-loop kinematical numerators for different representations,
\be
      n_i^{{\cal N}=2,{\rm fund}} = \overline{n}_i^{{\cal N}=2,{\rm fund}} = 
      n_i^{{\cal N}=2,{\rm adj}} ,
\label{fundadjN2}
\ee
which reduces the fundamental matter computation to the one with the adjoint matter.

Color-kinematics dual numerators for the component amplitude with external gluons
and adjoint ${\cal N}=2$ internal matter have already been constructed
in \rcites{Carrasco:2012ca,Bern:2013yya,Nohle:2013bfa} and \citePaper{3}.
The result presented here is a minor variation of these known forms.
The main novelties are that we use the compact ansatz~\eqn{ansatz},
embed the external gluons into their vector supermultiplets,
generalize the internal color representation
and give the complete results for the integrand, including the $\mu$-terms and the snail diagrams.

Using the ansatz in \eqn{ansatz} for $N=4-{\cal N}_{\rm eff}=2$,
we have 174 free parameters to solve for.
By combining the equation
\be
      n_{\rm box}=\overline{n}_{\rm box} ,
\label{CPTN2}
\ee
with the definition~\eqref{matterConjugation} of the antifundamental numerators,
we can reduce the ansatz by a half, leaving only 87 free parameters.

Imposing the dihedral symmetries from~\sec{sec:symmetriescuts} gives further constraints:
the cyclic symmetry~\eqref{rotatesymmetry} fixes 58 parameters.
A single further variable is fixed by the flip relation~\eqref{reverse},
which becomes a symmetry after imposing \eqn{CPTN2}. Enforcing the $d$-dimensional unitarity cuts in \fig{fig:singlecuts} on the cut integrand fixes 24 out of 28 remaining parameters.
At this point, the cubic diagram representation satisfies all conditions
to be the correct amplitude, along with the built-in color-kinematics duality.

We choose to fix the remaining four free parameters
by imposing aesthetic or practical constraints
to obtain compact and manageable numerator expressions.
Requiring that the snail numerator $n_{\rm snail}$
vanishes for any on-shell external momenta gives two additional conditions.
Finally, requiring that the parity-even part of the $s$-channel triangle
defined in \eqn{Jac1} is proportional to $s$ gives two more conditions.
The latter implicitly enforces no dependence on $\kappa_{12}$ and $\kappa_{34}$
in that triangle.\footnote{Alternatively, one can demand that the bubble numerator vanishes,
as was done in \rcites{Carrasco:2012ca,Nohle:2013bfa} and \citePaper{3}.
This also fixes the last two parameters,
but gives a more complicated expression for the triangle.}

Having thus solved for all free parameters, we obtain the following box numerator:
\beal
      n_{\rm box}^{{\cal N}=2,{\rm fund}}
      = & \, (\kappa_{12}+\kappa_{34}) \frac{(s-\ell_s)^2}{2s^2}
        + (\kappa_{23}+\kappa_{14}) \frac{\ell_t^2}{2t^2}
        + (\kappa_{13}+\kappa_{24}) \frac{st+(s+\ell_u)^2}{2u^2} \\
      & - 2i \epsilon(1,2,3,\ell)\frac{\kappa_{13}-\kappa_{24}}{u^2}
        + \mu^2 \Big( \frac{\kappa_{12}+\kappa_{34}}{s}
                     +\frac{\kappa_{23}+\kappa_{14}}{t}
                     +\frac{\kappa_{13}+\kappa_{24}}{u} \Big) ,
\label{N1evenBox}
\eeal
where the short-hand notation~\eqref{lproducts} for loop-momentum invariants is used and the parameters $\kappa_{ij}$ encoding the external multiplets
are defined in \eqn{kappa}.

The other numerators are given by the kinematic relations in \eqns{Jac1}{Jac2}.
To be explicit, the triangle numerator is
\beal
      n_{\rm tri}^{{\cal N}=2,{\rm fund}}
      = & \, (\kappa_{23}+\kappa_{14})\frac{s(t-2\ell_t)}{2t^2}
        - (\kappa_{13}+\kappa_{24})\frac{s(u-2\ell_u)}{2u^2} \\
      & + 2i \epsilon(1,2,3,\ell)\frac{\kappa_{23}-\kappa_{14}}{t^2}
        + 2i \epsilon(1,2,3,\ell)\frac{\kappa_{13}-\kappa_{24}}{u^2} ,
\eeal
and the internal bubble numerator is
\be
      n_{\rm bub}^{{\cal N}=2,{\rm fund}}
      = s \Big( \frac{\kappa_{23}+\kappa_{14}}{t}-\frac{\kappa_{13}+\kappa_{24}}{u}\Big) .
\ee

In principle, we could ignore any further contributions to the integrand of the amplitude,
since snail (external bubble) and tadpole diagrams should integrate to zero
in dimensional regularization. 
However, for completeness, we give the result for the only non-vanishing diagram
of this type -- the snail.
Because of ${\cal N}=2$ supersymmetry power counting, it must include an overall factor $p_4^2$ (in analogy to a one-loop propagator correction), which vanishes on shell.
This contribution is not visible in our ansatz,
because the external legs were placed on shell from the very start.
Nevertheless, a careful analysis of a singular two-particle unitarity cut reveals
that the numerator of the snail diagram shown in \fig{SnailTadpolesFigure} is given by
\be
      n_{\rm snail}^{{\cal N}=2,{\rm fund}}
      = -\frac{p_4^2}{2} \Big( \frac{\kappa_{23}+\kappa_{14}}{t}
                             - \frac{\kappa_{13}+\kappa_{24}}{u} \Big) .
\ee
As the snail graph has a propagator $1/p_4^2$,
the above numerator gives a finite contribution to the integrand,
after the $0/0$ is properly canceled out.
Once this is done the snail-diagram contribution can be included into the amplitude~\eqref{ampl}
as
\be
      \frac{1}{4} \sum_{S_4} {\cal I}_{\rm snail} ,
\label{snailampl}
\ee
where ${\cal I}_{\rm snail}$ is defined by eqs.~\eqref{Ifund} or~\eqref{Iadj}
with the respective color factors
\be
      c_{\rm snail} = \Tr([[T^{a_1},T^{a_2}],T^{a_3}]T^{a_4})
      \qquad \text{or} \qquad
      c_{\rm snail}^{\rm adj} = \tf^{a_1a_2c}\tf^{ca_3d}\tf^{bde}\tf^{ea_4b} .
\label{snailFundColorFactor}
\ee

Of course, the snail diagram~\eqref{snailampl} still integrates to zero,
so it would be justified to drop it.
Nevertheless, we choose to explicitly display the snail graph
because there is some potential interest in the planar ${\cal N}=2$ integrand,
in analogy with the recent advances with the planar ${\cal N}=4$ integrand~\cite{ArkaniHamed:2010kv,ArkaniHamed:2012nw}.
Lastly, the two remaining numerators $n_\text{tadpole}$ and $n_\text{xtadpole}$ are
manifestly zero, consistent with naive expectations in ${\cal N}=2$ supersymmetric theories.

If we now assemble the full amplitude~\eqref{ampl}
and recast it into the color-ordered form~\eqref{fundOrdering},
we obtain the two inequivalent partial amplitudes in $D=4-2\epsilon$,
which are most easily expressed as
\begin{subequations} \begin{align}
      A^{{\cal N}=2,{\rm fund}}_4(1^-\!,2^-\!,3^+\!,4^+) =
            \frac{i \braket{12}^2 [34]^2}{ (4\pi)^{D/2} } 
            \bigg\{\!-&\frac{1}{st} I_2(t) \bigg\} , \\
      A^{{\cal N}=2,{\rm fund}}_4(1^-\!,2^+\!,3^-\!,4^+) =
            \frac{i \braket{13}^2 [24]^2}{ (4\pi)^{D/2} }
            \bigg\{\!-&\frac{r_{\Gamma}}{2u^2}
                       \left( \ln^2\!\left( \frac{-s}{-t} \right)\!+ \pi^2 \right)
                     + \frac{1}{su} I_2(s) + \frac{1}{tu} I_2(t)
            \bigg\} ,
\end{align} \label{N2fund} \end{subequations}
\!\!\!\!\;where $I_2$ is the standard scalar bubble integral
\be
      I_2(t) = \frac{r_{\Gamma}}{\epsilon (1-2\epsilon)} (-t)^{-\epsilon} .
\label{I2}
\ee
Here and below, the integrated expressions are shown up to $O(\epsilon)$,
and the standard prefactor of dimensional regularization is
\begin{equation}
      r_{\Gamma} = \frac{ \Gamma(1+\epsilon) \Gamma^2(1-\epsilon) }
                        { \Gamma(1-2\epsilon) } .
\label{rgamma}
\end{equation}

Due to \eqn{fundadjN2}, the partial amplitudes~\eqref{N2fund} are the coefficients of both the clockwise and counterclockwise fundamental color traces in the color-dressed amplitude.
Moreover, they coincide with the primitive amplitudes $A^\text{1-loop,adj}_4$
with adjoint ${\cal N}=1$ matter in the loop (same as adjoint ${\cal N}=2$ in our convention, see \eqn{adjointNeq2}), which is the version that is best known in the literature~\cite{Bern:1994cg,Bern:1995db}.

\subsection{The amplitude with ${\cal N}=1$ fundamental matter}
\label{sec:N1}

In this section, we work out the numerators of the four-point one-loop amplitude
with a ${\cal N}=1$ fundamental matter multiplet $\Phi_{{\cal N}=1}$
circulating in the loop and adjoint vectors ${\cal V}_{{\cal N}=1}$
on the external legs.
The result is the first known color-kinematics representation of this amplitude.

At one loop, the amplitude, along with its numerators,
can be naturally decomposed into two simpler components:
the parity-even and parity-odd contributions:
\begin{subequations} \begin{align}
   n_{i}^{{\cal N}=1, {\rm fund}} & =
      \frac{1}{2}\, n_{i}^{{\cal N}=1, {\rm even}}
    + \frac{1}{2}\, n_{i}^{{\cal N}=1, {\rm odd}} ,
\label{nfund} \\
   \overline{n}_{i}^{{\cal N}=1, {\rm fund}} & =
      \frac{1}{2}\, n_{i}^{{\cal N}=1, {\rm even}}
    - \frac{1}{2}\, n_{i}^{{\cal N}=1, {\rm odd}} ,
\label{nantifund}
\end{align} \label{nfundboth} \end{subequations}
\!\!\!\!\;where we define
\begin{subequations} \begin{align}
      n_{i}^{{\cal N}=1, {\rm even}} & \equiv n_{i}^{{\cal N}=1, {\rm fund}}
                                     + \overline{n}_{i}^{{\cal N}=1, {\rm fund}} ,
\label{neven} \\
      n_{i}^{{\cal N}=1, {\rm odd}}  & \equiv n_{i}^{{\cal N}=1, {\rm fund}}
                                     - \overline{n}_{i}^{{\cal N}=1, {\rm fund}} .
\label{nodd}
\end{align} \label{nevenodd} \end{subequations}
\!\!\!\!\;According to \eqn{nplusnbar}, summing over the fundamental and antifundamental one-loop numerators effectively corresponds to promoting the ${\cal N}=1$ multiplet to the adjoint representation, which by \eqn{adjointNeq2} is equivalent to a contribution of a ${\cal N}=2$ multiplet. Thus we have the following equalities:
\be
      n_{i}^{{\cal N}=1, {\rm even}} = n_{i}^{{\cal N}=2, {\rm adj}}
                                     = n_{i}^{{\cal N}=2, {\rm fund}} .
\ee

Since we have already found compact expressions for the ${\cal N}=2$ numerators
in \sec{sec:N2}, we can now focus entirely on the parity-odd ${\cal N}=1$ numerators.
Unlike the parity-even ones, they have the expected amount of supersymmetry,
hence ${\cal N}_{\rm eff}=1$.
Using the ansatz~\eqref{ansatz} as before, except with $N=4-{\cal N}_{\rm eff}=3$,
gives us 546 free parameters to solve for.
Similarly to the procedure in the previous section,
we can immediately eliminate half of the parameters by imposing the defining property
that the box numerator should be odd under matter conjugation:
\be
      n_{\rm box}^{\rm odd}=-\overline{n}_{\rm box}^{\rm odd} ,
\label{CPTN1}
\ee
which reduces the problem to 273 undetermined parameters.

Next comes the dihedral symmetry of \sec{sec:symmetriescuts}:
the cyclic symmetry~\eqref{rotatesymmetry} fixes 208 parameters.
An additional 20 are constrained by the flip relation~\eqref{reverse},
which becomes a symmetry after imposing \eqn{CPTN1}. 

Out of the remaining 45 free parameters, 20 are fixed
by the four-dimensional unitarity cuts shown in \fig{fig:singlecuts}.
Requiring that the snail numerator vanishes on shell gives 17 additional constraints. Demanding that the power counting of the numerator
is at worst $\ell^m$ for one-loop $m$-gons gives 3 more conditions,
leaving only five parameters to fix.
One can check that four of them correspond to
the genuine freedom of the color-kinematics representations,
and one parameter is determined by $d$-dimensional unitarity cuts.
However, since it is difficult to analytically continue four-dimensional chiral fermions
to $d>4$, computing the correct unitarity cut is challenging.
Instead, in our final representation,
we denote this unfixed parameter by $a$
and manually choose the remaining four parameters
to the values that give more compact expressions.

The box numerator for the ${\cal N}=1$ odd contribution is then given by
\beal
      n_{\rm box}^{{\cal N}=1, {\rm odd}} \!=\,
            & (\kappa_{12}-\kappa_{34}) \frac{(\ell_s-s)^3}{2s^3}
            + (\kappa_{23}-\kappa_{14}) \frac{\ell_t^3}{2 t^3}
            + (\kappa_{13}-\kappa_{24}) \frac{1}{2}
              \Big( \frac{\ell_u^3}{u^3} + \frac{3s\ell_u^2}{u^3}
                                         - \frac{3s\ell_u}{u^2} + \frac{s}{u} \Big) \\
          & - 2i \epsilon(1,2,3,\ell)  (\kappa_{13}+\kappa_{24}) \frac{2\ell_u-u}{u^3}
            - a \mu^2 (\kappa_{13}-\kappa_{24})\frac{s-t}{u^2} ,
\label{N1oddBox}
\eeal
and the triangle and bubble numerators can be obtained using the definition~\eqref{Jac1},
see \rcitePaper{4} for explicit expressions.
As for the three remaining numerators
$n_{\rm snail}$, $n_{\rm tadpole}$ and $n_{\rm xtadpole}$, defined by \eqn{Jac2},
they are manifestly zero in this representation.

We note that the combinations of $n_i$ and $c_i$ that appear in \eqn{Ifund}
can be rewritten as follows:
\be
      n_i^{{\cal N}=1,{\rm fund}} c_i
    + \overline{n}_i^{{\cal N}=1,{\rm fund}} \overline{c}_i
    = \frac{1}{2}\, n_i^{{\cal N}=1,{\rm even}}(c_i+\overline{c}_i)
    + \frac{1}{2}\, n_i^{{\cal N}=1,{\rm odd}}(c_i-\overline{c}_i) .
\ee
This implies that we can write the following relation for the color-dressed amplitudes:
\be
      {\cal A}^{{\cal N}=1,{\rm fund}}_4
    = \frac{1}{2}\, {\cal A}^{{\cal N}=2,{\rm fund}}_4
    + \frac{1}{2}\, {\cal A}^{{\cal N}=1,{\rm odd}}_4 ,
\label{AN1}
\ee
where the last ``amplitude'' is constructed from parity-odd numerators and color factors,
\be
      c_i^{\rm odd}=c_i-\overline{c}_i .
\ee
Although chiral gauge anomalies are not the topic of the current work, it is interesting to note that this contribution contains all the anomalies of the ${\cal N}=1$ amplitude.

The two inequivalent color-ordered amplitudes for the odd part must coincide with
those for chiral fermions (see \sec{sec:N0fermion}).
These amplitudes were, for example, computed from four-dimensional unitarity cuts and locality conditions in \rcite{Huang:2013vha};
\begin{subequations} \begin{align}
      A^{{\cal N}=1,{\rm odd}}_4(1^-\!,2^-\!,3^+\!,4^+) & = 0 , \\
      A^{{\cal N}=1,{\rm odd}}_4(1^-\!,2^+\!,3^-\!,4^+) & =
            \frac{i r_{\Gamma} \braket{13}^2 [24]^2}{ (4\pi)^{D/2} }
            \bigg\{ \frac{s\!-\!t}{2u^3}
                    \left( \ln^2\!\left( \frac{-s}{-t} \right)\!+ \pi^2 \right)
                 +\!\frac{2}{u^2} \ln\!\left( \frac{-s}{-t} \right)
                 -  \frac{s\!-\!t}{2stu}
            \bigg\} .
    \label{N1odd} 
\end{align}\end{subequations}
After comparing to these results, the integration of the color-kinematics representation~\eqref{N1oddBox} fixes the last parameter to $a=3/2$.
Interestingly, the precise value of $a$ is irrelevant for the gravity amplitudes constructed in \sec{sec:gravity},
as the parameter $a$ drops out after the double-copy amplitudes are integrated.

\subsection{The MHV amplitude with fundamental ${\cal N}=0$ scalar matter}
\label{sec:N0scalar}

In this section, we construct the MHV amplitude with the fundamental scalar
$\Phi^{\rm scalar}_{{\cal N}=0}$ circulating in the loop and external adjoint vectors
${\cal V}_{{\cal N}=0}$, \ie gluons.
As in the ${\cal N}=2$ case, after ignoring the color representation the scalar matter is effectively CPT-invariant and thus the color-stripped amplitudes for $\Phi_{{\cal N}=0}^{\rm scalar}$,
$\overline{\Phi}_{{\cal N}=0}^{\rm scalar}$ and $\Phi_{{\cal N}=0}^{\rm adj\;scalar}$
are the same. Therefore, we can equate the numerators of the different multiplets:
\be
      n_i^{\rm scalar} = \overline{n}_i^{\rm scalar} = 
      n_i^{\rm adj\;scalar} .
\label{fundadjN0}
\ee
The color-kinematics representation of the amplitude with an adjoint scalar
has been previously constructed~\cite{Bern:2013yya,Nohle:2013bfa}
in a fully-covariant form (using formal polarization vectors).
The result presented here will be a helicity-based computation of this amplitude.
The main novelties are that we use the compact ansatz~\eqn{ansatz}
that relies on the four-dimensional notation for the external legs,
and that we trivially generalize the internal color representation.
Note that we only give the MHV amplitude, leaving out, for brevity,
the amplitudes exclusively present in non-supersymmetric theories:
the all-plus-helicity and one-minus-helicity amplitudes,
and their parity conjugates (see \rcite{Nohle:2013bfa} for these).

Using the ansatz~\eqref{ansatz} with ${\cal N}_{\rm eff}=0$
gives a parametrization with 1428 free variables.
Imposing that
\be
      n_{\rm box} = \overline{n}_{\rm box} ,
\label{CPTN0}
\ee
immediately reduces the undetermined parameters by a factor of two.
The cyclic symmetry~\eqref{rotatesymmetry} fixes
512 out of 714 remaining parameters,
and the flip relation~\eqref{reverse} eliminates 29 more.
The $d$-dimensional unitarity cuts in \fig{fig:singlecuts} fix 118 parameters,
leaving 55 free.

Next we impose practical constraints.
Similarly to \rcites{Bern:2013yya,Nohle:2013bfa},
we demand that the snail diagram in \fig{SnailTadpolesFigure}
gives a scaleless integral:
otherwise it will not necessarily integrate to zero.\footnote{Imposing
that the snail numerator vanish would be ideal,
but this is not consistent with the ansatz~\eqref{ansatz}
and the cuts in the ${\cal N}=0$ case.}
A scaleless snail integral is achieved
if its numerator is allowed to be a function of only one scalar product
between the external and internal momenta, namely $p_4 \cdot \ell$,
where $p_4$ is the external momentum directly entering the bubble subgraph of the snail.
This means that the integral is a function of only $p_4^2=0$,
so it must vanish in dimensional regularization.
This constraint fixes 25 parameters.

Continuing with aesthetic conditions, we require that the triangle numerator
has no dependence on $\kappa_{12}$ or $\kappa_{34}$, which gives 20 more conditions. Demanding that the power counting of each $m$-gon numerator is at most $\ell^m$
gives 5 more constraints.
The remaining 5 parameters are fixed manually to obtain more compact expressions for the numerators.

As a result, we obtain the following box numerator for the one-loop four-gluon amplitude with a scalar in the loop:
\beal
   n_{\rm box}^{\rm scalar} \! =
      & - (\kappa_{12}+\kappa_{34})
          \Big( \frac{\ell_s^4}{4 s^4} - \frac{\ell_s^2 (2 \ell^2 + 3 \ell_s)}{4 s^3}
              + \frac{2 \ell^2 \ell_s + \ell_s^2 - 2 \mu^4}{2 s^2}
              - \frac{2 \ell^2 - \ell_s + s}{4 s} \Big) \\
      & - (\kappa_{23}+\kappa_{14})
          \Big( \frac{\ell_t^4}{4 t^4}
              - \frac{\ell_t^2 (2 \ell^2 - \ell_s - \ell_u + t)}{4 t^3}
              - \frac{\mu^4}{t^2} \Big) \\
      & - (\kappa_{13}+\kappa_{24})
          \Big( \frac{\ell_u^3 (\ell_u + 3 s)}{4 u^4}
              - \frac{\ell_u( \ell_u (2 \ell^2 - \ell_s) - \ell_s^2 + \ell_t^2
                            + 4 s (\ell^2 + \ell_u + 2 \mu^2))}{4 u^3} \\
      & ~~~~~~~~~~~~~~~~~~~~~~~~~~~~~~~
              - \frac{\ell_s^2 - \ell_t^2 + 3 \ell_u^2 + 4 \ell^2 t
                     + 8 \mu^2  (\ell_u - s + \mu^2)}{8 u^2}
              - \frac{\ell_s - s}{4 u} \Big) \\ 
      & + 2i \epsilon(1,2,3,\ell) (\kappa_{13}-\kappa_{24})
          \frac{\ell_u^2 - u \ell_u - 2 \mu^2 u}{u^4} .
\label{N0scalarBox}
\eeal
The other numerators are given by the kinematic relations~\eqref{Jac1}
and~\eqref{Jac2} and can be found in \rcitePaper{4}.
There are also numerators for external bubbles and
external and internal tadpoles that can be obtained from \eqn{Jac2}.
They do not vanish but still integrate to zero. 

Assembling the pieces of the ${\cal N}=0$ amplitude for a fundamental
or an adjoint scalar is done along the lines of \sec{sec:amplitudeconstruction}.
In both cases, the primitive color-stripped amplitudes are the same,
only the color dressing will differ between the gauge-group representations.
The primitive color-stripped amplitudes for a single scalar contribution
in the loop are known~\cite{Bern:1995db} to be:
\begin{subequations} \begin{align}
      A^{{\cal N}=0,{\rm scalar}}_4(1^-\!,2^-\!,3^+\!,4^+) & =
            \frac{i \braket{12}^2 [34]^2}{ (4\pi)^{D/2} } 
            \bigg\{\!- \frac{1}{6st} I_2(t) - \frac{r_{\Gamma}}{9st} \bigg\} , \\
      A^{{\cal N}=0,{\rm scalar}}_4(1^-\!,2^+\!,3^-\!,4^+) & =
            \frac{i \braket{13}^2 [24]^2}{ (4\pi)^{D/2} }
            \bigg\{\!- \frac{r_{\Gamma} st}{2u^4}
                       \left( \ln^2\!\left( \frac{-s}{-t} \right)\!+ \pi^2 \right) \\ &
                     - \left( \frac{s\!-\!t}{2u^3} - \frac{1}{6su} \right) I_2(s)
                     - \left( \frac{t\!-\!s}{2u^3} - \frac{1}{6tu} \right) I_2(t)
                     + \frac{r_{\Gamma}}{2u^2} - \frac{r_{\Gamma}}{9st}
            \bigg\} \nn ,
\end{align} \label{N0scalar} \end{subequations}
\!\!\!\!\;and indeed our construction agrees with these.

\subsection{The MHV amplitude with fundamental ${\cal N}=0$ fermion matter}
\label{sec:N0fermion}

Here we present the MHV amplitude with a fundamental chiral fermion
$\Phi^{\rm fermion}_{{\cal N}=0}$ circulating in the loop
and external adjoint vectors ${\cal V}_{{\cal N}=0}$, \ie gluons.
This amplitude is actually a simple linear combination of the three amplitudes
discussed above, so no extra work is needed.
Note that, as for the scalar case, we only give the MHV amplitude.

The basic identity that we use is that a chiral fermion is obtained after subtracting the scalar $\Phi_{{\cal N}=0}^{\rm scalar}$ from the $\Phi_{{\cal N}=1}$ multiplet.
For the one-loop amplitudes this implies that
\be
   {\cal A}^{\rm fermion}_4 = {\cal A}^{{\cal N}=1,{\rm fund}}_4
                            - {\cal A}^{\rm scalar}_4 \\  
                            = \frac{1}{2}\,{\cal A}^{{\cal N}=2,{\rm fund}}_4
                            + \frac{1}{2}\,{\cal A}^{{\cal N}=1,{\rm odd}}_4
                            - {\cal A}^{\rm scalar}_4 ,
\ee
where we used \eqn{AN1}.
Hence we can define the fundamental fermion numerators as:
\begin{subequations} \begin{align}
   n_i^{\rm fermion} \equiv
        \frac{1}{2}\,n_i^{{\cal N}=2,{\rm fund}}
      + \frac{1}{2}\,n_i^{{\cal N}=1,{\rm odd}}
      - n_i^{\rm scalar} ,
\label{nfermion} \\
   \overline{n}_i^{\rm fermion} \equiv
        \frac{1}{2}\,n_i^{{\cal N}=2,{\rm fund}}
      - \frac{1}{2}\,n_i^{{\cal N}=1,{\rm odd}}
      - n_i^{\rm scalar} .
\label{nbarfermion}
\end{align} \label{nfermionboth} \end{subequations}
\!\!\!\!\;
Note that, just as in the ${\cal N}=1$ case, the parity-odd contributions
to the chiral fermion amplitude come entirely from the odd ${\cal N}=1$ sector.
Thus, interestingly, the chiral-anomalous part of the chiral fermion amplitude
effectively has ${\cal N}=1$ supersymmetry.

\section{One-loop four-point supergravity amplitudes}
\label{sec:gravity}

In this section, we assemble the duality-satisfying numerators
into various one-loop four-point supergravity amplitudes.
Similar to \eqn{ampl}, we write these amplitudes as
\be
{\cal M}^{\text{1-loop}}_4 = \Big(\frac{\kappa}{2}\Big)^4 \sum_{S_4}
      \Big( \frac{1}{8}\, {\cal I}_{\rm box}
          + \frac{1}{4}\, {\cal I}_{\rm tri}
          + \frac{1}{16}\,{\cal I}_{\rm bub} \Big) ,
\label{ampl2}
\ee
where the three integrals correspond to the canonically labeled graphs
in \fig{fig:ChiralBoxes}, and the rational prefactors compensate for overcount
in the sum over the set $S_4$ of 24 permutations of the four external legs, as well as overcount in the phase space of the bubble graph.

\subsection{Matter amplitudes}
\label{sec:mattergravity}

Here we check the double-copy construction
for gravity amplitudes with external graviton multiplets $H_{\cal N}$\footnote{More
precisely, to get $H_{\cal N}$ multiplets,
the external dilaton-axion multiplets $X$, $\overline{X}$ should be projected out,
which is straightforward for external states.}
and internal matter of the types shown in the right column of \tab{tab:DCconstructions}.

First of all, consider the following double copy of the ${\cal N}=2$ numerators
of \sec{sec:N2}:
\be
   {\cal I}_i^{{\cal N}=2+2,{\rm matter}} = \int\!\!\frac{d^{d}\ell}{(2\pi)^{d}} \,
      \frac{n_i^{{\cal N}=2,{\rm fund}}\,\overline{n}_i^{{\cal N}=2,{\rm fund}}
        \!+ \overline{n}_i^{{\cal N}=2,{\rm fund}}\,n_i^{{\cal N}=2,{\rm fund}}}{D_i} ,
\label{Neq2p2Integrals}
\ee
which is a precise implementation of the last line and right column of \tab{tab:DCconstructions}.
Due to the effective CPT invariance of the ${\cal N}=2$ matter numerators, this fundamental-representation double copy gives exactly twice the known~\cite{Carrasco:2012ca,Bern:2013yya,Nohle:2013bfa},\citePaper{3} adjoint double copy,
\be
   {\cal I}_i^{{\cal N}=2+2,{\rm matter}} = 2 \, {\cal I}_i^{{\cal N}=4,{\rm matter}}=2\!\int\!\!\frac{d^{d}\ell}{(2\pi)^{d}}
                  \frac{(n_i^{{\cal N}=2,{\rm adj}})^2}{D_i} .
\label{Neq2p2adjIntegrals}
\ee
The adjoint-representation double copy corresponds to a single ${\cal V}_{{\cal N}=4}$ multiplet contribution in the loop. Up to this factor of two,
we recover the ${\cal N}=4$ matter amplitude,
first computed by Dunbar and Norridge in \rcite{Dunbar:1994bn}:
\beal
   M^{{\cal N}=4,{\rm matter}}_4(1^{--}\!,2^{--}\!,3^{++}\!,4^{++})
    = \frac{i r_{\Gamma} \braket{12}^4 [34]^4}{ (4\pi)^{D/2} } \frac{1}{2 s^4}
      \bigg\{\!& - t u \left( \ln^2\!\left( \frac{-t}{-u} \right)\!+ \pi^2 \right) \\
               & + s(t\!-\!u) \ln\!\left( \frac{-t}{-u} \right)\!+ s^2
      \bigg\} ,
\label{N4gravitymatter}
\eeal
and the other component amplitudes are given by supersymmetry Ward identities.

Using the same relation for the ${\cal N}=2$ matter numerators,
and additionally taking into account that
\be
   n_i^{{\cal N}=1,{\rm fund}} + \overline{n}_i^{{\cal N}=1,{\rm fund}}
                               = n_i^{{\cal N}=2,{\rm adj}},
\label{N1even}
\ee
the fundamental-representation double copy
in the second-to-last line of \tab{tab:DCconstructions} becomes
\be
   {\cal I}_i^{{\cal N}=2+1,{\rm matter}} = \int\!\!\frac{d^{d}\ell}{(2\pi)^{d}} \,
      \frac{n_i^{{\cal N}=2,{\rm fund}}\,\overline{n}_i^{{\cal N}=1,{\rm fund}}
        \!+ \overline{n}_i^{{\cal N}=2,{\rm fund}}\,n_i^{{\cal N}=1,{\rm fund}}}{D_i}={\cal I}_i^{{\cal N}=4,{\rm matter}} .
\label{Neq2p1Integrals}
\ee
So these contributions also reduce to the adjoint-representation double copy~\eqref{Neq2p2adjIntegrals}
of the amplitude~\eqref{N4gravitymatter},
though this time without the factor of two.

The discussed two entries of \tab{tab:DCconstructions} can be regarded merely
as a new perspective
on the corresponding adjoint-representation double copies.
However, the remaining part of the right column of \tab{tab:DCconstructions}
is genuinely tied to the novel fundamental construction,
and not trivially related to adjoint double copies.
For instance, the $({\cal N}\!=\!1)^2$ matter double copy successfully exploits\footnote{Recall that ${\cal N}=1$ (chiral) adjoint matter works differently
since it effectively has ${\cal N}=2$ supersymmetry
and thus produces $({\cal N}\!=\!2)^2 = ({\cal N}\!=\!4)$ matter in the double copy.}  the chiral structure, giving
\be
   {\cal I}_i^{{\cal N}=1+1,{\rm matter}} = \int\!\!\frac{d^{d}\ell}{(2\pi)^{d}} \,
      \frac{n_i^{{\cal N}=1,{\rm fund}}\,\overline{n}_i^{{\cal N}=1,{\rm fund}}
        \!+ \overline{n}_i^{{\cal N}=1,{\rm fund}}\,n_i^{{\cal N}=1,{\rm fund}}}{D_i}=2 \, {\cal I}_i^{{\cal N}=2,{\rm matter}}  ,
\label{Neq1p1Integrals}
\ee
where the last identity indicates that this construction corresponds
to twice the contribution of a single non-chiral ${{\cal N}=2}$ matter multiplet.
The graviton component of this matter amplitude was computed
in \rcite{Dunbar:1994bn} using non-chiral building blocks,
its integrated form is
\beal
   M^{{\cal N}=2,{\rm matter}}_4(1^{--}\!,2^{--}\!,3^{++}\!,4^{++})
    = \frac{i r_{\Gamma} \braket{12}^4 [34]^4}{ (4\pi)^{D/2} }
      \bigg\{\! - \frac{t^2 u^2}{2s^6}
                  \left( \ln^2\!\left( \frac{-t}{-u} \right)\!+ \pi^2 \right) & \\
                + \frac{(t-u)(t^2 + 8tu + u^2)}{12s^5}
                  \ln\!\left( \frac{-t}{-u} \right)
                + \frac{t^2 + 14tu + u^2}{24s^4} &
      \bigg\} .
\label{N2gravitymatter}
\eeal
The fact that this amplitude agrees with our chiral double copy, up to the factor of two, is a highly nontrivial check of our construction.

The remaining three entries of \tab{tab:DCconstructions} provide
even more constraining checks of our procedure,
as they include the non-supersymmetric chiral-fermion numerators~\eqref{nfermionboth}.
We verified that the double copy,
\be
   {\cal I}_i^{{\cal N}=0+0,{\rm matter}} = \int\!\!\frac{d^{d}\ell}{(2\pi)^{d}} \,
      \frac{n_i^{\rm fermion}\,\overline{n}_i^{\rm fermion}
        \!+ \overline{n}_i^{\rm fermion}\,n_i^{\rm fermion}}{D_i} = 2\,{\cal I}_i^{{\cal N}=0,\rm scalar} ,
\label{Neq0p0Integrals}
\ee
reproduces twice the scalar-matter amplitude~\cite{Dunbar:1994bn}:
\beal
 & M^{{\cal N}=0,{\rm scalar}}_4(1^{--}\!,2^{--}\!,3^{++}\!,4^{++})
    = \frac{i r_{\Gamma} \braket{12}^4 [34]^4}{ (4\pi)^{D/2} } \frac{1}{2}
      \bigg\{\! - \frac{t^3 u^3}{s^8}
                  \left( \ln^2\!\left( \frac{-t}{-u} \right)\!+ \pi^2 \right) \\ &
                + \frac{(t\!-\!u)(t^4\!+\!9 t^3 u\!+\!46 t^2 u^2\!+\!9 t u^3\!+\!u^4)}
                       {30s^7}
                  \ln\!\left( \frac{-t}{-u} \right) \!
                + \frac{2t^4\!+\!23 t^3 u\!+\!222 t^2 u^2\!+\!23 t u^3\!+\!2u^4}
                       {180s^6}
      \bigg\} .
\label{N0gravityscalar}
\eeal
Replacing the left-copy fermion numerators by the chiral ${\cal N}=1$ numerators,
\be
   {\cal I}_i^{{\cal N}=1+0,{\rm matter}} = \int\!\!\frac{d^{d}\ell}{(2\pi)^{d}} \,
      \frac{n_i^{{\cal N}=1,{\rm fund}}\,\overline{n}_i^{\rm fermion}
        \!+ \overline{n}_i^{{\cal N}=1,{\rm fund}}\,n_i^{\rm fermion}}{D_i}= {\cal I}_i^{{\cal N}=2,{\rm matter}}  ,
\label{Neq1p0Integrals}
\ee
also reproduces the correct amplitude~\eqref{N2gravitymatter}.
Doing the same for ${\cal N}=2$ numerators,
\be
   {\cal I}_i^{{\cal N}=2+0,{\rm matter}} = \int\!\!\frac{d^{d}\ell}{(2\pi)^{d}} \,
      \frac{n_i^{{\cal N}=2,{\rm fund}}\,\overline{n}_i^{\rm fermion}
        \!+ \overline{n}_i^{{\cal N}=2,{\rm fund}}\,n_i^{\rm fermion}}{D_i}= {\cal I}_i^{{\cal N}=2,{\rm vector}}  ,
\label{Neq2p0Integrals}
\ee
results in the correct ${\cal V}_{{\cal N}=2}$-matter amplitude, which can be otherwise calculated as
\be
        M^{{\cal N}=2,\rm vector}_4(1,2,3,4) =  M^{{\cal N}=4,\rm matter}_4(1,2,3,4) -2M^{{\cal N}=2,\rm matter}_4(1,2,3,4) .
\label{N2gravityvector}
\ee

In addition, even if they are not included in \tab{tab:DCconstructions},
we note that the fundamental scalar numerators of \sec{sec:N0scalar}
also produce sensible double copies at one loop,
such as
\be
   {\cal I}_i^{{\cal N}=0'+0', {\rm \, scalar \, matter}} =
      \int\!\!\frac{d^{d}\ell}{(2\pi)^{d}} \,
      \frac{n_i^{\rm scalar} \overline{n}_i^{\rm scalar}
        \!+ \overline{n}_i^{\rm scalar} n_i^{\rm scalar}}{D_i}
      = 2\,{\cal I}_i^{{\cal N}=0,\rm scalar} ,
\label{Neq0p0IntegralsScalar}
\ee
which integrates to twice the scalar-matter amplitude~\eqref{N0gravityscalar}.
However, this is equivalent to the adjoint construction
of \rcites{Bern:2013yya,Nohle:2013bfa}.
\Eqn{N1even} also implies that the double copy
\be
   {\cal I}_i^{{\cal N}=1+0', {\rm matter}} = \int\!\!\frac{d^{d}\ell}{(2\pi)^{d}} \,
      \frac{n_i^{{\cal N}=1,{\rm fund}}\,\overline{n}_i^{\rm scalar}
        \!+ \overline{n}_i^{{\cal N}=1,{\rm fund}}\,n_i^{\rm scalar}}{D_i}= {\cal I}_i^{{\cal N}=2,\rm matter} ,
\label{Neq1p0IntegralsScalar}
\ee
trivially follows from the adjoint double copy 
$({\cal N}\!=\!2,{\rm adj})\times({\cal N}\!=\!0,{\rm scalar})$,
which integrates to the ${\cal N}=2$ matter amplitude~\eqref{N2gravitymatter}.

While the double copies involving scalars work flawlessly for one-loop amplitudes with external graviton multiplets, we expect similar constructions to be more problematic or ambiguous at higher loops. In particular, as far as we know, the scalars cannot be used to obtain pure gravity amplitudes at two loops and higher. Furthermore, in the absence of supersymmetry, the scalar self-interactions are not fixed by universal considerations.

\subsection{Pure gravity amplitudes}
\label{sec:puregravity}

\begin{figure}[t]
      \centering
      \includegraphics[scale=0.75]{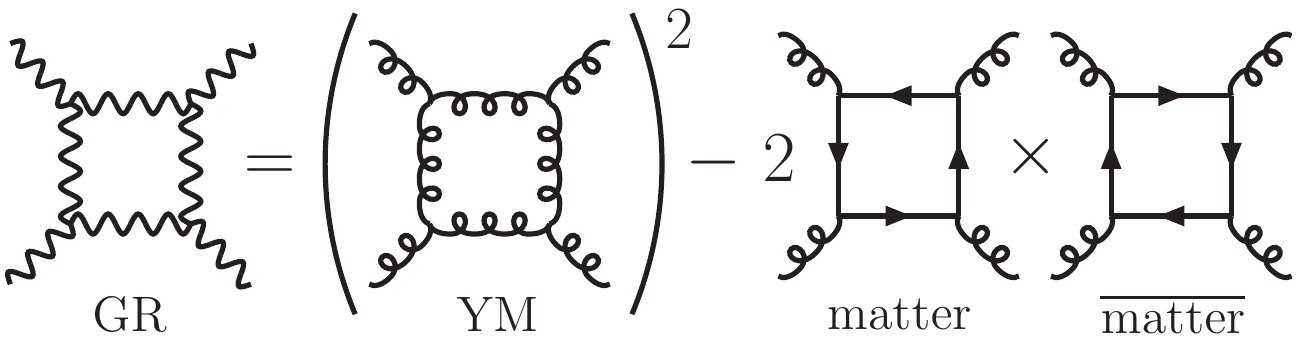}
      \vspace{-5pt}
\caption[a]{\small A pure gravity amplitude obtained by subtraction of dilaton and axion.
At one loop, the matter contribution can be from either a fermion or a scalar.}
\label{SubtractionFigure}
\end{figure}

In the previous section, we checked that our fundamental numerators produce all the matter-amplitude double copies from \tab{tab:DCconstructions}.
Based on that and the conventional adjoint-representation double copies,
we can now formulate what the recipe in \sec{sec:FundDoubleCopy} gives
for pure (super-)gravity four-point\footnote{In principle, the formulas in this subsection are valid for any multiplicity at one loop, once the corresponding numerators are known.} one-loop amplitudes. We will not write out the explicitly integrated forms of the one-loop four-point amplitudes,
as they can be found in \rcite{Dunbar:1994bn}.
After integration, all our amplitudes agree with the results therein.

As in illustrated in \fig{SubtractionFigure},
the integrals for pure Einstein gravity, or general relativity (GR),
in our framework are given by
\be
   {\cal I}_i^{\rm GR} = \int\!\!\frac{d^{d}\ell}{(2\pi)^{d}}
      \frac{(n_i^{\rm YM})^2 - 2\,n_i^{\rm fermion}\,\overline{n}_i^{\rm fermion}}
           {D_i} .
\label{Neq0fermIntegrals}
\ee
The pure Yang-Mills numerators
can be easily constructed from the numerators~\eqref{N4Box},
\eqref{N1evenBox} and \eqref{N0scalarBox} through the well-known~\cite{Bern:1993mq}
one-loop supersymmetry decomposition. This decomposition, for both numerators and amplitudes at one loop, is given by
\beal
   n_i^{\rm YM}& = n_i^{{\cal N}=4} - 4\,n_i^{{\cal N}=2,{\rm adj}}
                                   + 2\,n_i^{\rm adj\,scalar} , \\
   {\cal A}^{\rm YM}_m &={\cal A}^{{\cal N}=4}_m - 4\,{\cal A}^{{\cal N}=2,{\rm adj}}_m
                                   + 2\,{\cal A}^{\rm adj\,scalar}_m .
\eeal
The fermion ghosts in \eqn{Neq0fermIntegrals}
subtract the dilaton and the axion contributions.
However, due to the fact that at one loop double copies of anti-aligned chiral fermions are equal to scalar double copies, the same integrals can be reproduced simply as
\be
   {\cal I}_i^{\rm GR} = \int\!\!\frac{d^{d}\ell}{(2\pi)^{d}}
                  \frac{(n_i^{\rm YM})^2 -2(n_i^{\rm adj\,scalar})^2}{D_i} .
\label{Neq0scIntegrals}
\ee
Although the latter construction is nice,
the higher-loop cut checks from \sec{sec:cutchecks} suggest
that this equality is accidental.
The fundamental-fermion framework resulting in \eqn{Neq0fermIntegrals}
is the one that should be generally valid at higher loops.

The pure ${\cal N}=1$ supergravity amplitude is generated by the following integrals:
\be
   {\cal I}_i^{{\cal N}=1\,{\rm SG}} = \int\!\!\frac{d^{d}\ell}{(2\pi)^{d}}
      \frac{ n_i^{{\cal N}=1\,{\rm  SYM}}\,n_i^{\rm YM}
           - n_i^{{\cal N}=1, {\rm fund}}\,\overline{n}_i^{\rm fermion}
           - \overline{n}_i^{{\cal N}=1, {\rm fund}}\,n_i^{\rm fermion}}{D_i} ,
\label{Neq1fermIntegrals}
\ee
and the integrals for pure ${\cal N}=2$ supergravity are given by
\be
   {\cal I}_i^{{\cal N}=2\,{\rm SG}} = \int\!\!\frac{d^{d}\ell}{(2\pi)^{d}}
      \frac{ (n_i^{{\cal N}=1\,{\rm  SYM}})^2
           - 2\,n_i^{{\cal N}=1, {\rm fund}} \,
             \overline{n}_i^{{\cal N}=1, {\rm fund}} }{D_i} .
\label{Neq2Integrals}
\ee
The numerators for ${\cal N}=1$ SYM can be constructed from the numerators~\eqref{N4Box} and \eqref{N1evenBox} through the following one-loop supersymmetry decomposition~\cite{Bern:1993mq}:
\be
   n_i^{{\cal N}=1\, {\rm  SYM}} = n_i^{{\cal N}=4} - 3\,n_i^{{\cal N}=2,{\rm adj}} .
\ee
In \eqns{Neq1fermIntegrals}{Neq2Integrals}, the matter ghosts subtract the contributions
from one and two ${\cal N}=2$ matter multiplets, respectively. 

Alternatively, we can write the pure ${\cal N}=2$ supergravity using the integrals
\be
   {\cal I}_i^{{\cal N}=2\,{\rm SG}} = \int\!\!\frac{d^{d}\ell}{(2\pi)^{d}}
      \frac{ n_i^{{\cal N}=2\,{\rm  SYM}}\,n_i^{\rm YM}
           - n_i^{{\cal N}=2, {\rm fund}}\,\overline{n}_i^{\rm fermion}
           - \overline{n}_i^{{\cal N}=2, {\rm fund}}\,n_i^{\rm fermion}}{D_i} ,
\label{Neq2fermIntegrals}
\ee
where the ghosts cancel a ${\cal V}_{{\cal N}=2}$-matter multiplet,
\ie an abelian vector multiplet. 
The numerators for ${\cal N}=2$ SYM can be constructed
from the numerators~\eqref{N4Box} and~\eqref{N1evenBox}
through the following decomposition~\cite{Bern:1993mq}:
\be
   n_i^{{\cal N}=2\, {\rm  SYM}} = n_i^{{\cal N}=4} - 2\,n_i^{{\cal N}=2,{\rm adj}} .
\ee
The fact that the above two very different constructions of pure ${\cal N}=2$ supergravity give the same amplitude is highly nontrivial.
If our prescription does not encounter unexpected obstructions at higher loops,
this nontrivial equality should be true in all generality.

The pure ${\cal N}=3$ supergravity amplitude is obtained through the integrals
\be
   {\cal I}_i^{{\cal N}=3\,{\rm SG}} = \int\!\!\frac{d^{d}\ell}{(2\pi)^{d}}
      \frac{ n_i^{{\cal N}=2\,{\rm  SYM}}\,n_i^{{\cal N}=1\,{\rm  SYM}}
           - n_i^{{\cal N}=2, {\rm fund}}\,\overline{n}_i^{{\cal N}=1, {\rm fund}}
           - \overline{n}_i^{{\cal N}=2, {\rm fund}}\,n_i^{{\cal N}=1, {\rm fund}}}{D_i} ,
\label{Neq3Integrals}
\ee
where the ghosts subtract out a ${\cal V}_{{\cal N}=4}$-matter multiplet,
equal to the combination of two ${\cal N}=3$ chiral multiplets.

Although pure ${\cal N}=4$ supergravity is a factorizable theory, meaning that its amplitudes and integrals can be written as an adjoint double copy of pure YM theories, 
\be
   {\cal I}_i^{{\cal N}=4\,{\rm SG}} = \int\!\!\frac{d^{d}\ell}{(2\pi)^{d}}
      \frac{ n_i^{{\cal N}=4\,{\rm  SYM}}\,n_i^{\rm YM}}{D_i} ,
\label{Neq4Integrals}
\ee
they can also be obtained through a fundamental-representation double copy 
\be
   {\cal I}_i^{{\cal N}=4\,{\rm SG}} = \int\!\!\frac{d^{d}\ell}{(2\pi)^{d}}
      \frac{ n_i^{{\cal N}=2\,{\rm  SYM}}\,n_i^{{\cal N}=2\,{\rm  SYM}}
           - 2n_i^{{\cal N}=2, {\rm fund}}\,\overline{n}_i^{{\cal N}=2, {\rm fund}}}{D_i} .
\label{Neq4IntegralsFund}
\ee
Similarly to the ${\cal N}=2$ case, that these two very different constructions
of pure ${\cal N}=4$ supergravity give the same amplitude
is a nontrivial result~\cite{Tourkine:2012vx,Carrasco:2012ca}.

With that said, we have gone through all the cases listed in \tab{tab:DCconstructions}.
Our double-copy prescription produces the correct one-loop four-point amplitudes
for ${\cal N}=0,1,2,3,4$ pure supergravities,
as confirmed by comparing to the previously known results~\cite{Dunbar:1994bn}.

\section{Discussion}
\label{sec:bcjdiscussion}

In this chapter, we have extended the scope of color-kinematics duality to 
matter fields in the fundamental representation.
As we showed on various examples, this allows us to construct gravity scattering amplitudes in a broad range of (super-)gravity theories using the double-copy prescription. This includes Einstein gravity, pure ${\cal N}<4$ supergravity, and supergravities with arbitrary non-self-interacting matter. 

Our main focus is on the issue
of unwanted matter degrees of freedom propagating in the loops
that occur when one attempts to construct pure ${\cal N}<4$ supergravities
using the color-kinematics duality.
Such supersymmetric dilaton-axion matter is inherent
to the standard double copy of adjoint-representation gauge theories.
Using the fundamental representation for Yang-Mills matter gives us a means to differentiate these fields from the adjoint ones, so that the double copies can be performed separately.
If we promote the matter double copies to be ghosts,
they cancel the unwanted matter states in the double copy of the vector fields.
Moreover, the ghost double-copy prescription,
$c_i \rightarrow (-1)^{|i|} \bar{n}'_i$,
can be easily replaced by a tunable-matter prescription,
$c_i \rightarrow (N_X)^{|i|} \bar{n}'_i$,
thus producing theories with any number of matter multiplets coupled to gravity.

We have presented nontrivial evidence supporting our framework
using examples both for tree and loop-level amplitudes, as well as more general arguments.
At tree level, we have checked the construction of gravitational amplitudes
with non-self-interacting matter though seven points, with the most general external states.
At one loop, we have constructed the color-kinematics representation
of four-gluon amplitudes with general fundamental matter circulating in the loop.
Using the double-copy prescription, we have obtained the one-loop four-graviton amplitudes
in Einstein gravity, pure ${\cal N}<4$ supergravity and supergravity with generic matter.

Our explicit calculations give novel and simple forms for Yang-Mills numerators containing ${\cal N}=2,1$ supersymmetric fundamental matter, as well as non-supersymmetric fundamental scalars and fermions. Equivalent numerators that satisfy the color-kinematics duality were known~\cite{Carrasco:2012ca,Bern:2013yya,Nohle:2013bfa},\citePaper{3} for all but the odd part of ${\cal N}=1$ the matter contribution, which we give in \eqn{N1oddBox}. After integration, our new representations of one-loop gravity amplitudes are in full accord with the results of Dunbar and Norridge~\cite{Dunbar:1994bn}.
Moreover, they provide a direct check of the nontrivial equality of the symmetric
${\cal N}=1+1$ and the asymmetric ${\cal N}=2+0$ supergravity construction.

At two loops, we have checked the consistency of the construction of  Einstein gravity and pure ${\cal N}<4$ supergravity.  We considered four-dimensional unitarity cuts of two-loop amplitudes, and applied the ghost prescription to the matter double copies. The cuts show that, when using double copies of fundamental fermions, our construction correctly eliminates the unwanted dilaton-axion degrees of freedom naturally present in the double copy of pure Yang-Mills theory.
We have checked that the same cancellation happens in the supersymmetric generalizations. The two-loop checks also show that when using double copies of fundamental scalars the cancellation with the dilaton-axion states is incomplete. Table \ref{tab:DCconstructions} summarizes the valid double copies for the matter ghosts.

While the double copy of fundamental scalars is unsuitable for the construction of Einstein gravity, it is interesting to note that the fundamental-scalar amplitudes nontrivially satisfy the color-kinematics duality. The resulting double-copy amplitude should correspond to some gravitational amplitudes that are corrected by four-scalar terms and possibly higher-order interactions. We leave the details of the non-supersymmetric double copy of fundamental scalars for future work.

While our construction of pure supergravities passes many nontrivial checks, it is well known that higher-loop calculations can be plagued by subtleties coming from dimensional regularization. Thus it is important that our prescription is carefully scrutinized by explicit $L>1$ calculations in both pure and matter-coupled supergravities. Starting at two loops, the scheme-dependence of different types of dimensional-regularization prescriptions would be interesting to study.
It would be also interesting to see how our double-copy prescription for pure gravities
can be extended beyond $d=4-2\epsilon$ dimensions.

We hope that our results will open a new window into the study of pure ${\cal N}=0,1,2,3$ supergravity multiloop amplitudes and their ultraviolet properties.
Pure Einstein gravity is known to have a divergence in the two-loop effective action, as was proven in \rcites{Goroff:1985sz,Goroff:1985th,vandeVen:1991gw}
using computerized symbolic manipulations. The ${\cal N}=1,2,3$ theories are known to have no divergences at one and two loops due to supersymmetry~\cite{Grisaru:1976nn,Tomboulis:1977wd,Deser:1977nt,Howe:1980th},
but the status of these theories at three loops and beyond is unknown.
It would be interesting to reinvestigate the ${\cal N}<4$ theories from the modern point of view using new analytic methods and structures, including the current results, in hope of gaining better understanding of the structure of quantum gravity.

\chapter*{Conclusion}
\addcontentsline{toc}{chapter}{Conclusion}

In this thesis, we have given an account of some exemplary topics
in the field of scattering amplitudes.

In \chap{chap:treelevel}, we reviewed the BCFW recursion~\cite{Britto:2004ap,Britto:2005fq}
--- an on-shell method of calculating tree-level amplitudes
that had greatly boosted the understanding of the perturbation theory of gauge theories
over the last decade.
We have taken a significant step away from their established zone of applicability
by studying the on-shell recursion for tree-level objects
that are not entirely on-shell.
Surprisingly, we discovered that the standard obstacle for securing the recursion
poses no more problems than in the fully on-shell case.
Instead, we ran into a new issue --- that of the unphysical poles.
However, we found ways to evade them and thus use the on-shell recursion method
to find an infinite family of off-shell fermion currents.

Interestingly, solutions to infinite families of problems exist beyond tree level.
First such achievements~\cite{Bern:1994zx,Bern:1994cg} can be regarded as
the turning point for establishing the amplitudes field as it is today.
Since then, more impressive all-multiplicity results followed~\cite{Bern:2004bt,
Bedford:2004nh,Bedford:2004nh,Berger:2006vq,Bern:2005hh,BjerrumBohr:2007vu,Dunbar:2009ax,
Prygarin:2011gd,Brandhuber:2012wy}.
However, most of them relied on the simplicity
of the $n$-point tree-level MHV amplitude~\cite{Parke:1986gb},
which served as the indispensable input for analytic loop calculations.

In \chap{chap:oneloop}, we showed that
the tree-level NMHV amplitudes~\cite{Drummond:2008vq,Drummond:2008bq}
are also simple enough to use them for analytic one-loop calculations.
Remarkably, that simplicity can be uncovered via the supersymmetric rendition
of the BCFW recursion~\cite{Elvang:2008na,Dixon:2010ik}.
We took these tree-level NMHV amplitudes as the input
to compute all the one-loop NMHV amplitudes
for gluon scattering in ${\cal N}=1$ super-Yang-Mills theory.
For that, we employed the powerful method of spinor integration~\cite{Britto:2005ha,Anastasiou:2006jv},
for which we also provided a streamlined rederivation,
motivated by a new version
of the general formula for the coefficient of the bubble master integral.

Recently, there have been impressive advances in applying the BCFW recursion
to the four-dimensional integrands of the maximally-supersymmetric Yang-Mills theory
\cite{Mason:2009sa,ArkaniHamed:2010kv,ArkaniHamed:2012nw,
Arkani-Hamed:2013jha,Arkani-Hamed:2013kca} up to all loops.
In contrast to this, the applicability of recursive methods
to integrated loop-level amplitudes is very limited~\cite{Bern:2005hh}.
Interestingly, our analytic results for the NMHV amplitude
seem to suggest the possibility of non-trivial recursion relations
among various loop-integral coefficients.
Once this phenomenon is understood systematically,
it might extend the scope of the inductive approach
beyond the previously established cases.

Another big class of quantum field theories has seen rapid progress ---
the perturbation theory of gravity.
In principle, gravity can be considered as a gauge field theory,
the associated vector bundle of which was replaced by the cotangent bundle.
This corresponds to swapping the independent color vector space
by the Minkowski space cotangent to the spacetime itself,
this being done locally, at each spacetime point.

That intimate connection between Yang-Mills theories and gravities
was converted into a computation device by the discovery
of the BCJ color-kinematics duality and double copy~\cite{Bern:2008qj,Bern:2010ue},
according to which gravity amplitude integrands are obtained from the gauge theory ones
by replacing color factors by kinematic factors.
It is also seems natural that this recipe requires
the kinematic factors not to be arbitrary
but rather respect the gauge theory algebra.
The corresponding kinematic algebra was even made manifest
in the self-dual sector of the gauge theories~\cite{Monteiro:2011pc}.

The BCJ duality was originally formulated for gauge theories
in fields in the adjoint representation.
In the lower-supersymmetry cases,
the double copy of such theories results in gravities
with some hardwired matter content.
For example, the gravity that can be obtained from
the pure non-supersymmetric Yang-Mills theory
is Einstein's general relativity theory coupled to an antisymmetric tensor and dilaton.

In \chap{chap:bcj},
we transcended both these limitations
by extending the scope of the color-kinematics duality
to include particles in the fundamental representation.
Intuitively, this can be regarded as the ``complexification''
of the once ``real'' BCJ construction.
The (super-)gravity theories given by such a double copy
contain an arbitrary number of matter supermultiplets
which do not interact with each other.
Remarkably, this includes the pure gravity theories,
that are obtained from the adjoint-representation construction,
after the unwanted degrees of freedom are canceled
by the ghost-like double copies,
made possible by the novel fundamental-representation construction.

Even though the details of its $d$-dimensional realization are still to be clarified,
we expect that our generalization of the BCJ construction will open new venues
for the study of pure ${\cal N}=0,1,2,3$ supergravity multiloop amplitudes
and their ultraviolet properties.
For instance, pure Einstein gravity is known~\cite{Goroff:1985sz,Goroff:1985th,vandeVen:1991gw} to diverge at two loops,
whereas the UV behavior of the ${\cal N}=1,2,3$ theories is currently unclear
beyond two loops~\cite{Grisaru:1976nn,Tomboulis:1977wd,Deser:1977nt,Howe:1980th}.
Hopefully, investigating the ${\cal N}<4$ theories using the modern approaches
will shed more light on the structure of quantum gravity.

\appendix
\chapter{Spinor residues}
\label{app:residues}

Simple spinor residues are defined as
\be
      \Res_{\lambda=\zeta} \frac{F(\la,\lb)}{\braket{\zeta|\la}}
            = F(\zeta,\tilde{\zeta})
            =-\Res_{\lambda=\zeta} \frac{F(\la,\lb)}{\braket{\la|\zeta}} .
\label{simplespinorpole}
\ee
Multiple spinor poles can in principle be extracted using the following formula:
\be
      \Res_{\lambda=\zeta} \frac{F(\la,\lb)}{\braket{\zeta|\la}^k\!\!}
            = \frac{1}{\braket{\eta|\zeta}^{k-1}}
              \bigg\{ \frac{1}{(k-1)!}
                      \frac{\mathrm{d}^{(k-1)}}{\mathrm{d}t^{(k-1)}}
                      F(\zeta-t\eta,\tilde{\zeta})
              \bigg\} \bigg|_{t=0} ,
\label{multiplespinorpole}
\ee
where $\eta$ is an arbitrary auxiliary spinor not equal to the pole spinor $\zeta$.
However, in $\mathcal{N}=1$ SYM there are no multiple poles.

Care must be taken when dealing with poles of the form $\bra{\la}K|q]$
because it is equivalent to a pole $\braket{\zeta|\la}$ with the following value of $\zeta$:
\beal
   \bra{\zeta} & = [q|K| , ~~~~~~~~~~~~\:
   \ket{\zeta} = - |K|q] , \\
   [\tilde{\zeta}| & = - \bra{q}K| , ~~~~~~~~~~
   |\tilde{\zeta}] \;\! = |K\ket{q} .
\label{compositepole}
\eeal

\bibliographystyle{JHEP}
\bibliography{references}

\end{document}